\documentclass[preprint,12pt,authoryear]{elsarticle}

\usepackage{amssymb}
\usepackage{hyperref}
\usepackage{amsmath}
\usepackage{graphicx}%
\usepackage{multirow}%
\usepackage{amsmath,amssymb,amsfonts}%
\usepackage{amsthm}%
\usepackage{mathrsfs}%
\usepackage[title]{appendix}%
\usepackage{xcolor}%
\usepackage{textcomp}%
\usepackage{manyfoot}%
\usepackage{booktabs}%
\usepackage{algorithm}%
\usepackage{algorithmicx}%
\usepackage{algpseudocode}%
\usepackage{listings}%
\usepackage{mathtools}

\usepackage{color,xcolor,ucs}
\usepackage{mathtools}   \usepackage{tikz} 
\usepackage{ amssymb }
\usepackage{extarrows} 
\usepackage{pgf,tikz}
\usepackage{float}
\usetikzlibrary{positioning}
\usetikzlibrary{shapes.geometric}
\usetikzlibrary{shapes.misc}
\usetikzlibrary{arrows}
\usepackage{caption}
\usepackage{mathrsfs}
\usetikzlibrary{arrows,shapes,automata,backgrounds,petri,positioning}
\usetikzlibrary{decorations.pathmorphing}
\usetikzlibrary{decorations.shapes}
\usetikzlibrary{decorations.text}
\usetikzlibrary{decorations.fractals}
\usetikzlibrary{decorations.footprints}
\usetikzlibrary{shadows}
\usetikzlibrary{calc}
\usetikzlibrary{spy}
\usepackage{amsmath}
\usepackage{array}
\usepackage{ amssymb }
\usepackage{braket}
\usepackage{qcircuit}
\usepackage{soul}
\usepackage{braket} 
\usepackage{relsize}

\usepackage{amsmath}
\usepackage{ amssymb }
\usepackage{braket}
\usepackage{qcircuit}
\usepackage{soul}
\usepackage{braket}

\journal{Journal of Computational Physics}

\begin{document}

\begin{frontmatter}



\title{Exact solvability of an Ising-type model, and exact solvability of the 6-vertex, and 8-vertex, models} 


\author{Pete Rigas} 

\affiliation{ 
            city={Newport Beach},
            postcode={92625}, 
            state={CA},
            country={United States, pbr43@cornell.edu}}

\begin{abstract}
We compute the action-angle coordinates for an Ising type model whose L-operator has been previously studied in the literature by Bazhanov and Sergeev. In comparison to computations with such operators that have been examined previously by the author for the 4-vertex, 6-vertex, and 20-vertex, models, computations for asymptotically approximating a collection of sixteen identities with the Poisson bracket, which together constitute the Poisson structure of the Ising type model, exhibit dependencies upon nearest neighbor interactions. Inspite of the fact that L-operators for the 20-vertex model are defined in terms of combinatorial, and algebraic, constituents unlike such operators for the 6-vertex model which are defined in terms of projectors and Pauli basis elements, L-operators for the Ising-type model can be used for concluding that a model which interpolates between the 6-vertex, and 8-vertex, models is exactly solvable. (\textbf{MSC Class}: 34L25; 60K35)
\end{abstract}

\begin{graphicalabstract}
\end{graphicalabstract}

\begin{highlights}
\item The Quantum Inverse Scattering Method (QISM) approach, is based upon seminal work from Faddeev and Takhatajan, which asserts that Integrability of Hamiltonian systems are integrable, or equivalently, relating to exactly solvable structure. The method includes:

\begin{itemize}
\item[$\bullet$] Formulating well posed solutions to the Nonlinear Schrodinger equation,
Introducing expressions for the Poisson brackets,

\item[$\bullet$]
Formulating sets of conditions for action-angle variables, which approximate dynamics that are approximately linear,

\item[$\bullet$]
Characterizing integrability from approximating a collection of Poisson brackets.
\end{itemize}

\item The QISM approach has wide appeal within Mathematical and Statistical Physics, particularly for characterizing integrability of closely related systems. Of the systems that can be of interest, and amenable to the QISM approach, previous work of the author in this area includes:

\begin{itemize}
\item[$\bullet$] Characterizing exact solvability of the 6-vertex model, from integrability of inhomogeneous limit shapes and a Hamiltonian flow,
\item[$\bullet$]
Characterizing exact integrability, and solvability, of the 4-vertex and 20-vertex models,
\item[$\bullet$]
Formulating contour integral representations for the 20-vertex model under domain-wall boundary conditions,
\item[$\bullet$]
Establishing well-posed formulations of the Bethe ansatz equations for open boundary conditions of the $D^2_3$ spin chain.
\end{itemize}

\end{highlights}

\begin{keyword}
Statistical Physics, Mathematical Physics, eigenvalue problem, SOS model, 6-vertex model, 7-vertex model, 20-vertex model, 4-vertex model


\end{keyword}

\end{frontmatter}


\section{Introduction}

\subsection{Overview}

\noindent Exactly solvable, and integrable, models of Statistical Physics have gathered significant attention, whether from the perspective of connections with the Bethe ansatz and correlations {\color{blue}[12},{\color{blue}20},{\color{blue}26]}, continuum limits, {\color{blue}[23]}, limit shapes, and associated asymptotic phenomena, {\color{blue}[8},{\color{blue}9},{\color{blue}10},{\color{blue}11},{\color{blue}35},{\color{blue}36},{\color{blue}37},{\color{blue}38},{\color{blue}39},{\color{blue}43]}, amongst several other closely related topics {\color{blue}[1},{\color{blue}2},{\color{blue}4},{\color{blue}6},{\color{blue}17},{\color{blue}18},{\color{blue}19},{\color{blue}22},{\color{blue}24},{\color{blue}25},{\color{blue}27},{\color{blue}28},{\color{blue}29]}. To make use of rapidly emerging developments for many models, and settings, at the intersection of Mathematical Physics, Statistical Physics, and Quantum Physics, {\color{blue}[30},{\color{blue}31},{\color{blue}32},{\color{blue}33},{\color{blue}34]}, we propose a new perspective for applying the quantum inverse scattering method (QISM) to a model that has interactions which signficantly differ from those of vertex models. From previous work of the author, {\color{blue}[41},{\color{blue}45]}, properties of the Poisson bracket - anticommutativity, Leibniz' rule, bilinearity, and the Jacobi identity - one can determine whether integrable, or exactly solvable, properties of vertex models exist. Besides classes of vertex models, which are but one type of many, models appearing in Statistical Physics, Ising type models are another. Such classes of models are interesting due to the fact that, from some perspectives, they greatly differ between the behavior of the Gaussian free field. Despite such differences, there can still be an adaptation of the QISM framework to such models, which not only relies upon similar steps of performing computations with an L-operator for asymptotically representing the transfer and quantum  monodromy matrices, but also for formalizing notions of Poisson structures, which hold implications for accompanying Yang-Baxter algebras.

Further studying interactions between such themes, which span the fields of Discrete and Integrable Probability simultaneously, remains of interest. In determining whether integrable properties of models play roles in discrete structures, pursuing such a research program for vertex model, in comparison to Ising type models, predominantly relies upon the ice-rule. As a conversation rule for vertex models, in two dimensions the ice-rule states that there must be two incoming, and two outgoing arrows, surrounding each vertex of the square lattice $\textbf{Z}^2$, while in three dimensions it states that there must be three incoming, or three outgoing arrows, surrounding each vertex of the triangular lattice $\textbf{T}$. For Ising type models, in the absence of the ice rule, one instead defines a probability measure over finite volume with a nearest neighbor interactions from a Hamiltonian, as well as with an inverse temperature. Nonwithstanding of the differences in which vertex, and the Ising, models are defined, it is still possible to apply components of the QISM in each situation.

For vertex models, in comparison to Ising type models, asymptotic approximations for the transfer matrix, and quantum monodromy matrix, can be formulated from products of L-operators. That is, asymptotically in infinite volume, having a representation of the L-operator, whether in terms of Pauli basis elements and other closely related factors dictates which terms determine the approximation of each Poisson bracket, within the entire Poisson structure, in infinite volume. Moreover, whether a vertex model of Statistical Physics has \textit{isotropic} parameters, namely a set of parameters that are all set to be equal, also plays a significant role in determining macroscopic properties of vertex configurations in weak finite limits that are taken towards infinite volume. Ising type models, albeit being introduced in a completely different way, and hence with a completely separate procedure for taking the finite volume limit for obtaining probability measures supported over the entirety of $\textbf{Z}^2$, can still be analyzed with similar computations from those provided by the author {\color{blue}[41},{\color{blue}45]}. 

\subsection{The Poisson and Commutator brackets}

As an operation that takes two arguments, the Poisson bracket $\big\{ \cdot , \cdot \big\}$ satisfies the following set of properties, given test functions $f$, $g$ and $h$,

\begin{itemize}
    \item [$\bullet$] \textit{Anticommutativity}. $\big\{ f, g \big\}  =  - \big\{ g , f \big\} $

    \item[$\bullet$] \textit{Bilinearity}. For real $a,b$, $\big\{ af + bg , h \big\} = a \big\{ f ,h \big\} + b \big\{ g , h \big\},$ and $\big\{ h , af + bg \big\} = a \big\{ h , f \big\} + b \big\{ h , g \big\} $

    \item[$\bullet$] \textit{Leibniz' rule}. $\big\{ fg , h \big\} = \big\{ f , h \big\} g + f \big\{ g , h \big\}$, 

    \item[$\bullet$] \textit{Jacobi identity}. $\big\{ f , \big\{ g , h \big\} \big\} + \big\{ g , \big\{ h , f \big\} \big\} + \big\{ h , \big\{ f , g \big\} \big\} = 0$ \end{itemize}

While making use of the Poisson bracket, defined above, can be used to deduce that integrable and exactly solvable structures hold for \textit{Classical} L-operators, for \textit{Quantum} L-operators, one makes use of the commutator bracket,

\begin{align*}
    \bigg[ \bigg[ \cdot , \cdot \bigg] \bigg] .
\end{align*}

\noindent As a bilinear mapping that is dependent upon complex-valued linear operators, the commutator, as does the Poisson bracket mentioned above, satisfies the following properties. For the below bracket, denote,

\begin{align*}
 \textit{First complex-valued linear operator} \equiv  \mathscr{L}_1 \big( \textbf{C} \big)  = \mathscr{L}_1 , \\ \\  \textit{Second complex-valued linear operator} \equiv  \mathscr{L}_2 \big( \textbf{C} \big)  = \mathscr{L}_2 , \\ \\ \textit{Third complex-valued linear operator} \equiv  \mathscr{L}_3 \big( \textbf{C} \big)  = \mathscr{L}_3  , 
\end{align*}

\noindent from which the properties of the commutator include:

\begin{itemize}
    \item [$\bullet$] \textit{Anticommutativity}. $\bigg[ \bigg[ \mathscr{L}_1 , \mathscr{L}_2 \bigg] \bigg]   =  - \bigg[ \bigg[ \mathscr{L}_2 , \mathscr{L}_1 \bigg] \bigg] $

    \item[$\bullet$] \textit{Bilinearity}. For complex $A,B$, $\bigg[ \bigg[  A \mathscr{L}_1 + B \mathscr{L}_2 , \mathscr{L}_3  \bigg] \bigg]  = A \bigg[ \bigg[ \mathscr{L}_1 , \mathscr{L}_3 \bigg] \bigg]  + B \bigg[ \bigg[  \mathscr{L}_2 , \mathscr{L}_3 \bigg] \bigg] ,$ and $\bigg[ \bigg[  \mathscr{L}_3 , A \mathscr{L}_1 + B \mathscr{L}_2 \bigg] \bigg]  = A \bigg[ \bigg[ \mathscr{L}_3  , \mathscr{L}_1 \bigg] \bigg]  + B \bigg[ \bigg[ \mathscr{L}_3 , \mathscr{L}_2 \bigg] \bigg]  $

    \item[$\bullet$] \textit{Leibniz' rule}. $\bigg[ \bigg[  \mathscr{L}_1 \mathscr{L}_2 , \mathscr{L}_3 \bigg] \bigg]  = \bigg[ \bigg[ \mathscr{L}_1  , \mathscr{L}_3  \bigg] \bigg]  \mathscr{L}_2  +  \mathscr{L}_1  \bigg[ \bigg[ \mathscr{L}_2 , \mathscr{L}_3 \bigg] \bigg]$, 

    \item[$\bullet$] \textit{Jacobi identity}. $\bigg[  \bigg[ \mathscr{L}_1 , \bigg[ \bigg[ \mathscr{L}_2 , \mathscr{L}_3  \bigg] \bigg]  \bigg] \bigg]  + \bigg[ \bigg[  \mathscr{L}_2 , \bigg[ \bigg[  \mathscr{L}_3 , \mathscr{L}_1 \bigg] \bigg]  \bigg] \bigg]  + \bigg[ \bigg[  \mathscr{L}_3  , \bigg[ \bigg[ \mathscr{L}_1 , \mathscr{L}_2 \bigg] \bigg]  \bigg] \bigg]  = 0$ \end{itemize}

\noindent given the three test linear operators $\mathscr{L}_1$, $\mathscr{L}_2$ and $\mathscr{L}_3$. In comparison to the action-angle coordinates which one wishes to study in Classical Mechanics, in Quantum Mechanics one studies complex-valued linear functionals. With respect to the commutator bracket, if the bracket,

\begin{align*}
     \bigg[ \bigg[   \Phi^{\mathrm{Ising-type}}   \big[ \textbf{C} \big]        , \bar{\Phi^{\mathrm{Ising-type}}}        \big[ \textbf{C} \big]                \bigg] \bigg]  \equiv   \bigg[ \bigg[     \Phi^{\mathrm{Ising-type}}              , \bar{\Phi^{\mathrm{Ising-type}}}       \bigg] \bigg]   ,
\end{align*}

\noindent were to vanish, the family,

\[
\left\{\!\begin{array}{ll@{}>{{}}l} \underline{(1)}:      \bigg[ \bigg[ A^{\mathrm{Ising}} \big( u \big)        , A^{\mathrm{Ising}} \big( u^{\prime} \big)   \bigg] \bigg]  
\text{ , } \\  \underline{(2)}:  \bigg[ \bigg[         A^{\mathrm{Ising}} \big( u \big)        ,        B^{\mathrm{Ising}} \big( u^{\prime} \big)     \bigg] \bigg]   \text{ , } \\  \underline{(3)}:   \bigg[ \bigg[   A^{\mathrm{Ising}} \big( u \big)       ,  C^{\mathrm{Ising}} \big( u^{\prime} \big) \bigg] \bigg]  
 \text{, }   \\  \underline{(4)}: \bigg[ \bigg[   A^{\mathrm{Ising}} \big( u \big)       ,  D^{\mathrm{Ising}} \big( u^{\prime} \big) \bigg] \bigg]    \text{ , } \\ \underline{(5)} : \bigg[ \bigg[ B^{\mathrm{Ising}} \big( u \big) , A^{\mathrm{Ising}} \big( u^{\prime} \big) \bigg]  \bigg] \text{, } \end{array}\right.
\]

\[
\left\{\!\begin{array}{ll@{}>{{}}l}   \underline{(6)}: \bigg[ \bigg[  B^{\mathrm{Ising}} \big( u \big) , B^{\mathrm{Ising}} \big( u^{\prime} \big) \bigg] \bigg]  \text{, } \\ \underline{(7)}: \bigg[ \bigg[ B^{\mathrm{Ising}} \big( u \big) , C^{\mathrm{Ising}} \big( u^{\prime} \big) \bigg] \bigg] \text{, } \\ \underline{(8)}: \bigg[ \bigg[  B^{\mathrm{Ising}} \big( u \big)  , D^{\mathrm{Ising}} \big( u^{\prime} \big)   \bigg] \bigg] \text{, } \\   \underline{(9)}:  \bigg[ \bigg[ C^{\mathrm{Ising}} \big( u \big)  , A^{\mathrm{Ising}} \big( u^{\prime} \big)   \bigg] \bigg]  \text{, } \\ \underline{(10)}: \bigg[ \bigg[ C^{\mathrm{Ising}} \big( u \big) , B^{\mathrm{Ising}} \big( u^{\prime} \big) \bigg]  \bigg]  \text{, } \\ \underline{(11)}: \bigg[ \bigg[ C^{\mathrm{Ising}} \big( u \big) , C^{\mathrm{Ising}} \big( u^{\prime} \big) \bigg] \bigg]  \text{, }  \\ \underline{(12)}: \bigg[ \bigg[ C^{\mathrm{Ising}} \big( u \big) , D^{\mathrm{Ising}} \big( u^{\prime} \big) \bigg] \bigg]  \text{, } \\ \underline{(13)}: \bigg[ \bigg[  D^{\mathrm{Ising}} \big( u \big)  , A^{\mathrm{Ising}} \big( u^{\prime} \big)  \bigg] \bigg]  \text{, } \\ 
   \underline{(14)}: \bigg[ \bigg[   D^{\mathrm{Ising}} \big( u \big)  , B^{\mathrm{Ising}} \big( u^{\prime} \big)  \bigg] \bigg]  \text{, } \\ \underline{(15)}: \bigg[ \bigg[  D^{\mathrm{Ising}} \big( u \big)  , C^{\mathrm{Ising}} \big( u^{\prime} \big) \bigg] \bigg]   \text{, }  \\ \underline{(16)}: \bigg[ \bigg[  D^{\mathrm{Ising}} \big( u \big)  ,       D^{\mathrm{Ising}} \big( u^{\prime} \big)  \bigg] \bigg]  \text{, }
\end{array}\right.
\]

\noindent of commutator brackets could each be approximated with the constants provided above.

\subsection{Compact representations over Lie algebras }

\noindent In comparison to taking the finite weak volume limit of Poisson brackets, through the identification,

\begin{align*}
    \big\{ \hat{\textit{Classical Operator 1}} , \hat{\textit{Classical Operator 2}} \big\} \overset{\textit{support} \rightarrow \textbf{Z}^2}{\longrightarrow}   \big\{ {\textit{Classical Operator 1}} , \textit{Classical} \\ \textit{ Operator 2} \big\} , 
\end{align*}

\noindent for the commutator bracket one takes the finite weak volume limit through the identification,

\begin{align*}
  \frac{1}{i \hbar} \bigg[ \bigg[ \hat{\textit{Quantum Operator 1}} , \hat{\textit{Quantum Operator 2}} \bigg] \bigg] \overset{\hbar \rightarrow 0^{+}}{\longrightarrow}  \bigg[ \bigg[ \textit{Quantum Operator 1} , \textit{Quantum} \\ \textit{ Operator 2} \bigg] \bigg]  ,
\end{align*}

\noindent where,

\begin{align*}
   \hat{\textit{Classical Operator 1}} \equiv   \textit{Finite volume approximation of Classical Operator 1} , \\ \\  \hat{\textit{Classical Operator 2}} \equiv  \textit{Finite volume approximation of Classical Operator 2} , \\ \\   \hat{\textit{Quantum Operator 1}} \equiv \textit{Finite volume approximation of Quantum Operator 1} , \\ \\  \hat{\textit{Quantum Operator 2}} \equiv  \textit{Finite volume approximation of Quantum Operator 2} , 
\end{align*}

\noindent Fix a spectral parameter $\lambda$ over $\textbf{Z}^2$, and Lie algebra $\textbf{g}$. For the endomorphism space,

\begin{align*}
   \mathrm{End} \big( V_a \otimes \mathcal{H} \big)  ,
\end{align*}

\noindent supported over the tensor product space of a vector space $V_a$, for $a \in \textbf{R}$, with the Hilbert space $\mathcal{H}$, the Quantum L-operators for the Ising-type model are related to representations of the form,

\begin{align*}
 L^{\mathrm{Ising}}  \equiv   L^{\mathrm{Ising}}  \big( i , \lambda \big)  \equiv \underset{i \in \textbf{Z}}{\sum} f_i \big( \lambda \big) \rho_a \big( T_i \big) \otimes \mathcal{O}_i   ,
\end{align*}

\noindent where,

\begin{align*}
    \textit{Quantum Operator 1} \in   \mathrm{End} \big( V_a \otimes \mathcal{H} \big)  , \\ \\ \textit{Quantum Operator 2} \in   \mathrm{End} \big( V_a \otimes \mathcal{H} \big)  . 
\end{align*}

\noindent for,

\begin{align*}
       T_i  \equiv \textit{basis of } \textbf{g} , \\ \\  \rho_a \equiv \textit{finite dimensional representation associated with each } V_a ,  \\ \\ \mathcal{O}_i \equiv \textit{operators supported over } \mathcal{H} , 
\\ \\ 
   f_i \big( \lambda \big) \equiv \textit{functions of the spectral parameters } \lambda  .
\end{align*}

\noindent In forthcoming arguments, we make use of the following sequence of arguments for constructing a transfer matrix from Quantum L-operators:

\begin{itemize}
    \item[$\bullet$] \textit{Fixing a basis over the Lie algebra}. With a fixed basis over the algebra, as a generalization of a complex-valued vector space, a compact representation,

    \begin{align*}
      \rho \equiv \rho_1 \cup \rho_2 \cup \rho_3 \cup \rho_4   ,
    \end{align*}

    \noindent can be introduced for the following Quantum L-operators,

        \begin{align*}
     \rho_{1, \textbf{g}} \equiv   \rho_1 \equiv   \textit{compact representation Quantum Operator 1 over } \textbf{g}  , \\ \\     \rho_{2, \textbf{g}} \equiv \rho_2 \equiv   \textit{compact representation Quantum Operator 2 over } \textbf{g} , \\ \\     \rho_{3, \textbf{g}} \equiv \rho_3 \equiv   \textit{compact representation Quantum Operator 3 over }  \textbf{g} , \\ \\     \rho_{4, \textbf{g}} \equiv \rho_4 \equiv   \textit{compact representation Quantum Operator 4 over } \textbf{g}  , 
        \end{align*}

    \noindent which are used to construct the transfer matrix.

    \item[$\bullet$] \textit{Fixing irreducible representations from the Lie algebra with the highest weight}. Fix the two representations,

    \begin{align*}
      \mathcal{R}_1 \equiv \textit{representation for which } \underset{\textit{representations r}}{\mathrm{sup}} w \big( r \big) \textit{is maximized, corresponding to the} \\ \textit{representation of } V_a   , \\  \\   \mathcal{R}_2 \equiv \textit{representation for which } \underset{\textit{representations $r^{\prime}$} : r^{\prime} \neq r}{\mathrm{sup}} w \big( r^{\prime} \big) \textit{is maximized, corresponding to} \\ \textit{ the representation of } \mathcal{H} , 
    \end{align*}

\noindent over $\textbf{g}$.

    \item[$\bullet$] \textit{RTT and RLL relations}. For the Ising-type model, these two collection of relations for RTT and RLL are introduced in \textit{1.9}.

    \item[$\bullet$] \textit{Defining representations corresponding to those of the Quantum L-operator over the endomorphism space}. The Ising-type L-operator, $\mathcal{L}$, is introduced in \textit{1.4}, the next section.

    \item[$\bullet$] \textit{Imposing the RTT and RLL relations from the Ising-type L-operators}. Given the representations over $\textbf{g}$ introduced in the above items, over the same endomorphism space one can enforce the RTT and RLL relations from $\mathcal{L}$. Such relations are closely related to the braiding operations for the Ising-type $R$ matrix, which is also introduced in \textit{1.9}.

    \item[$\bullet$] \textit{Constructing the Ising-type transfer matrix}. One can construct a desired approximation for the transfer matrix by taking properties of L-operators $\mathcal{L}$.

    \item[$\bullet$] \textit{Computation of conserved quantities, along with representations for the Hamiltonian}. When possible, one computes conserved quantities, which have infinitely many degrees of freedom. Furthermore, when possible, one also looks to compute representations of the Hamiltonian from the logarithmic derivative of the transfer matrix, which has been achieved for spin chains as described in \textit{1.8}.

\end{itemize}

\subsection{Hybrid Ising L-operators, and their connections with L-operators for integrable vertex models}

\noindent For some $a \in \textbf{Z}$, introduce the L-operator for an Ising type model, {\color{blue}[3]},

\[
\mathcal{L}^{a^{\prime}}_a \equiv \bigg[ \mathcal{L} \big[   x , \frac{x^{\prime}}{\xi} , y \big] \bigg]^{a^{\prime}}_a  \equiv  \begin{bmatrix}
   \frac{xx^{\prime}}{y^2} \delta_{a,a^{\prime}+1} - \frac{y^2}{xx^{\prime}} \delta_{a,a^{\prime}-1}      & q^a \bigg[  \frac{x}{x^{\prime} } \delta_{a,a^{\prime}+1} - \frac{x^{\prime}}{x} \delta_{a,a^{\prime}-1}  \bigg]  \\ q^{-a} \bigg[ \frac{x}{x^{\prime}} \delta_{a,a^{\prime}-1} - \frac{x^{\prime}}{x} \delta_{a,a^{\prime}+1 } \bigg]  & \frac{xx^{\prime}}{y^2} \delta_{a,a^{\prime}-1} - \frac{y^2}{xx^{\prime}} \delta_{a,a^{\prime}+1 }  
\end{bmatrix} \text{, }
\]

\noindent where in the entries of the representation for the Ising type L-operator above, $x \neq x^{\prime}$ denote two positions over the square lattice, while $q$ parameters, which are widely studied in vertex models as "quantum" parameters, appear in other L-operators that have extensively been studied by the author in {\color{blue}[45]}, which take the form,

\[
\hat{L} \big( \xi \big) \equiv L^{3D}_1 =  \mathrm{exp} \big( \lambda_3 ( q^{-2 } \xi^s ) \big)    \bigg[ \begin{smallmatrix}
        q^{D_1}       &    q^{-2} a_1 q^{-D_1-D_2} \xi^{s-s_1}        &   a_1 a_2 q^{-D_1 - 3D_2} \xi^{s - s_1 - s_2}  \\ a^{\dagger}_1 q^{D_1} \xi^{s_1} 
             &      q^{-D_1 + D_2} - q^{-2} q^{D_1 -D_2} \xi^{s}     &     - a_2 q^{D_1 - 3D_2} \xi^{s-s_2}  \\ 0  &    a^{\dagger}_2 q^{D_2} \xi^{s_2} &  q^{-D_2} \\   
  \end{smallmatrix} \bigg] \text{, }         
\]

\noindent given a spectral parameter $\xi$ for the triangular lattice, unital associative mapping $\xi$, and differential operators,

\begin{align*}
D^j_k  \equiv \big(  D      \otimes \textbf{1} \big) \textbf{1}_{\{\textbf{r} \equiv e_k\}}  \text{, } D^j_{k+1}\equiv \big(  \textbf{1} \otimes D \big) \textbf{1}_{\{\textbf{r} \equiv e_{k+1}\}}  \text{. }  
\end{align*}

\noindent that was originally introduced in {\color{blue}[6]}, given basis vectors $e_k$ and $e_{k+1}$ of $\textbf{T}$. As a general observation, for adaptations of QISM for vertex models, one studies R-matrices, which satisfying the Yang Baxter equation. With several possible interpretations, such equations not only demonstrate the equality of partition functions for a vertex model in question, but also demonstrate how intertwinning relations can be formalized.

\subsection{This paper's contributions}

\noindent Quantum scattering methods have close relations with Integrability, and exact solvability. For vertex models in Statistical Mechanics, under a wide variety of boundary conditions, one can demonstrate that suitable \textit{action angle coordinates} exist, for which the dynamics of a system of interest are approximately linear. Action-angle coordinates have vanishing Poisson bracket, hence implying that there must exist a system constituting the underlying \textit{Poisson structure}. Albeit the fact that sevearl computations from previous work of the author have demonstrated that exact, or hybrid, integrable properties for vertex models holds, it is also of interest to determine whether similar properties hold for Ising type models.

Ising type models fundamentally differ from vertex models for the following reasons. First, vertex models assign probabilities to configurations that are determined by the ice rule, or generalizations of the ice rule, while Ising type models assign probabilities to configurations that are determined by coupling constants and, potentially, multiple sites of the lattice. This difference not only determines the probability measure, and its corresponding boundary conditions, but also the structure of L-operators. Whereas L-operators for vertex type models, including those for the 4-vertex, 6-vertex, and 20-vertex, models depend upon unital associate mappings, spectral parameters, amongst other components, L-operators for Ising type models depend upon the sites of the lattice, and $\delta$ function, which is dependent upon two strictly positive parameters. Depiste the fact that local interactions between vertex and Ising type models, in some regard, are encoded in completely opposing manners, being able to obtain explicit, closed form representations of correlation functions, and other quantities, with respective classes of boundary conditions, can be achieved for both classes of models.

To streamline computations that one encounters with the Poisson bracket, transfer matrices of vertex, and Ising type, models alike can be formulated from L-operators. While boundary conditions for Ising type models can be more restrictive, namely from the condition that boundary spins can be $+$, or $-$,  vertex configurations which have edges lying incident to a finite volume boundary can be taken to point inwards or outwards. In the case that one would like to determine whether exactly solvable structure of a model of interest holds, one would then have to perform several computations with the Poisson bracket. As arguments to the Poisson bracket, one would have to determine whether there are suitable, explicit, closed form representations of block representations of the transfer matrix. From L-operators of Ising type models, as defined in the next section, we demonstrate that one can equivalently perform computations with the Poisson bracket by formulating systems of relations of the form,

\begin{align*}
 \underset{i \neq  j \in V ( \textbf{Z}^2 ) }{\underset{a,a^{\prime} \in \textbf{R}}{\mathrm{span}}} \big\{ \mathcal{C}\mathcal{R}_1 \big(i , j , a,a^{\prime} \big) , \mathcal{C}\mathcal{R}_2 \big(i , j , a,a^{\prime} \big) , \mathcal{C}\mathcal{R}_3 \big(i , j , a,a^{\prime} \big) \big\}   \text{, }
\end{align*}

\noindent for $a \neq a^{\prime}$ as two strictly positive parameters, where 

\begin{align*}
  \mathcal{C}\mathcal{R} \big( i , j , a , a^{\prime} \big) \equiv \mathcal{C}\mathcal{R}_1 \big( i , j ,a,a^{\prime} \big) \cup  \mathcal{C}\mathcal{R}_2 \big(i , j , a,a^{\prime} \big) \cup \mathcal{C}\mathcal{R}_3 \big( i , j ,a,a^{\prime} \big)  \text{, }
\end{align*}

\noindent denotes the column representations spanned by the representation,

\begin{align*}
    \begin{bmatrix} A^{\text{Ising type}} \big( i , j , a , a^{\prime} \big) & B^{\text{Ising type}} \big( i , j , a , a^{\prime} \big)  \\ C^{\text{Ising type}} \big(i , j , a , a^{\prime} \big)  & D^{\text{Ising type}} \big( i , j , a , a^{\prime} \big)   
    \end{bmatrix}  \equiv   \begin{bmatrix} A^{\text{Ising type}}  & B^{\text{Ising type}}  \\ C^{\text{Ising type}}   & D^{\text{Ising type}}  
    \end{bmatrix}  \text{, }
\end{align*}    

\noindent of the Ising type transfer matrix. As described further in the next section, in comparison to previous approaches of the author, within the QISM, which have obtained series of relations for each operator appearing in representations of transfer matrices such as that of the Ising type model above, we describe how computations with the Poisson bracket can be implemented for Ising type models. In particular, given the greatly simplified nature of individual entries for each block of the representation of,

\begin{align*}
    \begin{bmatrix} A^{\text{Ising type}}  & B^{\text{Ising type}}  \\ C^{\text{Ising type}}   & D^{\text{Ising type}}  
    \end{bmatrix}  \text{, }
\end{align*}

\noindent which can be identified with the entries of the representation for,

\begin{align*}
t^{\text{Ising}} : \textbf{C}^2 \otimes \big( \textbf{C}^2 \big)^N  \longrightarrow  \textbf{C}^2 \otimes \big( \textbf{C}^2 \big)^N   \mapsto      \underset{1 \leq j \leq N}{\prod} \big( t^{\text{Ising Type}} \big)_j \propto \underset{1 \leq j \leq N}{\prod}   \mathcal{L}^{a^{\prime}_j}_{a_j}   \text{, }
\end{align*}

\noindent we formulate relations within the Ising type \textit{Poisson structure}, through brackets of the form,

\begin{align*}
  \bigg[\bigg[                       
  \underset{n \in \textbf{Z}}{\sum}            \bigg[    \frac{y^2}{x_1 x_{N-1}} \delta_{a,a^{\prime}-1} q^{-a} \frac{x_1}{x_{N-1}}     \delta_{a,a^{\prime}-1} + q^a \frac{X_{N-1}}{x_1} \delta_{a,a^{\prime}+1}  \frac{x_1 x_N}{y^2} \delta_{a,a^{\prime}-1} + \frac{y^2}{x_1 x_{N-2}}   \delta_{a,a^{\prime}-1} q^a \frac{x_1}{x_{N-1}} \\ \times \delta_{a,a^{\prime}+1}            +        q^a \frac{x_{N-2}}{x_1} \delta_{a,a^{\prime}-1} \frac{x_1 x_{N-1}}{y^2} \delta_{a,a^{\prime}-1 }         \bigg]                ,  \cdot    \bigg]\bigg]           \text{. } 
\end{align*}

\noindent from the entries of the Ising type L-operator,

\[
   \left\{\!\begin{array}{ll@{}>{{}}l} 
  x_i x_j \delta_{a,a^{\prime}+1}  - \frac{y^2}{x_i x_j} \delta_{a,a^{\prime}-1}  \text{, } \\  \\      q^a \frac{x_i}{x_j} \delta_{a,a^{\prime}+1} -  q^a \frac{x_i}{x_j} \delta_{a,a^{\prime}-1}    \text{, } \\  \\        q^{-a} \frac{x_i}{x_j} \delta_{a,a^{\prime}-1} -  q^{-a} \frac{x_i}{x_j} \delta_{a,a^{\prime}+1}     \text{, }  \\  \\      \frac{x_i x_j}{y^2} \delta_{a,a^{\prime}-1}  - \frac{y^2}{x_i x_j} \delta_{a,a^{\prime}+1}    \text{, } 
 \end{array}\right.
 \]

\noindent defined in the next section. Straightforwardly, from Poisson brackets of the above form, integrable, and exactly solvable, structures of the Ising type model can be deduced in nearly the same way that they are for vertex models. That is, albeit the fact that L-operators, as previously described, encode interactions through the vertices of the underlying lattice, in comparison to L-operators for vertex models which encode interactions along edges, suitable approximations for all Poisson brackets, within the \textit{entire} Poisson structure, can be deduced. In comparison to computations with the Poisson bracket for block operators of transfer matrices of vertex models, block operators for transfer matrices of the Ising type model are far more simple, and can be studied from a single operator alone. Explicitly, the Poisson structure for the Ising type model, given the block operator decomposition,

\[ \begin{bmatrix} A^{\text{Ising type}}  & B^{\text{Ising type}}  \\ C^{\text{Ising type}}   & D^{\text{Ising type}}  
    \end{bmatrix} \text{, } \]

\noindent of the transfer matrix, takes the form,

\[
   \left\{\!\begin{array}{ll@{}>{{}}l} 
 \big\{ A^{\text{Ising type}} \big( i , j , a , a^{\prime} \big) , A^{\text{Ising type}} \big( i^{\prime} , j^{\prime} , a^{\prime\prime} , a^{\prime\prime\prime}\big)  \big\} , \cdots, \big\{  A^{\text{Ising type}} \big( i , j , a , a^{\prime} \big) \\ , D^{\text{Ising type}} \big( i^{\prime} , j^{\prime} , a^{\prime\prime} , a^{\prime\prime\prime} \big)  \big\}    \text{, }  \\  \vdots \\  \big\{ A^{\text{Ising type}} \big( i , j , a , a^{\prime} \big) , A^{\text{Ising type}} \big( i^{\prime} , j^{\prime} , a^{\prime\prime} , a^{\prime\prime\prime}\big)  \big\} , \cdots, \big\{  A^{\text{Ising type}} \big( i , j , a , a^{\prime} \big) \\ , D^{\text{Ising type}} \big( i^{\prime} , j^{\prime} , a^{\prime\prime} , a^{\prime\prime\prime} \big)  \big\}  \text{. }  
 \end{array}\right.
 \]

 \noindent In the next section, we define operators for the Ising type model. 

\subsection{Quantum-inverse scattering type objects}

\noindent For the Ising type model with L-operator $\mathcal{L}$ introduced above, the fact that each entry of the two by two representation is the difference of two terms implies that it is equal to,

\begin{align*}
 \begin{bmatrix}  x_i x_j \delta_{a,a^{\prime}+1} & q^a \frac{x_i}{x_j} \delta_{a,a^{\prime}+1}   \\ q^{-a} \frac{x_i}{x_j} \delta_{a,a^{\prime}-1} & \frac{x_i x_j}{y^2} \delta_{a,a^{\prime}-1} 
 \end{bmatrix} - \begin{bmatrix}           \frac{y^2}{x_i x_j} \delta_{a,a^{\prime}-1}  &
 q^a \frac{x_i}{x_j} \delta_{a,a^{\prime}-1} \\ q^{-a} \frac{x_i}{x_j} \delta_{a,a^{\prime}+1} & \frac{y^2}{x_i x_j} \delta_{a,a^{\prime}+1} \end{bmatrix} \text{, }
\end{align*}

\noindent under the change of coordinates of $\mathcal{L}^{a^{\prime}}_a$, which is given by,

\begin{align*}
 \begin{bmatrix}  x_i x_j \delta_{a,a^{\prime}+1}  - \frac{y^2}{x_i x_j} \delta_{a,a^{\prime}-1}   &   q^a \frac{x_i}{x_j} \delta_{a,a^{\prime}+1} -  q^a \frac{x_i}{x_j} \delta_{a,a^{\prime}-1}    \\  q^{-a} \frac{x_i}{x_j} \delta_{a,a^{\prime}-1} -  q^{-a} \frac{x_i}{x_j} \delta_{a,a^{\prime}+1}  & \frac{x_i x_j}{y^2} \delta_{a,a^{\prime}-1}  - \frac{y^2}{x_i x_j} \delta_{a,a^{\prime}+1} 
 \end{bmatrix} \text{. }
\end{align*}

\noindent To apply computations of the type that have previously been introduced by the author in {\color{blue}[41},{\color{blue}45]}, it suffices to consider cross terms from neighboring L-operators. In the form introduced above for sites $x_i$ and $x_j$ over the square lattice, one would consider products of the form,

\begin{align*}
 \underset{x_i \neq x_j}{\underset{1 \leq i < j \leq N}{\prod}} \bigg\{  \begin{bmatrix}  x_i x_j \delta_{a,a^{\prime}+1} & q^a \frac{x_i}{x_j} \delta_{a,a^{\prime}+1}   \\ q^{-a} \frac{x_i}{x_j} \delta_{a,a^{\prime}-1} & \frac{x_i x_j}{y^2} \delta_{a,a^{\prime}-1} 
 \end{bmatrix} - \begin{bmatrix}           \frac{y^2}{x_i x_j} \delta_{a,a^{\prime}-1}  &
 q^a \frac{x_i}{x_j} \delta_{a,a^{\prime}-1} \\ q^{-a} \frac{x_i}{x_j} \delta_{a,a^{\prime}+1} & \frac{y^2}{x_i x_j} \delta_{a,a^{\prime}+1} \end{bmatrix} \bigg\}    \text{, }
\end{align*}

\noindent for some $N>0$. As the weak finite volume limit is taken as $N \longrightarrow + \infty$, products of the form above can also be taken over $j$ such that,

\begin{align*}
   \big\{ j \in \textbf{N} : j < i \big\}  \text{, }
\end{align*}

\noindent with,

\begin{align*}
 \underset{x_i \neq x_j}{\underset{1 \leq j < i \leq N}{\prod}} \bigg\{  \begin{bmatrix}  x_i x_j \delta_{a,a^{\prime}+1} & q^a \frac{x_i}{x_j} \delta_{a,a^{\prime}+1}   \\ q^{-a} \frac{x_i}{x_j} \delta_{a,a^{\prime}-1} & \frac{x_i x_j}{y^2} \delta_{a,a^{\prime}-1} 
 \end{bmatrix} - \begin{bmatrix}           \frac{y^2}{x_i x_j} \delta_{a,a^{\prime}-1}  &
 q^a \frac{x_i}{x_j} \delta_{a,a^{\prime}-1} \\ q^{-a} \frac{x_i}{x_j} \delta_{a,a^{\prime}+1} & \frac{y^2}{x_i x_j} \delta_{a,a^{\prime}+1} \end{bmatrix} \bigg\}    \text{, }
\end{align*}

\noindent Despite $\mathcal{L}$ being introduced as an L-operator for an Ising type model, which the authors of {\color{blue}[]} observed as being a distant relative of the 6-vertex, and 8-vertex, models, several components parallel the structure of correlations, in addition to interactions, which are encapsulated by the ice rule of vertex models. Fundamentally, as a conservation rule, the ice rule significantly impacts the geometry of vertex configurations of the 6-vertex, and 20-vertex, models, {\color{blue}[41},{\color{blue}45]}, as well as characteristics of integrability, and exact solvability, that one would hope to prove about such models.

In the following, to obtain asymptotic approximations of the transfer matrix, and quantum monodromy matrix, for the Ising type model, we discuss behaviors of the expansion when $N \equiv 2$ first. Under this choice of $N$, the terms that emerge from the product,

\begin{align*}
 \underset{x_i \neq x_j}{\underset{1 \leq i < j \leq 2}{\prod}} \bigg\{   \begin{bmatrix}  x_i x_j \delta_{a,a^{\prime}+1} & q^a \frac{x_i}{x_j} \delta_{a,a^{\prime}+1}   \\ q^{-a} \frac{x_i}{x_j} \delta_{a,a^{\prime}-1} & \frac{x_i x_j}{y^2} \delta_{a,a^{\prime}-1} 
 \end{bmatrix} - \begin{bmatrix}           \frac{y^2}{x_i x_j} \delta_{a,a^{\prime}-1}  &
 q^a \frac{x_i}{x_j} \delta_{a,a^{\prime}-1} \\ q^{-a} \frac{x_i}{x_j} \delta_{a,a^{\prime}+1} & \frac{y^2}{x_i x_j} \delta_{a,a^{\prime}+1} \end{bmatrix} \bigg\}   \text{, }
\end{align*}

\noindent or from the product,

\begin{align*}
 \underset{x_i \neq x_j}{\underset{1 \leq j < i \leq 2}{\prod}} \bigg\{ \begin{bmatrix}  x_i x_j \delta_{a,a^{\prime}+1} & q^a \frac{x_i}{x_j} \delta_{a,a^{\prime}+1}   \\ q^{-a} \frac{x_i}{x_j} \delta_{a,a^{\prime}-1} & \frac{x_i x_j}{y^2} \delta_{a,a^{\prime}-1} 
 \end{bmatrix} - \begin{bmatrix}           \frac{y^2}{x_i x_j} \delta_{a,a^{\prime}-1}  &
 q^a \frac{x_i}{x_j} \delta_{a,a^{\prime}-1} \\ q^{-a} \frac{x_i}{x_j} \delta_{a,a^{\prime}+1} & \frac{y^2}{x_i x_j} \delta_{a,a^{\prime}+1} \end{bmatrix} \bigg\}   \text{, }
\end{align*}

\noindent include,

\begin{align*}
\begin{bmatrix}
    x_1 x_2 \delta_{a,a^{\prime}+1} & q^a \frac{x_1}{x_2} \delta_{a,a^{\prime}+1}    \\ q^{-a}  \frac{x_1}{x_2} \delta_{a,a^{\prime}-1} & \frac{x_1 x_2}{y^2} \delta_{a,a^{\prime}-1}  \end{bmatrix}
 \begin{bmatrix}
    x_2 x_3 \delta_{a,a^{\prime}+1} & q^a \frac{x_2}{x_3} \delta_{a,a^{\prime}+1}     \\        q^{-a} \frac{x_2}{x_3} \delta_{a,a^{\prime}-1} & \frac{x_2 x_3}{y^2} \delta_{a,a^{\prime}-1}     \end{bmatrix} \text{, } \\ \\  -   \begin{bmatrix}
 \frac{y^2}{x_1 x_2} \delta_{a,a^{\prime}-1} & q^a        \frac{x_1}{x_2} \delta_{a,a^{\prime}-1}    \\    q^{-a} \frac{x_1}{x_2} \delta_{a,a^{\prime}+1}     &     \frac{y^2}{x_1 x_2 } \delta_{a,a^{\prime}+1}     \end{bmatrix} \begin{bmatrix}
   x_2 x_3 \delta_{a,a^{\prime}+1} & q^a \frac{x_2}{x_3} \delta_{a,a^{\prime}+1}    \\ q^{-a}  \frac{x_2}{x_3} \delta_{a,a^{\prime}-1} & \frac{x_2 x_3}{y^2} \delta_{a,a^{\prime}-1}   \end{bmatrix}   \text{, } \\  \\ - \begin{bmatrix}
    x_1 x_2 \delta_{a,a^{\prime}+1} & q^a \frac{x_1}{x_2} \delta_{a,a^{\prime}+1}    \\ q^{-a}  \frac{x_1}{x_2} \delta_{a,a^{\prime}-1} & \frac{x_1 x_2}{y^2} \delta_{a,a^{\prime}-1}  \end{bmatrix} \begin{bmatrix} \frac{y^2}{x_2 x_3} \delta_{a,a^{\prime}-1 } & q^a \frac{x_2}{x_3} \delta_{a,a^{\prime}-1} \\ q^{-a} \frac{x_2}{x_3} \delta_{a,a^{\prime}+1} &          \frac{y^2}{x_2 x_3} \delta_{a,a^{\prime}+1 }   
    \end{bmatrix} \text{, } \\   \\  \begin{bmatrix}
 \frac{y^2}{x_1 x_2} \delta_{a,a^{\prime}-1} & q^a        \frac{x_1}{x_2} \delta_{a,a^{\prime}-1}    \\    q^{-a} \frac{x_1}{x_2} \delta_{a,a^{\prime}+1}     &     \frac{y^2}{x_1 x_2 } \delta_{a,a^{\prime}+1}     \end{bmatrix}  \begin{bmatrix} \frac{y^2}{x_2 x_3} \delta_{a,a^{\prime}-1 } & q^a \frac{x_2}{x_3} \delta_{a,a^{\prime}-1} \\ q^{-a} \frac{x_2}{x_3} \delta_{a,a^{\prime}+1} &          \frac{y^2}{x_2 x_3} \delta_{a,a^{\prime}+1 }   
    \end{bmatrix}     \text{. } 
  \end{align*}

\noindent In each of the four terms above, minus signs accumulate in the middle two terms. In the finite weak volume limit as $N \longrightarrow + \infty$, for either one of the the infinite products,

\begin{align*}
 \underset{x_i \neq x_j}{\underset{1 \leq i < j \leq + \infty}{\prod}} \bigg\{  \begin{bmatrix}  x_i x_j \delta_{a,a^{\prime}+1} & q^a \frac{x_i}{x_j} \delta_{a,a^{\prime}+1}   \\ q^{-a} \frac{x_i}{x_j} \delta_{a,a^{\prime}-1} & \frac{x_i x_j}{y^2} \delta_{a,a^{\prime}-1} 
 \end{bmatrix} - \begin{bmatrix}           \frac{y^2}{x_i x_j} \delta_{a,a^{\prime}-1}  &
 q^a \frac{x_i}{x_j} \delta_{a,a^{\prime}-1} \\ q^{-a} \frac{x_i}{x_j} \delta_{a,a^{\prime}+1} & \frac{y^2}{x_i x_j} \delta_{a,a^{\prime}+1} \end{bmatrix} \bigg\}   \text{, } \\   \underset{x_i \neq x_j}{\underset{1 \leq j < i \leq + \infty}{\prod}}\bigg\{  \begin{bmatrix}  x_i x_j \delta_{a,a^{\prime}+1} & q^a \frac{x_i}{x_j} \delta_{a,a^{\prime}+1}   \\ q^{-a} \frac{x_i}{x_j} \delta_{a,a^{\prime}-1} & \frac{x_i x_j}{y^2} \delta_{a,a^{\prime}-1} 
 \end{bmatrix} - \begin{bmatrix}           \frac{y^2}{x_i x_j} \delta_{a,a^{\prime}-1}  &
 q^a \frac{x_i}{x_j} \delta_{a,a^{\prime}-1} \\ q^{-a} \frac{x_i}{x_j} \delta_{a,a^{\prime}+1} & \frac{y^2}{x_i x_j} \delta_{a,a^{\prime}+1} \end{bmatrix} \bigg\}     \text{, }
\end{align*}

\noindent of L-operators to asymptotically approximate the Ising type transfer matrix given by the mapping,

\begin{align*}
t^{\text{Ising Type}} : \textbf{C}^2 \otimes \big( \textbf{C}^2 \big)^N  \longrightarrow  \textbf{C}^2 \otimes \big( \textbf{C}^2 \big)^N   \mapsto      \underset{1 \leq j \leq N}{\prod} \big( t^{\text{Ising Type}} \big)_j \propto \underset{1 \leq j \leq N}{\prod}   \mathcal{L}^{a^{\prime}_j}_{a_j}   \text{, }
\end{align*}

\noindent where the points $a^{\prime}_j$, and $a_j$, are contained within the finite volume $\mathscr{V}_i \subsetneq \textbf{Z}^2$, for,

\begin{align*}
  \big( t^{\text{Ising Type}} \big)_j \cap j \textbf{Z}^2 \neq \emptyset  \text{, } \\       \mathrm{support} \big( \mathcal{L}^{a^{\prime}_j}_{a_j} \big)  \subsetneq \mathrm{support} \big( j \textbf{Z}^2 \big) \subsetneq \mathrm{support} \big( \textbf{Z}^2 \big) \subsetneq \textbf{Z}^2     \text{, }
\end{align*}

\noindent The quantum monodromy matrix given by the mapping, 

\begin{align*}
    T^{\text{Ising Type}} : \textbf{C}^2 \otimes \big( \textbf{C}^2 \big)^N  \longrightarrow  \textbf{C}^2 \otimes \big( \textbf{C}^2 \big)^N   \mapsto  \mathrm{tr} \bigg[     \underset{1 \leq j \leq N}{\prod} \big( t^{\text{Ising Type}} \big)_j \bigg]   \propto \mathrm{tr} \bigg[  \underset{1 \leq j \leq N}{\prod}   \mathcal{L}^{a^{\prime}_j}_{a_j}  \bigg]      \text{. }
\end{align*}

\noindent The two mappings above can be seen as lower dimensional objects in comparison to the three-dimensional quantum monodromy matrix for the 20-vertex model examined by the author in {\color{blue}[45]}, which takes the form,

{\small \begin{align*}
 T^{3D}_{a,b} \big(    \big\{ u_i \big\} , \big\{ v^{\prime}_j  \big\} , \big\{ w^{\prime\prime}_k \big\}     \big) :   \textbf{C}^3 \otimes \big( \textbf{C}^3 \big)^{\otimes ( |N| + ||M||_1 )}  \longrightarrow   \textbf{C}^3 \otimes \big( \textbf{C}^3 \big)^{\otimes ( |N| + ||M||_1 )} \\   \mapsto  \overset{-N}{\underset{j=0}{\prod}} \text{ }  \overset{\underline{M}}{\underset{k=0}{\prod}} \bigg[  \mathrm{diag} \big( \mathrm{exp} \big(  \alpha \big(i, j,k \big)  \big)   , \mathrm{exp} \big(  \alpha \big( i,j,k\big)  \big)   , \mathrm{exp} \big(  \alpha \big( i, j,k \big)  \big) \big) \\ \times  R_{ia,jb,kc} \big( u - u_i ,  u^{\prime} - v^{\prime}_j , w-w^{\prime\prime}_k \big) \bigg]  \text{, } 
\end{align*}}

\noindent where $R \equiv \mathcal{R}$ denotes the Universal R-matrix (see {\color{blue}[6]} for more details, including the factorization into contributions from the K-matrix). Beyond $N\equiv 3$ terms in the product of L-operators that has been previously discussed, a few of the terms that will be computed for the asymptotic expansion of the transfer matrix include,

\begin{align*}
\begin{bmatrix}
    x_1 x_2 \delta_{a,a^{\prime}+1} & q^a \frac{x_1}{x_2} \delta_{a,a^{\prime}+1}    \\ q^{-a}  \frac{x_1}{x_2} \delta_{a,a^{\prime}-1} & \frac{x_1 x_2}{y^2} \delta_{a,a^{\prime}-1}  \end{bmatrix} 
 \begin{bmatrix}
    x_2 x_3 \delta_{a,a^{\prime}+1} & q^a \frac{x_2}{x_3} \delta_{a,a^{\prime}+1}     \\        q^{-a} \frac{x_2}{x_3} \delta_{a,a^{\prime}-1} & \frac{x_2 x_3}{y^2} \delta_{a,a^{\prime}-1}     \end{bmatrix}  \begin{bmatrix}
    x_3 x_4 \delta_{a,a^{\prime}+1} & q^a \frac{x_3}{x_4} \delta_{a,a^{\prime}+1}     \\        q^{-a} \frac{x_3}{x_4} \delta_{a,a^{\prime}-1} & \frac{x_3 x_4}{y^2} \delta_{a,a^{\prime}-1}     \end{bmatrix}  \text{, } \\ \\ -   \begin{bmatrix}
 \frac{y^2}{x_1 x_2} \delta_{a,a^{\prime}-1} & q^a        \frac{x_1}{x_2} \delta_{a,a^{\prime}-1}    \\    q^{-a} \frac{x_1}{x_2} \delta_{a,a^{\prime}+1}     &     \frac{y^2}{x_1 x_2 } \delta_{a,a^{\prime}+1}     \end{bmatrix} \begin{bmatrix}
   x_2 x_3 \delta_{a,a^{\prime}+1} & q^a \frac{x_2}{x_3} \delta_{a,a^{\prime}+1}    \\ q^{-a}  \frac{x_2}{x_3} \delta_{a,a^{\prime}-1} & \frac{x_2 x_3}{y^2} \delta_{a,a^{\prime}-1}   \end{bmatrix}  \begin{bmatrix}
    x_3 x_4 \delta_{a,a^{\prime}+1} & q^a \frac{x_3}{x_4} \delta_{a,a^{\prime}+1}     \\        q^{-a} \frac{x_3}{x_4} \delta_{a,a^{\prime}-1} & \frac{x_3 x_4}{y^2} \delta_{a,a^{\prime}-1}     \end{bmatrix}               \text{, }    \end{align*}

    \begin{align*}                           \begin{bmatrix}
 \frac{y^2}{x_1 x_2} \delta_{a,a^{\prime}-1} & q^a        \frac{x_1}{x_2} \delta_{a,a^{\prime}-1}    \\    q^{-a} \frac{x_1}{x_2} \delta_{a,a^{\prime}+1}     &     \frac{y^2}{x_1 x_2 } \delta_{a,a^{\prime}+1}     \end{bmatrix}  \begin{bmatrix} \frac{y^2}{x_2 x_3} \delta_{a,a^{\prime}-1 } & q^a \frac{x_2}{x_3} \delta_{a,a^{\prime}-1} \\ q^{-a} \frac{x_2}{x_3} \delta_{a,a^{\prime}+1} &          \frac{y^2}{x_2 x_3} \delta_{a,a^{\prime}+1 }   
    \end{bmatrix}  \begin{bmatrix} \frac{y^2}{x_3 x_4} \delta_{a,a^{\prime}-1 } & q^a \frac{x_3}{x_4} \delta_{a,a^{\prime}-1} \\ q^{-a} \frac{x_3}{x_4} \delta_{a,a^{\prime}+1} &          \frac{y^2}{x_3 x_4} \delta_{a,a^{\prime}+1 }   
    \end{bmatrix}  \text{, } \\  \\    \begin{bmatrix}
    x_1 x_2 \delta_{a,a^{\prime}+1} & q^a \frac{x_1}{x_2} \delta_{a,a^{\prime}+1}    \\ q^{-a}  \frac{x_1}{x_2} \delta_{a,a^{\prime}-1} & \frac{x_1 x_2}{y^2} \delta_{a,a^{\prime}-1}  \end{bmatrix} \begin{bmatrix} \frac{y^2}{x_2 x_3} \delta_{a,a^{\prime}-1 } & q^a \frac{x_2}{x_3} \delta_{a,a^{\prime}-1} \\ q^{-a} \frac{x_2}{x_3} \delta_{a,a^{\prime}+1} &          \frac{y^2}{x_2 x_3} \delta_{a,a^{\prime}+1 }   
    \end{bmatrix} \begin{bmatrix} \frac{y^2}{x_2 x_3} \delta_{a,a^{\prime}-1 } & q^a \frac{x_2}{x_3} \delta_{a,a^{\prime}-1} \\ q^{-a} \frac{x_2}{x_3} \delta_{a,a^{\prime}+1} &          \frac{y^2}{x_2 x_3} \delta_{a,a^{\prime}+1 }   
    \end{bmatrix}   \text{. } \\  
    \end{align*}

\noindent Within the quantum inverse scattering framework, analyzing how interactions of either a vertex, or Ising type, model, are encoded through entries of the L-operator continues to remain of great interest. In L-operators for vertex models, one typically encounters Pauli basis elements, projectors, and possibly terms from a mapping into a unital associative algebra, whereas in L-operators for Ising type models, as indicated through the operator above, one encounters contributions from all possible pairwise contributions of two points within some finite volume. For an Ising type model, contributions from each entry of the representation for the L-operator differ from computations of L-operators for vertex models previously studied by the author in {\color{blue}[41},{\color{blue}45},{\color{blue}46]}. As one possible representative of an asymptotic expansion of the transfer matrix as the system size tends to that of $\textbf{Z}^2$, products of an L-operator can also be used for studying counterparts of the emptiness formation probability for the 20-vertex model, {\color{blue}[47]}, from previous computations with contour-integral representations under domain-wall boundary conditions for the 6-vertex model {\color{blue}[7]}. As such, determining how the dimensionality of the underlying state space of a vertex model, whether in two or three dimensions, in addition to the dependency of the model on boundary conditions, are of great significance. Furthermore, conjectures that have been raised with regards to the convergence of the scaling limit of the 6-vertex model to the Gaussian free field is also indicative of behaviors of other models in Statistical Physics - in which interactions interpolate between pointwise interactions such as those encoded through Hamiltonians of the Ising and Potts models, to interactions for the Gaussian free field which are not defined pointwise.

In the forthcoming sections, we provide computations for products of L-operators for the Ising type model. After having obtained a system of relations for each of the four entries of the product representation, as has already been investigated by the author for the 4-vertex, 6-vertex, and 20-vertex, models, we characterize integrability, and exact solvability, of the Ising-type model.

\subsection{Statement of Main Results}

\noindent We collect the two main results below.

\bigskip

\noindent \textbf{Lemma} (\textit{product representation of the transfer matrix from two, and three, L-operators of the Ising-type model}). Given the L-operator defined for the Ising-type formulation of the 6-vertex model, for the space $\mathscr{F}^{\prime}$ of functions,

\[
  \mathscr{F}^{\prime} \equiv  \left\{\!\begin{array}{ll@{}>{{}}l} 
  \mathcal{I}^1_1 \equiv \mathcal{I}^1_1\big( x_1 , \cdots, x_N, a , a+1, a^{\prime}, a^{\prime}-1
 , a^{\prime}+1  \big) , \\  \\  \mathcal{I}^2_1 \equiv \mathcal{I}^2_1\big( x_1 , \cdots, x_N, a , a+1, a^{\prime}, a^{\prime}-1
 , a^{\prime}+1  \big) , \\ \\   \mathcal{I}^3_1 \equiv \mathcal{I}^3_1\big( x_1 , \cdots, x_N, a , a+1, a^{\prime}, a^{\prime}-1
 , a^{\prime}+1  \big) , \\  \\  \mathcal{I}^4_1 \equiv \mathcal{I}^4_1\big( x_1 , \cdots, x_N, a , a+1, a^{\prime}, a^{\prime}-1
 , a^{\prime}+1  \big)  , \\ \\   \mathcal{I}^1_2 \equiv \mathcal{I}^1_2\big( x_1 , \cdots, x_N, a , a+1, a^{\prime}, a^{\prime}-1
 , a^{\prime}+1  \big) , \\ \\  \mathcal{I}^2_2 \equiv \mathcal{I}^2_2\big( x_1 , \cdots, x_N, a , a+1, a^{\prime}, a^{\prime}-1
 , a^{\prime}+1  \big) ,  \\ \\  \mathcal{I}^3_2 \equiv \mathcal{I}^3_2\big( x_1 , \cdots, x_N, a , a+1, a^{\prime}, a^{\prime}-1
 , a^{\prime}+1  \big) , \\ \\  \mathcal{I}^4_2 \equiv \mathcal{I}^4_2\big( x_1 , \cdots, x_N, a , a+1, a^{\prime}, a^{\prime}-1
 , a^{\prime}+1  \big) ,  \end{array}\right. \] \[  \left\{\!\begin{array}{ll@{}>{{}}l} \mathcal{I}^1_1 \mathcal{I}^1_2 + \mathcal{I}^3_1 \mathcal{I}^3_2    \equiv  \big( \mathcal{I}^1_1 \mathcal{I}^1_2 + \mathcal{I}^3_1 \mathcal{I}^3_2  \big) \big( x_1 , \cdots, x_N, a , a+1, a^{\prime}, a^{\prime}-1
 , a^{\prime}+1  \big) , \\  \\ \mathcal{I}^2_1 \mathcal{I}^2_2 + \mathcal{I}^4_1 \mathcal{I}^4_2    \equiv  \big( \mathcal{I}^2_1 \mathcal{I}^2_2 + \mathcal{I}^4_1 \mathcal{I}^4_2  \big) \big( x_1 , \cdots, x_N, a , a+1, a^{\prime}, a^{\prime}-1
 , a^{\prime}+1  \big) ,  \\ \\  \mathcal{I}^1_1 \mathcal{I}^3_2 + \mathcal{I}^3_1 \mathcal{I}^4_2    \equiv  \big( \mathcal{I}^1_1 \mathcal{I}^3_2 + \mathcal{I}^3_1 \mathcal{I}^4_2  \big) \big( x_1 , \cdots, x_N, a , a+1, a^{\prime}, a^{\prime}-1
 , a^{\prime}+1  \big) , \\  \\ \mathcal{I}^3_1 \mathcal{I}^2_2 + \mathcal{I}^4_1 \mathcal{I}^4_2    \equiv  \big( \mathcal{I}^3_1 \mathcal{I}^2_2 + \mathcal{I}^4_1 \mathcal{I}^4_2    \big) \big( x_1 , \cdots, x_N, a , a+1, a^{\prime}, a^{\prime}-1
 , a^{\prime}+1  \big) ,  \end{array}\right.
 \]

\noindent to asymptotically approximate the transfer matrix in infinite volume, one has the following two representations, the first two of which are,

\[  \bigg[ \begin{smallmatrix} \mathcal{I}^1_1  & \mathcal{I}^2_1   \\ \mathcal{I}^3_1   & \mathcal{I}^4_1   \end{smallmatrix} \bigg] \text{, }  \bigg[ \begin{smallmatrix} \mathcal{I}^1_2  & \mathcal{I}^2_2   \\ \mathcal{I}^3_2  & \mathcal{I}^4_2    \end{smallmatrix} \bigg] \text{, }    \]

\noindent and the third of which is,

\[   \bigg[ \begin{smallmatrix} \mathcal{I}^1_1 \mathcal{I}^1_2 + \mathcal{I}^3_1 \mathcal{I}^3_2   &  \mathcal{I}^2_1 \mathcal{I}^2_2 + \mathcal{I}^4_1 \mathcal{I}^4_2   \\ \mathcal{I}^1_1 \mathcal{I}^3_2 + \mathcal{I}^3_1 \mathcal{I}^4_2  &  \mathcal{I}^3_1 \mathcal{I}^2_2 + \mathcal{I}^4_1 \mathcal{I}^4_2    \end{smallmatrix} \bigg] ,  \]

\noindent of the Ising-type transfer matrix.

\bigskip

\noindent With the result above, the remaining main result below provides conditions on the action-angle coordinates. In the main result below, we state the conditions under which action-angle coordinates of the Ising-type model vanish. Specifically, in comparison to previous computations with L-operators for two, and three, dimensional vertex models, the forthcoming result below establishes an iff correspondence between the Poisson bracket of the \textit{lower} order expansion of action-angle coordinates, and the \textit{higher} order expansion of action-angle coordinates. With respect to the Poisson bracket, one performs computations of the form,

\begin{align*}
 \bigg[\bigg[ \underset{i \neq  j \in V ( \textbf{Z}^2 ) }{\underset{a,a^{\prime} \in \textbf{R}}{\mathrm{span}}} \big\{ \mathcal{I}^1_1 \mathcal{I}^1_2 + \mathcal{I}^3_1 \mathcal{I}^3_2   ,       \mathcal{I}^2_1 \mathcal{I}^2_2 + \mathcal{I}^4_1 \mathcal{I}^4_2  ,  \mathcal{I}^1_1 \mathcal{I}^3_2 + \mathcal{I}^3_1 \mathcal{I}^4_2  ,   \mathcal{I}^3_1 \mathcal{I}^2_2 + \mathcal{I}^4_1 \mathcal{I}^4_2   \big\}   \\  , \mathrm{log} \bigg[ \underset{i \neq  j \in V ( \textbf{Z}^2 ) }{\underset{a,a^{\prime} \in \textbf{R}}{\mathrm{span}}} \big\{ \mathcal{I}^1_1 \mathcal{I}^1_2 + \mathcal{I}^3_1   \mathcal{I}^3_2   ,       \mathcal{I}^2_1\mathcal{I}^2_2  + \mathcal{I}^4_1 \mathcal{I}^4_2  ,  \mathcal{I}^1_1 \mathcal{I}^3_2 + \mathcal{I}^3_1 \mathcal{I}^4_2  ,   \mathcal{I}^3_1 \mathcal{I}^2_2  + \mathcal{I}^4_1 \mathcal{I}^4_2   \big\} \bigg]  \bigg]\bigg] , \end{align*}

\noindent where,

\begin{align*}
     \underset{i \neq  j \in V ( \textbf{Z}^2 ) }{\underset{a,a^{\prime} \in \textbf{R}}{\mathrm{span}}} \big\{ \textit{First-order Ising-type block operators} \big\} =    \underset{i \neq  j \in V ( \textbf{Z}^2 ) }{\underset{a,a^{\prime} \in \textbf{R}}{\mathrm{span}}} \big\{   \mathcal{I}^1_1 , \mathcal{I}^2_1 , \mathcal{I}^3_1 , \mathcal{I}^4_1 \big\}  , \\  \underset{i \neq  j \in V ( \textbf{Z}^2 ) }{\underset{a,a^{\prime} \in \textbf{R}}{\mathrm{span}}} \big\{ \textit{Second-order Ising-type block operators} \big\} =    \underset{i \neq  j \in V ( \textbf{Z}^2 ) }{\underset{a,a^{\prime} \in \textbf{R}}{\mathrm{span}}} \big\{  \mathcal{I}^1_2 , \mathcal{I}^2_2 , \mathcal{I}^3_2 , \mathcal{I}^4_2 \big\}  , \\  \underset{i \neq  j \in V ( \textbf{Z}^2 ) }{\underset{a,a^{\prime} \in \textbf{R}}{\mathrm{span}}} \big\{ \textit{Third-order Ising-type block operators} \big\} =    \underset{i \neq  j \in V ( \textbf{Z}^2 ) }{\underset{a,a^{\prime} \in \textbf{R}}{\mathrm{span}}} \big\{     \mathcal{I}^1_3 , \mathcal{I}^2_3 , \mathcal{I}^3_3 , \mathcal{I}^4_3 \big\}      , \end{align*}

\noindent corresponding to the set of linear combinations of Ising-type block operators. For three-dimensional transfer matrices of vertex models, one can similarly express the finite=dimensiional representation of the transfer matrix with,

\begin{align*}
\textbf{T}^{3D} \equiv   \underset{\textbf{T}}{\mathrm{span}} \big\{    \mathcal{T}\mathcal{C}_1   ,  \mathcal{T}\mathcal{C}_2 ,  \mathcal{T}\mathcal{C}_3    \big\}  \text{, } 
\end{align*}

\noindent where,

\begin{align*}
 \underset{\textbf{T}}{\mathrm{span}} \big\{    \mathcal{T}\mathcal{C}_1   ,  \mathcal{T}\mathcal{C}_2 ,  \mathcal{T}\mathcal{C}_3           \big\} \approx    \underset{\underline{j} \in \textbf{R}^2  , k \in \textbf{N}}{\mathrm{span}} \big\{ \mathcal{B}_1 ,  \mathcal{B}_2 ,  \mathcal{B}_3 \big\}   \text{, }
\end{align*}

\noindent for,

{\small \begin{align*}
 \mathcal{B}_1 \equiv \begin{bmatrix}     q^{D^j_k + D^{j+1}_k}  + q^{-2} a^j_k q^{-D^j_k - D^j_{k+1}} \xi^{s-s^j_k}   \\  \\   \big( a^j_k \big)^{\dagger}  q^{D^j_k} \xi^{s^j_k} q^{D^{j+1}_k} + q^{-D^j_k + D^j_{k+1}} \big( a^{j+1}_k \big)^{\dagger} q^{D^j_{k+1}}  \xi^{s^{j+1}_k} - q^{-2} q^{D^j_k} \big( a^j_k \big)^{\dagger} \\   \\ \big( a^j_{k+1} \big)^{\dagger} q^{D^j_{k+1}} \xi^{s^j_{k+1}} \big( a^{j+1}_k \big)^{\dagger} q^{D^{j+1}_k} \xi^{s^{j+1}_k}  \\    \end{bmatrix}   \text{, }   
\\ \\ 
 \mathcal{B}_2 \equiv     \begin{bmatrix}    \mathcal{E}_1 \\     \\      \big( a^j_k \big)^{\dagger} q^{D^j_{k+1}} \xi^{s^j_{k+1}} q^{-D^{j+1}_k + D^{j+1}_{k+1}}  - \big(   a^j_{k+1}   \big)^{\dagger}  q^{D^j_{k+1}} \xi^{s^j_{k+1}} q^{-2} q^{D^{j+1}_k} + q^{-D^{j}_{k+1}} \big( a^{j+1}_k \big)^{\dagger} q^{D^{j+1}_{k+1}} \xi^{s^{j+1}_{k+1}} \\ \\   q^{D^j_k} a^{j+1}_k a^{j+1}_{k+1} q^{-D^{j+1}_k - 3 D^{j+1}_{k+1} } \xi^{s-s^{j+1}_k - s^{j+1}_{k+1}} + q^{-2} a^j_k q^{-D^j_k - D^j_{k+1}} \\ \times   \xi^{s-s^j_{k+1}} \big( - a^{j+1}_{k+1} \big)^{\dagger} q^{D^{j+1}_{k} - 3 D^{j+1}_{k+1}} \xi^{s-s^{j+1}_{k+1}} \\ \end{bmatrix} 
\text{, }   
\end{align*} }

\begin{align*}
    \mathcal{E}_1 \equiv   q^{D^j_k} q^{-2} a^{j+1}_k - q^{D^{j+1}_k - D^{j+1}_{k+1} } \xi^{s-s^{j+1}_k} + q^{-2} a^j_k q^{-D^j_k - D^j_{k+1}} \xi^{s-s^j_k} q^{-2} q^{D^{j+1}_k} \xi^s +  a^j_k a^j_{k+1} q^{-D^j_k - 3 D^j_{k+1}} \\ \times \xi^{s-s^j_k - s^j_{k+1}} \big( a^{j+1}_k \big)^{\dagger} q^{D^{j+1}_{k+1}} \xi^{s^{j+1}_{k+1}}   \big( a^j_k \big)^{\dagger} q^{D^j_k } \xi^{s^j_k} q^{-2} a^{j+1}_k q^{-D^{j+1}_k - D^{j+1}_{k+1}}  \text{, } \\ \\ 
  \mathcal{B}_3 \equiv   \bigg[  \begin{smallmatrix}   \mathcal{E}^{\prime}_1  \\    \mathcal{E}_2     \\      \big( a^{j}_{k+1} \big)^{\dagger} q^{D^j_{k+1}} \xi^{s^j_{k+1}} \big( - a^{j+1}_{k+1} \big)^{\dagger} q^{D^{j+1}_k - 3 D^{j+1}_{k+1}} \xi^{s-s^{j+1}_k} + \big( q^{-D^j_{k+1}} \big) \big( q^{-D^{j+1}_{k+1}}  \big)    \end{smallmatrix} \bigg]   \text{, }  
\\ \\ 
\mathcal{E}^{\prime}_1 \equiv   q^{D^j_k} a^{j+1}_k a^{j+1}_{k+1} q^{-D^{j+1}_k - 3 D^{j+1}_{k+1} } \xi^{s-s^{j+1}_k - s^{j+1}_{k+1}} + q^{-2}  a^j_k q^{-D^j_k - D^j_{k+1}} \big( - a^{j+1}_{k+1} \big)^{\dagger} q^{-D^{j+1}_k - 3 D^{j+1}_{k+1}}  \\ \times     \xi^{s-s^{j+1}_k} + q^{-2}  a^j_k q^{-D^j_k - D^j_{k+1}} \xi^{s-s^j_k} \big( a^{j+1}_{k+1} \big)^{\dagger}   q^{D^{j+1}_k - 3 D^{j+1}_{k+1}} \xi^{s-s^{j+1}_k}    + a^j_k a^j_{k+1}  q^{-D^j_k - 3 D^j_{k+1}}   \xi^{s-s^j_k - s^j_{k+1}} \\ \times   q^{-D^{j+1}_{k+1}}    \text{, } \\ \\ 
\mathcal{E}_2  \equiv  \big( a^{j}_{k} \big)^{\dagger}       q^{D^j_k} \xi^{s^j_k} a^{j+1}_k a^{j+1}_{k+1} q^{-D^{j+1}_k - 3 D^{j+1}_{k+1}} \xi^{s- s^{j+1}_k - s^{j+1}_{k+1}}   + q^{-D^j_k + D^j_{k+1}} \big( - a^{j+1}_{k+1} \big)^{\dagger}     q^{-D^{j+1}_k - 3 D^{j+1}_{k+1}} \\ \times \xi^{s-s^{j+1}_{k+1}} - q^{-2} q^{D^j_k} \xi^s        \big( - a^{j+1}_{k+1} \big)^{\dagger}    q^{D^{j+1}_k - 3 D^{j+1}_{k+1}} \xi^{s-s^{j+1}_k} -  q^{-2} q^{D^j_k} \xi^s  \big( - a^{j+1}_{k+1} \big)^{\dagger} \\ \times q^{D^{j+1}_k - 3 D^{j+1}_{k+1}} \xi^{s-s^{j+1}_k}   \text{, } 
\end{align*}

\noindent corresponding to the products of block operators of,

\[
\hat{L} \big( \xi \big) \equiv L^{3D}_1 =  \mathrm{exp} \big( \lambda_3 ( q^{-2 } \xi^s ) \big)    \bigg[ \begin{smallmatrix}
        q^{D_1}       &    q^{-2} a_1 q^{-D_1-D_2} \xi^{s-s_1}        &   a_1 a_2 q^{-D_1 - 3D_2} \xi^{s - s_1 - s_2}  \\ a^{\dagger}_1 q^{D_1} \xi^{s_1} 
             &      q^{-D_1 + D_2} - q^{-2} q^{D_1 -D_2} \xi^{s}     &     - a_2 q^{D_1 - 3D_2} \xi^{s-s_2}  \\ 0  &    a^{\dagger}_2 q^{D_2} \xi^{s_2} &  q^{-D_2} \\   
  \end{smallmatrix} \bigg] \text{, }         
\]

\noindent the L-operator for the 20-vertex model.

\bigskip

\noindent \textbf{Theorem} (\textit{integrability through suitable action-angle coordinates}). Denote lower-dimensional action-angle coordinates obtained by products of three Ising-type L-operators with $\Phi^{\mathrm{Isng}}_{\text{Lower order}}$. There exists Ising-type action-angle coordinates, $\Phi^{\mathrm{Ising}}$, for which $ \big[ \big[ \Phi^{\mathrm{Ising}}_{\text{Lower order}}, \bar{\Phi^{\mathrm{Ising}}}_{\text{Lower order}} \big] \big] \approx 0 \Longleftrightarrow \big[ \big[ \Phi^{\mathrm{Ising}}, \bar{\Phi^{\mathrm{Ising}}} \big] \big]  \approx 0 $.

\subsection{Paper organization}

\noindent To obtain the desired asymptotic expansions of the transfer, and quantum monodromy, matrices, we first compute all entries of the product representation for two L-operators. Beyond this first computation for products of L-operators, as in the case for adaptations of the quantum-inverse scattering method for the 4-vertex, 6-vertex, and 20-vertex, models, we demonstrate the existence of suitable action-angle variables for the Ising type model, $\Phi^{\mathrm{Ising}} \equiv \Phi$, for which,

\begin{align*}
  \bigg[ \bigg[ \Phi , \bar{\Phi} \bigg] \bigg]  \equiv 0  \text{. }
\end{align*}

\noindent The existence of action-angle variables, such as the one provided above, not only implies that there exists a linearization of the model that sufficiently approximates nonlinear dynamics, and behaviors, of systems with a large system size, but also that integrability, and exact solvability, are expected to hold. For vertex models, as the dimensionality of the underlying state space increases to three dimensions, properties of integrability, and exact solvability, do not hold in the same manner due to the fact that Poisson brackets of action-angle coordinates do not necessarily vanish. In spite of the fact that difficulties exist in adapting the quantum inverse scattering method to show that integrability holds in the context of the 20-vertex model, one can still perform computations with the L-operator which can then be used to characterize Poisson structures in three dimensions, predominantly from the fact that asymptotic approximations for each entry of the product representation for the transfer matrix, and hence for the quantum monodromy matrix, lead to computations with the Poisson bracket.

For the Ising type model, obtaining an asymptotic representation for the transfer, and quantum monodromy, matrices sheds further light on integrability and exactly solvability of the model through the set of Poisson brackets that one must approximate, which together constitute the Poisson structure. Such a collection of brackets is obtained by exhausting all possible options for each of the two arguments of the Poisson bracket, which can further be analyzed to conclude that integrability for the model holds. Such observations, in addition to computations performed by the author in several previous works described at length near the end of the previous subsection above, can be used for analyzing asymptotic behaviors, integrability, and exact solvability, of the Ising type model. As a means of complementing, from previous characterizations of the transfer, and quantum monodromy, matrices in asymptotically large finite volume,           To demonstrate the computations that one would expect to encounter with the Poisson bracket for demonstrating that integrability, and exact solvability, holds for the Ising type model within the quantum inverse scattering framework, consider the spanning set,

\begin{align*}
  \underset{i \neq  j \in V ( \textbf{Z}^2 ) }{\underset{a,a^{\prime} \in \textbf{R}}{\mathrm{span}}} \bigg[\bigg[    x_1 x_3 \delta_{a,a^{\prime}+1}      ,   q^{-a}  \frac{x_1}{x_3} \delta_{a,a^{\prime}-1} q^a \frac{x_4}{x_1} \delta_{a,a^{\prime}-1}      ,  x_1 x_2 \delta_{a,a^{\prime}+1} \frac{y^2}{x_1 x_3} \delta_{a,a^{\prime}-1} ,      q^{-a} \frac{x_1}{x_2} \delta_{a,a^{\prime}-1} q^a \frac{x_3}{x_1} \\ \times \delta_{a,a^{\prime}-1}    ,        q^a  \frac{x_1}{x_3} \delta_{a,a^{\prime}+1} \frac{y^2}{x_1 x_4} \delta_{a,a^{\prime}-1}   ,     \frac{x_1 x_3}{y^2} \delta_{a,a^{\prime}-1} q^a \frac{x_4}{x_1} \delta_{a,a^{\prime}-1} , x_1 x_2 \delta_{a,a^{\prime}+1} q^{-a} \frac{x_3}{x_1} \delta_{a,a^{\prime}+1} , q^{-a} \\ \times  \frac{x_1}{x_2} \delta_{a,a^{\prime}-1} \frac{y^2}{x_1 x_3} \delta_{a,a^{\prime}+1}                   \bigg]\bigg]   \text{, }
\end{align*}

\noindent corresponding to the interactions between $x_1$ and $x_3$. In comparison to L-operators for two-dimensional, and for three-dimensional, vertex models,  additional possible interactions from L-operators for the Ising-type model are also dependent upon,

\begin{align*}
       \underset{i \neq  j \in V ( \textbf{Z}^2 ) }{\underset{a,a^{\prime} \in \textbf{R}}{\mathrm{span}}}\bigg[\bigg[       x_1 x_2 \delta_{a,a^{\prime}+1} \frac{y^2}{x_1 x_3 } \delta_{a,a^{\prime}-1}       ,  q^{-a} \frac{x_1}{x_2} \delta_{a,a^{\prime}-1}  , x_1 x_3 \delta_{a,a^{\prime}+1} q^a \frac{x_4}{x_1}  , q^{-a} \frac{x_1}{x_3} \delta_{a,a^{\prime}-1} \frac{y^2}{x_1 x_4} \delta_{a,a^{\prime}+1} , x_1  \\ \times  x_2 \delta_{a,a^{\prime}+1} q^{-a} \frac{x_3}{x_1} \delta_{a,a^{\prime}+1}  ,    q^{-a} \frac{x_1}{x_2} \delta_{a,a^{\prime}-1}     \frac{y^2}{x_1 x_3 } \delta_{a,a^{\prime}+1}           ,        q^a \frac{x_1}{x_3} \delta_{a,a^{\prime}+1} q^{-a} \frac{x_4}{x_1} \delta_{a,a^{\prime}+1} , \frac{x_1 x_3}{y^2} \delta_{a,a^{\prime}-1} \\ \times  q^a \frac{x_4}{x_1} \delta_{a,a^{\prime}-1}   \bigg]\bigg]     \text{. }
\end{align*}

\noindent Besides $\delta$ factors appearing next to the product, $x_1 x_3$, of two positions, or to the ratio, $\frac{x_1}{x_3}$, of two positions, $q$ parameters parallel similar factors appearing in L-operators of vertex models,

\begin{align*}
   q^{D^j_k}     \text{, } \\  q^{-D^j_k -D^j_{k+1}}  \text{, } \\  q^{-D^j_k + D^j_{k+1}} \text{, } \\  q^{-D^j_k} \text{, } 
\end{align*}

\noindent in the second, first, and third, columns of the representation,

\[  \bigg[ \begin{smallmatrix}     q^{D^j_k}       &    q^{-2} a^j_k q^{-D^j_k -D^j_{k+1}} \xi^{s-s^k_j}        &  a^j_k a^j_{k+1} q^{-D^j_k - 3D^j_{k+1}} \xi^{s - s^j_k - s^j_{k+1}}  \\ \big( a^j_k \big)^{\dagger} q^{D^j_k} \xi^{s^j_k} 
             &      q^{-D^j_k + D^j_{k+1}} - q^{-2} q^{D^j_k -D^j_{k+1}} \xi^{s}     &     - a^j_k q^{D^j_k - 3D^j_{k+1}} \xi^{s-s^j_k}  \\ 0  &    a^{\dagger}_j q^{D^j_k} \xi^{s^j_k} &  q^{-D^j_k} \\    \end{smallmatrix} \bigg] \text{, }   \]

\noindent where the $q$ parameters are raised to operators that are defined over the tensor product of two vector spaces, in addition to mappings $\xi$ into a unital associative algebra. As $N \longrightarrow + \infty$, asymptotically the Ising-type model transfer matrix is also dependent upon several additional spanning sets rather than the first two introduced above.

\bigskip

\noindent One such relation is,

\begin{align*}
    \underset{i \neq  j \in V ( \textbf{Z}^2 ) }{\underset{a,a^{\prime} \in \textbf{R}}{\mathrm{span}}} \bigg[\bigg[ x_1 x_3 \delta_{a,a^{\prime}+1}  x_1 x_4 \delta_{a,a^{\prime}+1} + x_1 x_4 \delta_{a,a^{\prime}+1} x_1 x_5 \delta_{a,a^{\prime}+1}  + q^{-a} \frac{x_1}{x_4} \delta_{a,a^{\prime}-1} q^a \frac{x_1}{x_5} \delta_{a,a^{\prime}+1 } \\ +  x_1 x_3 \delta_{a,a^{\prime}+1}   x_1 x_6 \delta_{a,a^{\prime}+1}  +    q^{-a} \frac{x_1}{x_3} \delta_{a,a^{\prime}-1} q^a \frac{x_1}{x_6} \delta_{a,a^{\prime}+1}   +  x_1 x_3 \delta_{a,a^{\prime}+1} q^{-a} \frac{x_1}{x_5} \delta_{a,a^{\prime}-1}  \\ , \cdot        \bigg]\bigg]    \text{. }
\end{align*}

\noindent In comparison to action-angle variables of the 20-vertex model, $\Phi^{20-V}$, for which,

\begin{align*}
  \big\{ \Phi^{20-V} , \bar{\Phi^{20-V}} \big\} \neq 0  \text{, }
\end{align*}

\noindent stronger notions of Integrability follow from the vanishing Poisson bracket of the action-angle variables for the Ising type model with its complex conjugate. Terms appearing in the first, and second, spanning sets over elements of the Ising-type L-operator directly represent terms appearing in the Poisson bracket,

\begin{align*}
   \big\{ \cdot , \cdot \big\}  \text{, } 
\end{align*}

\noindent with,

\begin{align*}
  \bigg[\bigg[                       
\bigg\{   \underset{n \in \textbf{Z}}{\sum}            \bigg[    \frac{y^2}{x_1 x_{N-1}} \delta_{a,a^{\prime}-1} q^{-a} \frac{x_1}{x_{N-1}}     \delta_{a,a^{\prime}-1} + q^a \frac{X_{N-1}}{x_1} \delta_{a,a^{\prime}+1}  \frac{x_1 x_N}{y^2} \delta_{a,a^{\prime}-1} + \frac{y^2}{x_1 x_{N-2}}   \delta_{a,a^{\prime}-1} q^a  \\ \times \frac{x_1}{x_{N-1}} \delta_{a,a^{\prime}+1}            +        q^a \frac{x_{N-2}}{x_1} \delta_{a,a^{\prime}-1} \frac{x_1 x_{N-1}}{y^2} \delta_{a,a^{\prime}-1 }         \bigg]          \bigg\}       ,  \cdot    \bigg]\bigg]           \text{. } 
\end{align*}

\noindent The second argument of the Poisson bracket, designed with $\cdot$, could equal,

\begin{align*}
             x_1 x_2 \delta_{a,a^{\prime}+1}        q^{-a} \frac{x_3}{x_1}          \delta_{a,a^{\prime}+1}       \text{, } \\ q^{-a} \frac{x_1}{x_2}  \delta_{a,a^{\prime}-1} \frac{y^2}{x_1 x_3 } \delta_{a,a^{\prime}+1}        \text{, } \\   q^a \frac{x_1}{x_3} \delta_{a,a^{\prime}+1} \frac{y^2}{x_1 x_4 } \delta_{a,a^{\prime}-1}                                   \text{, } \\             q^a \frac{x_1}{x_2} \delta_{a,a^{\prime}+1} q^{-a} \frac{x_3}{x_1} \delta_{a,a^{\prime}+1} \text{, } \\     \frac{x_1 x_2}{y^2} \delta_{a,a^{\prime}-1} \frac{y^2}{x_1 x_3} \delta_{a,a^{\prime}+1}           \text{. }
\end{align*}

\noindent Over the integers, additional contributions emerge by replacing the first entry of the Poisson bracket above with,

\begin{align*}
   \underset{N \in \textbf{Z}}{\sum}    \bigg[      q^{-a} \frac{x_1}{x_{N-2}} \delta_{a,a^{\prime}-1} \frac{y^2}{x_1 x_N } \delta_{a,a^{\prime}+1} + q^a \frac{x_1}{x_{N-1}} \delta_{a,a^{\prime}+1} q^{-a} \frac{x_N}{x_1} \delta_{a,a^{\prime}+1} + q^a  \frac{x_1}{x_{N-1}} \delta_{a,a^{\prime}+1} q^{-a} \frac{x_N}{x_1}  \\ \times   \delta_{a,a^{\prime}+1}   + \frac{x_1 x_{N-1}}{y^2} \delta_{a,a^{\prime}-1} q^a \frac{x_N}{x_1} \delta_{a,a^{\prime}-1}     \bigg]      \text{. }
\end{align*}

\noindent Such computations parallel those of the 20-vertex model, in which the product of the vertex-type L-operator is proportional to,

\begin{align*}
  \big( a^j_k \big)^{\dagger} q^{D^j_k} \xi^{s^j_k} q^{D^{j+1}_k} + q^{-D^j_k + D^j_{k+1}} \big( a^{j+1}_k \big)^{\dagger} q^{D^{j+1}_k} \xi^{s^{j+1}_k} + \big( a^j_{k+1} \big)^{\dagger} q^{D^{j}_{k+1}}          \xi^{s^{j}_{k+1}} \big( a^{j+1}_k \big)^{\dagger} q^{D^{j+1}_k} \xi^{s^{j}_{k+1}}       \text{. }
\end{align*}

\noindent In the next section, we compute the first order approximation for products of L-operators from the Ising type model. Specifically, given the interactions from vertices $i$ and $j$ over the underlying lattice, block operators in representations for L-operators, and products of L-operators, determine computations with the Poisson bracket. Block operators for the Ising type transfer matrix are related to the following characteristics for other models in Statistical Mechanics:

\begin{itemize}
\item[$\bullet$] \textit{Stationarity in the weak finite volume limit}. Denote a finite volume $\mathscr{F} \subsetneq \textbf{Z}^2$. As $\mathscr{F} \longrightarrow \textbf{Z}^2$, the support of each block operator in the L-operator representation exhausts that of $\textbf{Z}^2$,

\begin{align*}
  A^{\text{Ising type}}_{\mathscr{F}}  , \\ B^{\text{Ising type}}_{\mathscr{F}}  , \\ C^{\text{Ising type}}_{\mathscr{F}}  , \\ D^{\text{Ising type}}_{\mathscr{F}}  ,
\end{align*}

\noindent the transfer matrix supported over finite volume approaches,

\begin{align*}
 \underset{i \neq  j \in V ( \mathscr{F} ) }{\underset{a,a^{\prime} \in \textbf{R}}{\mathrm{span}}} \big\{  A^{\text{Ising type}}_{\mathscr{F}}  ,  B^{\text{Ising type}}_{\mathscr{F}}  , C^{\text{Ising type}}_{\mathscr{F}}  ,  D^{\text{Ising type}}_{\mathscr{F}}  \big\} \\ \overset{\mathscr{F} \longrightarrow \textbf{Z}^2}{\longrightarrow}  \underset{i \neq  j \in V ( \textbf{Z}^2 ) }{\underset{a,a^{\prime} \in \textbf{R}}{\mathrm{span}}} \big\{  A^{\text{Ising type}}_{\textbf{Z}^2}  ,  B^{\text{Ising type}}_{\textbf{Z}^2}  , C^{\text{Ising type}}_{\textbf{Z}^2}   ,  D^{\text{Ising type}}_{\textbf{Z}^2}  \big\}  \text{, }
\end{align*}

\noindent through a suitable weak finite volume limit.

\item[$\bullet$] \textit{Computation of logarithmic derivatives of the transfer matrix}. With respect to the spectral parameters $u$ of the $D^{(2)}_3$ spin-chain, one has that,

\begin{align*}
   \frac{\mathrm{d}}{\mathrm{d} u} \bigg\{  \mathrm{log} \big( \textbf{T}^{\mathrm{Quasi-periodic}} \big( u \big) \big) \bigg\}   \bigg|_{u \equiv 0} \text{, } 
\end{align*}

\noindent for the transfer matrix under quasi-periodic boundary conditions,

\begin{align*}
   \textbf{T}^{\text{Quasi-periodic}} \big( u \big) \equiv \mathrm{tr}_0 \big[    \textbf{K}_0   \textbf{T}_0 \big( u \big)   \big] \equiv  \mathrm{tr} \bigg[ \textbf{K}_0     \prod_{1 \leq j \leq L}  \textbf{R}_{0j} \big( u \big)  \bigg]   \text{, }  
\end{align*}

\item[$\bullet$] \textit{Parity-time symmetry}. For objects associated with the $D^{(2)}_3$ spin-chain in the previous item, \textit{parity-time} symmetry of the R-matrix stipulates,

\begin{align*}
   R_{21} \big( u \big) \equiv \mathcal{P}_{12} \mathcal{R}_{12} \big( u \big) \mathcal{P}_{12} \equiv     R^{t_1 , t_2}_{12} \big( u \big)       \text{, }  
\end{align*}

\noindent for the permutation matrix $\mathcal{P}$, and strictly positive times $t_1 \neq t_2$.

\item[$\bullet$] \textit{Transformation of weights encoded in the R-matrix to weights encoded in Boltzmann weight matrices of Solid-on-Solid models}. Fix two external fields $H$ and $V$. Given the R-matrix,

\begin{align*}
  \begin{bmatrix}
      a \text{ }  \mathrm{exp} \big(  H + V \big)    & 0 & 0 & 0  \\
    0 & b \text{ } \mathrm{exp} \big( H - V \big) & c & 0  \\0 & c & b \text{ }  \mathrm{exp} \big( - H + V \big) & 0 \\ 0 & 0 & 0 & a \text{ }  \mathrm{exp} \big( - H - V \big) \\ 
  \end{bmatrix} \text{, }
\end{align*}

\noindent corresponding to $R \equiv R \big( u , H , V \big)$, with spectral parameters $u$ and $v$, one can determine entries of the Boltzmann weight matrix, $W$, from the intertwining relation,

\begin{align*}
 R \big( u - v \big) \bigg[  \psi \big( u \big)^a_b \otimes \psi \big( v \big)^b_c \bigg]   = \underset{b^{\prime}}{\sum}      \bigg[    \psi \big( v \big)^a_{b^{\prime}} \otimes \psi \big( u \big)^{b^{\prime}}_c \bigg]  W       \text{, }
\end{align*}

\noindent for intertwining vectors $\psi^a_b$, and $\psi^b_c$.

\item[$\bullet$] \textit{Nearest-neighbor decomposition of spin-chain Hamiltonian}. Under open-boundary conditions of the $D^{(2)}_3$ spin-chain, one has the following decomposition,

{\small\begin{align*}
   \mathcal{H} \big( k , \kappa , \textbf{K}_{-} , \textbf{K}_{+} \big)   \equiv \mathcal{H} \sim    \underset{1 \leq k \leq N-1}{\sum} h_{k,k+1} + \frac{1}{2\kappa} \bigg[ \textbf{K}^{-}_1  \big( 0 \big) \bigg]^{\prime}   +   \frac{1}{\mathrm{tr} \big( \textbf{K}_{+} \big( 0 \big) \big)  }  \mathrm{tr}_0 \textbf{K}_{0,+} \big( 0 \big) h_{N0} \text{, }  
\end{align*}}

\noindent in terms of the transfer matrix expanded about $u \equiv 0$.

\item[$\bullet$] \textit{Local Hamiltonian encoding via the logarithmic derivative of the spin-chain transfer matrix}. One has that,

\begin{align*}
     \textbf{H}^{\text{Quasi0periodic}} \approx - \mathrm{sinh} \big( 2 i \gamma \big) \bigg[ \frac{\mathrm{d}}{\mathrm{d} u} \bigg\{   \mathrm{log} \big( \textbf{T}^{\mathrm{Quasi-periodic}} \big( u \big)\big)  \bigg\}   \bigg|_{u \equiv 0}\bigg] + L \text{ } \mathrm{sinh} \big( 2 i \gamma \big) \\ \times  \big[             \mathrm{coth} \big( 2 i \gamma \big)  + \mathrm{coth} \big( 4 i \gamma \big)        \big] \textbf{I}^{\otimes L}          \text{, }  
\end{align*}

\noindent for anisotropy parameter $\gamma$ which is taken to be strictly positive.

\item[$\bullet$] \textit{Yang- Baxter equations}. The Yang-Baxter equations hold for a variety of models, which ultimately underlies integrable, and exactly solvable structure. Given the previously defined $R$-matrix for the 20-vertex model, in addition to $R$-matrix for the $6$-vertex model and $D^{(2)}_3$ spin-chains, the relations state,

\begin{align*}
  \underline{\text{20-vertex model}}: \mathcal{R}_{12} \big( u \big) \mathcal{R}_{13} \big( u + v \big) \mathcal{R}_{23} \big( v \big) =  \mathcal{R}_{23} \big( v \big) \mathcal{R}_{13} \big( u + v \big) \mathcal{R}_{12} \big( u \big)  \text{, } \\ \\ \underline{\text{6-vertex model}}:  R_{12} \big( u \big) R_{13} \big( u + v \big) R_{23} \big( v \big) =  R_{23} \big( v \big) R_{13} \big( u + v \big) R_{12} \big( u \big) \text{,} \\ \\ \underline{D^{(2)}_3 \text{ spin-chain}}:  \mathscr{R}_{12} \big( u - v \big) \mathscr{R}_{13} \big( u \big) \mathscr{R}_{23} \big( v \big) = \mathscr{R}_{23} \big( v \big) \mathscr{R}_{13} \big( u \big) \mathscr{R}_{12} \big( u - v \big)   \text{,}
\end{align*}

\noindent respectively.

\item[$\bullet$] \textit{Commutation relation between R, and K, matrices of the } $D^{(2)}_3$ \text{ spin-chain}. From previously defined objects of the spin-chain, one has that,

 \begin{align*}
       R_{12} \big( u - v \big) K_{1,-} \big( u \big) R_{21} \big( u + v \big) K_{2,-} \big( v \big) =  K_{2,-} \big( u \big) R_{12} \big( u + v \big) K_{1,-} \big( u \big) R_{21} \big( u - v \big)    \text{. }  
\end{align*}

\item[$\bullet$] \textit{Approximation of roots of the truncated Bethe equations with the root density approach}. Under quasi-periodic boundary conditions for the $D^{(2)}_3$ spin-chain, one has that the Bethe equations can be expressed as,

\begin{align*}
    \bigg[           \frac{\lambda_j - i}{\lambda_j + i}        \bigg]^L       \approx        \overset{m_1}{\underset{k \neq j}  {\prod}}  \text{ }     \overset{m_2}{\underset{k=1}{\prod} } \bigg\{  \bigg[         \frac{\lambda_j - \lambda_k - 2 i }{\lambda_j - \lambda_k + 2 i }              \bigg] \text{ } \bigg[          \frac{\lambda_j - \lambda_k + i }{\lambda_j - \lambda_k - i }                       \bigg]   \bigg\}                  \text{, }  
\end{align*}

\noindent for,

{\small \begin{align*}
      \textit{Eigenvalues of the first type, corresponding to Bethe roots of the first} \\ \textit{type}  \equiv \lambda_j   \text{,} \\  \\ \textit{Eigenvalues of the second type, corresponding to Bethe roots of the second} \\ \textit{type}  \equiv \lambda_k      \text{,} \\ \\   \textit{Length of the spin-chain} \equiv L      \text{. } 
\end{align*} }

\end{itemize}

\noindent The objects introduced in the above list are of interest to formalize, and further describe, for Ising type models from finite-dimensional representations of the transfer matrix obtained in the next section. Albeit the fact that one expects for more phases of possible behaviors, ranging from the presence of order, disorder, ferroelectricy, and antiferroelectricity for vertex models, in comparison to ferromagnetism and antiferromagnetism for Ising type models, suitable action-angle variables for each class of models still underlies integrable and exactly solvable structure. Moreover, while one can obtain information on action-angle variables from generalized families of brackets with the Poisson structure, performing computations with the bracket is extremely dependent upon encoding of local interactions, whether through stipulations on vertices or edges of the underlying lattice.

\subsection{Quantization of the Yang-Baxter equation for Quantum L-operators of the Ising-type model}

\noindent For the Quantum equivalent for the Yang Baxter equation - the Quantum Yang-Baxter equation - introduce,

\begin{align*}
   R^{\mathrm{Ising} } : V \otimes V \longrightarrow V \otimes V ,
\end{align*}

\noindent for a vector space, $V$. The equation is given by the statement,

\begin{align*}
  R^{\mathrm{Ising}}_{12} R^{\mathrm{Ising}}_{13} R^{\mathrm{Ising}}_{23} =    R^{\mathrm{Ising}}_{23}   R^{\mathrm{Ising}}_{13} R^{\mathrm{Ising}}_{12} ,
\end{align*}

\noindent over $V\otimes V \otimes V$. Given solutions to the Quantum Yang-Baxter equation, the Ising-type $R$-matrix, as a function of the Quantum L-operators introduced in \textit{1}, satisfies:

\begin{itemize}
\item[$\bullet$] \textit{RTT relation}. One has that,

\begin{align*}
  R^{\mathrm{Ising}}  \big(   t^{\text{Ising}}  \big)_{a_1}^{a^{\prime}_2}  \big(   t^{\text{Ising}}  \big)_{a^{\prime}_2}^{a_1}  \equiv   R^{\mathrm{Ising}} \big( a_1 - a^{\prime}_2 \big) \big(   t^{\text{Ising}}  \big)_{a_1}^{a^{\prime}_2}  \big(   t^{\text{Ising}}  \big)_{a^{\prime}_2}^{a_1}  =   \big(   t^{\text{Ising}}  \big)_{a^{\prime}_2}^{a_1}    \big(   t^{\text{Ising}}  \big)_{a_1}^{a^{\prime}_2} \\ \times R^{\mathrm{Ising}} \big( a_1 - a^{\prime}_2 \big)  \equiv  \big(   t^{\text{Ising}}  \big)_{a^{\prime}_2}^{a_1}    \big(   t^{\text{Ising}}  \big)_{a_1}^{a^{\prime}_2} R^{\mathrm{Ising}}             ,
\end{align*}

\noindent corresponds to the action of the Ising-type $R$ matrix, $R^{\mathrm{Ising}}$, on the product of two transfer matrices - given by the $t$ factors.

\item[$\bullet$] \textit{RLL relation}. One has that,

\begin{align*}
   R^{\mathrm{Ising}} \mathcal{L}_{a_1}^{a^{\prime}_2}  \mathcal{L}_{a^{\prime}_2}^{a_1}    \equiv   R^{\mathrm{Ising}} \big( a_1 - a^{\prime}_2 \big) \mathcal{L}_{a_1}^{a^{\prime}_2}  \mathcal{L}_{a^{\prime}_2}^{a_1}  =   \mathcal{L}_{a^{\prime}_2}^{a_1}    \mathcal{L}_{a_1}^{a^{\prime}_2} R^{\mathrm{Ising}} \big( a_1 - a^{\prime}_2 \big)  \equiv    \mathcal{L}_{a^{\prime}_2}^{a_1}    \mathcal{L}_{a_1}^{a^{\prime}_2} R^{\mathrm{Ising}}         ,
\end{align*}

\noindent corresponds to the action of the Ising-type $R$ matrix, $R^{\mathrm{Ising}}$, on the product of two L-operators -given by the $\mathcal{L}$ factors.

\item[$\bullet$] \textit{Commutation relation with respect to the bracket of two transfer matrices for the Ising-type model}. One has that,

\begin{align*}
 \bigg[ \bigg[      \big(   t^{\text{Ising}}  \big)_{a^{\prime}_2}^{a_1}  ,   \big(   t^{\text{Ising}}  \big)_{a_1}^{a^{\prime}_2}    \bigg] \bigg]   = 0  . 
\end{align*}

\item[$\bullet$] \textit{Quantum Bethe equations}. Fix $N$, a strictly positive parameter that is equal to the number of vertical spectral parameters for $\textbf{Z}^2$. From the Quantum Yang-Baxter equations, along with the $L$-operators for the Ising-type model, the Quantum Bethe equations take the form,

\[
   \left\{\!\begin{array}{ll@{}>{{}}l} 
  \bigg[   \frac{x_i x^{\prime}_i \delta_{a,a^{\prime}+i} - \frac{y}{2} x_i x^{\prime}_i \delta_{a_i,a^{\prime}_i - 1}}{q^{a_i} \frac{x_i}{x^{\prime}_i} \delta_{a_i , a^{\prime}_i + 1} - q^{a_i} \frac{x^{\prime}_i}{x_i} \delta_{a_i , a^{\prime}_i - 1}     } \bigg]^N   \\ =           \underset{1 \leq i^{\prime} \leq N}{\underset{1 \leq i \leq N}{\underset{i \neq i^{\prime}}{\prod}}}      \frac{q^a \frac{x}{x^{\prime}}  \delta_{a,a^{\prime}+1}     - q^{-a} \frac{x^{\prime}}{x} \delta_{a,a^{\prime}-1}     }{q^{-a} \frac{x}{x^{\prime}}   \delta_{a,a^{\prime}-1}   - q^{-a} \frac{x^{\prime}}{x} \delta_{a,a^{\prime}+1}   }                                        ,
 \end{array}\right.
 \]

\noindent which is related to the general form of the Quantum Bethe equations,

\[
   \left\{\!\begin{array}{ll@{}>{{}}l} 
 \bigg[ \frac{A^{\text{Ising type}} ( a , a^{\prime} )  }{D^{\text{Ising type}}( a , a^{\prime} )   } \bigg]^N   \equiv       \underset{1 \leq i^{\prime} \leq N}{\underset{1 \leq i \leq N}{\underset{i \neq i^{\prime}}{\prod}}}    \frac{f( a_i - a_{i^{\prime}} )}{f (  a^{\prime} - a )}      \\ \text{                  }  \text{                  }  \text{                  }  \Updownarrow  \\             \frac{\textit{First Quantum L-operator for the Ising-type model}}{\textit{Fourth Quantum L-operator for the Ising-type model}} \\  \equiv    \frac{A^{\text{Ising type}} }{D^{\text{Ising type}} } \equiv \frac{A^{\text{Ising type}} ( a , a^{\prime} )  }{D^{\text{Ising type}}( a , a^{\prime} )   }  \\     \text{                  }  \text{                  }  \text{                  }  \Updownarrow  \\   \frac{f( a_i - a_{i^{\prime}} )}{f (  a^{\prime} - a )}     \equiv    \frac{q^a \frac{x}{x^{\prime}} \delta_{a,a^{\prime}+1}    - q^{-a} \frac{x^{\prime}}{x} \delta_{a,a^{\prime}-1}     }{q^{-a} \frac{x}{x^{\prime}}   \delta_{a,a^{\prime}-1}   - q^{-a} \frac{x^{\prime}}{x} \delta_{a,a^{\prime}+1}   }       . 
 \end{array}\right.
 \]

\end{itemize}

\subsection{Enumeration of compact representations over the Lie group}

\noindent With respect to the commutator bracket, compact representations over the Lie group can be provided through the following correspondence:

\[
   \left\{\!\begin{array}{ll@{}>{{}}l} 
      \textit{First compact representation over $\textbf{g}$}         \equiv A^{\mathrm{Ising \text{ } type}}   , \\ \\   \textit{Second compact representation over $\textbf{g}$}         \equiv B^{\mathrm{Ising \text{ } type}}   ,  \\ \\  \textit{Third compact representation over $\textbf{g}$}         \equiv C^{\mathrm{Ising \text{ } type}}   ,   \\ \\   \textit{Fourth compact representation over $\textbf{g}$}         \equiv D^{\mathrm{Ising \text{ } type}}     .         
 \end{array}\right.
 \]

\noindent Finite dimensional, compact representations over $\textbf{g}$ are expected to satisfy:

\[
   \left\{\!\begin{array}{ll@{}>{{}}l} 
      \textit{First compact representation over $\textbf{g}$ has finite support for some $\Lambda$}         \equiv A^{\mathrm{Ising \text{ } type}}_{\Lambda}   , \\ \\   \textit{Second compact representation over $\textbf{g}$ has finite support for some $\Lambda$}         \equiv B^{\mathrm{Ising \text{ } type}}_{\Lambda}   ,  \\ \\  \textit{Third compact representation over $\textbf{g}$ has finite support for some $\Lambda$}         \equiv C^{\mathrm{Ising \text{ } type}}_{\Lambda}   ,   \\ \\   \textit{Fourth compact representation over $\textbf{g}$ has finite support for some $\Lambda$}         \equiv D^{\mathrm{Ising \text{ } type}}_{\Lambda}     .         
 \end{array}\right.
\] 

\begin{align*}
    \Updownarrow 
\end{align*}

\[
\underset{\Lambda \longrightarrow \textbf{Z}^2}{\mathrm{lim}}  \left\{\!\begin{array}{ll@{}>{{}}l} 
      \textit{First compact representation over $\textbf{g}$ has finite support for some $\Lambda$}         \equiv A^{\mathrm{Ising \text{ } type}}_{\Lambda}   , \\ \\   \textit{Second compact representation over $\textbf{g}$ has finite support for some $\Lambda$}         \equiv B^{\mathrm{Ising \text{ } type}}_{\Lambda}   ,  \\ \\  \textit{Third compact representation over $\textbf{g}$ has finite support for some $\Lambda$}         \equiv C^{\mathrm{Ising \text{ } type}}_{\Lambda}   ,   \\ \\   \textit{Fourth compact representation over $\textbf{g}$ has finite support for some $\Lambda$}         \equiv D^{\mathrm{Ising \text{ } type}}_{\Lambda}     .        
 \end{array}\right.
 \]

 \begin{align*}
     \textit{exists, and is well defined through the weak finite volume limit}.
 \end{align*}

\begin{align*}
    \Updownarrow 
\end{align*}

\[
\left\{\!\begin{array}{ll@{}>{{}}l} \underline{(1)}:      \bigg[ \bigg[ A^{\mathrm{Ising}}_{\Lambda} \big( u \big)        , A^{\mathrm{Ising}}_{\Lambda} \big( u^{\prime} \big)   \bigg] \bigg]  
\text{ , } \\  \underline{(2)}:  \bigg[ \bigg[         A^{\mathrm{Ising}}_{\Lambda} \big( u \big)        ,        B^{\mathrm{Ising}}_{\Lambda} \big( u^{\prime} \big)     \bigg] \bigg]   \text{ , } \\  \underline{(3)}:   \bigg[ \bigg[   A^{\mathrm{Ising}}_{\Lambda} \big( u \big)       ,  C^{\mathrm{Ising}}_{\Lambda} \big( u^{\prime} \big) \bigg] \bigg]  
 \text{, }   \\  \underline{(4)}: \bigg[ \bigg[   A^{\mathrm{Ising}}_{\Lambda} \big( u \big)       ,  D^{\mathrm{Ising}}_{\Lambda} \big( u^{\prime} \big) \bigg] \bigg]    \text{ , } \\ \underline{(5)} : \bigg[ \bigg[ B^{\mathrm{Ising}}_{\Lambda} \big( u \big) , A^{\mathrm{Ising}}_{\Lambda} \big( u^{\prime} \big) \bigg]  \bigg] \text{, } \\   \underline{(6)}: \bigg[ \bigg[  B^{\mathrm{Ising}}_{\Lambda} \big( u \big) , B^{\mathrm{Ising}}_{\Lambda} \big( u^{\prime} \big) \bigg] \bigg]  \text{, } \\ \underline{(7)}: \bigg[ \bigg[ B^{\mathrm{Ising}}_{\Lambda} \big( u \big) , C^{\mathrm{Ising}}_{\Lambda} \big( u^{\prime} \big) \bigg] \bigg] \text{, } \\ \underline{(8)}: \bigg[ \bigg[  B^{\mathrm{Ising}}_{\Lambda} \big( u \big)  , D^{\mathrm{Ising}}_{\Lambda} \big( u^{\prime} \big)   \bigg] \bigg] \text{, } \\   \underline{(9)}:  \bigg[ \bigg[ C^{\mathrm{Ising}}_{\Lambda} \big( u \big)  , A^{\mathrm{Ising}}_{\Lambda} \big( u^{\prime} \big)   \bigg] \bigg]  \text{, } 
\end{array}\right.
\]

 \[
\left\{\!\begin{array}{ll@{}>{{}}l} \underline{(10)}: \bigg[ \bigg[ C^{\mathrm{Ising}}_{\Lambda} \big( u \big) , B^{\mathrm{Ising}}_{\Lambda} \big( u^{\prime} \big) \bigg]  \bigg]  \text{, } \\ \underline{(11)}: \bigg[ \bigg[ C^{\mathrm{Ising}}_{\Lambda} \big( u \big) , C^{\mathrm{Ising}}_{\Lambda} \big( u^{\prime} \big) \bigg] \bigg]  \text{, }  \\ \underline{(12)}: \bigg[ \bigg[ C^{\mathrm{Ising}}_{\Lambda} \big( u \big) , D^{\mathrm{Ising}}_{\Lambda} \big( u^{\prime} \big) \bigg] \bigg]  \text{, } \\ \underline{(13)}: \bigg[ \bigg[  D^{\mathrm{Ising}}_{\Lambda} \big( u \big)  , A^{\mathrm{Ising}}_{\Lambda} \big( u^{\prime} \big)  \bigg] \bigg]  \text{, } \\ 
   \underline{(14)}: \bigg[ \bigg[   D^{\mathrm{Ising}}_{\Lambda} \big( u \big)  , B^{\mathrm{Ising}}_{\Lambda} \big( u^{\prime} \big)  \bigg] \bigg]  \text{, } \\ \underline{(15)}: \bigg[ \bigg[  D^{\mathrm{Ising}}_{\Lambda} \big( u \big)  , C^{\mathrm{Ising}}_{\Lambda} \big( u^{\prime} \big) \bigg] \bigg]   \text{, }  \\ \underline{(16)}: \bigg[ \bigg[  D^{\mathrm{Ising}}_{\Lambda} \big( u \big)  ,       D^{\mathrm{Ising}}_{\Lambda} \big( u^{\prime} \big)  \bigg] \bigg]  \text{, }
\end{array}\right.
\]

\begin{align*}
    \Updownarrow 
\end{align*}

\[
 \underset{\Lambda \longrightarrow \textbf{Z}^2}{\mathrm{lim}} \left\{\!\begin{array}{ll@{}>{{}}l} \underline{(1)}:      \bigg[ \bigg[ A^{\mathrm{Ising}}_{\Lambda} \big( u \big)        , A^{\mathrm{Ising}}_{\Lambda} \big( u^{\prime} \big)   \bigg] \bigg]  
\text{ , } \\  \underline{(2)}:  \bigg[ \bigg[         A^{\mathrm{Ising}}_{\Lambda} \big( u \big)        ,        B^{\mathrm{Ising}}_{\Lambda} \big( u^{\prime} \big)     \bigg] \bigg]   \text{ , } \\  \underline{(3)}:   \bigg[ \bigg[   A^{\mathrm{Ising}}_{\Lambda} \big( u \big)       ,  C^{\mathrm{Ising}}_{\Lambda} \big( u^{\prime} \big) \bigg] \bigg]  
 \text{, }   \\  \underline{(4)}: \bigg[ \bigg[   A^{\mathrm{Ising}}_{\Lambda} \big( u \big)       ,  D^{\mathrm{Ising}}_{\Lambda} \big( u^{\prime} \big) \bigg] \bigg]    \text{ , } \\ \underline{(5)} : \bigg[ \bigg[ B^{\mathrm{Ising}}_{\Lambda} \big( u \big) , A^{\mathrm{Ising}}_{\Lambda} \big( u^{\prime} \big) \bigg]  \bigg] \text{, } \\   \underline{(6)}: \bigg[ \bigg[  B^{\mathrm{Ising}}_{\Lambda} \big( u \big) , B^{\mathrm{Ising}}_{\Lambda} \big( u^{\prime} \big) \bigg] \bigg]  \text{, } \\ \underline{(7)}: \bigg[ \bigg[ B^{\mathrm{Ising}}_{\Lambda} \big( u \big) , C^{\mathrm{Ising}}_{\Lambda} \big( u^{\prime} \big) \bigg] \bigg] \text{, } \\ \underline{(8)}: \bigg[ \bigg[  B^{\mathrm{Ising}}_{\Lambda} \big( u \big)  , D^{\mathrm{Ising}}_{\Lambda} \big( u^{\prime} \big)   \bigg] \bigg] \text{, } \end{array}\right.
\]

 \[
\left\{\!\begin{array}{ll@{}>{{}}l}    \underline{(9)}:  \bigg[ \bigg[ C^{\mathrm{Ising}}_{\Lambda} \big( u \big)  , A^{\mathrm{Ising}}_{\Lambda} \big( u^{\prime} \big)   \bigg] \bigg]  \text{, } \\ \underline{(10)}: \bigg[ \bigg[ C^{\mathrm{Ising}}_{\Lambda} \big( u \big) , B^{\mathrm{Ising}}_{\Lambda} \big( u^{\prime} \big) \bigg]  \bigg]  \text{, } \\ \underline{(11)}: \bigg[ \bigg[ C^{\mathrm{Ising}}_{\Lambda} \big( u \big) , C^{\mathrm{Ising}}_{\Lambda} \big( u^{\prime} \big) \bigg] \bigg]  \text{, }  \\ \underline{(12)}: \bigg[ \bigg[ C^{\mathrm{Ising}}_{\Lambda} \big( u \big) , D^{\mathrm{Ising}}_{\Lambda} \big( u^{\prime} \big) \bigg] \bigg]  \text{, } \\ \underline{(13)}: \bigg[ \bigg[  D^{\mathrm{Ising}}_{\Lambda} \big( u \big)  , A^{\mathrm{Ising}}_{\Lambda} \big( u^{\prime} \big)  \bigg] \bigg]  \text{, } \\ 
   \underline{(14)}: \bigg[ \bigg[   D^{\mathrm{Ising}}_{\Lambda} \big( u \big)  , B^{\mathrm{Ising}}_{\Lambda} \big( u^{\prime} \big)  \bigg] \bigg]  \text{, } \\ \underline{(15)}: \bigg[ \bigg[  D^{\mathrm{Ising}}_{\Lambda} \big( u \big)  , C^{\mathrm{Ising}}_{\Lambda} \big( u^{\prime} \big) \bigg] \bigg]   \text{, }  \\ \underline{(16)}: \bigg[ \bigg[  D^{\mathrm{Ising}}_{\Lambda} \big( u \big)  ,       D^{\mathrm{Ising}}_{\Lambda} \big( u^{\prime} \big)  \bigg] \bigg]  \text{, }
\end{array}\right.
\]

\begin{align*}
    \Updownarrow 
\end{align*}

\[
   \left\{\!\begin{array}{ll@{}>{{}}l} 
    \textit{There exists $C_i$, where $1 \leq i \leq 16$, for which $C_i \big( \Lambda \big) \longrightarrow C_i $ as $\Lambda \longrightarrow \textbf{Z}^2$ in the finite} \\ \textit{weak volume limit (where  $\Lambda \longrightarrow \textbf{Z}^2 \Longleftrightarrow \mathrm{support} \big[ \Lambda \big]  \longrightarrow \mathrm{support} \big[ \textbf{Z}^2 \big]$ } .         
 \end{array}\right.
 \]

\section{Discrete Probabilistic objects}

\subsubsection{20-vertex model}

\begin{figure}
\begin{align*}
\includegraphics[width=1.05\columnwidth]{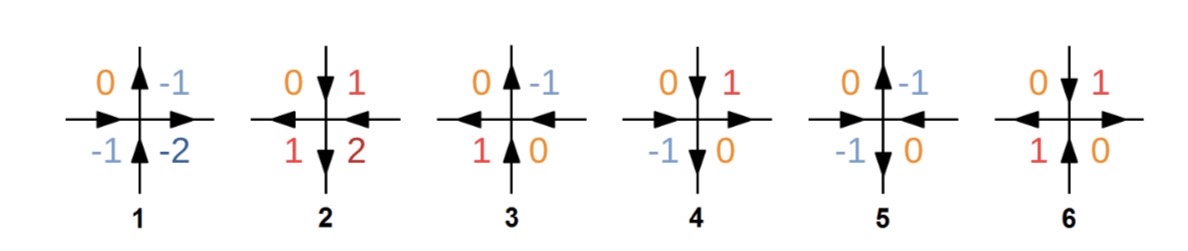}
\end{align*}
\caption{Each possible vertex for the six-vertex model, adapted from ${\color{blue}[8]}$.}
\end{figure}

\begin{figure}
\begin{align*}
\includegraphics[width=1.1\columnwidth]{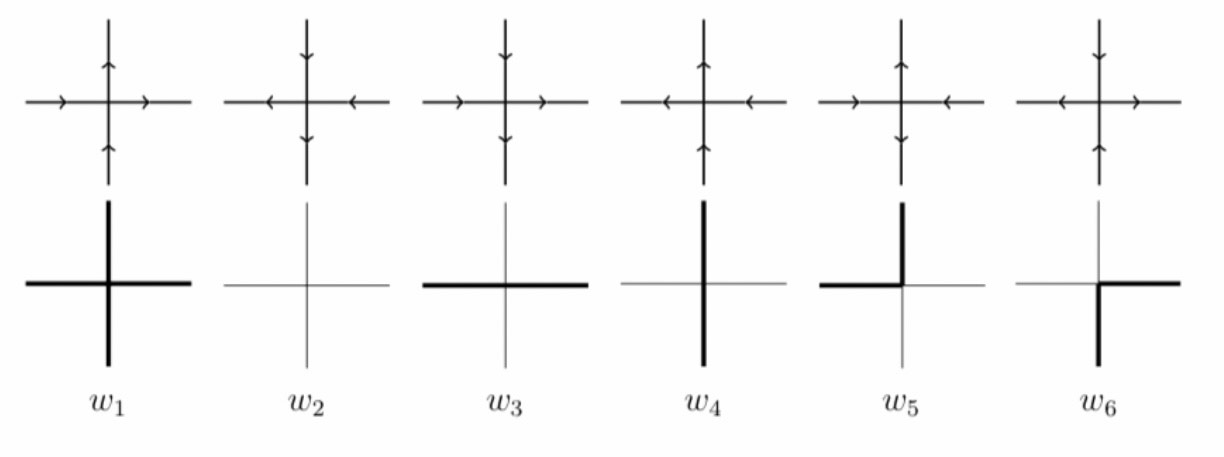}
\end{align*}
\caption{Another depiction of each possible vertex for the six-vertex model, adapted from {\color{blue}[24]}.}
\end{figure}

For boundary conditions $\xi$, either those with sufficiently flat slope, ie, flat, or domain walls, such as those introduced over the pentagonal lattice, {\color{blue}[15]}, the 20-vertex probability measure supported over the triangular lattice, $\textbf{T}$, takes the form,

\begin{align*}
   \textbf{P}^{20V , \xi}_{\textbf{T}} \big[ \cdot \big] \equiv  \textbf{P}^{20V}_{\textbf{T}} \big[ \cdot \big]   \text{, }
\end{align*}

\noindent which is explicitly given by the ratio of the vertex weight function and the partition function,

\begin{align*}
\textbf{P}^{20V}_{\textbf{T}}[      \omega         ]   \equiv \textbf{P}^{20V}[   \omega     ]     =  \frac{w_{20V}(\omega)}{Z^{20V}_{\textbf{T}}} \equiv \frac{w(\omega)}{Z_{\textbf{T}}} \text{, }
\end{align*}

\noindent for some vertex configuration $\omega \in \Omega^{20V}$ - the 20-vertex sample space, and weights similar to those introduced in the previous section for the 6-vertex model, namely, {\color{blue}[15]},

\begin{align*}
     w_0 \equiv   a_1 a_2 a_3   \text{, } 
\\
    w_1 \equiv   b_1 a_2 b_3   \text{, }  \end{align*}

    \begin{align*}     w_2    \equiv   b_1 a_2 c_3    \text{, } \\
   w_3 \equiv       a_1 b_2 b_3 + c_1 c_2 c_3  \text{, } \\     w_4 \equiv    c_1 a_2 a_3        \text{, } \\               w_5 \equiv   b_1 c_2 a_3   \text{, } \\  w_6 \equiv  b_1 b_2 a_3    \text{, } 
\end{align*}

\noindent and the partition function,

\begin{align*}
 Z_{\textbf{T}} \equiv \underset{\omega \in \Omega^{20V}}{\sum}  w \big( \omega \big) \text{. }
\end{align*}

\noindent As was the case for the two-dimensional vertex model, the 6-vertex model, introduced in the previous subsection, one can also introduce finite volume approximations to the transfer matrix before taking the weak infinite volume limit, from spectral parameters $u,v,w$, with,

{\small \[
T  \big(   \underline{M} , N ,  \underline{\lambda_{\alpha}} , u , v , w \big) \equiv    \overset{\underline{M}}{\underset{\underline{j}=0}{\prod}}  \text{ }  \overset{-N}{\underset{i=0}{\prod}} \bigg\{   \mathrm{exp} \big( \lambda_3 ( q^{-2} \xi^{s_i} ) \big) \bigg[     \begin{smallmatrix}      q^{D_i}       &    q^{-2} a_i q^{-D_i-D_j} \xi^{s-s_i}        &   a_i a_{j} q^{-D_i - 3D_j} \xi^{s - s_i - s_j}  \\ a^{\dagger}_i q^{D_i} \xi^{s_i} 
             &      q^{-D_i + D_j} - q^{-2} q^{D_i -D_j} \xi^{s}     &     - a_j q^{D_i - 3D_j} \xi^{s-s_j}  \\ 0  &    a^{\dagger}_j q^{D_j} \xi^{s_j} &  q^{-D_j} \\    \end{smallmatrix} \bigg]    \bigg\}  \text{. } 
\] }

\noindent As $M \longrightarrow + \infty$,  $N \longrightarrow - \infty$, the finite volume approximation above, in weak infinite volume, takes the form,

\begin{align*}
T \big(   \underline{\lambda}  \big) \equiv T\bigg( + \infty , - \infty ,   \underline{\lambda_{\alpha}} , \big\{ u_i \big\} , \big\{ v_j  \big\} , \big\{ w_k \big\} \bigg) =   \underset{\underline{M} \longrightarrow + \infty}{\mathrm{lim}} \text{ }  \underset{N \longrightarrow - \infty}{\mathrm{lim}}     T  \big(   M , N , \lambda_{\alpha} ,  v  ,  u    \big)    \text{, } 
\end{align*}

\noindent with,

\begin{align*}
 \textbf{T}^{3D} \big( \underline{\lambda} \big)   \equiv    \underset{ N \longrightarrow -  \infty}{\underset{\underline{M} \longrightarrow + \infty}{\mathrm{lim}}} \mathrm{tr} \bigg\{     \overset{\underline{M}}{\underset{j=0}{\prod}}  \text{ }  \overset{-N}{\underset{k=0}{\prod}} \mathrm{exp} \big( \lambda_3 ( q^{-2} \xi^{s^j_k} ) \big)    \bigg[ \begin{smallmatrix}     q^{D^j_k}       &    q^{-2} a^j_k q^{-D^j_k -D^j_{k+1}} \xi^{s-s^k_j}        &  *_2\\ \big( a^j_k \big)^{\dagger} q^{D^j_k} \xi^{s^j_k} 
             &     *_1   &     - a^j_k q^{D^j_k - 3D^j_{k+1}} \xi^{s-s^j_k}  \\ 0  &    a^{\dagger}_j q^{D^j_k} \xi^{s^j_k} &  q^{-D^j_k} \\    \end{smallmatrix} \bigg]      \bigg\}  
\text{, }
\end{align*}

\noindent for,

\begin{align*}
  *_1 \equiv  q^{-D^j_k + D^j_{k+1}} - q^{-2} q^{D^j_k -D^j_{k+1}} \xi^{s}    \text{, } \\   *_2 \equiv a^j_k a^j_{k+1} q^{-D^j_k - 3D^j_{k+1}} \xi^{s - s^j_k - s^j_{k+1}}   \text{. }
\end{align*}

\begin{figure}
\begin{align*}
\includegraphics[width=0.65\columnwidth]{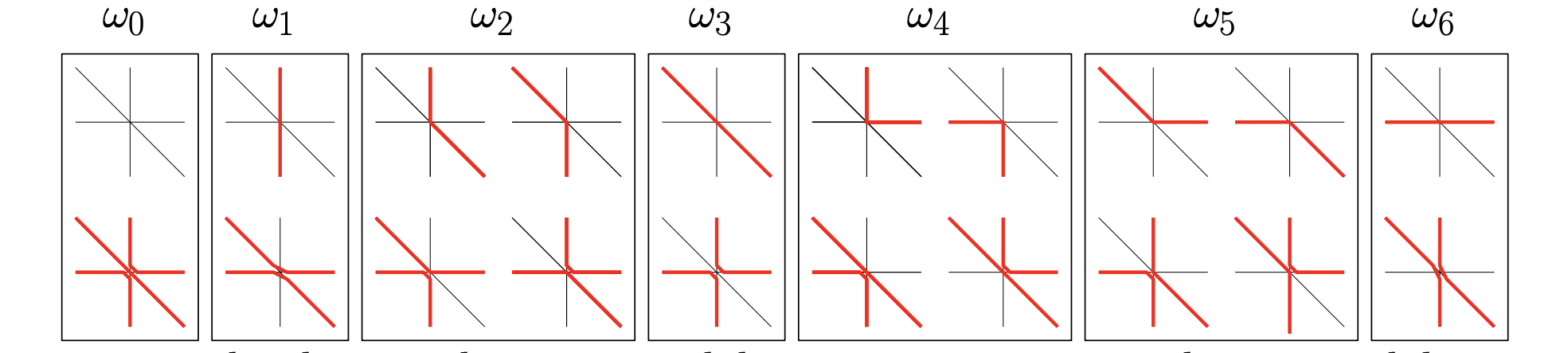}
\end{align*}
\caption{A depiction of each possible vertex for the triangular, or three dimensional, six-vertex model, adapted from {\color{blue}[15]}.}
\end{figure}

\begin{figure}
\begin{align*}
\includegraphics[width=0.85\columnwidth]{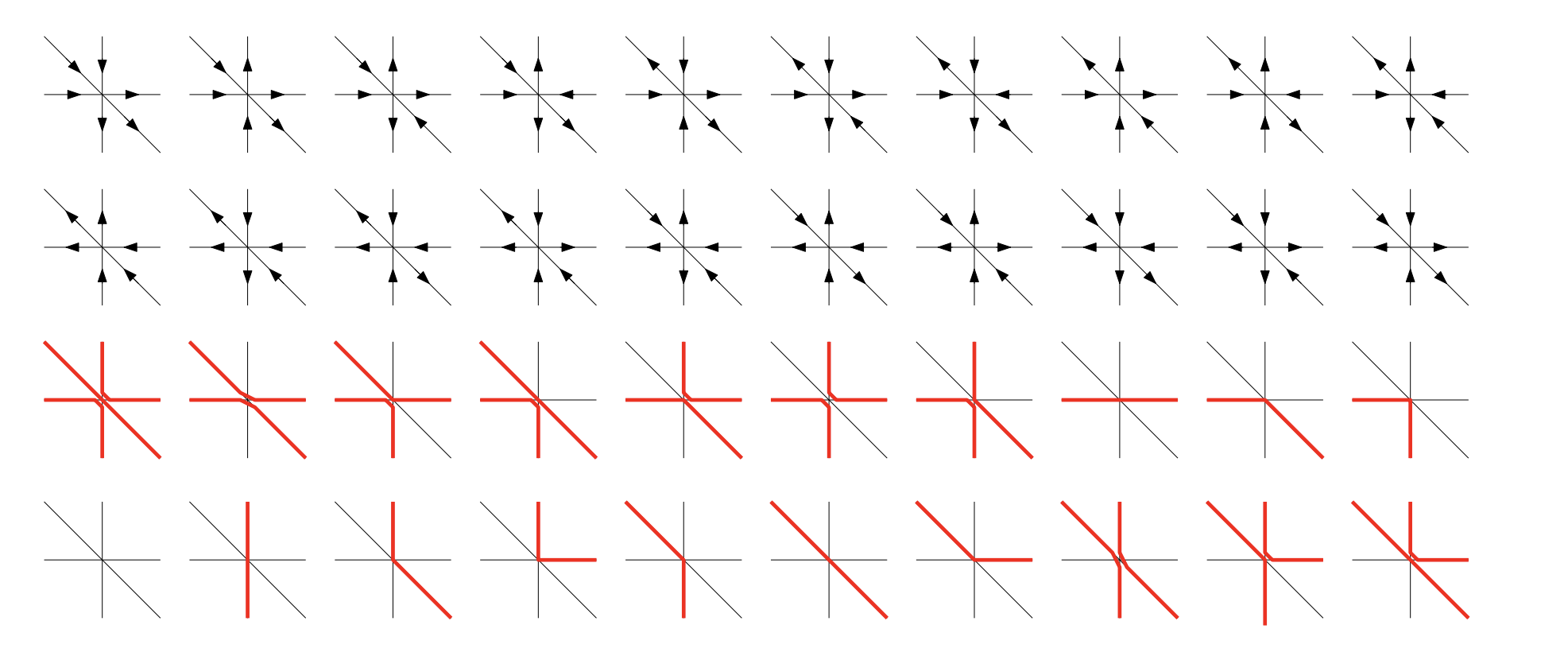}
\end{align*}
\caption{A depiction of each Boltzman weight for the triangular, or three dimensional, six-vertex model, also adapted from {\color{blue}[15]}.}
\end{figure}

\subsubsection{Solid-on-Solid model}

Despite the fact that previous adaptations of seminal work in {\color{blue}[17]} have studied objects relating to the Bethe ansatz, and several closely related objects, {\color{blue}[21},{\color{blue}22},{\color{blue}23},{\color{blue}24},{\color{blue}25},{\color{blue}26},{\color{blue}27},{\color{blue}28]}, adaptations of the quantum inverse scattering framework provided by the author in {\color{blue}[46]} are primarily reliant upon a higher-dimensional analog of computations with an L-operator of the 6-vertex model analyzed in {\color{blue}[42]}. As a byproduct of the quantum inverse scattering framework, it was conjectured in {\color{blue}[25]} that integrability of limit shapes for the 6-vertex model should imply integrability of a Hamiltonian flow under the presence of inhomogeneities, which was resolved by the author. For the rational 7-vertex model, given an R-matrix, the system of equations from which the intertwining vectors can be explicitly read off from takes the form,

\begin{align*}
 R \big( u - v \big) \bigg[   \psi \big( u \big)^a_b \otimes \psi \big( v \big)^b_c \bigg]  = \underset{b^{\prime}}{\sum}    \bigg[      \psi \big( v \big)^a_{b^{\prime}} \otimes \psi \big( u \big)^{b^{\prime}}_c \bigg]  W       \text{, }
\end{align*}

\noindent for two spectral parameters $u$ and $v$, intertwining vectors $\psi$, points $a,b,c,b^{\prime}$, and Boltzmann weight matrix $W$. In the equality of intertwining vectors above, the fact that the R-matrix for the rational 7-vertex model is dependent upon the difference of spectral parameters, $u-v$, rather than upon one, or two spectral parameters, $u$ and $v$ can be reflected through the combinatorial factors of the q-exponentials, in addition to indicates of the summation which depend upon $r$, a spectral parameter introduced along the rows, or columns, of some finite volume of the triangular lattice. With such observations, from factorization of the Universal R-matrix, it will be demonstrated that the system of relations takes the form,

\[
(*) \equiv \left\{\!\begin{array}{ll@{}>{{}}l} (1):  \mathcal{R}_1 \beta_l \big( u \big) \beta_{l+1} \big( u \big) =  \beta_l \big( v \big) \beta_{l+1} \big( u \big)   W_1 
\text{, } \\   (2):   \mathcal{R}_1 \beta_l \big( u \big) \gamma_{l+1} \big( u \big) =    \beta_l \big( v \big) \gamma_{l+1} \big( u \big)    W_1       \text{, } \\      (3):  \mathcal{R}_1 \beta_l \big( u \big) Z_{l+1} \big( u \big) =   \beta_l \big( v \big) Z_{l+1} \big( u \big)    W_1   \text{, } \\    (4):  \mathcal{R}_1 \beta_l \big( u \big) \beta_{l+1} \big( u \big) =   \gamma_l \big( v \big) \beta_{l+1} \big( u \big) W_1  \text{, } \\   (5):           \mathcal{R}_1 \gamma_l \big( u \big) \gamma_{l+1} \big( v \big) =  \gamma_l \big( v \big) \gamma_{l+1} \big) u \big)  W_1  \text{, } \\   (6): \mathcal{R}_1 \gamma_l \big( u \big) Z_{l+1} \big( v \big) =  \gamma_l \big( v \big) Z_{l+1} \big( u \big) W_1  \text{, } \\ (7): \mathcal{R}_1  Z_l \big( u \big) \beta_{l+1} \big( v \big) =  Z_l \big( v \big) \beta_{l+1} \big( u \big) W_1  \text{, } \\ (8): \mathcal{R}_1 Z_l \big( u \big) \gamma_{l+1} \big( v \big) =  Z_l \big( v \big) \gamma_{l+1} \big( u \big) W_1 \text{, } \\ (9): \mathcal{R}_1 Z_l \big( u \big) Z_{l+1} \big( v \big) =  Z_l \big( v \big) Z_{l+1} \big( u \big)      W_1 \text{, }       \end{array}\right.
\]

\noindent for the 20-vertex intertwining vectors,

\begin{align*}
  X_l \big( u \big) = \bigg[ \begin{smallmatrix}   \beta_l ( u )  \\ \gamma_l ( u ) \\ Z_l ( u ) 
  \end{smallmatrix} \bigg]  \text{, } \\ X_{l+1} \big( u \big) = \bigg[ \begin{smallmatrix}   \beta_{l+1} ( u )  \\ \gamma_{l+1} ( u ) \\ Z_{l+1} ( u ) 
  \end{smallmatrix} \bigg]  \text{. } 
\end{align*}

\noindent To distribute terms from the factorization of the Universal R-matrix, with $X_l \big( u \big)$ and $X_{l+1} \big( u \big)$ above, observe that from the first term of the R-matrix from the rational 7-vertex model,

\begin{align*}
 R \big( u - v \big)    \text{, }
\end{align*}

\noindent appears as a prefactor to the first intertwining vector $\psi \big( u \big)^a_b$, which appears in the tensor product,

\begin{align*}
  \psi \big( u \big)^a_b \otimes \psi \big( v \big)^b_c     \text{, }
\end{align*}

\noindent of intertwining vectors. For intertwining vectors of the 20-vertex model, in comparison to those of the rational 7-vertex model, the q-exponential factor,

\begin{align*}
 \underset{m \in \textbf{N}
}{\underset{\gamma \in \Delta+ ( A)}{\prod}}    \mathrm{exp}_q \big[ \big( q - q^{-1} \big)          s^{-1}_{m , \delta - \gamma} e_{\delta - \gamma + m \delta} \otimes f_{ \delta - \gamma + m \delta}              \big]  \text{, }
\end{align*}

\noindent appearing in the the Universal R-matrix factorization is distributed to $\psi \big( u \big)^a_b$ before taking the resultant tensor product with the remaining intertwining vector $\psi \big( v \big)^b_c$. From previous work on the rational 7-vertex model that has been discussed, each entry of the intertwining vectors, over the square lattice, has a $1$ either in the first or second entry, in addition to the other factor being dependent upon a product of strictly positive parameters, with contributions from $\alpha$, $n$, and $l$. From the system of equations for the rational 7-vertex model, the explicit form of the three entries appearing in the intertwining vectors of the 20-vertex model takes a similar form. In obtaining a general solution from the system of nine relations above, one distributes the tensor product from the intertwining vectors,

\begin{align*}
     X_l \big( u \big)    \text{, } \\ X_{l+1} \big( u \big) 
\text{. }
\end{align*}

\noindent In the system for the 20-vertex SOS model that is dependent upon entries of the Boltzmann weight matrix, the order in which spectral parameters are introduced into the tensor product, from each intertwining vector, is reversed. That is, from each of the nine relations listed above for each entry of the R-matrix, and of the Boltzmann weight matrix, the sequence in which the intertwining vectors, which are respectively dependent upon $u \equiv \underline{u}$ and $v\equiv \underline{v}$, comprise the reversed transformation that is applied to the Boltzmann weight matrix. Altogether, the universal R-matrix factorization into q-exponential, the K-matrix, and spectral parameters, implies that one must consider exponentials of the form,

\begin{align*}
   \mathscr{A}_1 \mathscr{A}_2 \mathscr{A}_3 \mathscr{A}_4 \text{, }
\end{align*}

\noindent where,

\begin{align*}
      \mathscr{A}_1 \equiv  \mathrm{exp} \bigg[         \hbar   \big( q - q^{-1} \big) s^{-1}_{m,\gamma} e_{\gamma+m \delta} \otimes f_{\gamma+ m \delta} \bigg] \text{, } \\  \mathscr{A}_2 \equiv  \mathrm{exp} \bigg[             \big( q - q^{-1} \big) \underset{m \in \textbf{Z}^+}{\sum}            \overset{r}{\underset{i^{\prime} \neq j^{\prime} \in \textbf{Z}}{\underset{i-j,i^{\prime}-j^{\prime}=1}{\sum}}}          u_m \big( i - j\big) \big( i^{\prime} - j^{\prime} \big)     e_{m\delta,\alpha_{i-j}} \otimes f_{m\delta , a_{j-j^{\prime}}}                       \bigg] \text{, } \end{align*}

      \begin{align*} \mathscr{A}_3 \equiv          \underset{m \in \textbf{N}}{\underset{\gamma \in \Delta_+ ( A)}{\prod}}         \mathrm{exp} \bigg[ \big( q - q^{-1} \big)        s^{-1}_{m,\delta-\gamma} e_{\delta - \gamma + m \delta }       \otimes   f_{\delta-\gamma+m \delta}    \bigg]  \text{, } \\ \mathscr{A}_4 \equiv          \mathrm{exp} \bigg[ \hbar    \overset{r}{\underset{i^{\prime}-j^{\prime}\in \textbf{Z}}{\underset{i-j \in \textbf{Z}}{\underset{i-j, i^{\prime}-j^{\prime}=1}{\sum} }}}     \beta_{(i-j)(i^{\prime}-j^{\prime})} h_{\alpha ( i-j)}  \otimes h_{\alpha(i^{\prime}-j^{\prime})}   \bigg]                     \text{. }
\end{align*}

\noindent In comparison to the first expression introduced for the exponential of the K-matrix that is dependent upon a \textit{single} spectral parameter, rather than the \textit{difference} of two spectral parameters, exponentials of the form above are introduced for boundary conditions to the first term $\mathcal{R} \big( u - v \big)$ appearing in the 20-vertex intertwining relation. Following the overview in the next subsection, we demonstrate how a system of relations, from the nine components, are obtained from the intertwining vectors. 

\bigskip

\noindent Under the presence of fixed boundary conditions for the 4-vertex model, for determining whether integrable, or hybrid integrable, properties of the model holds, one manipulates products of quantities of the form,

\begin{align*}
    L^{4V} \big(  n | u \big) \equiv L \big( n | u \big)    \equiv  \begin{bmatrix}
         L_{11} \big( n | u \big) & L_{12} \big( n | u \big) \\ L_{21} \big( n | u \big) & L_{22} \big( n | u \big) 
    \end{bmatrix}  = \begin{bmatrix}
       - u e_n & \sigma^{-}_n \\ \sigma^{+}_n & u^{-1} e_n 
    \end{bmatrix}
    \text{, }
\end{align*}

\noindent corresponding to the 4-vertex L-operator, with degrees of freedom,

\begin{align*}
 \mathrm{DOF\text{ }  1} \equiv \underset{n \in \textbf{Z}}{\bigcup}  \sigma^-_n     \text{, } \\ \mathrm{DOF\text{ }  2} \equiv  \underset{n \in \textbf{Z}}{\bigcup}  \sigma^+_n    \text{, } \\ \mathrm{DOF\text{ }  3} \equiv  \underset{n \in \textbf{Z}}{\bigcup}  e_n  \text{, }
\end{align*}

\noindent respectively spanned by the two standard Pauli basis elements, and the basis element $e_n$ of $\textbf{Z}$. For each vertex model, determining integrability under the presence of different boundary conditions amounts to determining closed form approximations to operators,

\begin{align*}
  A^{6V} , B^{6V} , C^{6V} , D^{6V}  \text{, } \\ A^{20V}, B^{20V}, C^{20V}, D^{20V}, E^{20V}, F^{20V}, G^{20V}, H^{20V}, I^{20V}  \text{, } \\  A^{4V} , B^{4V} , C^{4V} , D^{4V}  \text{, }
\end{align*}

\noindent appearing in,

\begin{align*}
\begin{bmatrix}
       A^{6V} \big( \lambda_{\alpha} , v_k \big)   & B^{6V} \big( \lambda_{\alpha} , v_k \big)   \\
    C^{6V} \big( \lambda_{\alpha}, v_k \big)  & D^{6V} \big( \lambda_{\alpha} , v_k \big)  \text{ }  
  \end{bmatrix} \text{, }
\\ \\ 
\begin{bmatrix}
 A^{20V} \big( \underline{u} 
 \big) & D^{20V} \big( \underline{u} \big)  & G^{20V} \big( \underline{u}  \big) \\ B^{20V} \big( \underline{u}  \big) & E^{20V} \big( \underline{u}  \big) & H^{20V} \big( \underline{u}  \big)  \\ C^{20V} \big( \underline{u}  \big)  &  F^{20V} \big( \underline{u}  \big) & I^{20V} \big( \underline{u}  \big) 
\end{bmatrix} \text{, }
\\ \\ 
    \begin{bmatrix}
        A^{4V} \big( \lambda_{\alpha} , v_k  \big) & B^{4V} \big(  \lambda_{\alpha} , v_k  \big) \\ C^{4V} \big(  \lambda_{\alpha} , v_k  \big) & D^{4V} \big(  \lambda_{\alpha} , v_k  \big) 
    \end{bmatrix}     \text{, }
\end{align*}

\noindent respectively corresponding to finite volume approximations of the 6-vertex, 20-vertex, and 4-vertex, models. From finite volume representations of transfer matrices for the 6-vertex, 20-vertex, and 4-vertex, models, one can study tensor products of Poisson brackets of transfer matrices, each of which take the form,

\begin{align*}
     \bigg\{  \begin{bmatrix}
        A^{6V} \big( \lambda_{\alpha} , v_k  \big) & B^{6V} \big(  \lambda_{\alpha} , v_k  \big) \\ C^{6V} \big(  \lambda_{\alpha} , v_k  \big) & D^{6V} \big(  \lambda_{\alpha} , v_k  \big) 
    \end{bmatrix}   \overset{\bigotimes}{,} \begin{bmatrix}
        A^{6V} \big( \lambda^{\prime}_{\alpha} , v^{\prime}_k  \big) & B^{6V} \big( \lambda^{\prime}_{\alpha} , v^{\prime}_k    \big) \\ C^{6V} \big(  \lambda^{\prime}_{\alpha} , v^{\prime}_k    \big) & D^{6V} \big( \lambda^{\prime}_{\alpha} , v^{\prime}_k   \big) 
    \end{bmatrix}   \bigg\}\textbf{}   \text{, }
\\ \\ 
\bigg\{    \begin{bmatrix}
 A \big( \underline{u} 
 \big) & D \big( \underline{u} \big)  & G \big( \underline{u}  \big) \\ B \big( \underline{u}  \big) & E \big( \underline{u}  \big) & H \big( \underline{u}  \big)  \\ C \big( \underline{u}  \big)  &  F \big( \underline{u}  \big) & I \big( \underline{u}  \big) 
\end{bmatrix}\overset{\bigotimes}{,}\begin{bmatrix}
 A \big( \underline{u^{\prime}} \big) & D \big( \underline{u^{\prime}} \big)  & G \big( \underline{u^{\prime}} \big) \\ B \big( \underline{u^{\prime}}\big) & E \big( \underline{u^{\prime}} \big) & H \big( \underline{u^{\prime}} \big)  \\ C \big( \underline{u^{\prime}} \big)  &  F \big( \underline{u^{\prime}} \big) & I \big( \underline{u^{\prime}} \big) 
\end{bmatrix} \bigg\} \text{, }
\\ \\ 
     \bigg\{  \begin{bmatrix}
        A^{4V} \big( \lambda_{\alpha} , v_k  \big) & B^{4V} \big(  \lambda_{\alpha} , v_k  \big) \\ C^{4V} \big(  \lambda_{\alpha} , v_k  \big) & D^{4V} \big(  \lambda_{\alpha} , v_k  \big) 
    \end{bmatrix}   \overset{\bigotimes}{,} \begin{bmatrix}
        A^{4V} \big( \lambda^{\prime}_{\alpha} , v^{\prime}_k   \big) & B^{4V} \big( \lambda^{\prime}_{\alpha} , v^{\prime}_k    \big) \\ C^{4V} \big( \lambda^{\prime}_{\alpha} , v^{\prime}_k   \big) & D^{4V} \big( \lambda^{\prime}_{\alpha} , v^{\prime}_k   \big) 
    \end{bmatrix}   \bigg\}      \text{, }
\end{align*}

\noindent respectively. From the finite volume representations of the transfer, and quantum monodromy matrices, of the 6-vertex, 20-vertex, and 4-vertex, models, one characterizes the relations,

\begin{align*}
         \big\{  T^{6V}_a \big( u , \big\{ v_k \big\}  \big)     \overset{\bigotimes}{,}   T^{6V}_a \big( u^{\prime} , \big\{ v^{\prime}_k \big\}  \big) \big\}   \text{, } \\  \\               \big\{  T^{20V}_{a,a^{\prime}} \big( u , \big\{ v_k \big\} , \big\{ v^{\prime}_k \big\} \big)     \overset{\bigotimes}{,}   T^{20V}_{a,a^{\prime}} \big( u , \big\{ \big( v_k \big)^{\prime} \big\} , \big\{ \big(  v^{\prime}_k \big)^{\prime} \big\}  \big)   \big\}      \text{, } \\    \\     \big\{  T^{4V}_a \big( u , \big\{ v_k \big\}  \big)     \overset{\bigotimes}{,}   T^{4V}_a \big( u^{\prime} , \big\{ v^{\prime}_k \big\}  \big) \big\}       \text{, } 
\end{align*}

\noindent which respectively equal,

\begin{align*}
     \bigg[   \bigg[  r^{6V}_{a,+}         \big( v_k - v^{\prime}_k \big)   \begin{bmatrix}
       A^{6V} \big( \lambda_{\alpha} , v_k \big)   & B^{6V} \big( \lambda_{\alpha} , v_k \big)   \\
    C^{6V} \big( \lambda_{\alpha}, v_k \big)  & D^{6V} \big( \lambda_{\alpha} , v_k \big)  \text{ }  
  \end{bmatrix} \bigg]   \bigotimes   \begin{bmatrix}
       A^{6V} \big( \lambda_{\alpha} , v_k \big)   & B^{6V} \big( \lambda_{\alpha} , v_k \big)   \\
    C^{6V} \big( \lambda_{\alpha}, v_k \big)  & D^{6V} \big( \lambda_{\alpha} , v_k \big)  \text{ }  
  \end{bmatrix} \bigg]  \\ -  \bigg[ \begin{bmatrix}
       A^{6V} \big( \lambda_{\alpha} , v_k \big)   & B^{6V} \big( \lambda_{\alpha} , v_k \big)   \\
    C^{6V} \big( \lambda_{\alpha}, v_k \big)  & D^{6V} \big( \lambda_{\alpha} , v_k \big)  \text{ }  
  \end{bmatrix} \bigotimes \bigg[  \begin{bmatrix}
       A^{6V} \big( \lambda_{\alpha} , v_k \big)   & B^{6V} \big( \lambda_{\alpha} , v_k \big)   \\
    C^{6V} \big( \lambda_{\alpha}, v_k \big)  & D^{6V} \big( \lambda_{\alpha} , v_k \big)  \text{ }  
  \end{bmatrix}   r^{6V}_{a,-}      \big( v_k - v^{\prime}_k  \big)              \bigg] \bigg]   \text{, } \end{align*}

  \begin{align*} \bigg[    \bigg[  r^{20V}_{+} \big( u_k - u^{\prime}_k , v_k - v^{\prime}_k , w_k - w^{\prime}_k \big)   \begin{bmatrix}
 A^{20V} \big( \underline{u} \big) & D^{20V} \big( \underline{u} \big)  & G^{20V} \big( \underline{u} \big) \\ B^{20V} \big( \underline{u} \big) & E^{20V} \big( \underline{u} \big) & H^{20V} \big( \underline{u} \big)  \\ C^{20V} \big( \underline{u} \big)  &  F^{20V} \big( \underline{u} \big) & I^{20V} \big( \underline{u} \big) 
\end{bmatrix}    \bigg] \\   \bigotimes  \begin{bmatrix}
 A^{20V} \big( \underline{u^{\prime}} \big) & D^{20V} \big( \underline{u^{\prime}} \big)  & G^{20V} \big( \underline{u^{\prime}} \big) \\ B^{20V} \big( \underline{u^{\prime}} \big) & E^{20V} \big( \underline{u^{\prime}} \big) & H^{20V} \big( \underline{u^{\prime}} \big)  \\ C^{20V} \big( \underline{u^{\prime}} \big)  &  F^{20V} \big( \underline{u^{\prime}} \big) & I^{20V} \big( \underline{u^{\prime}} \big) 
\end{bmatrix}   \bigg]   - \bigg[  \begin{bmatrix}
 A^{20V} \big( \underline{u} \big) & D^{20V} \big( \underline{u} \big)  & G^{20V} \big( \underline{u} \big) \\ B^{20V} \big( \underline{u} \big) & E^{20V} \big( \underline{u} \big) & H^{20V} \big( \underline{u} \big)  \\ C^{20V} \big( \underline{u} \big)  &  F^{20V} \big( \underline{u} \big) & I^{20V} \big( \underline{u} \big) 
\end{bmatrix} \\  \bigotimes  \bigg[ \begin{bmatrix}
 A^{20V} \big( \underline{u^{\prime}} \big) & D^{20V} \big( \underline{u^{\prime}} \big)  & G^{20V} \big( \underline{u^{\prime}} \big) \\ B^{20V} \big( \underline{u^{\prime}} \big) & E^{20V} \big( \underline{u^{\prime}} \big) & H^{20V} \big( \underline{u^{\prime}} \big)  \\ C^{20V} \big( \underline{u^{\prime}} \big)  &  F^{20V} \big( \underline{u^{\prime}} \big) & I^{20V} \big( \underline{u^{\prime}} \big) 
\end{bmatrix}  r^{20V}_{-} \big( u_k - u^{\prime}_k , v_k - v^{\prime}_k , w_k - w^{\prime}_k \big)          \bigg] \bigg] \text{, }  \\ \\  \bigg[    \bigg[  r^{4V}_{+} \big( u_k - u^{\prime}_k , v_k - v^{\prime}_k , w_k - w^{\prime}_k \big)   \begin{bmatrix}
        A^{4V} \big( \lambda^{\prime}_{\alpha} , v^{\prime}_k   \big) & B^{4V} \big( \lambda^{\prime}_{\alpha} , v^{\prime}_k    \big) \\ C^{4V} \big( \lambda^{\prime}_{\alpha} , v^{\prime}_k   \big) & D^{4V} \big( \lambda^{\prime}_{\alpha} , v^{\prime}_k   \big) 
    \end{bmatrix}   \bigg] \\   \bigotimes  \begin{bmatrix}
        A^{4V} \big( \lambda^{\prime}_{\alpha} , v^{\prime}_k   \big) & B^{4V} \big( \lambda^{\prime}_{\alpha} , v^{\prime}_k    \big) \\ C^{4V} \big( \lambda^{\prime}_{\alpha} , v^{\prime}_k   \big) & D^{4V} \big( \lambda^{\prime}_{\alpha} , v^{\prime}_k   \big) 
    \end{bmatrix} \bigg]   - \bigg[  \begin{bmatrix}
        A^{4V} \big( \lambda^{\prime}_{\alpha} , v^{\prime}_k   \big) & B^{4V} \big( \lambda^{\prime}_{\alpha} , v^{\prime}_k    \big) \\ C^{4V} \big( \lambda^{\prime}_{\alpha} , v^{\prime}_k   \big) & D^{4V} \big( \lambda^{\prime}_{\alpha} , v^{\prime}_k   \big) 
    \end{bmatrix} \\  \bigotimes  \bigg[ \begin{bmatrix}
        A^{4V} \big( \lambda^{\prime}_{\alpha} , v^{\prime}_k   \big) & B^{4V} \big( \lambda^{\prime}_{\alpha} , v^{\prime}_k    \big) \\ C^{4V} \big( \lambda^{\prime}_{\alpha} , v^{\prime}_k   \big) & D^{4V} \big( \lambda^{\prime}_{\alpha} , v^{\prime}_k   \big) 
    \end{bmatrix}  r^{4V}_{-} \big( u_k - u^{\prime}_k , v_k - v^{\prime}_k , w_k - w^{\prime}_k \big)          \bigg] \bigg]         \text{, } 
\end{align*}

\noindent where the quantities $r^{6V}_{\pm}$, $r^{20V}_{\pm}$, and $r^{4V}_{\pm}$, denote,

{\small \begin{align*}
  r^{6V}_{+} \big( u_k - u^{\prime}_k , v_k - v^{\prime}_k \big) \equiv  r^{6V}_{+ } \equiv     \underset{y \longrightarrow +\infty}{\mathrm{lim}}              \bigg[ E^{\mathrm{6V}} \big(   u^{\prime}  ,  v^{\prime}_k - v_k \big)      \bigotimes \bigg[                E^{\mathrm{6V}} \big(  u^{\prime}  ,  v^{\prime}_k - v_k \big)     r \big(  u_k - u^{\prime}_k   \\ , v_k - v^{\prime}_k       \bigg] \bigg]            \text{, } \\ \\  r^{6V}_{-} \big( u_k - u^{\prime}_k , v_k - v^{\prime}_k \big) \equiv  r^{6V}_{- } \equiv     \underset{y \longrightarrow -\infty}{\mathrm{lim}}              \bigg[ E^{\mathrm{6V}} \big(   u^{\prime}  ,  v^{\prime}_k - v_k \big)      \bigotimes \bigg[                E^{\mathrm{6V}} \big(  u^{\prime}  ,  v^{\prime}_k - v_k \big)     r \big(  u_k - u^{\prime}_k  \\  , v_k - v^{\prime}_k       \bigg] \bigg] \text{, } \\ \\ r^{20V}_{+} \big( u_k - u^{\prime}_k , v_k - v^{\prime}_k , w_k - w^{\prime}_k \big) \equiv  r^{3D}_{+ } \equiv     \underset{y \longrightarrow +\infty}{\mathrm{lim}}              \bigg[ E^{3D,\mathrm{6V}} \big(   \underline{u^{\prime}}  ,  v^{\prime}_k - v_k \big)      \bigotimes \bigg[                E^{3D,\mathrm{6V}} \big(  \underline{u^{\prime}}  ,  v^{\prime}_k - v_k \big)   \\ \times   r \big(  u_k - u^{\prime}_k  , v_k - v^{\prime}_k  ,  w_k - w^{\prime}_k \big)       \bigg] \bigg]            \text{, } \\ \\  r^{20V}_{-} \big( u_k - u^{\prime}_k , v_k - v^{\prime}_k , w_k - w^{\prime}_k \big) \equiv   r^{3D}_{-} \equiv      \underset{y \longrightarrow - \infty}{\mathrm{lim}}   \bigg[ E^{3D,\mathrm{6V}} \big(   \underline{u^{\prime}}  ,  v^{\prime}_k - v_k \big)      \bigotimes \bigg[                E^{3D,\mathrm{6V}} \big(  \underline{u^{\prime}}  ,  v^{\prime}_k - v_k \big) \\  \times     r \big( u_k - u^{\prime}_k   , v_k - v^{\prime}_k  ,  w_k - w^{\prime}_k   \big)       \bigg] \bigg]          \text{, } \\ \\  r^{4V}_{+} \big( u_k - u^{\prime}_k , v_k - v^{\prime}_k  \big) \equiv   r^{4V}_{+} \equiv      \underset{y \longrightarrow + \infty}{\mathrm{lim}}   \bigg[ E^{\mathrm{4V}} \big(   u^{\prime}  ,  v^{\prime}_k - v_k \big)      \bigotimes \bigg[                E^{\mathrm{4V}} \big(  u^{\prime}  ,  v^{\prime}_k - v_k \big)     r \big( u_k - u^{\prime}_k \\   , v_k - v^{\prime}_k    \big)       \bigg] \bigg]       \text{, } \\ \\ r^{4V}_{-} \big( u_k - u^{\prime}_k , v_k - v^{\prime}_k  \big) \equiv   r^{4V}_{-} \equiv      \underset{y \longrightarrow - \infty}{\mathrm{lim}}   \bigg[ E^{\mathrm{4V}} \big(   u^{\prime}  ,  v^{\prime}_k - v_k \big)      \bigotimes \bigg[                E^{\mathrm{4V}} \big(  u^{\prime}  ,  v^{\prime}_k - v_k \big)     r \big( u_k - u^{\prime}_k   \\ , v_k - v^{\prime}_k    \big)       \bigg] \bigg]  \text{, } 
\end{align*} }

\noindent for,

\begin{align*}
  E^{\mathrm{6V}} \big( x - v_k  , x \big) \equiv  \mathrm{exp} \big[             \mathrm{coth} \big( \frac{\eta}{2} + i \alpha_j - v_k \big)            \big]    \text{, } \\ \\  E^{3D,\mathrm{6V}} \big(   \underline{u^{\prime}}  ,  v^{\prime}_k - v_k ,  u^{\prime}_k - u_k  , w^{\prime}_k - w_k \big)  \equiv    \mathrm{exp} \bigg[  \frac{1}{2i} \begin{bmatrix} 1 & 0 & 0 \\ 0 & -1 & 0 \\ 0 & 0 & 0 \end{bmatrix}   +  \begin{bmatrix} 0 & 0 & \psi  \\ 0 & \bar{\psi}   & 0 \\ 0 & 0 & 0 \end{bmatrix}        \bigg]       \text{, } \\ \\     E^{\mathrm{4V}} \big( x - v_k  , x \big)  \equiv       \mathrm{exp} \bigg[  \frac{1}{2i} \begin{bmatrix} 1 & 0  \\ 0 & -1  \end{bmatrix}   +  \begin{bmatrix} 0 & 0   \\ 0 & \bar{\psi}      \end{bmatrix}        \bigg]                \text{. }
\end{align*}

\noindent With respect to the standard tensor product operation,

\begin{align*}
  \bigotimes \cdot  \text{, }
\end{align*}

\noindent one has the collection of relations,

\begin{align*}
      \big\{  T^{6V}_{-} \big( x , \underline{\lambda} \big) \overset{\bigotimes}{,} T^{6V}_{-} \big( x , \underline{\mu} \big)           \big\} = r^{6V} \big( \underline{\lambda} - \underline{\mu} \big) T^{6V}_{-} \big( x , \underline{\lambda} \big)\bigotimes T^{6V}_{-} \big( x , \underline{\mu} \big) - T^{6V}_{-} \big( x , \underline{\lambda} \big) \bigotimes T^{6V}_{-} \big( x , \underline{\mu} \big) \\ \times r^{6V}_{-} \big( \underline{\lambda} - \underline{\mu} \big)         \text{, } \\     \big\{  T^{6V}_{+} \big( x , \underline{\lambda} \big) \overset{\bigotimes}{,} T^{6V}_{+} \big( x , \underline{\mu} \big)           \big\} = T^{6V}_{+} \big( x , \underline{\lambda} \big) \bigotimes T^{6V}_{+} \big( x , \underline{\mu} \big) r^{6V}_{+} \big( \underline{\lambda} - \underline{\mu} \big) - r^{6V} \big( \underline{\lambda} - \underline{\mu} \big) T^{6V}_{+} \big( x , \underline{\lambda} \big) \\ \bigotimes T^{6V}_{+} \big( x , \underline{\mu} \big)      \\  \\ \big\{  T^{3D}_{-} \big( x , \underline{\lambda} \big) \overset{\bigotimes}{,} T^{3D}_{-} \big( x , \underline{\mu} \big)           \big\} = r \big( \underline{\lambda} - \underline{\mu} \big) T^{3D}_{-} \big( x , \underline{\lambda} \big)\bigotimes T^{3D}_{-} \big( x , \underline{\mu} \big) - T^{3D}_{-} \big( x , \underline{\lambda} \big) \bigotimes T^{3D}_{-} \big( x , \underline{\mu} \big) \\ \times r^{3D}_{-} \big( \underline{\lambda} - \underline{\mu} \big)         \text{, } \\     \big\{  T^{3D}_{+} \big( x , \underline{\lambda} \big) \overset{\bigotimes}{,} T^{3D}_{+} \big( x , \underline{\mu} \big)           \big\} = T^{3D}_{+} \big( x , \underline{\lambda} \big) \bigotimes T^{3D}_{+} \big( x , \underline{\mu} \big) r^{3D}_{+} \big( \underline{\lambda} - \underline{\mu} \big) - r^{3D} \big( \underline{\lambda} - \underline{\mu} \big) T^{3D}_{+} \big( x , \underline{\lambda} \big) \\ \bigotimes T^{3D}_{+} \big( x , \underline{\mu} \big)   \text{, }  \\ \\    \big\{  T^{4V}_{-} \big( x , \underline{\lambda} \big) \overset{\bigotimes}{,} T^{4V}_{-} \big( x , \underline{\mu} \big)           \big\} = r^{4V} \big( \underline{\lambda} - \underline{\mu} \big) T^{4V}_{-} \big( x , \underline{\lambda} \big)\bigotimes T^{4V}_{-} \big( x , \underline{\mu} \big) - T^{4V}_{-} \big( x , \underline{\lambda} \big) \bigotimes T^{4V}_{-} \big( x , \underline{\mu} \big) \\ \times r^{4V}_{-} \big( \underline{\lambda} - \underline{\mu} \big)         \text{, } \end{align*}

      \begin{align*}      \big\{  T^{4V}_{+} \big( x , \underline{\lambda} \big) \overset{\bigotimes}{,} T^{4V}_{+} \big( x , \underline{\mu} \big)           \big\} = T^{4V}_{+} \big( x , \underline{\lambda} \big) \bigotimes T^{4V}_{+} \big( x , \underline{\mu} \big) r^{4V}_{+} \big( \underline{\lambda} - \underline{\mu} \big) - r^{4V} \big( \underline{\lambda} - \underline{\mu} \big) T^{4V}_{+} \big( x , \underline{\lambda} \big) \\ \bigotimes T^{4V}_{+} \big( x , \underline{\mu} \big)   \text{, }
\end{align*}

\noindent for the transfer matrices of the 6-vertex, 20-vertex, and 4-vertex, models. Asymptotically, the transfer matrices satisfy,

\begin{align*}
   T^{6V}_{\pm} \big( x , {\lambda} \big) = \underset{y \longrightarrow \pm \infty}{\mathrm{lim}} T \big( x , y , {\lambda} \big) E^{6V} \big( y , {\lambda} \big) \text{, } \\ 
 T^{3D}_{\pm} \big( x , \underline{\lambda} \big) = \underset{y \longrightarrow \pm \infty}{\mathrm{lim}} T \big( x , y , \underline{\lambda} \big) E^{20V} \big( y , \underline{\lambda} \big)   
 \text{, } \\   T^{4V}_{\pm} \big( x , {\lambda} \big) = \underset{y \longrightarrow \pm \infty}{\mathrm{lim}} T \big( x , y , {\lambda} \big) E^{4V} \big( y , {\lambda} \big)  \text{, }
\end{align*}

\noindent while explicitly, each transfer matrix is respectively given by, for $j,k \in \textbf{N}$,

\begin{align*}
      T^{6V}_a  \big( u , \big\{ v_k \big\} , H , V \big)  \equiv 
  \begin{bmatrix}
       A \big( u \big)   & B \big( u \big)   \\
    C \big( u \big)  & D \big( u \big)  \text{ }  
  \end{bmatrix}   \text{, } \end{align*}

  \begin{align*} \textbf{T}^{3D} \big( \underline{\lambda} \big)   \equiv    \underset{ N \longrightarrow -  \infty}{\underset{\underline{M} \longrightarrow + \infty}{\mathrm{lim}}} \mathrm{tr} \bigg\{     \overset{\underline{M}}{\underset{j=0}{\prod}}  \text{ }  \overset{-N}{\underset{k=0}{\prod}} \mathrm{exp} \big( \lambda_3 ( q^{-2} \xi^{s^j_k} ) \big)     \bigg[ \begin{smallmatrix}     q^{D^j_k}       &    q^{-2} a^j_k q^{-D^j_k -D^j_{k+1}} \xi^{s-s^k_j}        & *_1 \\ \big( a^j_k \big)^{\dagger} q^{D^j_k} \xi^{s^j_k} 
             &     *_2 &   - a^j_k q^{D^j_k - 3D^j_{k+1}} \xi^{s-s^j_k}  \\ 0  &    a^{\dagger}_j q^{D^j_k} \xi^{s^j_k} &  q^{-D^j_k} \\    \end{smallmatrix} \bigg]      \bigg\} \text{, } \end{align*}

             \begin{align*} T^{4V} \big( \underline{u} \big) \equiv     \underset{0 \leq j \leq M}{\prod}  L^{4V} \big( j | u \big) \equiv    \begin{bmatrix}
        A \big( \underline{u} \big) & B \big( \underline{u} \big) \\ C \big( \underline{u} \big) & D \big( \underline{u} \big) 
    \end{bmatrix}     \text{, }
\end{align*}

\noindent for,

\begin{align*}
  *_1 \equiv  a^j_k a^j_{k+1} q^{-D^j_k - 3D^j_{k+1}} \xi^{s - s^j_k - s^j_{k+1}}    \text{, }\\  \\  *_2 \equiv  q^{-D^j_k + D^j_{k+1}} - q^{-2} q^{D^j_k -D^j_{k+1}} \xi^{s}       \text{. }
\end{align*}

\noindent An adaptation of such an integrability property was also demonstrated by the author in later works, {\color{blue}[]} for the 20-vertex model, and {\color{blue}[]} for the 4-vertex model, with the L-operators, integrable structures of vertex models is ultimately dependent upon whether Poisson brackets,

\begin{align*}
    \big\{  \Phi^{6V} \big( {\lambda} \big)  ,   \bar{\Phi^{6V} \big( {\lambda} \big) } \big\}      \text{, } \\  \big\{  \Phi^{20V} \big( \underline{\lambda} \big)  ,   \bar{\Phi^{20V} \big( \underline{\lambda} \big) } \big\}       \text{, }  \\  \big\{  \Phi^{4V} \big( {\lambda} \big)  ,   \bar{\Phi^{4V} \big( {\lambda} \big) } \big\}       \text{, } 
\end{align*}

\noindent of action-angle variables, with respect to their complex conjugates, vanish. The existence of such a vanishing Poisson bracket would imply that, with respect to the action-angle coordinates, the associated, potentially time-dependent, dynamics of the system would be linear. In comparison to other vertex models described in this article, the 20-vertex model, as the highest-dimensional counterpart, exhibits a Poisson structure with the most number of brackets. That is, in comparison to the following collection,

\[
\left\{\!\begin{array}{ll@{}>{{}}l} \underline{(1)}:      \big\{  A \big( u \big)        , A \big( u^{\prime} \big)   \big\} 
\text{ , } \\  \underline{(2)}:  \big\{         A \big( u \big)        ,        B \big( u^{\prime} \big)     \big\}  \text{ , } \\  \underline{(3)}:   \big\{   A \big( u \big)       ,  C \big( u^{\prime} \big) \big\} 
 \text{, }   \\  \underline{(4)}: \big\{   A \big( u \big)       ,  D \big( u^{\prime} \big) \big\}   \text{ , } \\ \underline{(5)} : \big\{ B \big( u \big) , A \big( u^{\prime} \big) \big\} \text{, } \\ \underline{(6)}: \big\{ B \big( u \big) , B \big( u^{\prime} \big) \big\} \text{, } \\ \underline{(7)}: \big\{ B \big( u \big) , C \big( u^{\prime} \big) \big\} \text{, } \\ \underline{(8)}: \big\{ B \big( u \big)  , D \big( u^{\prime} \big)   \big\} \text{, } \\ \underline{(9)}:  \big\{ C \big( u \big)  , A \big( u^{\prime} \big)   \big\} \text{, } \\ \underline{(10)}: \big\{ C \big( u \big) , B \big( u^{\prime} \big) \big\} \text{, } \\ \underline{(11)}: \big\{ C \big( u \big) , C \big( u^{\prime} \big) \big\} \text{, }  \\ \underline{(12)}: \big\{ C \big( u \big) , D \big( u^{\prime} \big) \big\} \text{, } \\ \underline{(13)}: \big\{ D \big( u \big)  , A \big( u^{\prime} \big)  \big\} \text{, } \\ 
   \underline{(14)}: \big\{  D \big( u \big)  , B \big( u^{\prime} \big)  \big\} \text{, } \\ \underline{(15)}: \big\{  D \big( u \big)  , C \big( u^{\prime} \big) \big\}  \text{, }  \\ \underline{(16)}: \big\{  D \big( u \big)  ,       D \big( u^{\prime} \big)  \big\} \text{, }
\end{array}\right.
\]



\noindent of 16 Poisson brackets from block representations of the transfer matrix for the 6-vertex model, the transfer for the 20-vertex model having more block entry representations, from the fact that,

\[
\begin{bmatrix}
 A \big( \underline{u} \big) & D \big( \underline{u} \big)  & G \big( \underline{u} \big) \\ B \big( \underline{u} \big) & E \big( \underline{u} \big) & H \big( \underline{u} \big)  \\ C \big( \underline{u} \big)  &  F \big( \underline{u} \big) & I \big( \underline{u} \big) 
\end{bmatrix}  \text{, } 
\]

\noindent implies that the Poisson structure has $81$ brackets, which take the form,

\[
\left\{\!\begin{array}{ll@{}>{{}}l} \boxed{(1)}:      \big\{  A \big( \underline{u} \big)        , A \big( \underline{u^{\prime}} \big)   \big\} 
\text{, } &  \boxed{(22)}: \big\{ C \big( \underline{u} \big) , D \big( \underline{u^{\prime}} \big) \big\} \text{, }  &  \boxed{(43)}: \big\{ E \big( \underline{u} \big) , G \big( \underline{u^{\prime}} \big) \big\} \text{, }  \\  \boxed{(2)}:  \big\{         A \big( \underline{u} \big)        ,        B \big( \underline{u^{\prime}} \big)     \big\}  \text{
, } &  \boxed{(23)}:   \big\{  C \big( \underline{u} \big) , E \big( \underline{u^{\prime}} \big)  \big\}  \text{, } &  \boxed{(44)}: \big\{ E \big( \underline{u} \big) , H \big( \underline{u^{\prime}} \big) \big\} \text{, }   \\  \boxed{(3)}:   \big\{   A \big( \underline{u} \big)       ,  C \big( \underline{u^{\prime}} \big) \big\} 
 \text{, } &  \boxed{(24)}: \big\{ C \big( \underline{u} \big) , F \big( \underline{u^{\prime}} \big) \big\} \text{, } & \boxed{(45)}: \big\{ E \big( \underline{u} \big) , I \big( \underline{u^{\prime}} \big) \big\} \text{, } \\ \boxed{ (4)}: \big\{   A \big( \underline{u} \big)       ,  D \big( \underline{u^{\prime}} \big) \big\}   \text{, } & \boxed{(25)}: \big\{ C \big( \underline{u} \big) , G \big( \underline{u^{\prime} } \big) \big\} \text{, } &  \boxed{(46)}: \big\{ F \big( \underline{u} \big) ,     A \big( \underline{u^{\prime }} \big) \big\} \text{, } \\ \boxed{(5)}: \big\{ A \big( \underline{u} \big) , E \big( \underline{u^{\prime}} \big) \big\}  \text{,  } &   \boxed{(26)}: \big\{ C \big( \underline{u} \big) , H \big( \underline{u^{\prime}} \big) \big\} \text{, } & \boxed{(47)}:  \big\{ F \big( \underline{u} \big) , B \big( \underline{u^{\prime}}  \big) \big\} \text{, }  \\ \boxed{(6)} : \big\{ A \big( \underline{u} \big) , F \big( \underline{u^{\prime} } \big) \big\}  \text{ , }   & \boxed{(27)}: \big\{ C \big( \underline{u} \big) , I \big( \underline{u^{\prime}}  \big) \big\} \text{, } &     \boxed{(48)}: \big\{ F \big( \underline{u} \big) , C \big( \underline{u^{\prime}} \big) \big\} \text{, }     \\ \boxed{(7)}: \big\{ A \big( \underline{u} \big) , G \big( \underline{u^{\prime}} \big) \big\} \text{, }  &    \boxed{(28)}: \big\{ D \big( \underline{u} \big)  , A \big(\underline{u^{\prime}} \big)  \big\} \text{, } & \boxed{(49)}:   \big\{ F \big( \underline{u} \big) , D \big( \underline{u^{\prime}} \big) \big\} \text{, }  
\\  \boxed{(8)}: \big\{ A \big( \underline{u} \big) , H \big( \underline{u^{\prime}} \big) \big\} &  \boxed{(29)}: \big\{  D \big( \underline{u} \big)  , B \big( \underline{u^{\prime}} \big)  \big\} \text{, } &   \boxed{(50)}: \big\{ F \big( \underline{u} \big) , E \big( \underline{u^{\prime}} \big) \big\} \text{, }  \\ \boxed{(9)}: \big\{ A \big( \underline{u} \big) , I \big( \underline{u^{\prime}} \big) \big\}  \text{, } &    \boxed{(30)}: \big\{  D \big( \underline{u} \big)  , C \big( \underline{u^{\prime}} \big) \big\} \text{, } &  \boxed{(51)}:   \big\{ F \big( \underline{u} \big) , F \big( \underline{u^{\prime}} \big) \big\} \text{, }  \\   \boxed{(10)} : \big\{ B \big( \underline{u} \big) , A \big( \underline{u^{\prime}} \big) \big\} \text{, } &    \boxed{(31)}: \big\{  D \big( \underline{u} \big)  ,       D \big( \underline{u^{\prime}} \big)  \big\} \text{, } &  \boxed{(52)}: \big\{ F \big( \underline{u} \big) , G \big( \underline{u^{\prime}} \big) \big\} \text{, }
\\ \boxed{(11)}: \big\{ B \big( \underline{u} \big) , B \big( \underline{u^{\prime}} \big) \big\} \text{, } &   \boxed{(32)} : \big\{ D \big( \underline{u} \big) , E \big( \underline{u^{\prime}}\big) \big\} \text{, } &   \boxed{(53)}: \big\{ F \big( \underline{u} \big) , H \big( \underline{u^{\prime}} \big) \big\} \text{, } \\  \boxed{(12)}: \big\{ B \big( \underline{u} \big) , C \big( \underline{u^{\prime}} \big) \big\} \text{, }  &  \boxed{(33)}: \big\{ D \big( \underline{u} \big) , F \big( \underline{u^{\prime}} \big) \big\} \text{, } &   \boxed{(54)}:  \big\{ F \big( \underline{u} \big) , I \big( \underline{u^{\prime}} \big) \big\} \text{, }  \\ \boxed{(13)}: \big\{ B \big( \underline{u} \big)  , D \big( \underline{u^{\prime}} \big)   \big\} \text{, } & \boxed{(34)}: \big\{ D \big( \underline{u} \big) , G \big( \underline{u^{\prime}} \big) \big\} \text{, }  &  \boxed{(55)}: \big\{ G \big( \underline{u} \big) , A \big( \underline{u^{\prime}} \big) \big\} \text{, }   \\ \boxed{(14)}: \big\{ B \big( \underline{u} \big) , E \big( \underline{u^{\prime}} \big) \big\}  \text{, } &   \boxed{(35)}: \big\{ D \big( \underline{u} \big) , H \big( \underline{u^{\prime}} \big) \big\} \text{, } &   \boxed{(56)}: \big\{ G \big( \underline{u} \big) , B \big( \underline{u^{\prime}} \big) \big\} \text{, } \\ \boxed{(15)}: \big\{ B \big( \underline{u} \big) , F \big( \underline{u^{\prime}} \big) \big\}  \text{, } &   \boxed{(36)}: \big\{ D \big( \underline{u} \big) , I \big( \underline{u^{\prime}} \big) \big\} \text{, } &  \boxed{(57)}:  \big\{ G \big( \underline{u} \big) , C \big( \underline{u^{\prime}} \big) \big\} \text{, } \\ \boxed{(16)}: \big\{ B \big( \underline{u} \big) , G \big( \underline{u^{\prime}} \big) \big\} \text{, } & \boxed{(37)}: \big\{ E \big( \underline{u} \big) , A \big( \underline{u^{\prime}} \big) \big\}  \text{, } &  \boxed{(58)}: \big\{ G \big( \underline{u} \big) , D \big( \underline{u^{\prime} }\big) \big\} \text{, } \\ \boxed{(17)}: \big\{ B \big( \underline{u} \big) , H \big( \underline{u^{\prime}} \big) \big\} \text{, } & \boxed{(38)}: \big\{ E \big( \underline{u} \big) , B \big( \underline{u^{\prime}} \big) \big\} \text{, } & \boxed{(59)} : \big\{ G \big( \underline{u} \big) , E \big( \underline{u^{\prime}} \big) \big\} \text{, } \\  \boxed{(18)}: \big\{ B \big( \underline{u} \big) , I \big( \underline{u^{\prime}} \big) \big\}   \text{, }  &    \boxed{(39)}: \big\{ E \big( \underline{u} \big) , C \big( \underline{u^{\prime}} \big) \big\} \text{, } & \boxed{(60)}:  \big\{ G \big( \underline{u} \big) , F \big( \underline{u^{\prime}} \big) \big\} \text{, } \\  \boxed{(19)}:  \big\{ C \big( \underline{u} \big)  , A \big( \underline{u^{\prime}} \big)   \big\} \text{, } &    \boxed{(40)}:   \big\{ E \big( \underline{u} \big) , D \big( \underline{u^{\prime}} \big) \big\} \text{, } & \boxed{(61)}: \big\{ G \big( \underline{u} \big) , G \big( \underline{u^{\prime}} \big) \big\} \text{, } \\ \boxed{(20)}: \big\{ C \big( \underline{u} \big) , B \big( \underline{u^{\prime}} \big) \big\} \text{, } & \boxed{(41)}: \big\{ E \big( \underline{u} \big) , E \big( \underline{u^{\prime}} \big) \big\} \text{, } & \boxed{(62)}: \big\{ G \big( \underline{u} \big) , H \big( \underline{u^{\prime}} \big) \big\} \text{, } \\ \boxed{(21)}: \big\{ C \big( \underline{u} \big) , C \big( \underline{u^{\prime}} \big) \big\} \text{, }  &  \boxed{(42)}: \big\{ E \big( \underline{u} \big) , F \big( \underline{u^{\prime}} \big) \big\} \text{, } &  \boxed{(63)}: \big\{ G \big( \underline{u} \big) , I \big( \underline{u^{\prime}} \big) \big\} \text{, }
 \end{array}\right.
\]



\noindent corresponding to the first $63$ brackets within the structure, and,

\[\left\{\!\begin{array}{ll@{}>{{}}l}   \boxed{(64)}: \big\{ H \big( \underline{u} \big) , A \big( \underline{u^{\prime}} \big) \big\} \text{, } & \boxed{(70)}: \big\{ H \big(\underline{u} \big) , G \big( \underline{u^{\prime}}  \big) \big\} \text{, } & \boxed{(76)}: \big\{ I \big( \underline{u} \big) , D \big( \underline{u^{\prime}} \big) \big\} \text{, }  \\ \boxed{(65)}: \big\{ H \big( \underline{u} \big) , B \big( \underline{u^{\prime}} \big) \big\} \text{, } & \boxed{(71)}:  \big\{ H \big( \underline{u} \big) , H \big( \underline{u^{\prime} }\big) \big\} \text{, }      &   \boxed{(77)}: \big\{ I \big( \underline{u} \big) , E \big( \underline{u^{\prime}} \big) \big\} \text{, }   \\ \boxed{(66)}: \big\{ H \big( \underline{u} \big) , C \big( \underline{u^{\prime}} \big) \big\} \text{, } &  \boxed{(72)}: \big\{ H \big( \underline{u} \big) , I \big( \underline{u^{\prime}} \big) \big\} \text{, }         &       \boxed{(78)}: \big\{ I \big( \underline{u} \big) , F \big( \underline{u^{\prime}} \big) \big\} \text{, }    \\ \boxed{(67)} : \big\{ H \big( \underline{u} \big) , D \big( \underline{u^{\prime}} \big) \big\} \text{ , }  &  \boxed{(73)}: \big\{ I \big( \underline{u} \big) , A \big( \underline{u^{\prime}} \big) \big\} \text{, }        &   \boxed{(79)}: \big\{ I \big( \underline{u} \big) , G \big( \underline{u^{\prime}} \big) \big\} \text{, }  \\ \boxed{(68)}: \big\{ H \big( \underline{u} \big) , E \big( \underline{u^{\prime}} \big) \big\} \text{, }   &    \boxed{(74)}: \big\{ I \big( \underline{u} \big) , B \big( \underline{u^{\prime}} \big) \big\} \text{, }        &         \boxed{(80)}:  \big\{ I \big( \underline{u} \big) , H \big( \underline{u^{\prime}} \big) \big\} \text{, }          \\ \boxed{(69)}: \big\{ H \big( \underline{u} \big) , F \big( \underline{u^{\prime}} \big) \big\} \text{, }  &  \boxed{(75)}: \big\{ I \big( \underline{u} \big) , C \big( \underline{u^{\prime}} \big) \big\} \text{, }     & \boxed{(81)}: \big\{ I \big( \underline{u} \big) , I \big( \underline{u^{\prime}} \big) \big\} \text{, }   
\end{array}\right.
\]

\noindent corresponding to the remaining $17$ brackets within the structure. To quantify asymptotic properties of the limit,

\[ \textbf{T}^{3D} \big( \underline{\lambda} \big)   \equiv    \underset{ N \longrightarrow -  \infty}{\underset{\underline{M} \longrightarrow + \infty}{\mathrm{lim}}} \mathrm{tr} \bigg\{      \overset{\underline{M}}{\underset{j=0}{\prod}}  \text{ }  \overset{-N}{\underset{k=0}{\prod}} \mathrm{exp} \big( \lambda_3 ( q^{-2} \xi^{s^j_k} ) \big)     \bigg[ \begin{smallmatrix}     q^{D^j_k}       &    q^{-2} a^j_k q^{-D^j_k -D^j_{k+1}} \xi^{s-s^k_j}        & *_1 \\ \big( a^j_k \big)^{\dagger} q^{D^j_k} \xi^{s^j_k} 
             &     *_2 &   - a^j_k q^{D^j_k - 3D^j_{k+1}} \xi^{s-s^j_k}  \\ 0  &    a^{\dagger}_j q^{D^j_k} \xi^{s^j_k} &  q^{-D^j_k} \\    \end{smallmatrix} \bigg]      \bigg\}  \text{, }
\]

\noindent over $\textbf{T}$, we compute products of L-operators by taking into account contributions from the following groups of terms. First, we consider powers of $q$, which can be determined by taking the power to be various operations acting over the lattice,

\[
   \text{Powers of } q     \equiv  \left\{\!\begin{array}{ll@{}>{{}}l} 
     q^{D^{j+2}_k}  \text{, }  \\ q^{-2} q^{-D^j_k} \text{, }  \\ 
  q^{-2} q^{-D^{j+1}_k - D^{j+2}_k} \text{ , } \\ q^{-D^j_k} \text{ , } \\ q^{-D^{j+2}_k} \text{, }   \\  q^{-D^{j+2}_{k+1}} \text{, } \\ q^{-D^{j+2}_k - D^{j+1}_k}
 \text{, } \\ 
 q^{-D^{j+2}_{k+1} - D^{j+1}_{k+1}}   \text{, } 
 \end{array}\right.
 \]

\noindent or, images of unital associative mappings,

\[
   \text{Images of the unital associative mapping } \xi     \equiv  \left\{\!\begin{array}{ll@{}>{{}}l} 
 \xi^s  \text{, }  \\  \xi^{s-s^j_k} \text{, } \\ q^{-2} \xi^{s-s^{j+1}_k} \text{, } \\ q^{-2} \xi^{s-s^{j+1}_{k+1}} \text{, } \\   \xi^s \text{, } \\ \xi^{s-s^j_k} \text{, } \\ q^{-2} \xi^{s-s^{j+1}_k} \text{, } \\ q^{-2} \xi^{s-s^{j+1}_{k+1}} \text{. }  \end{array}\right.
 \]

 \noindent or, finally, powers of $q$ and unital associative mappings simultaneously,

 \[
   \text{Powers of } q  \text{, and images of the unital associative mapping } \xi   \equiv  \left\{\!\begin{array}{ll@{}>{{}}l} 
       q^{-2} q^{-D^j_k} \xi^s \text{ , } \\ q^{-2} q^{-D^j_k} \xi^{s-s^j_k} \text{ , } \\  q^{-2} q^{-D^j_k} \xi^{s-s^j_k - s^j_{k-1}} \text{ , }  \\  q^{-2} q^{-D^j_k} \xi^{s-s^j_k - s^j_{k-1}} \text{ , }  \\ q^{-2} q^{-D^j_k} \xi^{s-s^{j+1}_k - s^{j+1}_{k-1}} \text{ , }  \\ 
   \end{array}\right.
 \] 

\noindent From the superposition for each Poisson bracket provided above, from previous computations with the bracket in the 6-vertex, and 20-vertex, models  {\color{blue}[42},{\color{blue}46]}, the following terms are approximately,

\begin{align*}
  \big\{ \mathscr{I}^1_1 \big( u \big) , \mathscr{I}^1_1 \big( u^{\prime} \big) \big\} \approx \frac{1}{u-u^{\prime}} \equiv C^1_1 \propto \mathscr{C}^1_1 \text{, } \\ \big\{ \mathscr{I}^1_2 \big( u \big), \mathscr{I}^1_2 \big( u^{\prime} \big) \big\} \approx \frac{1}{u-u^{\prime}}  \equiv C^1_2 \propto \mathscr{C}^1_2 \text{, }  \\ 
 \big\{ \mathscr{I}^1_3 \big( u \big), \mathscr{I}^1_3 \big( u^{\prime} \big) \big\} \approx \frac{1}{u-u^{\prime}} \equiv C^1_3  \propto \mathscr{C}^1_3 \text{, }  \\  \big\{ \mathscr{I}^2_1 \big( u \big), \mathscr{I}^2_1 \big( u^{\prime} \big) \big\} \approx \frac{1}{u-u^{\prime}} \equiv C^2_1 \propto \mathscr{C}^2_1 
 \text{, }  \\  \big\{  \mathscr{I}^2_2 \big( u \big), \mathscr{I}^2_2 \big( u^{\prime} \big) \big\} \approx \frac{1}{u-u^{\prime}}  \equiv C^2_2   \propto \mathscr{C}^2_2  \text{, } \\ \big\{ \mathscr{I}^2_3 \big( u \big), \mathscr{I}^2_3 \big( u^{\prime} \big) \big\} \approx \frac{1}{u-u^{\prime}}   \equiv C^2_3   \propto \mathscr{C}^2_3 \text{, }   \end{align*}
 
\begin{align*}
  \big\{ \mathscr{I}^2_4 \big( u \big), \mathscr{I}^2_4  \big( u^{\prime} \big) \big\} \approx \frac{1}{u-u^{\prime}}  \equiv C^2_4   \propto \mathscr{C}^2_4 \text{, } \\  \big\{ \mathscr{I}^3_1 \big( u \big), \mathscr{I}^3_1 \big( u^{\prime} \big) \big\} \approx \frac{1}{u-u^{\prime}}  \equiv C^3_1  \propto \mathscr{C}^3_1  \text{, }  \\ \big\{ \mathscr{I}^3_2 \big( u \big), \mathscr{I}^3_2 \big( u^{\prime} \big) \big\} \approx \frac{1}{u-u^{\prime}}   \equiv C^3_2   \propto \mathscr{C}^3_2 
 \text{, } \\ \big\{ \mathscr{I}^3_3 \big( u \big), \mathscr{I}^3_3 \big( u^{\prime} \big) \big\} \approx \frac{1}{u-u^{\prime}}  \equiv C^3_3  \propto \mathscr{C}^3_3  \text{, } \\  \big\{ \mathscr{I}^3_4 \big( u \big), \mathscr{I}^3_4 \big( u^{\prime} \big) \big\} \approx \frac{1}{u-u^{\prime}}  \equiv C^3_4  \propto \mathscr{C}^3_4  \text{, }    \\ 
 \big\{ \mathscr{I}^4_1 \big( u \big) , \mathscr{I}^4_1 \big( u^{\prime} \big) \big\} \approx  \frac{1}{u-u^{\prime}}   \equiv C^4_1  \propto \mathscr{C}^4_1 \text{, } \\  \big\{ \mathscr{I}^4_2 \big( u \big) , \mathscr{I}^4_2 \big( u^{\prime} \big) \big\} \approx  \frac{1}{u-u^{\prime}}   \equiv C^4_2  \propto \mathscr{C}^4_2 \text{, } \\  \big\{ \mathscr{I}^4_3 \big( u \big) , \mathscr{I}^4_3 \big( u^{\prime} \big) \big\} \approx  \frac{1}{u-u^{\prime}}  \equiv C^4_3 \propto \mathscr{C}^4_3  \text{, } \\ 
 \big\{ \mathscr{I}^4_4 \big( u \big) , \mathscr{I}^4_4 \big( u^{\prime} \big) \big\} \approx  \frac{1}{u-u^{\prime}}  \equiv C^4_4  \propto \mathscr{C}^4_4  \text{, } \\ \big\{ \mathscr{I}^4_5 \big( u \big) , \mathscr{I}^4_5 \big( u^{\prime} \big) \big\} \approx  \frac{1}{u-u^{\prime}}  \equiv C^4_5   \propto \mathscr{C}^4_5 \text{. }
\end{align*}

\noindent The collection $\mathscr{C}$ are used to approximate, asymptotically, the approximation to each Poisson bracket of the 4-vertex model. In addition to approximating which Poisson brackets, within the structure, asymptotically behave like $\big( u - u^{\prime} \big)^{-1}$, products of blocks representations in transfer matrices for the 20-vertex model satisfy,

\begin{figure}
\begin{align*}
\includegraphics[width=1.22\columnwidth]{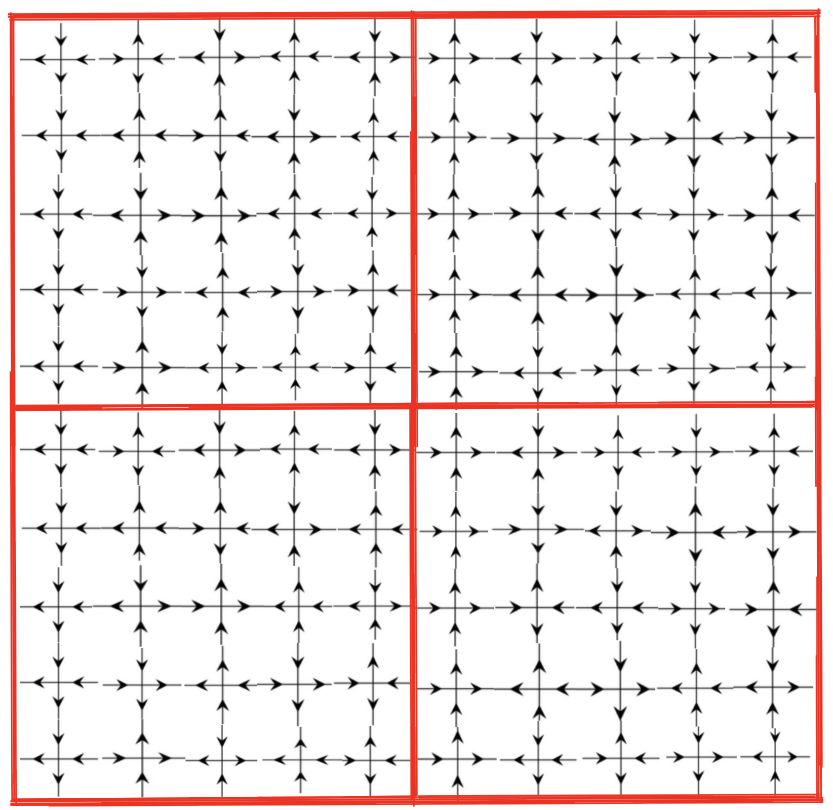}
\end{align*}
\caption{A depiction of a two-dimensional vertex configuration of the 6-vertex model sampled over $\textbf{Z}^2$. The box, whose boundary is outlined in red, is comprised of four equal boxes whose boundaries are also outlined in red within the interior.}
\end{figure} 

\begin{align*}
         \underline{G \big( \underline{u} \big) E \big( \underline{u^{\prime}} \big) C \big( \underline{u^{\prime\prime}} \big) } =       f \big( \lambda_{\alpha} , \lambda_r , \lambda_{r^{\prime}} \big)   f \big( \lambda , \lambda^{\prime} \big)     C \big( \underline{u^{\prime\prime}} \big) E \big( \underline{u^{\prime}} \big)  G \big( \underline{u} \big)     +    f \big( \lambda_{\alpha} , \lambda_r , \lambda_{r^{\prime}} \big) g \big( \lambda^{\prime} , \lambda \big)   \\ \times   C \big( \underline{u^{\prime}}    \big) E \big( \underline{u^{\prime\prime}} \big)    G \big( \underline{u} \big)                            +        g \big( \lambda_{\alpha} , \lambda_r , \lambda_{r^{\prime}} \big)      f \big(   \lambda , \lambda^{\prime} \big)  C \big( \underline{u^{\prime}} \big) E \big( \underline{u} \big)         G \big( \underline{u^{\prime\prime} } \big)              +  g \big( \lambda_{\alpha} , \lambda_r , \lambda_{r^{\prime}} \big)      g \big( \lambda^{\prime} , \lambda \big) \\ \times   C \big(  \underline{u}     \big) E \big( \underline{u^{\prime}}       \big)                      G \big( \underline{u^{\prime\prime} } \big)                \text{, } \\ \\ \underline{ I \big( \underline{u} \big) H \big( \underline{u^{\prime}} \big) G \big( \underline{u^{\prime\prime}} \big)}  =           f \big( \lambda_{\alpha} , \lambda_r , \lambda_{r^{\prime}} \big)       f \big( \lambda , \lambda^{\prime} \big)                G \big( \underline{u^{\prime\prime}} \big) H \big( \underline{u^{\prime}} \big)        I \big( \underline{u} \big)     +   f \big( \lambda_{\alpha} , \lambda_r , \lambda_{r^{\prime}} \big)   g \big( \lambda^{\prime} , \lambda \big)  \\ \times   G \big( \underline{u^{\prime}} \big) H \big( \underline{u^{\prime\prime}} \big)                          I \big( \underline{u} \big)            +  g \big( \lambda_{\alpha} , \lambda_r  , \lambda_{r^{\prime}} \big)    f \big( \lambda , \lambda^{\prime} \big) G \big( \underline{u^{\prime\prime}} \big) H \big(   \underline{u^{\prime}} \big) I \big( \underline{u^{\prime}}  \big)   +  g \big( \lambda_{\alpha} , \lambda_r  , \lambda_{r^{\prime}} \big)          g \big( \lambda^{\prime} , \lambda \big)              \end{align*}
         
         \begin{align*} \times          G \big( \underline{u^{\prime}} \big)  H \big( \underline{u^{\prime\prime}} \big) I \big( \underline{u^{\prime}}  \big)        \text{, } \\   \\      \underline{  A \big( \underline{u }  \big) D \big( \underline{u^{\prime}} \big) G \big( \underline{u^{\prime\prime}} \big) }   =          f \big( \lambda_{\alpha}  , \lambda_r , \lambda_{r^{\prime}} \big)               f \big( \lambda , \lambda^{\prime} \big)                       G \big(    \underline{u^{\prime\prime}}   \big)    D \big( \underline{u^{\prime}} \big)       A \big( \underline{u} \big)   +  f \big( \lambda_{\alpha}  , \lambda_r , \lambda_{r^{\prime}} \big)    g \big( \lambda^{\prime} , \lambda \big)  \\ \times   G \big( \underline{u^{\prime}} \big) D \big( \underline{u^{\prime\prime}} \big)                               A \big( \underline{u} \big)     +   g \big( \lambda_{\alpha} , \lambda_r , \lambda_{r^{\prime}} \big)           f \big( \lambda , \lambda^{\prime} \big)   G \big( \underline{u} \big)  D \big( \underline{u^{\prime\prime}} \big)      A \big( \underline{u} \big)      +  g \big( \lambda_{\alpha} , \lambda_r , \lambda_{r^{\prime}} \big)   g \big( \lambda^{\prime} , \lambda \big) \\   \times  G \big( \underline{u^{\prime\prime}} \big)  D \big( \underline{u} \big)          A \big( \underline{u} \big)     \text{, } \\ \\ 
         \underline{A \big( \underline{u } \big) E \big( \underline{u^{\prime}} \big) I \big( \underline{u^{\prime\prime}} \big)}       =           f \big( \lambda_{\alpha}  , \lambda_r , \lambda_{r^{\prime}} \big)        f \big( \lambda , \lambda^{\prime} \big)     I \big( \underline{u^{\prime\prime}} \big) E \big( \underline{u^{\prime}} \big)  A \big( \underline{u} \big)    +    f \big( \lambda_{\alpha}  , \lambda_r , \lambda_{r^{\prime}} \big)    g \big( \lambda^{\prime} , \lambda \big)          \\  \times      I \big( \underline{u^{\prime}} \big) E \big( \underline{u^{\prime\prime}} \big)           A \big( \underline{u} \big)     +   g \big( \lambda_{\alpha} , \lambda_r , \lambda_{r^{\prime}} \big)        f \big( \lambda , \lambda^{\prime} \big)     I \big( \underline{u}  \big) E \big( \underline{u^{\prime\prime}} \big)               A \big( \underline{u^{\prime}} \big)      + g \big( \lambda_{\alpha} , \lambda_r , \lambda_{r^{\prime}} \big)  g \big( \lambda^{\prime} , \lambda \big)       \\ \times               I \big( \underline{u^{\prime\prime}} \big)           E \big( \underline{u}  \big)      A \big( \underline{u^{\prime}} \big)             \text{, } 
\end{align*}

\noindent where,

\begin{align*}
   f \big( \lambda_{\alpha} , \lambda_{r} , \lambda_{r^{\prime}} \big) \equiv     \frac{\mathrm{sin} \big( \lambda_{r^{\prime}} - \lambda_r - \lambda_{\alpha} + 2 \eta  \big)}{\mathrm{sin} \big( \lambda_{r^{\prime}} - \lambda_r - \lambda_{\alpha } \big) }                    \text{, } \\ 
g \big( \lambda_{\alpha} , \lambda_{r} , \lambda_{r^{\prime}} \big) \equiv   \frac{\mathrm{sin} \big( 2 \eta \big)}{\mathrm{sin} \big( \lambda_{r^{\prime}} - \lambda_r - \lambda_{\alpha} \big)}                  \text{. } 
\end{align*}

\noindent The above expressions for the product of block representations of the 20-vertex transfer matrix is generated from the embedded Poisson bracket,

{\small \[\
\bigg\{    \begin{bmatrix}
 A \big( \underline{u} 
 \big) & D \big( \underline{u} \big)  & G \big( \underline{u}  \big) \\ B \big( \underline{u}  \big) & E \big( \underline{u}  \big) & H \big( \underline{u}  \big)  \\ C \big( \underline{u}  \big)  &  F \big( \underline{u}  \big) & I \big( \underline{u}  \big) 
\end{bmatrix}\overset{\bigotimes}{,} \bigg\{ \begin{bmatrix}
 A \big( \underline{u^{\prime}} \big) & D \big( \underline{u^{\prime}} \big)  & G \big( \underline{u^{\prime}} \big) \\ B \big( \underline{u^{\prime}}\big) & E \big( \underline{u^{\prime}} \big) & H \big( \underline{u^{\prime}} \big)  \\ C \big( \underline{u^{\prime}} \big)  &  F \big( \underline{u^{\prime}} \big) & I \big( \underline{u^{\prime}} \big) 
\end{bmatrix}  \overset{\bigotimes}{,} \begin{bmatrix}
 A \big( \underline{u^{\prime\prime}} \big) & D \big( \underline{u^{\prime\prime}} \big)  & G \big( \underline{u^{\prime\prime}} \big) \\ B \big( \underline{u^{\prime\prime}}\big) & E \big( \underline{u^{\prime\prime }} \big) & H \big( \underline{u^{\prime\prime}} \big)  \\ C \big( \underline{u^{\prime\prime}} \big)  &  F \big( \underline{u^{\prime\prime}} \big) & I \big( \underline{u^{\prime\prime}} \big) 
\end{bmatrix}  \bigg\}  \text{ } \bigg\} \text{, } 
\]  }

\noindent where each transfer matrix is dependent upon a different collection of spectral parameters, whether it be $u, u^{\prime}$, or $u^{\prime\prime}$. The QISM approach can also be adapted for spin chains, whether with periodic open boundary conditions, or for open boundary conditions entirely. The corresponding transfer matrix, as a product of $K$ and $R$ matrices, takes the form,

\begin{align*}
   \textbf{T} \big( u \big) \equiv \mathrm{tr}_0 \big(    \textbf{K}_0   \textbf{T}_0 \big( u \big)   \big) \equiv  \mathrm{tr} \big( \textbf{K}_0     \prod_{1 \leq j \leq L}  \textbf{R}_{0j} \big( u \big)  \big)  \text{, }  
\end{align*}

\noindent for,the $36 \times 36$ R-matrix,

\begin{align*}
  R \big( u \big)  \equiv  \mathrm{exp} \big( - 2 u - 6 \eta \big) R_J \big( x \big)  \text{, }  
\end{align*}

\noindent along with closely associated objects,

\begin{align*}
    \textbf{K} \equiv \mathrm{diag} \big( \mathrm{exp} \big( i \phi_1 \big) , \mathrm{exp} \big( i \phi_2 \big) , 1 , 1 , \mathrm{exp} \big( - i \phi_2 \big) , \mathrm{exp} \big( - i \phi_1 \big)   \big)   \text{,} \\   \widetilde{R} \big( u \big) \propto B_{12} B_{34} 
         \textbf{R}^{\prime}_{12,34}\big( u \big)    B_{12} B_{34} \equiv B_{12} B_{34} 
           \bigg(  R_{14} \big( u \big) R_{13} \big( u \big) R_{24} \big( u \big) \\ \times R_{23} \big( u \big)        \bigg)  B_{12} B_{34}         \text{, } \\        \textbf{R}^{\prime}_{12,34} \big( u \big) = R_{43} \big( - \theta \big) R_{13} \big( u \big) R_{14} \big( u + \theta \big) R_{23} \big( u - \theta \big) R_{24} \big( u \big) R_{34} \big( \theta \big)         \text{, } \\        \\ 
 B \equiv  \begin{bmatrix}
    1 & 0 & 0    & 0 \\ 0 & \frac{\mathrm{cosh} \big( \frac{\eta}{2}\big) }{\sqrt{\mathrm{cosh} \big( \eta \big) }} &   - \frac{\mathrm{sinh} \big( \frac{\eta}{2}\big)}{\sqrt{\mathrm{cosh} \big( \eta \big)}}    &  0 \\ 0 & - \frac{\mathrm{sinh} \big( \frac{\eta}{2} \big)}{\sqrt{\mathrm{cosh} \big( \eta \big)}} & - \frac{\mathrm{cosh} \big( \frac{\eta}{2} \big) }{\sqrt{\mathrm{cosh}\big( \eta \big)}}  & 0 \\ 0 & 0 & 0 & 1
\end{bmatrix}  \Longleftrightarrow B^2 = \textbf{I} \text{, } \\  \\  K_{-} \big( \lambda \big) \equiv \begin{bmatrix}
 - \mathrm{exp} \big( - \lambda \big) \big( \mathrm{exp} \big( 2 \lambda \big) + k \big) & 0   &  0 & 0  \\ 0 & (**)_1  & (**)_2 & 0 \\ 0 & (*)_1  & (*)_2 & 0 \\ 0 & 0 & 0 & (***)_1 
\end{bmatrix}  \text{, } \\ \\  (**)_1 \equiv - \frac{1}{2} \big( 1 + \mathrm{exp} \big( 2 \lambda \big) \big) \mathrm{exp} \big( \lambda \big) \big( 1 + k \big) \text{, } \\    (**)_2 \equiv  \frac{1}{2} \big( \mathrm{exp} \big( 2 \lambda \big) - 1 \big) \big( 1 - k \big) \mathrm{exp} \big( \lambda \big) \text{, }  \\  (*)_1 \equiv \frac{1}{2} \big( \mathrm{exp} \big( 2 \lambda \big) - 1 \big) \big( 1 - k \big) \mathrm{exp} \big( \lambda \big)\text{, } \end{align*}

\begin{align*}  (*)_2 \equiv - \frac{1}{2} \big( 1 + \mathrm{exp} \big( 2 \lambda \big) \big) \mathrm{exp} \big( \lambda \big) \big( 1 + k \big) \text{, } \\ (***)_1  \equiv - \mathrm{exp} \big( 3 \lambda \big) \big( \mathrm{exp} \big( 2 \lambda \big) + k \big) \text{, } \\        
\end{align*}

\noindent The higher rank transfer matrix, from a suitable encoding of boundary conditions from its own respective class of $K$ matrices, takes the form,

{\small \begin{align*}
   \textbf{T}^{\mathrm{Quasi-periodic}} \big( u \big) \equiv \mathrm{tr}_0 \big[    \textbf{K}_{+,0} \big( u \big)   \textbf{T}_{+,0} \big( u \big) \textbf{K}_{-,0} \big( u \big)       \textbf{T}_{-,0} \big( u \big)        \big] \equiv  \mathrm{tr}_0 \bigg[ \textbf{K}_{+,0}  \big( u \big)    \prod_{1 \leq j \leq L}  \textbf{R}_{+,0j} \big( u \big)   \\ \times   \textbf{K}_{-,0} \big( u \big)  \underset{1 \leq j^{\prime} \leq L}{\prod}      \textbf{R}_{-,j^{\prime}0}  \big( u \big)     \bigg]  \equiv \mathrm{tr}_0 \bigg[ \textbf{K}^{\mathrm{open}}_{+,0}  \big( u \big)    \prod_{1 \leq j \leq L}  \textbf{R}_{+,a0} \big( u \big)     \textbf{K}^{\mathrm{open}}_{-,0} \big( u \big) \underset{1 \leq j^{\prime} \leq L}{\prod}      \textbf{R}_{-,j^{\prime}0}  \big( u \big)     \bigg]  \text{, }  
\end{align*}} 

\bigskip

\noindent for the higher rank spin chain transfer matrix,

\begin{align*}
  \textbf{T}^{\mathrm{Quasi-periodic}}_{D^{(2)}_3} \big( u \big) \equiv \textbf{T}^{\mathrm{Quasi-periodic}} \big( u \big)   \text{, }  
\end{align*}

\noindent with open boundary conditions enforced through the K matrix,

\[
\textbf{K}^{\mathrm{Quasi-periodic}}_{-} \big( u \big) \equiv \textbf{K}_{-} \big( u \big)  \equiv \begin{bmatrix}
k_0 \big( u \big)  & 0 & 0 & 0 & 0 & 0  \\ 0 & k_0 \big( u \big) & 0 & 0 & 0 & 0  \\ 0 & 0  & k_1 \big( u \big) & k_2 \big( u \big)   & 0 & 0 \\ 0 & 0 & k_3 \big( u \big)  & k_4 \big( u \big)  &  0 & 0  \\ 0 & 0 & 0 & 0 & k_5 \big( u \big) & 0  \\  0 & 0 & 0 & 0 & 0 & k_5 \big( u \big)  
\end{bmatrix} \text{. }  
\]

\noindent After having obtained a system of relations from the product of two-dimensional L-operators expressed in terms of Pauli bases, for the inhomogeneous 6-vertex model one obtains the following set of $16$ relations for two-dimensional Poisson structure. In order to systematically evaluate each Poisson bracket within the two-dimensional structure, one introduces a reparameterization of each Poisson bracket in terms of lower-dimensional quantities, which can in turn be used to approximate each bracket with higher-dimensional terms from products of L-operators. Specifically, the system takes the below form,

\[
\left\{\!\begin{array}{ll@{}>{{}}l} (1):      \bigg\{   \bigg[      A_3 \big( u \big) + B_3 \big( u\big) \bigg]  \bigg[   \big( \mathrm{sin} \big( 2 \eta \big) \big)^{n-3} \mathscr{A}_1  + \mathscr{A}_2 +  \mathscr{A}_3 \bigg]       ,  \bigg[      A_3 \big( u^{\prime} \big) + B_3 \big( u^{\prime} \big) \bigg]  \bigg[   \big( \mathrm{sin} \big( 2 \eta \big) \big)^{n-3} \mathscr{A}^{\prime}_1  \\   + \mathscr{A}^{\prime}_2 +  \mathscr{A}^{\prime}_3 \bigg]    \bigg\} 
\text{, } \\  (2):  \bigg\{       \bigg[      A_3 \big( u \big) + B_3 \big( u\big) \bigg]  \bigg[   \big( \mathrm{sin} \big( 2 \eta \big) \big)^{n-3} \mathscr{A}_1  + \mathscr{A}_2 +  \mathscr{A}_3 \bigg]        ,    \bigg[      A_3 \big( u^{\prime} \big) + B_3 \big( u^{\prime}  \big) \bigg]  \bigg[  \big( \mathrm{sin} \big( 2 \eta \big) \big)^{n-4} \mathscr{B}^{\prime}_1   \\ + \mathscr{B}^{\prime}_2 + \mathscr{B}^{\prime}_3  \bigg]   \bigg\}  \text{, } \\  (3):   \bigg\{    \bigg[      A_3 \big( u \big) + B_3 \big( u\big) \bigg]  \bigg[   \big( \mathrm{sin} \big( 2 \eta \big) \big)^{n-3} \mathscr{A}_1  + \mathscr{A}_2 +  \mathscr{A}_3 \bigg]        ,  \bigg[         C_3 \big( u^{\prime} \big) + D_3 \big( u^{\prime} \big)       \bigg]  \bigg[  \big( \mathrm{sin} \big( 2 \eta \big) \big)^{n-3} \mathscr{C}^{\prime}_1 \\  + \mathscr{C}^{\prime}_2 + \mathscr{C}^{\prime}_3  \bigg]  \bigg\} 
 \text{, }    \\ (4): \bigg\{     \bigg[     A_3 \big( u \big) + B_3 \big( u\big) \bigg]  \bigg[   \big( \mathrm{sin} \big( 2 \eta \big) \big)^{n-3} \mathscr{A}_1  + \mathscr{A}_2 +  \mathscr{A}_3 \bigg]         ,   \bigg[  C_3 \big( u^{\prime} \big) + D_3 \big( u^{\prime} \big)    \bigg]  
  \bigg[  \big( \mathrm{sin} \big( 2 \eta \big) \big)^{n-4}\mathscr{D}^{\prime}_1 \\   + \mathscr{D}^{\prime}_2 + \mathscr{D}^{\prime}_3   \bigg] \bigg\}   \text{, } \\ (5) : \bigg\{ \bigg[      A_3 \big( u \big) + B_3 \big( u \big) \bigg]  \bigg[  \big( \mathrm{sin} \big( 2 \eta \big) \big)^{n-4} \mathscr{B}_1 + \mathscr{B}_2 + \mathscr{B}_3  \bigg] ,  \bigg[      A_3 \big( u^{\prime} \big) + B_3 \big( u^{\prime} \big) \bigg]   \bigg[   \big( \mathrm{sin} \big( 2 \eta \big) \big)^{n-3} \mathscr{A}^{\prime}_1 \\   + \mathscr{A}^{\prime}_2 +  \mathscr{A}^{\prime}_3 \bigg]  \bigg\} \text{, }  \\ (6): \bigg\{ \bigg[      A_3 \big( u \big) + B_3 \big( u \big) \bigg]  \bigg[  \big( \mathrm{sin} \big( 2 \eta \big) \big)^{n-4} \mathscr{B}_1 + \mathscr{B}_2 + \mathscr{B}_3  \bigg] ,   \bigg[      A_3 \big( u^{\prime} \big) + B_3 \big( u^{\prime}  \big) \bigg]  \bigg[  \big( \mathrm{sin} \big( 2 \eta \big) \big)^{n-4} \mathscr{B}^{\prime}_1   \\   + \mathscr{B}^{\prime}_2 + \mathscr{B}^{\prime}_3  \bigg] \bigg\} \text{, } \\ (7): \bigg\{ \bigg[      A_3 \big( u \big) + B_3 \big( u \big) \bigg]  \bigg[  \big( \mathrm{sin} \big( 2 \eta \big) \big)^{n-4} \mathscr{B}_1 + \mathscr{B}_2 + \mathscr{B}_3  \bigg] , \bigg[         C_3 \big( u^{\prime} \big) + D_3 \big( u^{\prime} \big)       \bigg]  \bigg[  \big( \mathrm{sin} \big( 2 \eta \big) \big)^{n-3} \mathscr{C}^{\prime}_1  \\  + \mathscr{C}^{\prime}_2 + \mathscr{C}^{\prime}_3  \bigg]  \bigg\} \text{, }  \\  (8): \bigg\{ \bigg[      A_3 \big( u \big) + B_3 \big( u \big) \bigg]  \bigg[  \big( \mathrm{sin} \big( 2 \eta \big) \big)^{n-4} \mathscr{B}_1 + \mathscr{B}_2 + \mathscr{B}_3  \bigg]  ,  \bigg[  C_3 \big( u^{\prime} \big) + D_3 \big( u^{\prime} \big)    \bigg] 
  \bigg[  \big( \mathrm{sin} \big( 2 \eta \big) \big)^{n-4}\mathscr{D}^{\prime}_1   \\   + \mathscr{D}^{\prime}_2 + \mathscr{D}^{\prime}_3   \bigg]   \bigg\} \text{, } \\ (9):  \bigg\{ \bigg[         C_3 \big( u \big) + D_3 \big( u \big)       \bigg]  \bigg[  \big( \mathrm{sin} \big( 2 \eta \big) \big)^{n-3} \mathscr{C}_1 + \mathscr{C}_2 + \mathscr{C}_3  \bigg]  ,  \bigg[      A_3 \big( u^{\prime} \big) + B_3 \big( u^{\prime} \big) \bigg]  \bigg[   \big( \mathrm{sin} \big( 2 \eta \big) \big)^{n-3} \mathscr{A}^{\prime}_1 \end{array}\right.
\]

\[
\left\{\!\begin{array}{ll@{}>{{}}l}     + \mathscr{A}^{\prime}_2 +  \mathscr{A}^{\prime}_3 \bigg]  \bigg\} \text{, } 
\\  (10): \bigg\{ \bigg[         C_3 \big( u \big) + D_3 \big( u \big)       \bigg]  \bigg[  \big( \mathrm{sin} \big( 2 \eta \big) \big)^{n-3} \mathscr{C}_1 + \mathscr{C}_2 + \mathscr{C}_3  \bigg]  ,   \bigg[      A_3 \big( u^{\prime} \big) + B_3 \big( u^{\prime}  \big) \bigg]  \bigg[ \big( \mathrm{sin} \big( 2 \eta \big) \big)^{n-4} \mathscr{B}^{\prime}_1  \\  + \mathscr{B}^{\prime}_2 + \mathscr{B}^{\prime}_3  \bigg] \bigg\} \text{, } \\ (11): \bigg\{ \bigg[         C_3 \big( u \big) + D_3 \big( u \big)       \bigg]   \bigg[  \big( \mathrm{sin} \big( 2 \eta \big) \big)^{n-3} \mathscr{C}_1 + \mathscr{C}_2 + \mathscr{C}_3  \bigg]  , \bigg[         C_3 \big( u^{\prime} \big) + D_3 \big( u^{\prime} \big)       \bigg]  \bigg[  \big( \mathrm{sin} \big( 2 \eta \big) \big)^{n-3} \mathscr{C}^{\prime}_1 \\   + \mathscr{C}^{\prime}_2 + \mathscr{C}^{\prime}_3  \bigg] \bigg\} \text{, }  \\ (12): \bigg\{ \bigg[         C_3 \big( u \big) + D_3 \big( u \big)       \bigg]  \bigg[  \big( \mathrm{sin} \big( 2 \eta \big) \big)^{n-3} \mathscr{C}_1 + \mathscr{C}_2 + \mathscr{C}_3  \bigg]  , \bigg[  C_3 \big( u^{\prime} \big) + D_3 \big( u^{\prime} \big)    \bigg] 
 \bigg[  \bigg(  \big( \mathrm{sin} \big( 2 \eta \big) \big)^{n-4}\mathscr{D}^{\prime}_1\\  + \mathscr{D}^{\prime}_2 + \mathscr{D}^{\prime}_3   \bigg] \bigg\} \text{, } \\   (13): \bigg\{\bigg[  C_3 \big( u \big) + D_3 \big( u \big)    \bigg] 
 \bigg[   \big( \mathrm{sin} \big( 2 \eta \big) \big)^{n-4}\mathscr{D}_1  + \mathscr{D}_2 + \mathscr{D}_3   \bigg]  , \bigg[      A_3 \big( u^{\prime} \big) + B_3 \big( u^{\prime} \big) \bigg]  \bigg[   \big( \mathrm{sin} \big( 2 \eta \big) \big)^{n-3} \mathscr{A}^{\prime}_1 \\   + \mathscr{A}^{\prime}_2 +  \mathscr{A}^{\prime}_3 \bigg]  \bigg\} \text{, }   \\  (14): \bigg\{  \bigg[  C_3 \big( u \big) + D_3 \big( u \big)    \bigg] 
 \bigg[   \big( \mathrm{sin} \big( 2 \eta \big) \big)^{n-4}\mathscr{D}_1 + \mathscr{D}_2 + \mathscr{D}_3   \bigg]  ,  \bigg[      A_3 \big( u^{\prime} \big) + B_3 \big( u^{\prime} \big) \bigg]  \bigg[  \big( \mathrm{sin} \big( 2 \eta \big) \big)^{n-4} \mathscr{B}^{\prime}_1 \\  + \mathscr{B}^{\prime}_2 + \mathscr{B}^{\prime}_3  \bigg]   \bigg\} \text{, } \\ (15): \bigg\{ \bigg[  C_3 \big( u \big) + D_3 \big( u \big)    \bigg] 
 \bigg[  \big( \mathrm{sin} \big( 2 \eta \big) \big)^{n-4}\mathscr{D}_1 + \mathscr{D}_2 + \mathscr{D}_3   \bigg] , \bigg[         C_3 \big( u^{\prime} \big) + D_3 \big( u^{\prime} 
 \big)       \bigg]  \bigg[  \big( \mathrm{sin} \big( 2 \eta \big) \big)^{n-3} \mathscr{C}^{\prime}_1  \\   + \mathscr{C}^{\prime}_2 + \mathscr{C}^{\prime}_3  \bigg] \bigg\} \text{, }  \\ (16): \bigg\{  \bigg[  C_3 \big( u \big) + D_3 \big( u \big)    \bigg] 
   \bigg[  \big( \mathrm{sin} \big( 2 \eta \big) \big)^{n-4}\mathscr{D}_1 + \mathscr{D}_2 + \mathscr{D}_3   \bigg] ,    \bigg[ C_3 \big( u^{\prime} \big) + D_3 \big( u^{\prime} \big)    \bigg] 
  \bigg[  \big( \mathrm{sin} \big( 2 \eta \big) \big)^{n-4}\mathscr{D}^{\prime}_1 \\  + \mathscr{D}^{\prime}_2 + \mathscr{D}^{\prime}_3   \bigg]  \bigg\} \text{. }
\end{array}\right.
\]




\noindent for,

\begin{align*}
  \mathscr{A}^{\prime}_1 \equiv  \mathscr{C}^{\prime}_1 \equiv \underset{1 \leq i \leq n - (i-3 )}{\prod} \sigma^{-,+}_{n-i} \text{, } \end{align*} 

  \begin{align*}
  \mathscr{A}^{\prime}_2 \equiv \mathscr{C}^{\prime}_2 \equiv \underset{1 \leq i \leq n - (i-3)}{\prod}    \mathrm{sin} \big( \lambda_{\alpha} - v_{n-i} + \eta \sigma^z_{n-i} \big)     \text{, } \end{align*}
  
  \begin{align*} \mathscr{A}^{\prime}_3 \equiv \mathscr{C}^{\prime}_3 \equiv  \underset{m+n^{\prime} = n-(i-3)}{\underset{1\leq j \leq  n^{\prime}}{\underset{1 \leq i \leq m}{\sum} }} \bigg\{       \bigg[ \text{ }   \underset{1 \leq i \leq m}{\prod} \mathrm{sin} \big( \lambda_{\alpha} - v_{n-i} \pm \eta \sigma^z_{n-i} \big)   \bigg] \text{ } \big( \mathrm{sin} \big( 2 \eta \big) \big)^{n^{\prime}-1}   \\ \times  \bigg[ \text{ }   \underset{1 \leq j \leq n^{\prime}}{ \prod}  \sigma^{-,+}_{n-j}     \bigg]         \bigg\}      \text{, } 
\end{align*}

\noindent and,

\begin{align*}
    \mathscr{B}^{\prime}_1 \equiv  \mathscr{D}^{\prime}_1 \equiv    \text{ }     \underset{2 \leq i \leq n-(i-3)}{\prod}     \sigma^{-,+}_{n-i}  \text{, } \\   \mathscr{B}^{\prime}_2 \equiv  \mathscr{D}^{\prime}_2 \equiv               \underset{2 \leq i \leq n-(i-3)}{\prod}   \mathrm{sin} \big( \lambda_{\alpha} - v_{n-i} +    \eta \sigma^z_{n-i}     \big)       \text{, } \end{align*}

    \begin{align*}  \mathscr{B}^{\prime}_3  \equiv  \mathscr{D}^{\prime}_3 \equiv    \text{ }  \underset{m+n^{\prime} = n-(i-3)}{\underset{2\leq j \leq  n^{\prime}}{\underset{2 \leq i \leq m}{\sum} }} \bigg\{       \bigg[ \text{ }   \underset{2 \leq i \leq m}{\prod} \mathrm{sin} \big( \lambda_{\alpha} - v_{n-i} \pm \eta \sigma^z_{n-i} \big)   \bigg] \text{ } \big( \mathrm{sin} \big( 2 \eta \big) \big)^{n^{\prime}-1}  \\ \times  \bigg[ \text{ }   \underset{2 \leq j \leq n^{\prime}}{ \prod}  \sigma^{-,+}_{n-j}     \bigg]         \bigg\}    \text{. } 
\end{align*}

\noindent The computations to obtain closed form representation for each block of a finite dimensional transfer matrix representation has been thoroughly characterized in previous work of the author. For the 4-vertex model, one obtains expressions for lower-dimensional representation of the transfer matrix from products of spins. Such a representation takes the form:

\bigskip

\noindent \textbf{Lemma}  (\textbf{Lemma} \textit{1}, {\color{blue}[46]}, \textit{adding on another term to the product representation of L-operators}, \textbf{Lemma} \textit{1}, {\color{blue}[42]}, \textbf{Lemma} \textit{2}, {\color{blue}[45]}, \textbf{Lemma} \textit{3}, {\color{blue}[45]}). The second representation corresponding to the product of the previous relation obtained in \textbf{Lemma} \textit{1}, with the next L-operator, can be expressed as,

\[
\begin{bmatrix}
 \underline{\mathcal{I}^{\prime}_1}   & \underline{\mathcal{I}^{\prime}_2}  \\ \underline{\mathcal{I}^{\prime}_3}   &  \underline{\mathcal{I}^{\prime}_4} 
\end{bmatrix}
\]

\noindent which is equivalent to the union of the set of linear combinations of,

{\tiny \[
\begin{bmatrix}
 i u \sigma^+_0 \sigma^-_0    i u \sigma^+_1 \sigma^-_1 i u \sigma^+_1 \sigma^-_1 i u \sigma^+_2 \sigma^-_2 + \sigma^-_0 \sigma^+_1 i u \sigma^+_1 \sigma^-_1 i u \sigma^+_2 \sigma^-_2 + i u \sigma^+_0 \sigma^-_0 i u \sigma^+_1  \sigma^-_1 \sigma^-_1 \sigma^+_2 + \sigma^-_0 \sigma^+_1 \sigma^-_1 \sigma^+_2 \\    \\            i u \sigma^+_0 \sigma^-_0 i u \sigma^+_1 \sigma^-_1 i u \sigma^+_1 \sigma^-_1 \sigma^+_2  + \sigma^-_0 \sigma^+_1 i u \sigma^+_1 \sigma^-_1 \sigma^+_2 - i u \sigma^+_0 \sigma^-_0 i u \sigma^+_1 \sigma^-_1 \sigma^+_1 i u \sigma^+_2 \sigma^-_2 + \sigma^-_0 \sigma^+_1 \sigma^+_1 i u \sigma^+_2 \sigma^-_2 \\ + i u \sigma^+_0 \sigma^+_1 \sigma^+_1 \sigma^-_2 + i u  \sigma^+_0 \sigma^+_1 i u \sigma^+_1 \sigma^-_1 i u \sigma^+_2 \sigma^-_2 +  \sigma^+_0 i u \sigma^+_1 \sigma^-_1 \sigma^+_1 \sigma^-_2 +    \sigma^+_0 i u \sigma^+_1 \sigma^-_1 i u \sigma^+_1 \sigma^-_1 \\ \times i u \sigma^+_2 \sigma^-_2          
\end{bmatrix} \text{, }
\] }

\noindent corresponding to the first row of the product representation, and of,

{\tiny \[
\begin{bmatrix}
 i u \sigma^+_0 \sigma^-_0 i u \sigma^+_1 \sigma^-_1 i u \sigma^+_1 \sigma^-_1 \sigma^-_2 + i u \sigma^+_0 \sigma^-_0 i u \sigma^+_1 \sigma^-_1 \sigma^-_1 i u \sigma^+_2 \sigma^-_2  + \sigma^-_0 \sigma^+_1 i u \sigma^+_1 \sigma^-_1 \sigma^-_2 +          \sigma^-_0 \sigma^+_1 \sigma^-_1 i u \sigma^+_2 \sigma^-_2 \\ +     i u \sigma^+_0 \sigma^-_0 \sigma^-_1 \sigma^+_2 \sigma^-_2 + i u      \sigma^+_0 \sigma^-_0 \sigma^-_1 i u \sigma^+_1 \sigma^-_1 i u \sigma^+_2 \sigma^-_2 + \sigma^-_0 i u \sigma^+_1 \sigma^-_1 \sigma^+_1 \sigma^-_2 + \sigma^-_0 i u \sigma^+_1 \sigma^-_1 i u \\ \times \sigma^+_1 \sigma^-_1 i u \sigma^+_2 \sigma^-_2          \\  \\                                      i u \sigma^+_0 \sigma^-_0 i u \sigma^+_1 \sigma^-_1 i u \sigma^+_1 \sigma^-_1 \sigma^-_2 + i u \sigma^+_0       \sigma^-_0 i u \sigma^+_1 \sigma^-_1 \sigma^-_1 i u \sigma^+_2 \sigma^-_2    + \sigma^-_0 \sigma^+_1 i u \sigma^+_1 \sigma^-_1 \sigma^-_2 + \sigma^-_0 \sigma^+_1 \sigma^-_1 i u \sigma^+_2 \sigma^-_2   \\ +                i u \sigma^+_0 \sigma^-_0 \sigma^-_1 \sigma^+_1 \sigma^-_2 + i u \sigma^+_0 \sigma^-_0 \sigma^-_1 i u \sigma^+_1 \sigma^-_1 i u \sigma^+_2 \sigma^-_2 + \sigma^-_0 i u \sigma^+_1 \sigma^-_1 \sigma^+_1 \sigma^-_2 + \sigma^-_0 i u \sigma^+_1 \sigma^-_1 i u \\ \times \sigma^+_1 \sigma^-_1          i u \sigma^+_2 \sigma^-_2        
\end{bmatrix} \text{, }
\] }

\noindent corresponding to the second row of the product representation.

\bigskip

\noindent For the inhomogeneous 6-vertex model with domain-walls, one performs computations with L-operators from the third order approximation, namely, the approximation of the inhomogeneous transfer matrix with spectral parameter $\lambda_{\alpha}$. Such a representation takes the form:

\bigskip

\noindent \textbf{Lemma}  (\textbf{Lemma} \textit{5}, {\color{blue}[41]}, \textit{iteratively obtaining the entries of the n th L-operator from the entries of the third L-operator}). The first entry of the $n$ th L-operator can be expressed in terms of, 

\begin{align*}
          A_n \big( \lambda_{\alpha} \big) =   \bigg[      A_3 \big( \lambda_{\alpha} \big) + B_3 \big( \lambda_{\alpha} \big) \bigg]   \big( \mathrm{sin} \big( 2 \eta \big) \big)^{n-3} \bigg[  \text{ }     \underset{1 \leq i \leq n-3}{\prod}     \sigma^{-,+}_{n-i}   \bigg] \\ +  \bigg[  A_3 \big( \lambda_{\alpha} \big)  + B_3 \big( \lambda_{\alpha } \big) \bigg]  \bigg[ \text{ } \underset{1 \leq i \leq n-3}{\prod}   \mathrm{sin} \big( \lambda_{\alpha} - v_{n-i} +    \eta \sigma^z_{n-j}     \big) \bigg]  \\ +   \bigg[ A_3 \big( \lambda_{\alpha}\big)  +  B_3 \big( \lambda_{\alpha} \big) \bigg] \bigg[ \text{ }  \underset{m,n^{\prime}: m+n^{\prime} = n-3}{\sum} \bigg[    \text{ }   \bigg[  \text{ }   \underset{1 \leq i \leq m}{\prod} \mathrm{sin} \big( \lambda_{\alpha} - v_{n-i} \pm \eta \sigma^z_{n-j} \big)   \bigg] \text{ } \\ \times \big( \mathrm{sin} \big( 2 \eta \big) \big)^{n^{\prime}-1}   \bigg[ \text{ }   \underset{1 \leq j \leq n^{\prime}}{ \prod}  \sigma^{-,+}_{n-j}     \bigg] \text{ }          \bigg]   \text{ }      \bigg]                                           \text{. } 
\end{align*}

\noindent The second entry of the $n$ th L-operator can be expressed in terms of,

\begin{align*}
 B_n \big( \lambda_{\alpha} \big) =     \bigg[      A_3 \big( \lambda_{\alpha} \big) + B_3 \big( \lambda_{\alpha} \big) \bigg]   \big( \mathrm{sin} \big( 2 \eta \big) \big)^{n-4} \bigg[ \text{ }     \underset{2 \leq i \leq n-3}{\prod}     \sigma^{-,+}_{n-i}   \bigg] \\ 
 +  \bigg[  A_3 \big( \lambda_{\alpha} \big)  + B_3 \big( \lambda_{\alpha } \big) \bigg] \bigg[ \text{ } \underset{2 \leq i \leq n-3}{\prod}   \mathrm{sin} \big( \lambda_{\alpha} - v_{n-i} +    \eta \sigma^z_{n-j}     \big) \bigg]  \\ +   \bigg[ A_3 \big( \lambda_{\alpha}\big)  +  B_3 \big( \lambda_{\alpha} \big) \bigg] \bigg[ \text{ }   \underset{m,n^{\prime}: m+n^{\prime} = n-3}{\sum}   \bigg[    \text{ }   \bigg[ \text{ }   \underset{2 \leq i \leq m}{\prod} \mathrm{sin} \big( \lambda_{\alpha} - v_{n-i} \pm \eta \sigma^z_{n-j} \big)   \bigg] \\ \times \text{ } \big( \mathrm{sin} \big( 2 \eta \big) \big)^{n^{\prime}-1}   \bigg[  \text{ }   \underset{2 \leq j \leq n^{\prime}}{ \prod}  \sigma^{-,+}_{n-j}     \bigg]    \text{ }      \bigg]     \text{ }   \bigg]                       \text{. } 
\end{align*}
\noindent The third entry of the $n$ th L-operator can be expressed in terms of,
\begin{align*}
          C_n \big( \lambda_{\alpha} \big) =   \bigg[      C_3 \big( \lambda_{\alpha} \big) + D_3 \big( \lambda_{\alpha} \big) \bigg]   \big( \mathrm{sin} \big( 2 \eta \big) \big)^{n-3} \bigg[ \text{ }     \underset{1 \leq i \leq n-3}{\prod}     \sigma^{-,+}_{n-i}   \bigg] \\ +  \bigg[  C_3 \big( \lambda_{\alpha} \big)  + D_3 \big( \lambda_{\alpha } \big) \bigg] \bigg[ \text{ } \underset{1 \leq i \leq n-3}{\prod}   \mathrm{sin} \big( \lambda_{\alpha} - v_{n-i} +    \eta \sigma^z_{n-j}     \big) \bigg] \\ +   \bigg[ C_3 \big( \lambda_{\alpha}\big)  +  D_3 \big( \lambda_{\alpha} \big) \bigg] \bigg[ \text{ }  \underset{m,n^{\prime}: m+n^{\prime} = n-3}{\sum}   \bigg[   \text{ }    \bigg[ \text{ }   \underset{1 \leq i \leq m}{\prod} \mathrm{sin} \big( \lambda_{\alpha} - v_{n-i} \pm \eta \sigma^z_{n-j} \big)   \bigg] \text{ } \\ \times \big( \mathrm{sin} \big( 2 \eta \big) \big)^{n^{\prime}-1}   \bigg[ \text{ }   \underset{1 \leq j \leq n^{\prime}}{ \prod}  \sigma^{-,+}_{n-j}     \bigg] \text{ }         \bigg]    \text{ }    \bigg]                            \text{. } 
\end{align*}

\noindent The fourth entry of the $n$ th L-operator can be expressed in terms of,

\begin{align*}
 D_n \big( \lambda_{\alpha} \big) =     \bigg[      C_3 \big( \lambda_{\alpha} \big) + D_3 \big( \lambda_{\alpha} \big) \bigg]   \big( \mathrm{sin} \big( 2 \eta \big) \big)^{n-4}  \bigg[ \text{ }     \underset{2 \leq i \leq n-3}{\prod}     \sigma^{-,+}_{n-i}   \bigg] 
  \\ + \bigg[  C_3 \big( \lambda_{\alpha} \big)  + D_3 \big( \lambda_{\alpha } \big) \bigg]    \bigg[ \text{ } \underset{2 \leq i \leq n-3}{\prod}   \mathrm{sin} \big( \lambda_{\alpha} - v_{n-i} +    \eta \sigma^z_{n-j}     \big) \bigg]    \\ +   \bigg[ C_3 \big( \lambda_{\alpha}\big)  +  D_3 \big( \lambda_{\alpha} \big) \bigg] \bigg[ \text{ }   \underset{m,n^{\prime}: m+n^{\prime} = n-3}{\sum}   \bigg[  \text{ }     \bigg[ \text{ }   \underset{2 \leq i \leq m}{\prod} \mathrm{sin} \big( \lambda_{\alpha} - v_{n-i} \pm \eta \sigma^z_{n-j} \big)   \bigg] \text{ } \\ \times  \big( \mathrm{sin} \big( 2 \eta \big) \big)^{n^{\prime}-1}   \bigg[ \text{ }   \underset{2 \leq j \leq n^{\prime}}{ \prod}  \sigma^{-,+}_{n-j}     \bigg]   \text{ }       \bigg]   \text{ }     \bigg]    \text{. } 
\end{align*}

\noindent For the 20-vertex model, given the fact that previously defined L-operators depend upon unital associative mappings, root systems, and suitable finite dimensional-representations, the product of L-operators takes the following representation:

\bigskip

\noindent \textbf{Lemma} (\textbf{Lemma} \textit{1}, {\color{blue}[45]}, \textit{the product representation when varying one spectral parameter of the twenty-vertex model, the base case induction step for lower order terms of the three-dimensional expansion}). One has that the product,

{\small \begin{align*}
    \bigg[  \begin{smallmatrix}      q^{D^j_2}              &  q^{-2} a^j_3 q^{-D^j_3 - D^j_1}     \xi^{s-s^j_3}        &     a^j_3 a^j_1 q^{-D^j_3 - 3 D^j_1} \xi^{s-s^j_3 - s^j_1}         \\    \big( a^j_3  \big)^{\dagger} q^{D^j_3} \xi^{s^j_3}        &   q^{-D^j_3 + D^j_1} - q^{-2} q^{D_3 - D_1} \xi^s    &       - a_1 q^{D_3 - 3 D_1} \xi^{s-s_3 - s_1}  \\   0  &    a^{\dagger}_1 q^{D_3} \xi^{s_3}   &    - q^{D_1}    \\       \end{smallmatrix} \bigg]  \\ \times \bigg[ \begin{smallmatrix}     q^{D_3}     &    q^{-2} a_3 q^{-D_3 - D_3}  \xi^{s-s_3}    &    a^2_3 q^{-D_3 - 3 D_3 }    \\             a^{\dagger}_3 q^{D_3} \xi^{s_3}           &   q^{-D_3 + D_2} - q^{-2} q^{D_3 - D_2}      &    -a_3 q^{D_3 - D_3} \xi^{s-s_3}  \\ 0   &     a^{\dagger}_3 q^{D_3} \xi^{s_3}     &   q^{-D_3}         \end{smallmatrix} \bigg]  \text{, }     \end{align*} }

\noindent can be expressed in terms of the union of the span of the three subspaces,

{\tiny \[
\begin{bmatrix}
 q^{D^j_1 + D^j_2} + \big( a^j_1 \big)^{\dagger} q^{D^j_1 } \xi^{s^j_1} q^{-2} a^j_2 q^{-D^j_2 - D^j_3 } \xi^{s-s^j_2}   \\ \\   q^{D^j_1} \big( a^j_2 \big)^{\dagger} q^{D^j_2} \xi^{s^j_2} + \big( a^j_1 \big)^{\dagger}   q^{D^j_1} \xi^{s^j_1} q^{-D^j_2 + D^j_3}  - \big( a^j_1 \big)^{\dagger} q^{D^j_1 } \xi^{s^j_1} q^{-2} q^{D^j_2 - 3 d^j_3 } \xi^s           \\   \\   \big( a^j_1 \big)^{\dagger} q^{D^j_1} \xi^{s^j_1} \big( a^j_3 \big)^{\dagger} q^{D^j_3 } \xi^{s^j_3 }  \end{bmatrix} \text{, } 
\]

\begin{align*}\begin{bmatrix}
      q^{-2} a^j_1 q^{-D^j_1 - D^j_2} \xi^{s-s^j_1} q^{D^j_2} + q^{-D^j_1 + D^j_2} q^{-2} a^j_2 q^{-D^j_2 - D^j_3 } \xi^{s-s^j_2} - q^{-2} q^{D^j_1 - 3 D^j_2 } \xi^s q^{-2} a^j_2 q^{-D^j_2 - D^j_3} \\ \times  \xi^{s-s^j_2}  + \big( a^j_2 \big)^{\dagger} q^{D^j_2} \xi^{s^j_2} a^j_2 a^j_3  q^{-D^j_2 - 3 D^j_3} \xi^{s-s^j_2 - s^j_3}        \\   \\ q^{-2} a^j_1 q^{-D^j_1 - D^j_2} \xi^{s-s^j_1} \big( a^j_2 \big)^{\dagger} q^{D^j_2} \xi^{s^j_2}   + q^{-D^j_1 + D^j_2 } q^{-D^j_2 + D^j_3} - q^{-2} q^{D^j_1 - 3 D^j_2 } \xi^s  q^{-D^j_2 + D^j_3} - q^{-D^j_1 + D^j_2} \\ \times  q^{-2} q^{D^j_2 - 3 D^j_3} \xi^s   + q^{-2} q^{D^j_1 - 3 D^j_2} \xi^s q^{-2} q^{D^j_2 - 3 D^j_3} \xi^s   +  \big( a^j_2 \big)^{\dagger} q^{D^j_2} \xi^{s^j_2} \\ \times \big( - a^j_3 \big) q^{D^j_3 - 3 D^j_4} \xi^{s-s^j_3}         \\      \\        q^{-2} a^j_1 q^{-D^j_1 - D^j_2} \xi^{s-s^j_1} \big( a^j_2 \big)^{\dagger} q^{D^j_2} \xi^{s^j_2}   + q^{-D^j_1 + D^j_2 } q^{-D^j_2 + D^j_3} - q^{-2} q^{D^j_1 - 3 D^j_2 } \xi^s  q^{-D^j_2 + D^j_3} - q^{-D^j_1 + D^j_2} \\ \times  q^{-2} q^{D^j_2 - 3 D^j_3} \xi^s   + q^{-2} q^{D^j_1 - 3 D^j_2} \xi^s q^{-2} q^{D^j_2 - 3 D^j_3} \xi^s   +  \big( a^j_2 \big)^{\dagger} q^{D^j_2} \xi^{s^j_2} \big( - a^j_3 \big) q^{D^j_3 - 3 D^j_4} \xi^{s-s^j_3}      
\end{bmatrix}\end{align*}

\begin{align*}\begin{bmatrix}
              a^j_1 a^j_2 q^{-D^j_1 - 3 D^j_2} \xi^{s-s^j_1 - s^j_2} q^{D^j_2} - a^j_2 q^{D^j_2 - 3 D^j_3} \xi^{s-s^j_2} q^{-2} a^j_2 q^{-D^j_2 - D^j_3} \xi^{s-s^j_2} + q^{-D^j_2} a^j_2 a^j_3 \\ \times  q^{-D^j_2 - 3 D^j_2} \xi^{s-s^j_2 - s^j_3}       \\      \\      a^j_1 a^j_2 q^{-D^j_1 - 3 D^j_2} \xi^{s-s^j_1 - s^j_2} \big( a^j_2 \big)^{\dagger} q^{D^j_2} \xi^{s^j_2} - a^j_2 q^{D^j_2 - 3 D^j_3} \xi^{s-s^j_2} \big(  q^{-D^j_2 + D^j_3} -  q^{-2 } q^{D^j_2 - 3 D^j_3} \xi^s \big)  - q^{-D^j_2}  a^j_3 \\ \times q^{D^j_3 - 3 D^j_4} \xi^{s-s^j_3}  \\   \\      - a^j_2 q^{D^j_2 - 3 D^j_3} \xi^{s-s^j_2} \big( a^j_3 \big)^{\dagger} q^{D^j_3} \xi^{s^j_3} + q^{-D^j_2 - D^j_3}       
\end{bmatrix} \text{. } \end{align*} }

\noindent The QISM over $\textbf{T}$ for the 20-vertex model under domain-wall boundary conditions, in comparison to the counterpart of the QISM for other vertex models, is also dependent upon the asymptotic behavior of the following, possibly infinite, products over three-dimensional representations,

\begin{align*}
\underset{\underline{M} \longrightarrow + \infty}{\mathrm{lim}}  \underset{ 0 \leq \underline{j} \leq \underline{M}} {\prod}\begin{bmatrix}
  \textbf{1}^{\underline{j}} & \textbf{4}^{\underline{j}} & \textbf{7}^{\underline{j}}     \\   \textbf{2}^{\underline{j}} & \textbf{5}^{\underline{j}} & \textbf{8}^{\underline{j}}   \\ \textbf{3}^{\underline{j}} & \textbf{6}^{\underline{j}} & \textbf{9}^{\underline{j}} 
    \end{bmatrix} 
    \text{, } \\ \\  \underset{N \longrightarrow + \infty}{\mathrm{lim}} \text{ } \underset{\underline{M} \longrightarrow + \infty}{\mathrm{lim}}  \underset{ 0 \leq \underline{j} \leq \underline{M}} {\prod}  \bigg\{ \underset{ - N \leq i \leq 0} {\prod} 
 \begin{bmatrix}
  \textbf{1}^{i,\underline{j}} & \textbf{4}^{i,\underline{j}} & \textbf{7}^{i,\underline{j}}     \\   \textbf{2}^{i,\underline{j}} & \textbf{5}^{i,\underline{j}} & \textbf{8}^{i,\underline{j}}   \\ \textbf{3}^{i,\underline{j}} & \textbf{6}^{i,\underline{j}} & \textbf{9}^{i,\underline{j}} 
    \end{bmatrix} \bigg\} 
    \text{, }  \\ 
 \begin{bmatrix}
  \textbf{1}^i & \textbf{4}^i & \textbf{7}^i     \\   \textbf{2}^i & \textbf{5}^i & \textbf{8}^i   \\ \textbf{3}^i & \textbf{6}^i & \textbf{9}^i 
    \end{bmatrix}
  \begin{bmatrix}
  \textbf{1}^{i+1} & \textbf{4}^{i+1} & \textbf{7}^{i+1}     \\   \textbf{2}^{i+1} & \textbf{5}^{i+1} & \textbf{8}^{i+1}   \\ \textbf{3}^{i+1} & \textbf{6}^{i+1} & \textbf{9}^{i+1} 
    \end{bmatrix}  \text{, }  \\ \\     \begin{bmatrix}
  \textbf{1}^i & \textbf{4}^i & \textbf{7}^i     \\   \textbf{2}^i & \textbf{5}^i & \textbf{8}^i   \\ \textbf{3}^i & \textbf{6}^i & \textbf{9}^i 
    \end{bmatrix}\bigg\{ 
  \begin{bmatrix}
  \textbf{1}^{i+1} & \textbf{4}^{i+1} & \textbf{7}^{i+1}     \\   \textbf{2}^{i+1} & \textbf{5}^{i+1} & \textbf{8}^{i+1}   \\ \textbf{3}^{i+1} & \textbf{6}^{i+1} & \textbf{9}^{i+1} 
    \end{bmatrix}  \begin{bmatrix}
  \textbf{1}^{\underline{j}} & \textbf{4}^{\underline{j}} & \textbf{7}^{\underline{j}}     \\   \textbf{2}^{\underline{j}} & \textbf{5}^{\underline{j}} & \textbf{8}^{\underline{j}}   \\ \textbf{3}^{\underline{j}} & \textbf{6}^{\underline{j}} & \textbf{9}^{\underline{j}} 
    \end{bmatrix} \bigg\}  \text{, } \\  \\ \begin{bmatrix}
  \textbf{1}^i & \textbf{4}^i & \textbf{7}^i     \\   \textbf{2}^i & \textbf{5}^i & \textbf{8}^i   \\ \textbf{3}^i & \textbf{6}^i & \textbf{9}^i 
    \end{bmatrix} \bigg\{ 
  \begin{bmatrix}
  \textbf{1}^{i+1} & \textbf{4}^{i+1} & \textbf{7}^{i+1}     \\   \textbf{2}^{i+1} & \textbf{5}^{i+1} & \textbf{8}^{i+1}   \\ \textbf{3}^{i+1} & \textbf{6}^{i+1} & \textbf{9}^{i+1} 
    \end{bmatrix}  \begin{bmatrix}
  \textbf{1}^{\underline{j}} & \textbf{4}^{\underline{j}} & \textbf{7}^{\underline{j}}     \\   \textbf{2}^{\underline{j}} & \textbf{5}^{\underline{j}} & \textbf{8}^{\underline{j}}   \\ \textbf{3}^{\underline{j}} & \textbf{6}^{\underline{j}} & \textbf{9}^{\underline{j}} 
    \end{bmatrix} \bigg\}  \propto    \underset{0 \leq \underline{j} \leq \underline{M}}{\prod} \bigg\{  \underset{ -N \leq i \leq 0} {\prod} 
 \begin{bmatrix}
  \textbf{1}^{i,\underline{j}} & \textbf{4}^{i,\underline{j}} & \textbf{7}^{i,\underline{j}}     \\   \textbf{2}^{i,\underline{j}} & \textbf{5}^{i,\underline{j}} & \textbf{8}^{i,\underline{j}}   \\ \textbf{3}^{i,\underline{j}} & \textbf{6}^{i,\underline{j}} & \textbf{9}^{i,\underline{j}} 
    \end{bmatrix}  \bigg\}  \text{, } \\   \underset{ -N \leq i \leq 0} {\prod}\begin{bmatrix}
  \textbf{1}^i & \textbf{4}^i & \textbf{7}^i     \\   \textbf{2}^i & \textbf{5}^i & \textbf{8}^i   \\ \textbf{3}^i & \textbf{6}^i & \textbf{9}^i 
    \end{bmatrix} \equiv \begin{bmatrix}
  \textbf{1}^1 & \textbf{4}^1 & \textbf{7}^1     \\   \textbf{2}^1 & \textbf{5}^1 & \textbf{8}^1   \\ \textbf{3}^1 & \textbf{6}^1 & \textbf{9}^1 
    \end{bmatrix} \bigg\{  \cdots \times \begin{bmatrix}
  \textbf{1}^N & \textbf{4}^N & \textbf{7}^N     \\   \textbf{2}^N & \textbf{5}^N & \textbf{8}^N   \\ \textbf{3}^N & \textbf{6}^N & \textbf{9}^N 
    \end{bmatrix} \bigg\}  \overset{N \longrightarrow + \infty}{\longrightarrow}  \begin{bmatrix}
  \textbf{1}^{+\infty} & \textbf{4}^{+\infty} & \textbf{7}^{+\infty}     \\   \textbf{2}^{+\infty} & \textbf{5}^{+\infty} & \textbf{8}^{+\infty}   \\ \textbf{3}^{+\infty} & \textbf{6}^{+\infty} & \textbf{9}^{+\infty} 
    \end{bmatrix} \text{, } \\ \begin{bmatrix}
  \textbf{1}^1 & \textbf{4}^1 & \textbf{7}^1     \\   \textbf{2}^1 & \textbf{5}^1 & \textbf{8}^1   \\ \textbf{3}^1 & \textbf{6}^1 & \textbf{9}^1 
    \end{bmatrix} \bigg\{  \cdots \times \begin{bmatrix}
  \textbf{1}^N & \textbf{4}^N & \textbf{7}^N     \\   \textbf{2}^N & \textbf{5}^N & \textbf{8}^N   \\ \textbf{3}^N & \textbf{6}^N & \textbf{9}^N 
    \end{bmatrix} \bigg\}  \equiv \begin{bmatrix}
  \textbf{1}_N & \textbf{4}_N & \textbf{7}_N     \\   \textbf{2}_N & \textbf{5}_N & \textbf{8}_{N}   \\ \textbf{3}_{N} & \textbf{6}_{N} & \textbf{9}_{N} 
    \end{bmatrix} \text{, }  \\ \\  \underset{ 0 \leq \underline{j} \leq \underline{M}} {\prod}\begin{bmatrix}
  \textbf{1}^{\underline{j}} & \textbf{4}^{\underline{j}} & \textbf{7}^{\underline{j}}     \\   \textbf{2}^{\underline{j}} & \textbf{5}^{\underline{j}} & \textbf{8}^{\underline{j}}   \\ \textbf{3}^{\underline{j}} & \textbf{6}^{\underline{j}} & \textbf{9}^{\underline{j}} 
    \end{bmatrix} \equiv \begin{bmatrix}
  \textbf{1}^1 & \textbf{4}^1 & \textbf{7}^1     \\   \textbf{2}^1 & \textbf{5}^1 & \textbf{8}^1   \\ \textbf{3}^1 & \textbf{6}^1 & \textbf{9}^1 
    \end{bmatrix} \bigg\{  \cdots \times \begin{bmatrix}
  \textbf{1}^N & \textbf{4}^N & \textbf{7}^N     \\   \textbf{2}^N & \textbf{5}^N & \textbf{8}^N   \\ \textbf{3}^N & \textbf{6}^N & \textbf{9}^N 
    \end{bmatrix}  \bigg\} \overset{\underline{M} \longrightarrow + \infty}{\longrightarrow}  \begin{bmatrix}
  \textbf{1}^{+\infty} & \textbf{4}^{+\infty} & \textbf{7}^{+\infty}     \\   \textbf{2}^{+\infty} & \textbf{5}^{+\infty} & \textbf{8}^{+\infty}   \\ \textbf{3}^{+\infty} & \textbf{6}^{+\infty} & \textbf{9}^{+\infty} 
    \end{bmatrix}  \text{, } \\ \\   \underset{ 0 \leq \underline{j} \leq \underline{M}} {\prod} \bigg\{ \underset{ -N \leq i \leq 0} {\prod} 
 \begin{bmatrix}
  \textbf{1}^{i,\underline{j}} & \textbf{4}^{i,\underline{j}} & \textbf{7}^{i,\underline{j}}     \\   \textbf{2}^{i,\underline{j}} & \textbf{5}^{i,\underline{j}} & \textbf{8}^{i,\underline{j}}   \\ \textbf{3}^{i,\underline{j}} & \textbf{6}^{i,\underline{j}} & \textbf{9}^{i,\underline{j}} 
    \end{bmatrix} \bigg\}  \overset{N \longrightarrow + \infty}{\overset{\underline{M} \longrightarrow + \infty}{\longrightarrow} } \begin{bmatrix}
  \textbf{1}^{-\infty, + \infty} & \textbf{4}^{-\infty, + \infty} & \textbf{7}^{-\infty, + \infty}    \\   \textbf{2}^{-\infty, + \infty} & \textbf{5}^{-\infty, + \infty} & \textbf{8}^{-\infty, + \infty}  \\ \textbf{3}^{-\infty, + \infty} & \textbf{6}^{-\infty, + \infty} & \textbf{9}^{-\infty, + \infty}
    \end{bmatrix} \propto \textbf{T} \big( \underline{\lambda} \big) \text{. }         \end{align*}

\noindent For performing computations with the Poisson bracket from L-operators of the 4-vertex, 6-vertex, and 20-vertex, models alike, one can make use of the following sequences of steps in the computation:

\begin{itemize}
    \item[$\bullet$] \boxed{(1)} : one application of the \underline{bilinearity}, (BL), property, to separate terms from $\mathcal{T}$, $A$, $B$, $C$, $D$, $E$, $F$, $G$, $H$, or $I$, factors,

    \item[$\bullet$] \boxed{(2)} : one application of the \underline{Leibniz rule}, (LR), to simplify the number of terms appearing in the first argument of the Poisson bracket from two to one,

    \item[$\bullet$] \boxed{(3)} : one application of the \underline{anticommutativity}, (AC), property, to reverse the order in which the arguments appear in the superposition of two Poisson brackets obtained after an application of (LR) in the previous step,

    \item[$\bullet$] \boxed{(4)} : one application of the \underline{Leibniz rule}, (LR), to simplify the number of arguments appearing in the first argument of the first Poisson bracket from the superposition of Poisson brackets obtained in the second step,

    \item[$\bullet$] \boxed{(5)} : a final application of the \underline{Leibniz rule}, (LR), to simplify the number of arguments appearing in the first argument of the second Poisson bracket from the superposition of Poisson brackets obtained in the second step,

    \item[$\bullet$] \boxed{(6)} : taking a product over constants obtained for each Poisson bracket, to obtain each oif the $81$ constants for approximating each Poisson bracket from the three-dimensional Poisson structure.
\end{itemize}

\noindent Specifically, in the case of block representations that were previously obtained by the author for the 20-vertex model with domain-walls, {\color{blue}[45]}, computations with the Poisson bracket take the following general form:

\begin{align*}
  \bigg\{     \bigg[ \underset{k \in \textbf{N}: 1 \leq k \leq 6}{\sum}  \underline{\mathcal{T}^k_{(1,1)}}     \bigg]  \bigg[ \underset{j \in \textbf{N}: 1 \leq j \leq 5}{\sum}    \bigg[   \underset{i \in \textbf{N} :  |  i    | \leq  4 \lceil \frac{m-4}{3} \rceil }{\prod}   \underline{A_{j,i} \big( \underline{u}\big) } \bigg]   \bigg]     ,      \bigg[  \underset{k^{\prime} \in \textbf{N}: 1 \leq k \leq 6}{\sum}   
 \big( \underline{\mathcal{T}^{k^{\prime}}_{(1,1)}}\big)^{\prime}    \bigg]  \\ \times \bigg[ \underset{j^{\prime} \in \textbf{N}: 1 \leq j^{\prime} \leq 5}{\sum}    \bigg[    \underset{i \in \textbf{N} :  |  i    | \leq  4 \lceil \frac{m-4}{3} \rceil }{\prod}   \underline{A_{j^{\prime},i} \big( \underline{u}\big) } \bigg] \bigg]     \bigg\}  \\ \\ \overset{(\mathrm{BL})}{=} \underset{j^{\prime} \in \textbf{N} : 1 \leq j^{\prime} \leq 5}{\underset{k^{\prime} \in \textbf{N}: 1 \leq k \leq 6}{\underset{j \in \textbf{N}: 1 \leq j \leq 5}{\underset{k \in \textbf{N}: 1 \leq k \leq 6}{\sum}}}}  \bigg\{     \underline{\mathcal{T}^k_{(1,1)}}      \bigg[   \underset{i \in \textbf{N} :  |  i    | \leq  4 \lceil \frac{m-4}{3} \rceil }{\prod}   \underline{A_{j,i} \big( \underline{u}\big) } \bigg]        ,      
 \big( \underline{\mathcal{T}^{k^{\prime}}_{(1,1)}}\big)^{\prime}     \bigg[    \underset{i \in \textbf{N} :  |  i    | \leq  4 \lceil \frac{m-4}{3} \rceil }{\prod}   \underline{A_{j^{\prime},i} \big( \underline{u}\big) } \bigg]     \bigg\} \\  \vdots \end{align*}
 
\begin{align*}
     \overset{(\mathrm{LR})}{=} -          \underset{j^{\prime} \in \textbf{N} : 1 \leq j^{\prime} \leq 5}{\underset{k^{\prime} \in \textbf{N}: 1 \leq k \leq 6}{\underset{j \in \textbf{N}: 1 \leq j \leq 5}{\underset{k \in \textbf{N}: 1 \leq k \leq 6}{\sum}}}}                  \bigg[  \bigg\{            \big( \mathcal{T}^{i^{\prime\prime}}_{(1,1)}  \big)^{\prime}                     , \underline{\mathcal{T}^i_{(1,1)}}           \bigg\} \bigg[  {\underset{j \neq j^{\prime} \in \textbf{N} :  |  i    | , |j |  , |j^{\prime} |  \leq  4 \lceil \frac{m-4}{3} \rceil }{\prod}}                             \underline{A_{j,i} \big( \underline{u}\big) } \text{ }   \underline{A_{j^{\prime},i} \big( \underline{u}\big) }       \bigg]   +   \bigg( \big( \mathcal{T}^{i^{\prime\prime}}_{(1,1)}  \big)^{\prime}   \bigg)  \\   \times   \bigg\{                                         \underset{j \neq j^{\prime} \in \textbf{N}: |j | , | j^{\prime}| \leq 4 \lceil \frac{m-4}{3} \rceil }{\underset{i \in \textbf{N} :  |  i    | \leq  4 \lceil \frac{m-4}{3} \rceil }{\prod}}                             \underline{A_{j,i} \big( \underline{u}\big) }      , \underline{\mathcal{T}^i_{(1,1)}}           \bigg\}  \bigg[   {\underset{i,j \in \textbf{N} :  |  i    | \leq  4 \lceil \frac{m-4}{3} \rceil }{\prod}}                             \underline{A_{j,i} \big( \underline{u}\big) }  \bigg]  +   \bigg(  \underline{\mathcal{T}^i_{(1,1)}}  \bigg)   \end{align*}

\begin{figure}
\begin{align*}
\includegraphics[width=0.63\columnwidth]{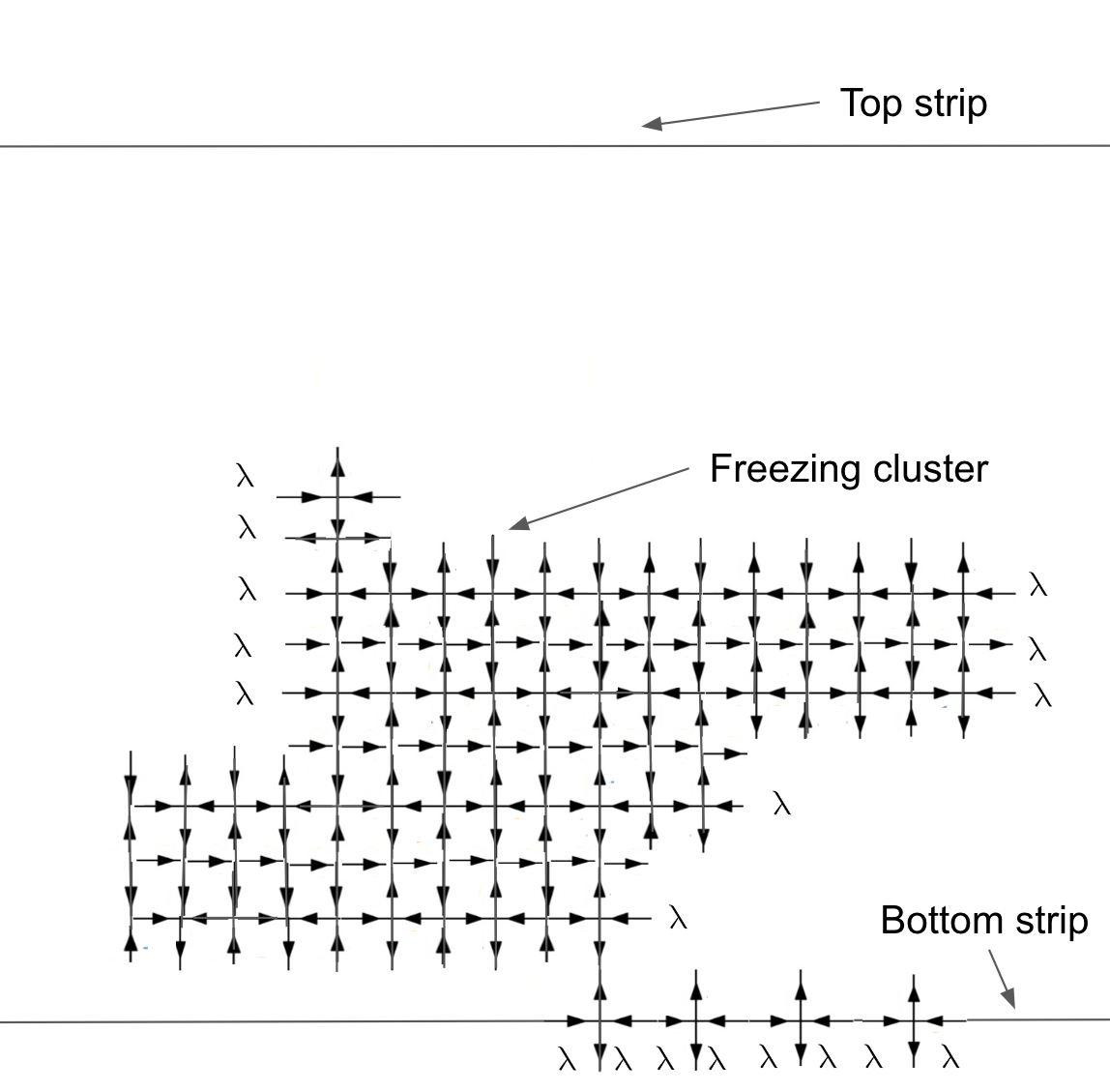}
\end{align*}
\caption{A depiction of a two-dimensional vertex configuration, reproduced from {\color{blue}[36]}, of the sloped boundary condition six-vertex model over a strip of $\textbf{Z}^2$. $\lambda$ spectral parameters are assigned to ingoing, and outgoing, edges of the configuration along the cluster of frozen faces of the height function.}
\end{figure}

    \begin{align*}  \times   \bigg[   \bigg\{   \big( \mathcal{T}^{i^{\prime\prime}}_{(1,1)}  \big)^{\prime}           ,           \underset{j \in \textbf{N} :  |  i    | \leq  4 \lceil \frac{m-4}{3} \rceil }{\prod}              \underline{A_{j,i} \big( \underline{u}\big) }            \bigg\}  \bigg[     \underset{j^{\prime} \in \textbf{N} :  |  i    | \leq  4 \lceil \frac{m-4}{3} \rceil }{\prod}              \underline{A_{j^{\prime},i} \big( \underline{u}\big) }         \bigg]   
  +    \bigg( \big( \mathcal{T}^{i^{\prime\prime}}_{(1,1)}  \big)^{\prime}   \bigg) \\ \times 
{\underset{i \neq j \neq j^{\prime} \in \textbf{N} :  |  i    | , |j |  , |j^{\prime} |  \leq  4 \lceil \frac{m-4}{3} \rceil }{\prod}}  \bigg\{                            \underline{A_{j,i} \big( \underline{u}\big) } ,       \underline{A_{j^{\prime},i} \big( \underline{u}\big) }              \bigg\}  \bigg]        \bigg]     \text{. }           
\end{align*}

\noindent The objects introduced as arguments to the Poisson brackets for the 20-vertex model introduced above include the following expressions for three-dimensional L-operators,

\begin{align*}
  \hat{L} \big( \xi \big) \equiv L^{3D}_1 =  \mathrm{exp} \big( \lambda_3 ( q^{-2 } \xi^s ) \big)    \bigg[ \begin{smallmatrix}
        q^{D_1}       &    q^{-2} a_1 q^{-D_1-D_2} \xi^{s-s_1}        &   a_1 a_2 q^{-D_1 - 3D_2} \xi^{s - s_1 - s_2}  \\ a^{\dagger}_1 q^{D_1} \xi^{s_1} 
             &      q^{-D_1 + D_2} - q^{-2} q^{D_1 -D_2} \xi^{s}     &     - a_2 q^{D_1 - 3D_2} \xi^{s-s_2}  \\ 0  &    a^{\dagger}_2 q^{D_2} \xi^{s_2} &  q^{-D_2} \\   
  \end{smallmatrix} \bigg]  \text{,} \\ \\     L^{3D}_2 =     \frac{\mathrm{exp} \big( - \lambda_3 ( q^{-2} \xi^{-s} ) \big)    }{1 - \xi^s}   \bigg[ \begin{smallmatrix}
     q^2 q^{D_1} - q^{-D_1} \xi^s   &  a_1 q^{D_1} \xi^{s_1}  &  q^{-1} a_1 a_2 \xi^{s_1 + s_2} \\                 
         a^{\dagger}_1 q^{-D_1 - D_2} \xi^{s-s_1}    &    - q^{D_1- D_2} \xi^s &    - a_2 q^{-D_2} \xi^{s_2} \\ - a_1^{\dagger} a^{\dagger}_2 q^{-D_1 - D_2} \xi^{s-s_1 - s_2} &    a^{\dagger}_2 q^{D_1 - D_2} \xi^{s-s_2} &  q^{-D_2} - q^{D_2} \xi^s \\   
  \end{smallmatrix} \bigg]       \text{,}      \\ \\                \overset{\underline{M}}{\underset{\underline{j}=0}{\prod}}  \text{ }  \overset{0}{\underset{i=-N}{\prod}}  \bigg\{ \mathrm{exp} \big( \lambda_3 ( q^{-2} \xi^{s_i} ) \big)  \bigg[   \begin{smallmatrix}     q^{D_i}       &    q^{-2} a_i q^{-D_i-D_j} \xi^{s-s_i}        &   a_i a_{j} q^{-D_i - 3D_j} \xi^{s - s_i - s_j}  \\ a^{\dagger}_i q^{D_i} \xi^{s_i} 
             &      q^{-D_i + D_j} - q^{-2} q^{D_i -D_j} \xi^{s}     &     - a_j q^{D_i - 3D_j} \xi^{s-s_j}  \\ 0  &    a^{\dagger}_j q^{D_j} \xi^{s_j} &  q^{-D_j} \\    \end{smallmatrix} \bigg]    \bigg\}                 \text{,} \\ \\          \overset{0}{\underset{i=-N}{\prod}}     \begin{bmatrix}
     \mathrm{sin} \big( \lambda_{\alpha} - v_{N-i} + \eta \sigma^z_{N-i} \big)       &    \mathrm{sin} \big( 2 \eta \big) \sigma^{-}_{N-i}    \\
      \mathrm{sin} \big( 2 \eta \big) \sigma^{+}_{N-i}     &   \mathrm{sin}  \big( \lambda_{\alpha} - v_{N-i} - \eta \sigma^z_{N-i} \big)     
  \end{bmatrix} \\    =        \overset{0}{\underset{i=-(N-1)}{\prod}}     \bigg\{  \begin{bmatrix}
         \mathrm{sin} \big( \lambda_{\alpha} - v_{N} + \eta \sigma^z_{N} \big)       &    \mathrm{sin} \big( 2 \eta \big) \sigma^{-}_{N}    \\
      \mathrm{sin} \big( 2 \eta \big) \sigma^{+}_{N}     &   \mathrm{sin}  \big( \lambda_{\alpha} - v_{N} - \eta \sigma^z_{N} \big)      
  \end{bmatrix} \\ \times   \begin{bmatrix}
         \mathrm{sin} \big( \lambda_{\alpha} - v_{N} + \eta \sigma^z_{N} \big)       &    \mathrm{sin} \big( 2 \eta \big) \sigma^{-}_{N}    \\
      \mathrm{sin} \big( 2 \eta \big) \sigma^{+}_{N}     &   \mathrm{sin}  \big( \lambda_{\alpha} - v_{N} - \eta \sigma^z_{N} \big)      
  \end{bmatrix}            \bigg\}            \text{,} \\ \\      \bigg\{    \bigg[  r^{3D}_{+} \big( u_k - u^{\prime}_k , v_k - v^{\prime}_k , w_k - w^{\prime}_k \big)   \begin{bmatrix}
 A \big( \underline{u} \big) & D \big( \underline{u} \big)  & G \big( \underline{u} \big) \\ B \big( \underline{u} \big) & E \big( \underline{u} \big) & H \big( \underline{u} \big)  \\ C \big( \underline{u} \big)  &  F \big( \underline{u} \big) & I \big( \underline{u} \big) 
\end{bmatrix}    \bigg]   \bigotimes  \begin{bmatrix}
 A \big( \underline{u^{\prime}} \big) & D \big( \underline{u^{\prime}} \big)  & G \big( \underline{u^{\prime}} \big) \\ B \big( \underline{u^{\prime}} \big) & E \big( \underline{u^{\prime}} \big) & H \big( \underline{u^{\prime}} \big)  \\ C \big( \underline{u^{\prime}} \big)  &  F \big( \underline{u^{\prime}} \big) & I \big( \underline{u^{\prime}} \big) 
\end{bmatrix}   \bigg\} \\  - \bigg\{  \begin{bmatrix}
 A \big( \underline{u} \big) & D \big( \underline{u} \big)  & G \big( \underline{u} \big) \\ B \big( \underline{u} \big) & E \big( \underline{u} \big) & H \big( \underline{u} \big)  \\ C \big( \underline{u} \big)  &  F \big( \underline{u} \big) & I \big( \underline{u} \big) 
\end{bmatrix}  \bigotimes  \bigg[ \begin{bmatrix}
 A \big( \underline{u^{\prime}} \big) & D \big( \underline{u^{\prime}} \big)  & G \big( \underline{u^{\prime}} \big) \\ B \big( \underline{u^{\prime}} \big) & E \big( \underline{u^{\prime}} \big) & H \big( \underline{u^{\prime}} \big)  \\ C \big( \underline{u^{\prime}} \big)  &  F \big( \underline{u^{\prime}} \big) & I \big( \underline{u^{\prime}} \big) 
\end{bmatrix} r^{3D}_{-} \big( u_k - u^{\prime}_k , v_k - v^{\prime}_k , w_k - w^{\prime}_k \big)          \bigg] \bigg\}       \text{,} \\ \\       \underset{\underline{M} \longrightarrow + \infty}{\underset{N \longrightarrow + \infty}{\mathrm{lim}}}  \textbf{T}^{3D}_{\underline{M},N} \big( \underline{\lambda} \big) \equiv \underset{\underline{M} \longrightarrow + \infty}{\underset{N \longrightarrow + \infty}{\mathrm{lim}}}      \bigg\{     E^{3D,\mathrm{6V}} \big(  \underline{M}  ,  v^{\prime}_k - v_k ,  u^{\prime}_k - u_k , w^{\prime}_k - w_k \big)        \textbf{T}^{3D} \big( \underline{M} , N  \big)   \\ \times   E^{3D,\mathrm{6V}} \big(  N ,  v^{\prime}_k  - v_k  ,  u^{\prime}_k - u_k   , w^{\prime}_k - w_k   \big) \bigg\}            \propto  \underset{\underline{M} \longrightarrow + \infty}{\underset{N \longrightarrow + \infty}{\mathrm{lim}}}  T^{3D}_a \big( \underline{M} , N  \big) = T^{3D}_a           \text{,} \end{align*}

\begin{align*}          \mathscr{V} \equiv    \bigg\{ v \in \textbf{T} : v \in  \underset{\textbf{T}}{\mathrm{span}} \big\{    \mathcal{T}\mathcal{C}_1   ,  \mathcal{T}\mathcal{C}_2 ,  \mathcal{T}\mathcal{C}_3           \big\} , v \not\in  \underset{\underline{j} \in \textbf{R}^2  , k \in \textbf{N}}{\mathrm{span}} \big\{ \mathcal{B}_1 ,  \mathcal{B}_2 ,  \mathcal{B}_3 \big\}  \bigg\}         \text{,} \\       \\    \Big\Updownarrow \\       \underset{1 \leq i \leq 3}{\bigcup} \mathrm{span} \big\{ \mathcal{T}\mathcal{C}_i    \big\} \subsetneq   \underset{1 \leq i \leq 3}{\bigcup} \mathrm{span} \big\{ \mathcal{R}_i    \big\}    \Longleftrightarrow       \begin{bmatrix} \vdots & \vdots & \vdots \\ \mathcal{R}_1 & \mathcal{R}_2 & \mathcal{R}_3 \\ \vdots & \vdots & \vdots
\end{bmatrix} \overset{n \longrightarrow + \infty}{\longrightarrow} \begin{bmatrix}
 A \big( \underline{u} \big) & D \big( \underline{u} \big)  & G \big( \underline{u} \big) \\ B \big( \underline{u} \big) & E \big( \underline{u} \big) & H \big( \underline{u} \big)  \\ C \big( \underline{u} \big)  &  F \big( \underline{u} \big) & I \big( \underline{u} \big) 
\end{bmatrix}  
\text{.} 
\end{align*}

\section{Computations with the Ising type L-operator}

\noindent Below, we obtain the entries of the two-dimensional representation for the transfer matrix. By making use of several observations from previous works of the author, {\color{blue}[41},{\color{blue}42},{\color{blue}45},{\color{blue}46},{\color{blue}47},{\color{blue}48]}, one may conclude that the Ising type model formulation of the 6-vertex model is exactly solvable. As in previous works of the author on two, and three, dimensional vertex models, the finite dimensional representation for the transfer matrix determines the lowest order modes of the action-angle coordinates. Besides the connections which the action-angle coordinates share with the underlying Poisson structure, the action-angle coordinates themselves also share connections with symplectic structures. Such structure incorporate, simultaneously, conditions on the Poisson bracket of the action-angle coordinates vanishing, in addition to additional Poisson brackets vanishing. The additional Poisson brackets vanishing, rather than those corresponding to the action-angle coordinates, has several geometric consequences for any model under consideration. While limit shapes, as the underlying lattice of $\textbf{Z}^2$ or $\textbf{T}$ is exhausted in weak infinite volume, can exhibit different mixing properties within the bulk, determining which universality classes to which limit shapes of a given model belong is closely related to integrable, and exactly solvable, structures. For example, the Ashkin-Teller model, which can obtained from an association, along a self-dual line of strictly positive parameters $J,U$, has $++$ boundary conditions, similar to how boundary spins can be encoded with $+$ spins in the Ising model. Such a class of boundary conditions, $\textbf{B}\textbf{C}_{++}$, is depicted in \textit{Figure 7} above.

\bigskip

\begin{figure}
\begin{align*}
\includegraphics[width=0.75\columnwidth]{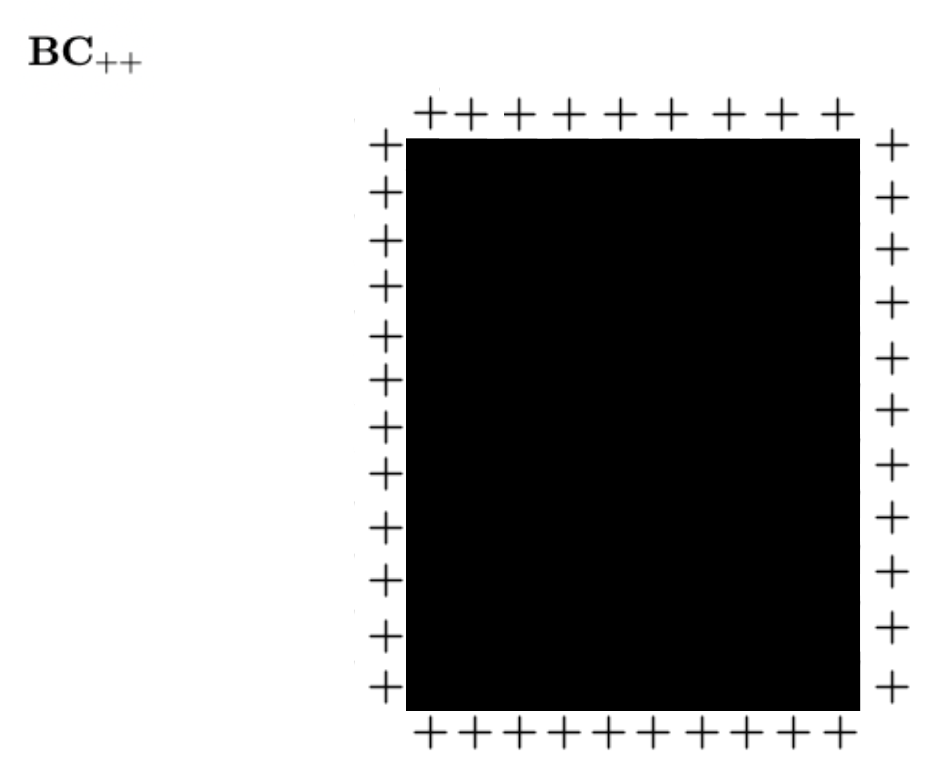}
\end{align*}
\caption{A depiction of one possible class of boundary conditions, $++$, for the Ashkin-Teller model. $+$ spins are placed adjacent to all of the vertices to the boundary of finite volumes over $\textbf{Z}^2$.}
\end{figure}

\noindent Below, we provide exact expressions for the operators in finite-dimensional representations of the transfer matrix. For convenience, we restate the following result again below:

\bigskip

\noindent \textbf{Lemma} (\textit{product representation of the transfer matrix from two, and three, L-operators of the Ising-type model}). Given the L-operator defined for the Ising-type formulation of the 6-vertex model, for the space $\mathscr{F}^{\prime}$ of functions,

\[
  \mathscr{F}^{\prime} \equiv  \left\{\!\begin{array}{ll@{}>{{}}l} 
  \mathcal{I}^1_1 \equiv \mathcal{I}^1_1\big( x_1 , \cdots, x_N, a , a+1, a^{\prime}, a^{\prime}-1
 , a^{\prime}+1  \big) , \\  \\  \mathcal{I}^2_1 \equiv \mathcal{I}^2_1\big( x_1 , \cdots, x_N, a , a+1, a^{\prime}, a^{\prime}-1
 , a^{\prime}+1  \big) , \\ \\   \mathcal{I}^3_1 \equiv \mathcal{I}^3_1\big( x_1 , \cdots, x_N, a , a+1, a^{\prime}, a^{\prime}-1
 , a^{\prime}+1  \big) , \\  \\  \mathcal{I}^4_1 \equiv \mathcal{I}^4_1\big( x_1 , \cdots, x_N, a , a+1, a^{\prime}, a^{\prime}-1
 , a^{\prime}+1  \big)  , \\ \\    \mathcal{I}^1_2 \equiv \mathcal{I}^1_2\big( x_1 , \cdots, x_N, a , a+1, a^{\prime}, a^{\prime}-1
 , a^{\prime}+1  \big) , \\ \\  \mathcal{I}^2_2 \equiv \mathcal{I}^2_2\big( x_1 , \cdots, x_N, a , a+1, a^{\prime}, a^{\prime}-1
 , a^{\prime}+1  \big) ,  \\ \\  \mathcal{I}^3_2 \equiv \mathcal{I}^3_2\big( x_1 , \cdots, x_N, a , a+1, a^{\prime}, a^{\prime}-1
 , a^{\prime}+1  \big) , \\ \\  \mathcal{I}^4_2 \equiv \mathcal{I}^4_2\big( x_1 , \cdots, x_N, a , a+1, a^{\prime}, a^{\prime}-1
 , a^{\prime}+1  \big) , \\ \\  \mathcal{I}^1_1 \mathcal{I}^1_2 + \mathcal{I}^3_1 \mathcal{I}^3_2    \equiv  \big( \mathcal{I}^1_1 \mathcal{I}^1_2 + \mathcal{I}^3_1 \mathcal{I}^3_2  \big) \big( x_1 , \cdots, x_N, a , a+1, a^{\prime}, a^{\prime}-1
 , a^{\prime}+1  \big) , \\ \\   \mathcal{I}^2_1 \mathcal{I}^2_2 + \mathcal{I}^4_1 \mathcal{I}^4_2    \equiv  \big( \mathcal{I}^2_1 \mathcal{I}^2_2 + \mathcal{I}^4_1 \mathcal{I}^4_2  \big) \big( x_1 , \cdots, x_N, a , a+1, a^{\prime}, a^{\prime}-1
 , a^{\prime}+1  \big) ,  \\ \\  \mathcal{I}^1_1 \mathcal{I}^3_2 + \mathcal{I}^3_1 \mathcal{I}^4_2    \equiv  \big( \mathcal{I}^1_1 \mathcal{I}^3_2 + \mathcal{I}^3_1 \mathcal{I}^4_2  \big) \big( x_1 , \cdots, x_N, a , a+1, a^{\prime}, a^{\prime}-1
 , a^{\prime}+1  \big) , \\  \\ \mathcal{I}^3_1 \mathcal{I}^2_2 + \mathcal{I}^4_1 \mathcal{I}^4_2    \equiv  \big( \mathcal{I}^3_1 \mathcal{I}^2_2 + \mathcal{I}^4_1 \mathcal{I}^4_2    \big) \big( x_1 , \cdots, x_N, a , a+1, a^{\prime}, a^{\prime}-1
 , a^{\prime}+1  \big) ,  \end{array}\right.
 \]

 \noindent one has the following two representations, the first two of which are,

\[  \bigg[ \begin{smallmatrix} \mathcal{I}^1_1  & \mathcal{I}^2_1   \\ \mathcal{I}^3_1   & \mathcal{I}^4_1   \end{smallmatrix} \bigg] \text{, }  \bigg[ \begin{smallmatrix} \mathcal{I}^1_2  & \mathcal{I}^2_2   \\ \mathcal{I}^3_2  & \mathcal{I}^4_2    \end{smallmatrix} \bigg] \text{, }    \]

\noindent and the third of which is,

\[   \bigg[ \begin{smallmatrix} \mathcal{I}^1_1 \mathcal{I}^1_2 + \mathcal{I}^3_1 \mathcal{I}^3_2   &  \mathcal{I}^2_1 \mathcal{I}^2_2 + \mathcal{I}^4_1 \mathcal{I}^4_2   \\ \mathcal{I}^1_1 \mathcal{I}^3_2 + \mathcal{I}^3_1 \mathcal{I}^4_2  &  \mathcal{I}^3_1 \mathcal{I}^2_2 + \mathcal{I}^4_1 \mathcal{I}^4_2    \end{smallmatrix} \bigg] ,  \]

\noindent where the entries of each representation equal,

\begin{align*}
 \mathcal{I}^1_1  \equiv  \mathcal{T}_1 + \mathcal{T}_2 \in \mathrm{span} \bigg\{ \mathcal{I}^1_1 \mathcal{I}^1_2 + \mathcal{I}^3_1 \mathcal{I}^3_2   ,       \mathcal{I}^2_1 \mathcal{I}^2_2 + \mathcal{I}^4_1 \mathcal{I}^4_2  ,  \mathcal{I}^1_1 \mathcal{I}^3_2 + \mathcal{I}^3_1 \mathcal{I}^4_2  ,   \mathcal{I}^3_1 \mathcal{I}^2_2 + \mathcal{I}^4_1 \mathcal{I}^4_2   \bigg\}  ,
\end{align*}

\noindent for,

\begin{align*}
 \mathcal{T}_1 \equiv   x_1 x_3 \delta_{a,a^{\prime}+1} x_1 x_{N-1}       \delta_{a,a^{\prime}+1} x_1 x_2 \delta_{a,a^{\prime}+1} x_1 x_N \delta_{a,a^{\prime}+1} + x_1 x_3 \delta_{a,a^{\prime}+1} x_1 x_{N-1} \delta_{a,a^{\prime}+1} q^{-a} \frac{x_1}{x_2} \\ \times  \delta_{a,a^{\prime}-1} q^a   \frac{x_1}{x_N} \delta_{a,a^{\prime}+1} \end{align*}

   \begin{align*}  \mathcal{T}_2 \equiv q^{-a} \frac{x_1}{x_3} \delta_{a,a^{\prime}-1} q^a \frac{x_1}{x_{N-1}} \delta_{a,a^{\prime}+1} x_1 x_2 \delta_{a,a^{\prime}+1}  x_1 x_N \delta_{a,a^{\prime}+1} + q^{-a} \frac{x_1}{x_3}         \delta_{a,a^{\prime}-1} q^a \frac{x_1}{x_{N-1}} \delta_{a,a^{\prime}+1} q^{-a} \frac{x_1}{x_2} \\ \times \delta_{a,a^{\prime}-1} q^a  \frac{x_1}{x_N} \delta_{a,a^{\prime}+1}       \text{, } \end{align*}

\noindent where,

\begin{align*}
  \underset{\sigma( j ) : 3 \leq j \leq N}{\prod} \bigg[  x_1 x_{\sigma ( j ) } \delta_{a,a^{\prime}+1} x_1 x_{\sigma( j ) } \delta_{a,a^{\prime}+1} x_1 x_{\sigma ( j ) } \delta_{a,a^{\prime}+1} + x_1 x_{\sigma( j ) } \delta_{a,a^{\prime}+1}  x_1 x_{\sigma ( j ) } \delta_{a,a^{\prime}+1} q^{-a} \frac{x_1}{x_{\sigma(j)}} \\ \times \delta_{a,a^{\prime}-1} q^a      \frac{x_1}{x_{\sigma(j)}} \delta_{a,a^{\prime}+1}   \bigg] \equiv \mathcal{T}_1  \text{, }
\end{align*}

\noindent and,

\begin{align*}
    \underset{\sigma( j ) : 3 \leq j \leq N}{\prod}  \bigg[ q^{-a} \frac{x_1}{x_{\sigma(j)}} \delta_{a,a^{\prime}-1} q^a \frac{x_1}{x_{\sigma(j)}} \delta_{a,a^{\prime}+1}      x_1 x_{\sigma( j )} \delta_{a,a^{\prime}+1} x_1 x_{\sigma(j)} \delta_{a,a^{\prime}+1} + q^{-a}  \frac{x_1}{x_{\sigma(j)}} \delta_{a,a^{\prime}-1} q^a \frac{x_1}{x_{\sigma(j)}}   \\ \times \delta_{a,a^{\prime}+1} q^{-a} \frac{x_1}{x_{\sigma(j)}} \delta_{a,a^{\prime}-1} q^a \frac{x_1}{x_{\sigma(j)}} \delta_{a,a^{\prime}+1}    \bigg] \equiv \mathcal{T}_2   \text{, }
\end{align*}

\noindent corresponding to the first entry,

\begin{align*}
    \mathcal{I}^2_1   \equiv \mathcal{T}_3 + \mathcal{T}_4 \in \mathrm{span} \bigg\{ \mathcal{I}^1_1 \mathcal{I}^1_2 + \mathcal{I}^3_1 \mathcal{I}^3_2   ,       \mathcal{I}^2_1 \mathcal{I}^2_2 + \mathcal{I}^4_1 \mathcal{I}^4_2  ,  \mathcal{I}^1_1 \mathcal{I}^3_2 + \mathcal{I}^3_1 \mathcal{I}^4_2  ,   \mathcal{I}^3_1 \mathcal{I}^2_2 + \mathcal{I}^4_1 \mathcal{I}^4_2   \bigg\}   ,
\end{align*}

\noindent for,

   \begin{align*}
  \mathcal{T}_3 \equiv    x_1 x_3 \delta_{a,a^{\prime}+1} q^{-a} \frac{x_1}{x_{N-1}} \delta_{a,a^{\prime}-1}     x_1 x_2 \delta_{a,a^{\prime}+1} x_1 x_N \delta_{a,a^{\prime}+1} + x_1 x_3 \delta_{a,a^{\prime}+1}    q^{-a} \frac{x_1}{x_{N-1} } \delta_{a,a^{\prime}-1}      q^{-a} \frac{x_1}{x_2} \delta_{a,a^{\prime}-1} \\ \times  q^a \frac{x_1}{x_N} \delta_{a,a^{\prime}+1} \text{,} 
  \end{align*}

  \begin{align*}
\mathcal{T}_4 \equiv   q^{-a}  \frac{x_1}{x_3} \delta_{a,a^{\prime}+1} \frac{x_1 x_{N-1}}{y^2} \delta_{a,a^{\prime}-1} x_1 x_2 \delta_{a,a^{\prime}+1} x_1 x_N \delta_{a,a^{\prime}+1}   + q^{-a} \frac{x_1}{x_3} \delta_{a,a^{\prime}+1} \frac{x_1 x_{N-1}}{y^2} \delta_{a,a^{\prime}-1} q^{-a} \frac{x_1}{x_2} \delta_{a,a^{\prime}-1} \\ \times q^a \frac{x_1}{x_N}  \delta_{a,a^{\prime}+1}           \text{, }
\end{align*}

\noindent where,

\begin{align*}
    \underset{\sigma(j): 3 \leq j \leq N}{\prod}        \bigg[  x_1 x_{\sigma(j)} \delta_{a,a^{\prime}+1 } q^{-a} \frac{x_1}{x_{\sigma(j)}}  \delta_{a,a^{\prime}-1} x_1 x_{\sigma(j)} \delta_{a,a^{\prime}+1} x_1 x_{\sigma(j)} \delta_{a,a^{\prime}+1} + x_1 x_{\sigma(j)}  \delta_{a,a^{\prime}+1} q^{-a} \frac{x_1}{x_{\sigma(j)}} \delta_{a,a^{\prime}-1} \\ \times  q^{-a} \frac{x_1}{x_{\sigma(j)}} \delta_{a,a^{\prime}-1}     q^{-a} \frac{x_1}{x_{\sigma(j)}} \delta_{a,a^{\prime}-1} q^a \frac{x_1}{x_{\sigma(j)}} \delta_{a,a^{\prime}+1}     \bigg] \equiv \mathcal{T}_3 \text{, }
\end{align*}

\noindent and,

\begin{align*}
      \underset{\sigma(j): 3 \leq j \leq N}{\prod}    \bigg[   q^{-a} \frac{x_1}{x_{\sigma(j)}} \delta_{a,a^{\prime}+1} \frac{x_1 x_{\sigma(j)}}{y^2} \delta_{a,a^{\prime}-1}   x_1 x_{\sigma(j)} \delta_{a,a^{\prime}+1} x_1 x_{\sigma(j)} \delta_{a,a^{\prime}+1}    + q^{-a} \frac{x_1}{x_{\sigma(j)}} \delta_{a,a^{\prime}+1} \frac{x_1 x_{\sigma(j)}}{y^2} \\ \times \delta_{a,a^{\prime}-1} q^{-a} \frac{x_1}{x_{\sigma(j)}} \delta_{a,a^{\prime}-1} q^a \frac{x_1}{x_{\sigma(j)}} \delta_{a,a^{\prime}+1}         \bigg] \equiv \mathcal{T}_4      \text{, }
\end{align*}

\noindent corresponding to the second entry,

\begin{align*}
 \mathcal{I}^3_1   \equiv \mathcal{T}_5 + \mathcal{T}_6 \in \mathrm{span} \bigg\{ \mathcal{I}^1_1 \mathcal{I}^1_2 + \mathcal{I}^3_1 \mathcal{I}^3_2   ,       \mathcal{I}^2_1 \mathcal{I}^2_2 + \mathcal{I}^4_1 \mathcal{I}^4_2  ,  \mathcal{I}^1_1 \mathcal{I}^3_2 + \mathcal{I}^3_1 \mathcal{I}^4_2  ,   \mathcal{I}^3_1 \mathcal{I}^2_2 + \mathcal{I}^4_1 \mathcal{I}^4_2   \bigg\}   ,
\end{align*}

\noindent for,

\begin{align*}
  \mathcal{T}_5 \equiv   x_1 x_3 \delta_{a,a^{\prime}+1} q^a \frac{x_1}{x_{N-1}}  \delta_{a,a^{\prime}+1} x_1 x_2 \delta_{a,a^{\prime}+1} x_1 x_N \delta_{a,a^{\prime}+1} + x_1 x_3 \delta_{a,a^{\prime}+1} q^a \frac{x_1}{x_{N-1} } \delta_{a,a^{\prime}+1} q^{-a} \frac{x_1}{x_2} \delta_{a,a^{\prime}-1} \\ \times q^a \frac{x_1}{x_N} \delta_{a,a^{\prime}+1} , \end{align*}

  \begin{align*} \mathcal{T}_6 \equiv q^a  \frac{x_1}{x_3} \delta_{a,a^{\prime}+1} \frac{x_1 x_{N-1}}{y^2} \delta_{a,a^{\prime}-1} x_1 x_2 \delta_{a,a^{\prime}+1} x_1 x_N \delta_{a,a^{\prime}+1}  + q^a \frac{x_1}{x_3} \delta_{a,a^{\prime}+1} \frac{x_1 x_{N-1}}{y^2} \delta_{a,a^{\prime}-1} q^{-a} \frac{x_1}{x_2} \delta_{a,a^{\prime}-1} \\ \times q^a \frac{x_1}{x_N} \delta_{a,a^{\prime}+1}   \text{, }
\end{align*}

\noindent where,

\begin{align*}
  \underset{\sigma(j): 3 \leq j \leq N}{\prod}\bigg[  x_1 x_{\sigma(j)} \delta_{a,a^{\prime}+1} q^a \frac{x_1}{x_{\sigma(j) }}  \delta_{a,a^{\prime}+1} x_1 x_{\sigma(j)} \delta_{a,a^{\prime}+1} x_1 x_{\sigma(j)}  \delta_{a,a^{\prime}+1} + x_1 x_{\sigma(j)}  \delta_{a,a^{\prime}+1} q^a \frac{x_1}{x_{\sigma(j)}  } \delta_{a,a^{\prime}+1}  \\ \times  q^{-a} \frac{x_1}{x_{\sigma(j)} }\delta_{a,a^{\prime}-1}  q^a \frac{x_1}{x_{\sigma(j)} } \delta_{a,a^{\prime}+1}  \bigg] \equiv \mathcal{T}_5   \text{, }
\end{align*}

\noindent and,

\begin{align*}
   \underset{\sigma(j): 3 \leq j \leq N}{\prod} \bigg[   q^a  \frac{x_1}{x_{\sigma(j)}} \delta_{a,a^{\prime}+1} \frac{x_1 x_{\sigma(j)}}{y^2} \delta_{a,a^{\prime}-1} x_1 x_{\sigma(j)} \delta_{a,a^{\prime}+1} x_1 x_{\sigma(j)} \delta_{a,a^{\prime}+1}  + q^a \frac{x_1}{x_{\sigma(j)}} \delta_{a,a^{\prime}+1} \frac{x_1 x_{\sigma(j)}}{y^2} \delta_{a,a^{\prime}-1} \\ \times q^{-a}  \frac{x_1}{x_{\sigma(j)}} \delta_{a,a^{\prime}-1}q^a \frac{x_1}{x_{\sigma(j)}} \delta_{a,a^{\prime}+1}\bigg] \equiv \mathcal{T}_6    \text{, }
\end{align*}

\noindent corresponding to the third entry, and, lastly,

\begin{align*}
    \mathcal{I}^4_1  \equiv  \mathcal{T}_7 + \mathcal{T}_8     \in \mathrm{span} \bigg\{ \mathcal{I}^1_1 \mathcal{I}^1_2 + \mathcal{I}^3_1 \mathcal{I}^3_2   ,       \mathcal{I}^2_1 \mathcal{I}^2_2 + \mathcal{I}^4_1 \mathcal{I}^4_2  ,  \mathcal{I}^1_1 \mathcal{I}^3_2 + \mathcal{I}^3_1 \mathcal{I}^4_2  ,   \mathcal{I}^3_1 \mathcal{I}^2_2 + \mathcal{I}^4_1 \mathcal{I}^4_2   \bigg\}     \text{, }
\end{align*}

\noindent for,

\begin{align*}
      \mathcal{T}_7 \equiv   x_1 x_3 \delta_{a,a^{\prime}+1} q^a \frac{x_1}{x_{N-1}} \delta_{a,a^{\prime}+1 } x_1 x_2 \delta_{a,a^{\prime}+1} q^{-a} \frac{x_1}{x_N} \delta_{a,a^{\prime}-1}   + x_1 x_3 \delta_{a,a^{\prime}+1} q^a \frac{x_1}{x_{N-1}} \delta_{a,a^{\prime}+1} q^{-a} \frac{x_1}{x_2} \delta_{a,a^{\prime}-1} \\ \times  \frac{x_1 x_N}{y^2} \delta_{a,a^{\prime}-1}  \end{align*}

        \begin{align*}\mathcal{T}_8 \equiv  q^a \frac{x_1}{x_3} \delta_{a,a^{\prime}+1} \frac{x_1 x_{N-1}}{y^2} \delta_{a,a^{\prime}-1} x_1 x_2 \delta_{a,a^{\prime}+1} q^{-a} \frac{x_1}{x_N} \delta_{a,a^{\prime}-1} + q^a \frac{x_1}{x_3} \delta_{a,a^{\prime}+1} \frac{x_1 x_{N-1}}{y^2} \delta_{a,a^{\prime}-1}  q^{-a} \frac{x_1}{x_2} \delta_{a,a^{\prime}-1} \\ \times   \frac{x_1 x_N}{y^2} \delta_{a,a^{\prime}-1}    \text{, }
\end{align*}

\noindent where,

\begin{align*}
    \underset{\sigma(j ) : 1 \leq j \leq N}{\prod} \bigg[ x_1 x_{\sigma(j)} \delta_{a,a^{\prime}+1} q^a \frac{x_1}{x_{\sigma(j)}} \delta_{a,a^{\prime}+1 } x_1 x_{\sigma(j)} \delta_{a,a^{\prime}+1} q^{-a} \frac{x_1}{x_{\sigma(j)}} \delta_{a,a^{\prime}-1}   + x_1 x_{\sigma(j)} \delta_{a,a^{\prime}+1} q^a \frac{x_1}{x_{\sigma(j)}} \delta_{a,a^{\prime}+1} q^{-a} \\ \times  \frac{x_1}{x_{\sigma(j)}} \delta_{a,a^{\prime}-1}  \frac{x_1 x_{\sigma(j)}}{y^2} \delta_{a,a^{\prime}-1} \bigg] \equiv \mathcal{T}_7 \text{, }
\end{align*}

\noindent and,

\begin{align*}
  \underset{\sigma(j ) : 1 \leq j \leq N}{\prod}  \bigg[  q^a \frac{x_1}{x_{\sigma(j)}} \delta_{a,a^{\prime}+1} \frac{x_1 x_{\sigma(j)}}{y^2} \delta_{a,a^{\prime}-1} x_1 x_{\sigma(j)} \delta_{a,a^{\prime}+1} q^{-a} \frac{x_1}{x_{\sigma(j)}} \delta_{a,a^{\prime}-1} + q^a \frac{x_1}{x_{\sigma(j)}} \delta_{a,a^{\prime}+1} \frac{x_1 x_{\sigma(j)}}{y^2} \delta_{a,a^{\prime}-1} \\ \times   q^{-a} \frac{x_1}{x_{\sigma(j)}} \delta_{a,a^{\prime}-1}   \frac{x_1 x_{\sigma(j)}}{y^2} \delta_{a,a^{\prime}-1} \bigg] \equiv  \mathcal{T}_8  \text{, }
\end{align*}

\noindent corresponding to the fourth entry. The collection of block operators in the second finite-dimensional representation, is spanned by,

\begin{align*}
 \mathcal{I}^1_2  \equiv  \mathcal{T}_9 + \mathcal{T}_{10} \in \mathrm{span} \bigg\{ \mathcal{I}^1_1 \mathcal{I}^1_2 + \mathcal{I}^3_1 \mathcal{I}^3_2   ,       \mathcal{I}^2_1 \mathcal{I}^2_2 + \mathcal{I}^4_1 \mathcal{I}^4_2  ,  \mathcal{I}^1_1 \mathcal{I}^3_2 + \mathcal{I}^3_1 \mathcal{I}^4_2  ,   \mathcal{I}^3_1 \mathcal{I}^2_2 + \mathcal{I}^4_1 \mathcal{I}^4_2   \bigg\}  ,
\end{align*}

\noindent is composed of,

\begin{align*}
   \mathcal{T}_9 \equiv  x_1 x_3 \delta_{a,a^{\prime}+1} q^a \frac{x_1}{x_{N-1}} \delta_{a,a^{\prime}+1} x_1 x_2 \delta_{a,a^{\prime}+1} x_1 x_2 \delta_{a,a^{\prime}+1 }  q^{-a} \frac{x_1}{x_N}  \delta_{a,a^{\prime}-1} + x_1 x_3 \delta_{a,a^{\prime}+1} q^a \frac{x_1}{x_{N-1}} \delta_{a,a^{\prime}+1} q^{-a}  \\ \times  \frac{x_1}{x_2}  \delta_{a,a^{\prime}-1}  \frac{x_1 x_N}{y^2} \delta_{a,a^{\prime}-1} \text{, } \\     \mathcal{T}_{10} \equiv  q^a \frac{x_1}{x_3} \delta_{a,a^{\prime}+1} \frac{x_1 x_{N-1}}{y^2} \delta_{a,a^{\prime}-1} x_1 x_2 \delta_{a,a^{\prime}+1}  q^{-a} \frac{x_1}{x_N} \delta_{a,a^{\prime}-1} + q^a \frac{x_1}{x_3} \delta_{a,a^{\prime}+1} \frac{x_1 x_{N-1}}{y^2} \delta_{a,a^{\prime}-1} q^{-a} \frac{x_1}{x_2} \delta_{a,a^{\prime}-1} \\ \times  \frac{x_1 x_N}{y^2} \delta_{a,a^{\prime}-1}      \text{, }
\end{align*}

\noindent where,

\begin{align*}
    \underset{\sigma(j ) : 1 \leq j \leq N}{\prod} \bigg[ q^a \frac{x_1}{x_{\sigma(j)}} \delta_{a,a^{\prime}+1} \frac{x_1 x_{\sigma(j)}}{y^2} \delta_{a,a^{\prime}-1} x_1 x_{\sigma(j)} \delta_{a,a^{\prime}+1}  q^{-a} \frac{x_1}{x_{\sigma(j)}} \delta_{a,a^{\prime}-1} + q^a \frac{x_1}{x_{\sigma(j)}} \delta_{a,a^{\prime}+1} \frac{x_1 x_{\sigma(j)}}{y^2}  \\ \times \delta_{a,a^{\prime}-1} q^{-a} \frac{x_1}{x_{\sigma(j)}} \delta_{a,a^{\prime}-1}   \frac{x_1 x_{\sigma(j)}}{y^2} \delta_{a,a^{\prime}-1}   \bigg] \equiv \mathcal{T}_{9}   \text{, } 
\end{align*}

\noindent and,

\begin{align*}
    \underset{\sigma(j ) : 1 \leq j \leq N}{\prod} \bigg[ q^a \frac{x_1}{x_3} \delta_{a,a^{\prime}+1} \frac{x_1 x_{N-1}}{y^2} \delta_{a,a^{\prime}-1} x_1 x_2 \delta_{a,a^{\prime}+1}  q^{-a} \frac{x_1}{x_N} \delta_{a,a^{\prime}-1} + q^a \frac{x_1}{x_3} \delta_{a,a^{\prime}+1} \frac{x_1 x_{N-1}}{y^2} \delta_{a,a^{\prime}-1} q^{-a} \frac{x_1}{x_2}  \\ \times  \delta_{a,a^{\prime}-1} \frac{x_1 x_N}{y^2} \delta_{a,a^{\prime}-1}  \bigg] \equiv \mathcal{T}_{10}    \text{, } 
\end{align*}

\noindent corresponding to the first term,

\begin{align*}
 \mathcal{I}^2_2  \equiv  \mathcal{T}_{11} + \mathcal{T}_{12} \in \mathrm{span} \bigg\{ \mathcal{I}^1_1 \mathcal{I}^1_2 + \mathcal{I}^3_1 \mathcal{I}^3_2   ,       \mathcal{I}^2_1 \mathcal{I}^2_2 + \mathcal{I}^4_1 \mathcal{I}^4_2  ,  \mathcal{I}^1_1 \mathcal{I}^3_2 + \mathcal{I}^3_1 \mathcal{I}^4_2  ,   \mathcal{I}^3_1 \mathcal{I}^2_2 + \mathcal{I}^4_1 \mathcal{I}^4_2   \bigg\}  ,
\end{align*}

\noindent for,

\begin{align*}
    \mathcal{T}_{11 } \equiv    q^{-a} \frac{x_1}{x_3} \delta_{a,a^{\prime}-1} q^a \frac{x_1}{x_{N-1}} \delta_{a,a^{\prime}+1} x_1 x_2 \delta_{a,a^{\prime}+1} q^{-a} \frac{x_1}{x_N} \delta_{a,a^{\prime}-1} + q^{-a} \frac{x_1}{x_3} \delta_{a,a^{\prime}-1} q^a \frac{x_1}{x_{N-1}} \delta_{a,a^{\prime}+1} q^{-a} \\ \times \frac{x_1}{x_2} \delta_{a,a^{\prime}-1} \frac{x_1 x_N}{y^2} \delta_{a,a^{\prime}-1 } , \\  \mathcal{T}_{12} \equiv \frac{x_1 x_3}{y^2} \delta_{a,a^{\prime}-1} \frac{x_1 x_{N-1}}{y^2} \delta_{a,a^{\prime}-1} x_1 x_2 \delta_{a,a^{\prime}+1 }  q^{-a} \frac{x_1}{x_N} \delta_{a,a^{\prime}-1} + \frac{x_1 x_3 }{y^2}      \delta_{a,a^{\prime}-1} \frac{x_1 x_{N-1}}{y^2} \delta_{a,a^{\prime}-1} q^{-a} \\ \times \frac{x_1}{x_2} \delta_{a,a^{\prime}-1} \frac{x_1 x_N}{y^2} \delta_{a,a^{\prime}-1}       \text{, }
\end{align*}

\noindent where,

\begin{align*}
  \underset{\sigma(j) : 1 \leq j \leq N}{\prod}  \bigg[          q^{-a} \frac{x_1}{x_{\sigma(j)}} \delta_{a,a^{\prime}-1} q^a \frac{x_1}{x_{\sigma(j)}} \delta_{a,a^{\prime}+1} x_1 x_{\sigma(j)} \delta_{a,a^{\prime}+1} q^{-a} \frac{x_1}{x_N} \delta_{a,a^{\prime}-1} + q^{-a} \frac{x_1}{x_{\sigma(j)}} \delta_{a,a^{\prime}-1} q^a \frac{x_1}{x_{\sigma(j)}} \\ \times \delta_{a,a^{\prime}+1} q^{-a}  \frac{x_1}{x_{\sigma(j)}} \delta_{a,a^{\prime}-1} \frac{x_1 x_{\sigma(j)}}{y^2} \delta_{a,a^{\prime}-1 } \bigg] \equiv \mathcal{T}_{11}  \text{, }
\end{align*}

\noindent and,

\begin{align*}
\underset{\sigma(j) : 1 \leq j \leq N}{\prod}  \bigg[  \frac{x_1 x_{\sigma(j)}}{y^2} \delta_{a,a^{\prime}-1} \frac{x_1 x_{\sigma(j)}}{y^2} \delta_{a,a^{\prime}-1} x_1 x_{\sigma(j)} \delta_{a,a^{\prime}+1 }  q^{-a} \frac{x_1}{x_{\sigma(j)}} \delta_{a,a^{\prime}-1}  + \frac{x_1 x_{\sigma(j)} }{y^2}      \delta_{a,a^{\prime}-1} \frac{x_1 x_{\sigma(j)}}{y^2} \\ \times \delta_{a,a^{\prime}-1} q^{-a} \frac{x_1}{x_{\sigma(j)}} \delta_{a,a^{\prime}-1} \frac{x_1 x_{\sigma(j)}}{y^2} \delta_{a,a^{\prime}-1}    \bigg] \equiv \mathcal{T}_{12} \text{, }
\end{align*}

\noindent corresponding to the second term,

\begin{align*}
 \mathcal{I}^3_2  \equiv  \mathcal{T}_{13} + \mathcal{T}_{14} \in \mathrm{span} \bigg\{ \mathcal{I}^1_1 \mathcal{I}^1_2 + \mathcal{I}^3_1 \mathcal{I}^3_2   ,       \mathcal{I}^2_1 \mathcal{I}^2_2 + \mathcal{I}^4_1 \mathcal{I}^4_2  ,  \mathcal{I}^1_1 \mathcal{I}^3_2 + \mathcal{I}^3_1 \mathcal{I}^4_2  ,   \mathcal{I}^3_1 \mathcal{I}^2_2 + \mathcal{I}^4_1 \mathcal{I}^4_2   \bigg\}  ,
\end{align*}

\noindent for,

\begin{align*}
       \mathcal{T}_{13} \equiv  q^{-a} \frac{x_1}{x_3} \delta_{a,a^{\prime}-1} q^a \frac{x_1}{x_{N-1}} \delta_{a,a^{\prime}+1} x_1 x_2 \delta_{a,a^{\prime}+1} x_1 x_2 \delta_{a,a^{\prime}+1 } q^a \frac{x_1}{x_N} \delta_{a,a^{\prime}+1} + q^{-a} \frac{x_1}{x_3} \delta_{a,a^{\prime}-1} q^a \frac{x_1}{x_{N-1}}  \\ \times  \delta_{a,a^{\prime}+1} q^a \frac{x_1}{x_2}  \delta_{a,a^{\prime}+1} \frac{x_1 x_N}{y^2} \delta_{a,a^{\prime}-1} , \\ \mathcal{T}_{14} \equiv  \frac{x_1 x_3}{y^2} \delta_{a,a^{\prime}-1} \frac{x_1 x_{N-1}}{y^2} \delta_{a,a^{\prime}-1} x_1 x_2 \delta_{a,a^{\prime}+1  } q^a \frac{x_1}{x_N}  \delta_{a,a^{\prime}+1} + \frac{x_1 x_3}{y^2} \delta_{a,a^{\prime}-1} \frac{x_1 x_{N-1}}{y^2} \delta_{a,a^{\prime}-1} q^a    \frac{x_1}{x_2}  \\ \times \delta_{a,a^{\prime}+1}  \frac{x_1 x_N}{y^2} \delta_{a,a^{\prime}-1}      \text{, }
\end{align*}

\noindent where,

\begin{align*}
     \underset{\sigma(j) : 1 \leq j \leq N}{\prod}  \bigg[ q^{-a} \frac{x_1}{x_{\sigma(j)}} \delta_{a,a^{\prime}-1} q^a \frac{x_1}{x_{N-1}} \delta_{a,a^{\prime}+1} x_1 x_{\sigma(j)} \delta_{a,a^{\prime}+1} x_1 x_{\sigma(j)} \delta_{a,a^{\prime}+1 } q^a \frac{x_1}{x_{\sigma(j)}} \delta_{a,a^{\prime}+1} + q^{-a} \frac{x_1}{x_{\sigma(j)}}  \\ \times  \delta_{a,a^{\prime}-1} q^a \frac{x_1}{x_{\sigma(j)}} \delta_{a,a^{\prime}+1} q^a  \frac{x_1}{x_2} \delta_{a,a^{\prime}+1} \frac{x_1 x_{\sigma(j)}}{y^2} \delta_{a,a^{\prime}-1} \bigg] \equiv \mathcal{T}_{13}  , \end{align*}

\noindent and,

\begin{align*}
   \underset{\sigma(j) : 1 \leq j \leq N}{\prod}   \bigg[  \frac{x_1 x_{\sigma(j)}}{y^2} \delta_{a,a^{\prime}-1} \frac{x_1 x_{\sigma(j)}}{y^2} \delta_{a,a^{\prime}-1} x_1 x_{\sigma(j)} \delta_{a,a^{\prime}+1  } q^a \frac{x_1}{x_{\sigma(j)}}  \delta_{a,a^{\prime}+1} + \frac{x_1 x_{\sigma(j)}}{y^2} \delta_{a,a^{\prime}-1} \frac{x_1 x_{\sigma(j)}}{y^2} \delta_{a,a^{\prime}-1} \\ \times  q^a \frac{x_1}{x_{\sigma(j)}} \delta_{a,a^{\prime}+1} \frac{x_1 x_{\sigma(j)}}{y^2} \delta_{a,a^{\prime}-1}   \bigg] \equiv \mathcal{T}_{14}     \text{, }
\end{align*}

\noindent corresponding to the third term, and,

\begin{align*}
 \mathcal{I}^4_2  \equiv  \mathcal{T}_{15} + \mathcal{T}_{16} \in \mathrm{span} \bigg\{ \mathcal{I}^1_1 \mathcal{I}^1_2 + \mathcal{I}^3_1 \mathcal{I}^3_2   ,       \mathcal{I}^2_1 \mathcal{I}^2_2 + \mathcal{I}^4_1 \mathcal{I}^4_2  ,  \mathcal{I}^1_1 \mathcal{I}^3_2 + \mathcal{I}^3_1 \mathcal{I}^4_2  ,   \mathcal{I}^3_1 \mathcal{I}^2_2 + \mathcal{I}^4_1 \mathcal{I}^4_2   \bigg\}  ,
\end{align*}

\noindent for,

\begin{align*}
     \mathcal{T}_{15} \equiv  q^{-a} \frac{x_1}{x_3} \delta_{a,a^{\prime}-1} q^a \frac{x_1}{x_{N-1}} \delta_{a,a^{\prime}+1} q^{-a} \frac{x_1}{x_2} \delta_{a,a^{\prime}+1} q^a \frac{x_1}{x_N} \delta_{a,a^{\prime}+1} + q^{-a} \frac{x_1}{x_3} \delta_{a,a^{\prime}-1} q^a \frac{x_1}{x_{N-1}} \delta_{a,a^{\prime}+1} \\ \times \frac{x_1 x_2}{y^2} \delta_{a,a^{\prime}-1}   \frac{x_1 x_N}{y^2} \delta_{a,a^{\prime}-1} ,  \\  \mathcal{T}_{16} \equiv \frac{x_1 x_3}{y^2} \delta_{a,a^{\prime}-1} \frac{x_1 x_{N-1}}{y^2} \delta_{a,a^{\prime}-1} q^{-a} \frac{x_1}{x_2} \delta_{a,a^{\prime}-1} q^a \frac{x_1}{x_N} \delta_{a,a^{\prime}+1} + \frac{x_1 x_3}{y^2} \delta_{a,a^{\prime}-1} \frac{x_1 x_{N-1}}{y^2} \delta_{a,a^{\prime}-1} \\ \times    \frac{x_1 x_2 }{y^2} \delta_{a,a^{\prime}-1}   \frac{x_1 x_N}{y^2} \delta_{a,a^{\prime}-1}   \text{, }
\end{align*}

\noindent where,

\begin{align*}
    \underset{\sigma( j ) : 1 \leq j \leq N}{\prod} \bigg[   q^{-a} \frac{x_1}{x_{\sigma( j )}} \delta_{a,a^{\prime}-1} q^a \frac{x_1}{x_{\sigma( j )}} \delta_{a,a^{\prime}+1} q^{-a} \frac{x_1}{x_{\sigma( j )}} \delta_{a,a^{\prime}+1} q^a \frac{x_1}{x_{\sigma( j )}} \delta_{a,a^{\prime}+1} + q^{-a} \frac{x_1}{x_{\sigma( j )}} \delta_{a,a^{\prime}-1} q^a \frac{x_1}{x_{\sigma( j )}}  \\ \times  \delta_{a,a^{\prime}+1}\frac{x_1 x_{\sigma( j )}}{y^2} \delta_{a,a^{\prime}-1}   \frac{x_1 x_{\sigma( j )}}{y^2} \delta_{a,a^{\prime}-1} \bigg] \equiv  \mathcal{T}_{15}   , \end{align*}

     \noindent and,

     \begin{align*} \underset{\sigma( j ) : 1 \leq j \leq N}{\prod} \bigg[ \frac{x_1 x_{\sigma( j )}}{y^2} \delta_{a,a^{\prime}-1} \frac{x_1 x_{\sigma( j )}}{y^2} \delta_{a,a^{\prime}-1} q^{-a} \frac{x_1}{x_2} \delta_{a,a^{\prime}-1} q^a \frac{x_1}{x_{\sigma( j )}} \delta_{a,a^{\prime}+1} + \frac{x_1 x_{\sigma( j )}}{y^2} \delta_{a,a^{\prime}-1} \frac{x_1 x_{\sigma( j )}}{y^2}  \\ \times  \delta_{a,a^{\prime}-1}   \frac{x_1 x_{\sigma( j )} }{y^2} \delta_{a,a^{\prime}-1}   \frac{x_1 x_{\sigma( j )}}{y^2} \delta_{a,a^{\prime}-1} \bigg] \equiv  \mathcal{T}_{16}   , 
     \end{align*}

\noindent corresponding to the fourth term.

\bigskip

\noindent In the final set of expressions below, we demonstrate, straightforwardly, how previous operators are multiplied together to obtain the entries of the final desired representation for the transfer matrix. The entries of the remaining representation, beginning with the first entry,

\begin{align*}
 \mathcal{I}^1_1 \mathcal{I}^1_2 + \mathcal{I}^3_1 \mathcal{I}^3_2 \equiv \mathcal{T}_{17} + \mathcal{T}_{18} + \mathcal{T}_{19} + \mathcal{T}_{20}    \text{, }
\end{align*}

\noindent includes,

\begin{align*}
 \mathcal{T}_{17} \equiv  \frac{y^2}{x_1 x_2} \delta_{a,a^{\prime}-1} x_1 x_3 \delta_{a,a^{\prime}+1} \frac{y^2}{x_1 x_3} \delta_{a,a^{\prime}-1} x_1 x_4 \delta_{a,a^{\prime}+1}   + \frac{y^2}{x_1 x_2} \delta_{a,a^{\prime}-1} x_1 x_3 \delta_{a,a^{\prime}+1} q^{-a} \frac{x_3}{x_1} \delta_{a,a^{\prime}+1}  q^a \\ \times \frac{x_1 }{x_4} \delta_{a,a^{\prime}+1}  \\  \mathcal{T}_{18} \equiv   q^{-a} \frac{x_2}{x_1} \delta_{a,a^{\prime}+1}q^a \frac{x_1}{x_3} \delta_{a,a^{\prime}+1} \frac{y^2}{x_1 x_3} \delta_{a,a^{\prime}-1} x_1 x_4 \delta_{a,a^{\prime}+1} + q^{-a} \frac{x_2}{x_1} \delta_{a,a^{\prime}+1} q^a \frac{x_1}{x_3} \delta_{a,a^{\prime}+1} q^{-a} \frac{x_3}{x_1} \delta_{a,a^{\prime}+1} q^a  \\ \times  \frac{x_1}{x_4} \delta_{a,a^{\prime}+1}  \end{align*}

 \noindent and,
 
 \begin{align*} \mathcal{T}_{19} \equiv   \frac{y^2}{x_1 x_3} \delta_{a,a^{\prime}-1} q^{-a} \frac{x_1}{x_4} \delta_{a,a^{\prime}-1} \frac{y^2}{x_1 x_2} \delta_{a,a^{\prime}-1} q^a \frac{x_1}{x_3} \delta_{a,a^{\prime}+1} + \frac{y^2}{x_1 x_3} \delta_{a,a^{\prime}-1} q^{-a} \frac{x_1}{x_4} \delta_{a,a^{\prime}-1} q^a \frac{x_2}{x_1} \delta_{a,a^{\prime}-1}  \\ \times  \frac{x_1 x_3}{y^2} \delta_{a,a^{\prime}-1} \\ \mathcal{T}_{20}  \equiv q^{-a} \frac{x_3}{x_1} \delta_{a,a^{\prime}+1} \frac{x_1 x_4}{y^2} \delta_{a,a^{\prime}-1} \frac{y^2}{x_1 x_2} \delta_{a,a^{\prime}-1} q^a \frac{x_1}{x_3} \delta_{a,a^{\prime}+1} + q^{-a} \frac{x_3}{x_1} \delta_{a,a^{\prime}+1} \frac{x_1 x_4}{y^2} \delta_{a,a^{\prime}-1} q^a \frac{x_2}{x_1} \delta_{a,a^{\prime}-1}  \\ \times  \frac{x_1 x_3}{y^2} \delta_{a,a^{\prime}-1}     \text{, }
\end{align*}

\noindent for,

\begin{align*}
\underset{\sigma(j) :  1 \leq j \leq N}{\prod}  \bigg[ \frac{y^2}{x_1 x_2} \delta_{a,a^{\prime}-1} x_1 x_3 \delta_{a,a^{\prime}+1} \frac{y^2}{x_1 x_3} \delta_{a,a^{\prime}-1} x_1 x_4 \delta_{a,a^{\prime}+1}   + \frac{y^2}{x_1 x_2} \delta_{a,a^{\prime}-1} x_1 x_3 \delta_{a,a^{\prime}+1} q^{-a} \frac{x_3}{x_1} \delta_{a,a^{\prime}+1}  q^a \\ \times \frac{x_1 }{x_4} \delta_{a,a^{\prime}+1} \bigg] \equiv \mathcal{T}_{17} ,  \end{align*}

\begin{align*}  \underset{\sigma(j) :  1 \leq j \leq N}{\prod} \bigg[ q^{-a} \frac{x_2}{x_1} \delta_{a,a^{\prime}+1}q^a \frac{x_1}{x_3} \delta_{a,a^{\prime}+1} \frac{y^2}{x_1 x_3} \delta_{a,a^{\prime}-1} x_1 x_4 \delta_{a,a^{\prime}+1} + q^{-a} \frac{x_2}{x_1} \delta_{a,a^{\prime}+1} q^a \frac{x_1}{x_3} \delta_{a,a^{\prime}+1} q^{-a} \frac{x_3}{x_1} \delta_{a,a^{\prime}+1} \\ \times  q^a   \frac{x_1}{x_4} \delta_{a,a^{\prime}+1}\bigg] \equiv \mathcal{T}_{18} , \end{align*}

\noindent and,

\begin{align*} \underset{\sigma(j) :  1 \leq j \leq N}{\prod} \bigg[  \frac{y^2}{x_1 x_3} \delta_{a,a^{\prime}-1} q^{-a} \frac{x_1}{x_4} \delta_{a,a^{\prime}-1} \frac{y^2}{x_1 x_2} \delta_{a,a^{\prime}-1} q^a \frac{x_1}{x_3} \delta_{a,a^{\prime}+1} + \frac{y^2}{x_1 x_3} \delta_{a,a^{\prime}-1} q^{-a} \frac{x_1}{x_4} \delta_{a,a^{\prime}-1} q^a \frac{x_2}{x_1}  \\ \times    \delta_{a,a^{\prime}-1}   \frac{x_1 x_3}{y^2} \delta_{a,a^{\prime}-1}\bigg] \equiv \mathcal{T}_{19} ,  \\  \underset{\sigma(j) :  1 \leq j \leq N}{\prod} \bigg[  q^{-a} \frac{x_3}{x_1} \delta_{a,a^{\prime}+1} \frac{x_1 x_4}{y^2} \delta_{a,a^{\prime}-1} \frac{y^2}{x_1 x_2} \delta_{a,a^{\prime}-1} q^a \frac{x_1}{x_3} \delta_{a,a^{\prime}+1} + q^{-a} \frac{x_3}{x_1} \delta_{a,a^{\prime}+1} \frac{x_1 x_4}{y^2} \delta_{a,a^{\prime}-1} q^a \frac{x_2}{x_1} \\ \times  \delta_{a,a^{\prime}-1}   \frac{x_1 x_3}{y^2} \delta_{a,a^{\prime}-1} \bigg] \equiv \mathcal{T}_{20}     \text{, }   \end{align*}

\noindent The second entry,

\begin{align*}        \mathcal{I}^2_1 \mathcal{I}^2_2 + \mathcal{I}^4_1 \mathcal{I}^4_2  \equiv \mathcal{T}_{21} + \mathcal{T}_{22} + \mathcal{T}_{23} + \mathcal{T}_{24}  \text{, }
\end{align*}

\noindent includes,

\begin{align*}
   \mathcal{T}_{21} 
 \equiv   \frac{y^2}{x_1 x_2} \delta_{a,a^{\prime}-1} x_1 x_3 \delta_{a,a^{\prime}+1} \frac{y^2}{x_1 x_3} \delta_{a,a^{\prime}-1} q^{-a} \frac{x_1}{x_4} \delta_{a,a^{\prime}-1} + \frac{y^2}{x_1 x_2} \delta_{a,a^{\prime}-1} x_1 x_3 \delta_{a,a^{\prime}+1} q^{-a} \frac{x_3}{x_1} \delta_{a,a^{\prime}+1} \\ \times \frac{x_1 x_4}{y^2} \delta_{a,a^{\prime}-1} \\  \mathcal{T}_{22} 
 \equiv   q^{-a} \frac{x_2}{x_1} \delta_{a,a^{\prime}+1} q^a \frac{x_1}{x_3} \delta_{a,a^{\prime}+1} \frac{y^2}{x_1 x_3} \delta_{a,a^{\prime}-1} q^{-a} \frac{x_1}{x_4} \delta_{a,a^{\prime}-1} + q^{-a} \frac{x_2}{x_1} \delta_{a,a^{\prime}+1} q^a \frac{x_1}{x_3} \delta_{a,a^{\prime}+1} q^{-a} \frac{x_3}{x_1} \delta_{a,a^{\prime}+1} \\ \times  \frac{x_1 x_4}{y^2} \delta_{a,a^{\prime}-1} \end{align*}

 \noindent and, 
 
 \begin{align*} \mathcal{T}_{23} \equiv  \frac{y^2}{x_1 x_2} \delta_{a,a^{\prime}-1} q^{-a} \frac{x_1}{x_3} \delta_{a,a^{\prime}-1} q^{-a} \frac{x_3}{x_1} \delta_{a,a^{\prime}+1} q^a \frac{x_1}{x_4} \delta_{a,a^{\prime}+1}   + \frac{y^2}{x_1 x_2} \delta_{a,a^{\prime}-1} q^{-a} \frac{x_1}{x_3} \delta_{a,a^{\prime}-1} \frac{y^2}{x_1 x_4} \delta_{a,a^{\prime}+1} \end{align*}

 \begin{align*} \times  \frac{x_1 x_4}{y^2} \delta_{a,a^{\prime}-1} \\ \mathcal{T}_{24} \equiv  q^{-a} \frac{x_2}{x_1} \delta_{a,a^{\prime}+1} \frac{x_1 x_3}{y^2} \delta_{a,a^{\prime}-1} q^{-a} \frac{x_3}{x_1} \delta_{a,a^{\prime}+1 }     q^a \frac{x_1}{x_4} \delta_{a,a^{\prime}+1} + q^{-a} \frac{x_2}{x_1} \delta_{a,a^{\prime}+1} \frac{x_1 x_3}{y^2} \delta_{a,a^{\prime}-1} \frac{y^2}{x_1 x_4}   \delta_{a,a^{\prime}+1} \\ \times  \frac{x_1 x_4}{y^2} \delta_{a,a^{\prime}-1}        \text{, }
\end{align*}

\noindent for,

\begin{align*}
\underset{\sigma(j) :  1 \leq j \leq N}{\prod}    \bigg[ \frac{y^2}{x_1 x_2} \delta_{a,a^{\prime}-1} x_1 x_3 \delta_{a,a^{\prime}+1} \frac{y^2}{x_1 x_3} \delta_{a,a^{\prime}-1} q^{-a} \frac{x_1}{x_4} \delta_{a,a^{\prime}-1} + \frac{y^2}{x_1 x_2} \delta_{a,a^{\prime}-1} x_1 x_3 \delta_{a,a^{\prime}+1} q^{-a} \frac{x_3}{x_1} \\ \times \delta_{a,a^{\prime}+1}  \frac{x_1 x_4}{y^2} \delta_{a,a^{\prime}-1} \bigg]     \equiv \mathcal{T}_{21} ,
  \\ \underset{\sigma(j) :  1 \leq j \leq N}{\prod}   \bigg[  q^{-a} \frac{x_2}{x_1} \delta_{a,a^{\prime}+1} q^a \frac{x_1}{x_3} \delta_{a,a^{\prime}+1} \frac{y^2}{x_1 x_3} \delta_{a,a^{\prime}-1} q^{-a} \frac{x_1}{x_4} \delta_{a,a^{\prime}-1} + q^{-a} \frac{x_2}{x_1} \delta_{a,a^{\prime}+1} q^a \frac{x_1}{x_3} \delta_{a,a^{\prime}+1} q^{-a} \frac{x_3}{x_1} \\ \times  \delta_{a,a^{\prime}+1}  \frac{x_1 x_4}{y^2} \delta_{a,a^{\prime}-1} \bigg]  \equiv \mathcal{T}_{22} ,
 \end{align*}
 
 \noindent and,
 
 \begin{align*} \underset{\sigma(j) :  1 \leq j \leq N}{\prod}   \bigg[ \frac{y^2}{x_1 x_2} \delta_{a,a^{\prime}-1} q^{-a} \frac{x_1}{x_3} \delta_{a,a^{\prime}-1} q^{-a} \frac{x_3}{x_1} \delta_{a,a^{\prime}+1} q^a \frac{x_1}{x_4} \delta_{a,a^{\prime}+1}   + \frac{y^2}{x_1 x_2} \delta_{a,a^{\prime}-1} q^{-a} \frac{x_1}{x_3} \delta_{a,a^{\prime}-1} \frac{y^2}{x_1 x_4} \\ \times  \delta_{a,a^{\prime}+1}  \frac{x_1 x_4}{y^2} \delta_{a,a^{\prime}-1} \bigg]  \equiv  \mathcal{T}_{23} , \\ \underset{\sigma(j) :  1 \leq j \leq N}{\prod}   \bigg[  q^{-a} \frac{x_2}{x_1} \delta_{a,a^{\prime}+1} \frac{x_1 x_3}{y^2} \delta_{a,a^{\prime}-1} q^{-a} \frac{x_3}{x_1} \delta_{a,a^{\prime}+1 }     q^a \frac{x_1}{x_4} \delta_{a,a^{\prime}+1} + q^{-a} \frac{x_2}{x_1} \delta_{a,a^{\prime}+1} \frac{x_1 x_3}{y^2} \delta_{a,a^{\prime}-1} \frac{y^2}{x_1 x_4}   \\ \times  \delta_{a,a^{\prime}+1}  \frac{x_1 x_4}{y^2} \delta_{a,a^{\prime}-1}     \bigg] \equiv \mathcal{T}_{24}     \text{. }
\end{align*}

\noindent The third entry,

\begin{align*}
\mathcal{I}^1_1 \mathcal{I}^3_2 + \mathcal{I}^3_1 \mathcal{I}^4_2 \equiv         \mathcal{T}_{25} + \mathcal{T}_{26} + \mathcal{T}_{27} + \mathcal{T}_{28}   \text{, }
\end{align*}

\noindent includes,

\begin{align*}
       \mathcal{T}_{25}  \equiv   \frac{y^2}{x_1 x_2} \delta_{a,a^{\prime}-1} q^a \frac{x_1 }{x_3} \delta_{a,a^{\prime}+1} \frac{y^2}{x_1 x_3} \delta_{a,a^{\prime}-1} q^{-a} \frac{x_1}{x_4} \delta_{a,a^{\prime}-1} +  \frac{y^2}{x_1 x_2} \delta_{a,a^{\prime}-1} q^a \frac{x_1}{x_3} \delta_{a,a^{\prime}+1} q^{-a} \frac{x_3}{x_1}   \delta_{a,a^{\prime}+1} \\ \times \frac{x_1 x_4}{y^2}  \delta_{a,a^{\prime}-1} \\  \mathcal{T}_{26} \equiv   q^a \frac{x_2}{x_1} \delta_{a,a^{\prime}-1} \frac{x_1 x_3}{y^2} \delta_{a,a^{\prime}-1} \frac{y^2}{x_1 x_3} \delta_{a,a^{\prime}-1} q^{-a} \frac{x_1}{x_4} \delta_{a,a^{\prime}-1}      + q^a \frac{x_2}{x_1} \delta_{a,a^{\prime}-1} \frac{x_1 x_3}{y^2} \delta_{a,a^{\prime}-1} q^{-a} \frac{x_3}{x_1} \delta_{a,a^{\prime}+1} \\ \times  \frac{x_1 x_4}{y^2} \delta_{a,a^{\prime}-1}    \end{align*}

       \noindent and,

       \begin{align*} \mathcal{T}_{27} \equiv  q^{-a} \frac{x_2}{x_1} \delta_{a,a^{\prime}+1} q^a \frac{x_1}{x_3} \delta_{a,a^{\prime}+1} q^{-a} \frac{x_3}{x_1} \delta_{a,a^{\prime}+1} q^a \frac{x_1}{x_4} \delta_{a,a^{\prime}+1} + q^{-a} \frac{x_2}{x_1} \delta_{a,a^{\prime}+1} q^a \frac{x_1}{x_3} \delta_{a,a^{\prime}+1}    \frac{y^2}{x_1 x_4} \delta_{a,a^{\prime}+1}\\ \times  \frac{x_1 x_4}{y^2} \delta_{a,a^{\prime}-1} \\ \mathcal{T}_{28} \equiv   \frac{y^2}{x_1 x_3} \delta_{a,a^{\prime}_+1} \frac{x_1 x_3}{y^2} \delta_{a,a^{\prime}-1} q^{-a} \frac{x_3}{x_1} \delta_{a,a^{\prime}+1} q^a \frac{x_1}{x_4} \delta_{a,a^{\prime}+1}  +    \frac{y^2}{x_1 x_3} \delta_{a,a^{\prime}+1} \frac{x_1 x_3}{y^2} \delta_{a,a^{\prime}-1} \frac{y^2}{x_1 x_4 } \delta_{a,a^{\prime}+1} \\ \times \frac{x_1 x_4}{y^2} \delta_{a,a^{\prime}-1}          \text{, }
\end{align*}

\noindent for,

\begin{align*}
       \underset{\sigma(j) :  1 \leq j \leq N}{\prod}   \bigg[   \frac{y^2}{x_1 x_2} \delta_{a,a^{\prime}-1} q^a \frac{x_1 }{x_3} \delta_{a,a^{\prime}+1} \frac{y^2}{x_1 x_3} \delta_{a,a^{\prime}-1} q^{-a} \frac{x_1}{x_4} \delta_{a,a^{\prime}-1} +  \frac{y^2}{x_1 x_2} \delta_{a,a^{\prime}-1} q^a \frac{x_1}{x_3} \delta_{a,a^{\prime}+1} q^{-a} \frac{x_3}{x_1}   \delta_{a,a^{\prime}+1} \\ \times \frac{x_1 x_4}{y^2}  \delta_{a,a^{\prime}-1}  \bigg]  \equiv  \mathcal{T}_{25} ,   \\  \underset{\sigma(j) :  1 \leq j \leq N}{\prod}   \bigg[  q^a \frac{x_2}{x_1} \delta_{a,a^{\prime}-1} \frac{x_1 x_3}{y^2} \delta_{a,a^{\prime}-1} \frac{y^2}{x_1 x_3} \delta_{a,a^{\prime}-1} q^{-a} \frac{x_1}{x_4} \delta_{a,a^{\prime}-1}      + q^a \frac{x_2}{x_1} \delta_{a,a^{\prime}-1} \frac{x_1 x_3}{y^2} \delta_{a,a^{\prime}-1} q^{-a} \frac{x_3}{x_1} \delta_{a,a^{\prime}+1} \\ \times  \frac{x_1 x_4}{y^2} \delta_{a,a^{\prime}-1}   \bigg]  \equiv   \mathcal{T}_{26}, \end{align*}
       
\noindent and,

       \begin{align*}   \underset{\sigma(j) :  1 \leq j \leq N}{\prod}   \bigg[   q^{-a} \frac{x_2}{x_1} \delta_{a,a^{\prime}+1} q^a \frac{x_1}{x_3} \delta_{a,a^{\prime}+1} q^{-a} \frac{x_3}{x_1} \delta_{a,a^{\prime}+1} q^a \frac{x_1}{x_4} \delta_{a,a^{\prime}+1} + q^{-a} \frac{x_2}{x_1} \delta_{a,a^{\prime}+1} q^a \frac{x_1}{x_3} \delta_{a,a^{\prime}+1}    \frac{y^2}{x_1 x_4} \end{align*}

       \begin{align*} \times  \delta_{a,a^{\prime}+1} \frac{x_1 x_4}{y^2} \delta_{a,a^{\prime}-1} \bigg]  \equiv \mathcal{T}_{27} \\ \underset{\sigma(j) :  1 \leq j \leq N}{\prod}   \bigg[     \frac{y^2}{x_1 x_3} \delta_{a,a^{\prime}_+1} \frac{x_1 x_3}{y^2} \delta_{a,a^{\prime}-1} q^{-a} \frac{x_3}{x_1} \delta_{a,a^{\prime}+1} q^a \frac{x_1}{x_4} \delta_{a,a^{\prime}+1}  +    \frac{y^2}{x_1 x_3} \delta_{a,a^{\prime}+1} \frac{x_1 x_3}{y^2} \delta_{a,a^{\prime}-1} \frac{y^2}{x_1 x_4 } \delta_{a,a^{\prime}+1} \\ \times \frac{x_1 x_4}{y^2} \delta_{a,a^{\prime}-1}   \bigg] \equiv  \mathcal{T}_{28}      \text{. }
\end{align*}

\noindent The fourth entry, 

\begin{align*}
\mathcal{I}^3_1 \mathcal{I}^2_2 + \mathcal{I}^4_1 \mathcal{I}^4_2 \equiv \mathcal{T}_{29} + \mathcal{T}_{30} + \mathcal{T}_{31} + \mathcal{T}_{32} \text{, }
\end{align*}

\noindent includes,

\begin{align*}
        \mathcal{T}_{29}  \equiv   \frac{y^2}{x_1 x_2} \delta_{a,a^{\prime}-1} \frac{y^2}{x_1 x_3} \delta_{a,a^{\prime}-1} \frac{y^2}{x_1 x_3} \delta_{a,a^{\prime}-1} \frac{y^2}{x_1 x_4} \delta_{a,a^{\prime}-1} +      \frac{y^2}{x_1 x_2} \delta_{a,a^{\prime}-1} \frac{y^2}{x_1 x_3} \delta_{a,a^{\prime}-1} q^{-a} \frac{x_3}{x_1} \delta_{a,a^{\prime}+1} q^a \\ \times \frac{x_4}{x_1} \delta_{a,a^{\prime}-1} \\  \mathcal{T}_{30} \equiv  q^{-a } \frac{x_2}{x_1} \delta_{a,a^{\prime}+1} q^a \frac{x_3}{x_1} \delta_{a,a^{\prime}-1}   \frac{y^2}{x_1 x_3} \delta_{a,a^{\prime}-1} \frac{y^2}{x_1 x_4} \delta_{a,a^{\prime}-1} + q^{-a} \frac{x_2}{x_1} \delta_{a,a^{\prime}+1} q^a \frac{x_3}{x_1} \delta_{a,a^{\prime}-1} q^{-a} \frac{x_3}{x_1} \delta_{a,a^{\prime}+1} q^a \\ \times   \frac{x_4}{x_1} \delta_{a,a^{\prime}-1} , \end{align*}

        \noindent and,

        \begin{align*} \mathcal{T}_{31}  \equiv    \frac{y^2}{x_1 x_2} \delta_{a,a^{\prime}-1} q^{-a} \frac{x_3}{x_1 } \delta_{a,a^{\prime}+1} \frac{y^2}{x_1 x_3} \delta_{a,a^{\prime}-1} q^a \frac{x_4}{x_1 } \delta_{a,a^{\prime}+1} + \frac{y^2}{x_1 x_2} \delta_{a,a^{\prime}-1} q^{-a} \frac{x_3}{x_1} \delta_{a,a^{\prime}+1}     q^a \frac{x_3}{x_1} \delta_{a,a^{\prime}-1} \\ \times  \frac{y^2}{x_1 x_4} \delta_{a,a^{\prime}+1} \\  \mathcal{T}_{32}  \equiv q^{-a} \frac{x_2}{x_1} \delta_{a,a^{\prime}+1} \frac{y^2}{x_1 x_3} \delta_{a,a^{\prime}+1}    \frac{y^2}{x_1 x_3} \delta_{a,a^{\prime}-1} q^a \frac{x_4}{x_1} \delta_{a,a^{\prime}+1} + q^{-a} \frac{x_2}{x_1} \delta_{a,a^{\prime}+1 } \frac{y^2}{x_1 x_3} \delta_{a,a^{\prime}+1} q^a \frac{x_3}{x_1} \delta_{a,a^{\prime}-1}  \\ \times  \frac{y^2}{x_1 x_4} \delta_{a,a^{\prime}+1}                 \text{, }
\end{align*}

\noindent for,

\begin{align*}
  \underset{\sigma(j) :  1 \leq j \leq N}{\prod}   \bigg[    \frac{y^2}{x_1 x_2} \delta_{a,a^{\prime}-1} \frac{y^2}{x_1 x_3} \delta_{a,a^{\prime}-1} \frac{y^2}{x_1 x_3} \delta_{a,a^{\prime}-1} \frac{y^2}{x_1 x_4} \delta_{a,a^{\prime}-1} +      \frac{y^2}{x_1 x_2} \delta_{a,a^{\prime}-1} \frac{y^2}{x_1 x_3} \delta_{a,a^{\prime}-1} q^{-a} \frac{x_3}{x_1}  \\ \times \delta_{a,a^{\prime}+1} q^a  \frac{x_4}{x_1} \delta_{a,a^{\prime}-1} \bigg] \equiv       \mathcal{T}_{29}    ,  \\ \underset{\sigma(j) :  1 \leq j \leq N}{\prod}   \bigg[    q^{-a } \frac{x_2}{x_1} \delta_{a,a^{\prime}+1} q^a \frac{x_3}{x_1} \delta_{a,a^{\prime}-1}   \frac{y^2}{x_1 x_3} \delta_{a,a^{\prime}-1} \frac{y^2}{x_1 x_4} \delta_{a,a^{\prime}-1} + q^{-a} \frac{x_2}{x_1} \delta_{a,a^{\prime}+1} q^a \frac{x_3}{x_1} \delta_{a,a^{\prime}-1} q^{-a} \\ \ \times  \frac{x_3}{x_1} \delta_{a,a^{\prime}+1} q^a   \frac{x_4}{x_1} \delta_{a,a^{\prime}-1} \bigg] \equiv \mathcal{T}_{30}   , \end{align*}

  \noindent and,

  \begin{align*} \underset{\sigma(j) :  1 \leq j \leq N}{\prod}   \bigg[   \frac{y^2}{x_1 x_2} \delta_{a,a^{\prime}-1} q^{-a} \frac{x_3}{x_1 } \delta_{a,a^{\prime}+1} \frac{y^2}{x_1 x_3} \delta_{a,a^{\prime}-1} q^a \frac{x_4}{x_1 } \delta_{a,a^{\prime}+1} + \frac{y^2}{x_1 x_2} \delta_{a,a^{\prime}-1} q^{-a} \frac{x_3}{x_1} \delta_{a,a^{\prime}+1}   \\ \times   q^a \frac{x_3}{x_1} \delta_{a,a^{\prime}-1}  \frac{y^2}{x_1 x_4} \delta_{a,a^{\prime}+1} \bigg] \equiv \mathcal{T}_{31} \\  \underset{\sigma(j) :  1 \leq j \leq N}{\prod}   \bigg[    q^{-a} \frac{x_2}{x_1} \delta_{a,a^{\prime}+1} \frac{y^2}{x_1 x_3} \delta_{a,a^{\prime}+1}    \frac{y^2}{x_1 x_3} \delta_{a,a^{\prime}-1} q^a \frac{x_4}{x_1} \delta_{a,a^{\prime}+1} + q^{-a} \frac{x_2}{x_1} \delta_{a,a^{\prime}+1 } \frac{y^2}{x_1 x_3} \delta_{a,a^{\prime}+1} q^a  \\ \times  \frac{x_3}{x_1} \delta_{a,a^{\prime}-1}   \frac{y^2}{x_1 x_4} \delta_{a,a^{\prime}+1}     \bigg] \equiv \mathcal{T}_{32}               \text{, }
\end{align*}

\noindent \textit{Proof of Lemma}. The desired entries for each term in the two representations for the transfer matrix can be obtained from straightforward computations. To begin, observe that the first desired entry from the representation,

\[  \bigg[ \begin{smallmatrix} \mathcal{I}^1_1  & \mathcal{I}^2_1   \\ \mathcal{I}^3_1   & \mathcal{I}^4_1   \end{smallmatrix} \bigg] \text{, }   \]

\noindent can be obtained from,

\begin{align*}
  \big[ x_1 x_3 \delta_{a,a^{\prime}+1} \frac{y^2}{x_1 x_4} \delta_{a,a^{\prime}-1}  + q^{-a} \frac{x_1}{x_3} \delta_{a,a^{\prime}-1}  q^a \frac{x_4}{x_1} \delta_{a,a^{\prime}-1 } \big] \big[    x_1 x_2 \delta_{a,a^{\prime}+1 } \frac{y^2}{x_1 x_3} \delta_{a,a^{\prime}-1} + q^{-a} \frac{x_1}{x_2} \delta_{a,a^{\prime}-1} q^a \\ \times  \frac{x_3}{x_1} \delta_{a,a^{\prime}-1}         \big]  \text{. }
\end{align*}

\noindent Along similar lines, one can obtain remaining entries of the first representation from the product of two L-operators for the Ising-type model from,

\begin{align*}
     \big[ q^a \frac{x_1}{x_3} \delta_{a,a^{\prime}+1} \frac{y^2}{x_1 x_4} \delta_{a,a^{\prime}-1} + \frac{x_1 x_3}{y^2}               \delta_{a,a^{\prime}-1}  q^a \frac{x_4}{x_1} \delta_{a,a^{\prime}-1}     \big]  \big[ x_1 x_2 \delta_{a,a^{\prime}+1}  q^{-a}    \frac{x_3}{x_1} \delta_{a,a^{\prime}+1} + q^{-a} \frac{x_1}{x_2} \delta_{a,a^{\prime}-1} \\ \times \frac{y^2}{x_1 x_3} \delta_{a,a^{\prime}+1}   \big] \text{, }  \\ \\   \big[      x_1 x_2 \delta_{a,a^{\prime}+1}  \frac{y^2}{x_1 x_3} \delta_{a,a^{\prime}-1}  + q^{-a} \frac{x_1}{x_2}  \delta_{a,a^{\prime}-1} q^a \frac{x_3}{x_1} \delta_{a,a^{\prime}-1}   \big] \big[  x_1 x_3 \delta_{a,a^{\prime}+1}  q^a \frac{x_4}{x_1} \delta_{a,a^{\prime}+1}  +  q^{-a} \frac{x_1}{x_3} \delta_{a,a^{\prime}-1} \end{align*}
     
     \begin{align*} \times  \frac{y^2}{x_1 x_4 } \delta_{a,a^{\prime}+1 }   \big]     \text{, } \\ \big[   x_1 x_2 \delta_{a,a^{\prime}+1} q^{-a} \frac{x_3}{x_1} \delta_{a,a^{\prime}+1}      + q^{-a} \frac{x_1}{x_3} \delta_{a,a^{\prime}-1} \frac{y^2}{x_1 x_3} \delta_{a,a^{\prime}+1}  \big] \big[ q^a \frac{x_1}{x_3} \delta_{a,a^{\prime}+1} q^{-a} \frac{x_4}{x_1} \delta_{a,a^{\prime}+1}  +  \frac{x_1 x_3}{y^2} \delta_{a,a^{\prime}-1}   \\ \times  q^a   \frac{x_4}{x_1} \delta_{a,a^{\prime}-1}       \big] \text{, } \\ \\  \big[       x_1 x_2 \delta_{a,a^{\prime}+1} \frac{y^2}{x_1 x_3} \delta_{a,a^{\prime}-1}   +  q^{-a} \frac{x_1}{x_2} \delta_{a,a^{\prime}-1} q^a \frac{x_3}{x_1} \delta_{a,a^{\prime}+1}       \big] \big[ q^a \frac{x_1}{x_3} \delta_{a,a^{\prime}+1}  \frac{y^2}{x_1 x_4} \delta_{a,a^{\prime}-1}   +  \frac{x_1 x_3}{y^2} \delta_{a,a^{\prime}-1}    q^a \\ \times  \frac{x_4}{x_1} \delta_{a,a^{\prime}-1}     \big]           \text{, } \\  \\      \big[    q^a \frac{x_1}{x_2} \delta_{a,a^{\prime}+1} \frac{y^2}{x_1 x_3} \delta_{a,a^{\prime}-1}   +  \frac{x_1 x_2}{y^2} \delta_{a,a^{\prime}_1} q^a \frac{x_3}{x_1} \delta_{a,a^{\prime}-1}     \big]  \big[    q^a \frac{x_1}{x_3} \delta_{a,a^{\prime}+1} q^{-a} \frac{x_4}{x_1} \delta_{a,a^{\prime}+1}    + \frac{x_1 x_3}{y^2} \delta_{a,a^{\prime}-1} \\ \times \frac{y^2}{x_1 x_4} \delta_{a,a^{\prime}+1}      \big]     \text{, }   
\\ \\ 
        \big[ x_1 x_2 \delta_{a,a^{\prime}+1} q^{-a} \frac{x_3}{x_1 } \delta_{a,a^{\prime}+1}  + q^{-a}   \frac{x_1}{x_2}  \delta_{a,a^{\prime}-1} \frac{y^2}{x_1 x_3} \delta_{a,a^{\prime}+1}     \big] \big[    q^a \frac{x_1}{x_3} \delta_{a,a^{\prime}+1}  \frac{y^2}{x_1 x_4} \delta_{a,a^{\prime}-1}  +  \frac{x_1 x_3}{y^2} \\ \times \delta_{a,a^{\prime}-1} q^a   \frac{x_4}{x_1} \delta_{a,a^{\prime}-1}      \big]  \text{, } \\ \\              \big[ q^a \frac{x_1}{x_2} \delta_{a,a^{\prime}+1} q^{-a} \frac{x_3}{x_1} \delta_{a,a^{\prime}+1} + \frac{x_1 x_2}{y^2} \delta_{a,a^{\prime}-1} \frac{y^2}{x_1 x_3} \delta_{a,a^{\prime}+1} \big] \big[     q^a \frac{x_1}{x_3} \delta_{a,a^{\prime}+1} q^{-a} \frac{x_4}{x_1} \delta_{a,a^{\prime}+1}         +       \frac{x_1 x_3}{y^2} \delta_{a,a^{\prime}-1} \\ \times \frac{y^2}{x_1 x_4} \delta_{a,a^{\prime}+1}     \big]      \text{, } \\ \\          \big[   x_1 x_2 \delta_{a,a^{\prime}+1} q^{-a} \frac{x_3}{x_1} \delta_{a,a^{\prime}+1}    +         q^{-a} \frac{x_1}{x_2} \delta_{a,a^{\prime}-1} \frac{y^2}{x_1 x_3} \delta_{a,a^{\prime}+1}     \big] \big[   q^a \frac{x_1}{x_3} \delta_{a,a^{\prime}+1} \frac{y^2}{x_1 x_4} \delta_{a,a^{\prime}-1}         +    \frac{x_1 x_3}{y^2} \delta_{a,a^{\prime}-1} q^a \\ \times \frac{x_4}{x_1} \delta_{a,a^{\prime}-1}    \big]      \text{, } \\ \\        \big[ q^a \frac{x_1}{x_2} \delta_{a,a^{\prime}+1} q^{-a} \frac{x_3}{x_1} \delta_{a,a^{\prime}+1}  +    \frac{x_1 x_2}{y^2} \delta_{a,a^{\prime}-1} \frac{y^2}{x_1 x_3 } \delta_{a,a^{\prime}+1}   \big]   \big[     q^a \frac{x_1}{x_3} \delta_{a,a^{\prime}+1} q^{-a} \frac{x_4}{x_1} \delta_{a,a^{\prime}+1}         + \frac{x_1 x_3}{y^2} \delta_{a,a^{\prime}-1} \end{align*}
     
     \begin{align*}      \times \frac{y^2}{x_1 x_4} \delta_{a,a^{\prime}+1}   \big]        \text{, } \\ \\       \big[  \frac{y^2}{x_1 x_2} \delta_{a,a^{\prime}-1} x_1 x_3 \delta_{a,a^{\prime}+1}      +     q^{-a} \frac{x_2}{x_1} \delta_{a,a^{\prime}+1} q^a \frac{x_1}{x_3} \delta_{a,a^{\prime}+1}   \big]      \big[ \frac{y^2}{x_1 x_3} \delta_{a,a^{\prime}-1} x_1 x_4 \delta_{a,a^{\prime}+1}       +  q^{-a} \frac{x_3}{x_1} \delta_{a,a^{\prime}+1} q^a \\ \times \frac{x_1}{x_4} \delta_{a,a^{\prime}+1 } \big]   \text{, } \\ \\     \big[    \frac{y^2}{x_1 x_3} \delta_{a,a^{\prime}-1} q^{-a} \frac{x_1}{x_4} \delta_{a,a^{\prime}-1}     +       q^{-a} \frac{x_3}{x_1} \delta_{a,a^{\prime}+1} \frac{x_1 x_4}{y^2} \delta_{a,a^{\prime}-1}     \big]      \big[            \frac{y^2}{x_1 x_2} \delta_{a,a^{\prime}-1} q^a \frac{x_1}{x_3} \delta_{a,a^{\prime}+1}          +    q^a \frac{x_2}{x_1} \delta_{a,a^{\prime}-1}  \\  \times \frac{x_1 x_3}{y^2} \delta_{a,a^{\prime}-1}             \big]   \text{, }      \\ \\ \big[    \frac{y^2}{x_1 x_2} \delta_{a,a^{\prime}-1} x_1 x_3 \delta_{a,a^{\prime}+1}       +      q^{-a} \frac{x_2}{x_1} \delta_{a,a^{\prime}+1} q^a \frac{x_1}{x_3} \delta_{a,a^{\prime}+1}          \big]  \big[     \frac{y^2}{x_1 x_3} \delta_{a,a^{\prime}-1} q^{-a}    \frac{x_1}{x_4} \delta_{a,a^{\prime}-1}     +              q^{-a} \frac{x_3}{x_1} \delta_{a,a^{\prime}+1} \\ \times \frac{x_1 x_4}{y^2} \delta_{a,a^{\prime}-1}            \big]  \text{, }   \\ \\  \big[     \frac{y^2}{x_1 x_2} \delta_{a,a^{\prime}-1} q^{-a} \frac{x_1}{x_3} \delta_{a,a^{\prime}-1}     +      q^{-a} \frac{x_2}{x_1} \delta_{a,a^{\prime}+1} \frac{x_1 x_3}{y^2} \delta_{a,a^{\prime}-1}     \big] \big[  q^{-a} \frac{x_3}{x_1} \delta_{a,a^{\prime}+1}    q^a \frac{x_1}{x_4} \delta_{a,a^{\prime}+1 }     + \frac{y^2}{x_1 x_4} \delta_{a,a^{\prime}+1} \\ \times \frac{x_1 x_4}{y^2} \delta_{a,a^{\prime}-1}  \big] \text{, } \end{align*}

        \begin{align*}       \big[ \frac{y^2}{x_1 x_2} \delta_{a,a^{\prime}-1} q^a \frac{x_1}{x_3} \delta_{a,a^{\prime}+1}  + q^a \frac{x_2}{x_!} \delta_{a,a^{\prime}-1} \frac{x_1 x_3}{y^2} \delta_{a,a^{\prime}-1} \big] \big[  \frac{y^2}{x_1 x_3} \delta_{a,a^{\prime}-1} q^{-a} \frac{x_1 }{x_4} \delta_{a,a^{\prime}-1} + q^{-a} \frac{x_3}{x_1} \delta_{a,a^{\prime}+1} \\ \times  \frac{x_1 x_4}{y^2}\delta_{a,a^{\prime}-1}  \big] \text{, } \\ \\ \big[ q^{-a} \frac{x_2}{x_1} \delta_{a,a^{\prime}+1} q^a \frac{x_1}{x_3} \delta_{a,a^{\prime}+1} + \frac{y^2}{x_1 x_3} \delta_{a,a^{\prime}+1} \frac{x_1 x_3}{y^2} \delta_{a,a^{\prime}-1} \big] \big[ q^{-a} \frac{x_3}{x_1} \delta_{a,a^{\prime}+1} q^a \frac{x_1}{x_4 } \delta_{a,a^{\prime}+1} + \frac{y^2}{x_1 x_4} \delta_{a,a^{\prime}+1} \\ \times \frac{x_1 x_4}{y^2} \delta_{a,a^{\prime}-1}  \big] \text{, }
\end{align*}

\noindent from which we conclude the argument, after having demonstrated that the entries of each product representation for the Ising type transfer matrix can be readily obtained from the products of finite-dimensional operators above. \boxed{}

\bigskip

\noindent Equipped with the two representations from the transfer matrix in the previous result, below we conclude that the Ising-type model is integrable. Following previous works of the author previously cited in other works, we demonstrate how arguments from each one of the Poisson bracket can be obtained, from which the existence of suitable action-angle variables can be inferred from the Poisson structure. We denote the Poisson bracket of the Ising type action-angle coordinates with,

\begin{align*}
    \big\{ \Phi^{\mathrm{Ising-type}} , \bar{\Phi^{\mathrm{Ising-type}}} \big\} . 
\end{align*}

\noindent With the system of expressions obtained from the third-order expansion of the Ising-type transfer matrix, we perform computations with the Poisson bracket in the forthcoming arguments below.

\bigskip

\noindent ($\Longleftarrow$) Suppose that $\big\{ \Phi^{\text{Ising-type}} , \bar{\Phi^{\text{Ising-type}}} \big\} \approx 0$. As has been examined for vertex models, spin-chains, and related models, introduced in previous sections, action-angle variables, for integrable and exactly solvable models, would satisfy,

\begin{align*}
 \big\{-  \mathcal{I}^2_1 \mathcal{I}^2_2 - \mathcal{I}^4_1 \mathcal{I}^4_2  ,  - \mathrm{log} \big[  \mathcal{I}^2_1 \mathcal{I}^2_2 + \mathcal{I}^4_1 \mathcal{I}^4_2      \big] \big\}  \approx 0  \Longleftrightarrow          \big\{ - \mathcal{I}^2_1   ,  \mathrm{log} \mathcal{I}^2_1  \big\} , \big\{ - \mathcal{I}^2_2   ,  \mathrm{log} \mathcal{I}^2_2  \big\} \approx 0      \text{. }
\end{align*}

\noindent The Ising-type transfer matrix, along with action-angle coordinates which can be determined from block operators in representations of the transfer matrix, determine the interactions that one would expect to observe in finite volume. Hence, from the previously obtained block operators for the third finite-dimensional representation of the Ising-type transfer matrix, it suffices to argue,

{\small \begin{align*}
 \bigg[\bigg[ \underset{i \neq  j \in V ( \textbf{Z}^2 ) }{\underset{a,a^{\prime} \in \textbf{R}}{\mathrm{span}}} \big\{ \mathcal{I}^1_1 \mathcal{I}^1_2 + \mathcal{I}^3_1 \mathcal{I}^3_2   ,       \mathcal{I}^2_1 \mathcal{I}^2_2 + \mathcal{I}^4_1 \mathcal{I}^4_2  ,  \mathcal{I}^1_1 \mathcal{I}^3_2 + \mathcal{I}^3_1 \mathcal{I}^4_2  ,   \mathcal{I}^3_1 \mathcal{I}^2_2 + \mathcal{I}^4_1 \mathcal{I}^4_2   \big\}    , \mathrm{log} \bigg[ \underset{i \neq  j \in V ( \textbf{Z}^2 ) }{\underset{a,a^{\prime} \in \textbf{R}}{\mathrm{span}}} \big\{ \mathcal{I}^1_1 \mathcal{I}^1_2 + \mathcal{I}^3_1  \\ \times  \mathcal{I}^3_2   ,       \mathcal{I}^2_1\mathcal{I}^2_2  + \mathcal{I}^4_1 \mathcal{I}^4_2  ,  \mathcal{I}^1_1 \mathcal{I}^3_2 + \mathcal{I}^3_1 \mathcal{I}^4_2  ,   \mathcal{I}^3_1 \mathcal{I}^2_2  + \mathcal{I}^4_1 \mathcal{I}^4_2   \big\} \bigg]  \bigg]\bigg] \end{align*}

 \begin{align*} \propto \underset{i \neq  j \in V ( \textbf{Z}^2 ) }{\underset{a,a^{\prime} \in \textbf{R}}{\mathrm{span}}}   \bigg[\bigg[  \big\{ \mathcal{I}^1_1 \mathcal{I}^1_2 + \mathcal{I}^3_1 \mathcal{I}^3_2  +        \mathcal{I}^2_1 \mathcal{I}^2_2 + \mathcal{I}^4_1 \mathcal{I}^4_2  +   \mathcal{I}^1_1 \mathcal{I}^3_2 + \mathcal{I}^3_1 \mathcal{I}^4_2  +   \mathcal{I}^3_1 \mathcal{I}^2_2 + \mathcal{I}^4_1 \mathcal{I}^4_2   \big\}    , \mathrm{log} \big[  \big\{ \mathcal{I}^1_1 \mathcal{I}^1_2 + \mathcal{I}^3_1 \\  \times  \mathcal{I}^3_2   ,       \mathcal{I}^2_1 \mathcal{I}^2_2  + \mathcal{I}^4_1 \mathcal{I}^4_2  ,  \mathcal{I}^1_1 \mathcal{I}^3_2 + \mathcal{I}^3_1 \mathcal{I}^4_2  ,   \mathcal{I}^3_1 \mathcal{I}^2_2  + \mathcal{I}^4_1 \mathcal{I}^4_2   \big\} \big]  \bigg]\bigg] \\ \\ \overset{(\mathrm{BL})}{=}  \underset{i \neq  j \in V ( \textbf{Z}^2 ) }{\underset{a,a^{\prime} \in \textbf{R}}{\mathrm{span}}}  \bigg[\bigg[  \mathcal{I}^1_1 \mathcal{I}^1_2  ,   \mathrm{log} \big[  \big\{ \mathcal{I}^1_1 \mathcal{I}^1_2 + \mathcal{I}^3_1 \mathcal{I}^3_2   +        \mathcal{I}^2_1 \mathcal{I}^2_2  + \mathcal{I}^4_1 \mathcal{I}^4_2 +  \mathcal{I}^1_1 \mathcal{I}^3_2 + \mathcal{I}^3_1 \mathcal{I}^4_2  ,   \mathcal{I}^3_1 \mathcal{I}^2_2  + \mathcal{I}^4_1 \mathcal{I}^4_2   \big\} \big]  \bigg]\bigg] \\ +   \underset{i \neq  j \in V ( \textbf{Z}^2 ) }{\underset{a,a^{\prime} \in \textbf{R}}{\mathrm{span}}}  \bigg[\bigg[  \mathcal{I}^3_1 \mathcal{I}^3_2 ,  \mathrm{log} \big[  \big\{ \mathcal{I}^1_1 \mathcal{I}^1_2 + \mathcal{I}^3_1 \mathcal{I}^3_2  +        \mathcal{I}^2_1  \mathcal{I}^2_2  + \mathcal{I}^4_1 \mathcal{I}^4_2 +   \mathcal{I}^1_1 \mathcal{I}^3_2 + \mathcal{I}^3_1 \mathcal{I}^4_2  +   \mathcal{I}^3_1 \mathcal{I}^2_2  + \mathcal{I}^4_1 \mathcal{I}^4_2   \big\} \big]  \bigg]\bigg] \\  +  \underset{i \neq  j \in V ( \textbf{Z}^2 ) }{\underset{a,a^{\prime} \in \textbf{R}}{\mathrm{span}}} \bigg[\bigg[           \mathcal{I}^2_1 \mathcal{I}^2_2  ,  \mathrm{log} \big[  \big\{ \mathcal{I}^1_1 \mathcal{I}^1_2 + \mathcal{I}^3_1 \mathcal{I}^3_2   +        \mathcal{I}^2_1 \mathcal{I}^2_2  + \mathcal{I}^4_1 \mathcal{I}^4_2  +  \mathcal{I}^1_1 \mathcal{I}^3_2 + \mathcal{I}^3_1 \mathcal{I}^4_2  +   \mathcal{I}^3_1 \mathcal{I}^2_2  + \mathcal{I}^4_1 \mathcal{I}^4_2   \big\} \big]  \bigg]\bigg] \\  +  \underset{i \neq  j \in V ( \textbf{Z}^2 ) }{\underset{a,a^{\prime} \in \textbf{R}}{\mathrm{span}}}  \bigg[\bigg[   \mathcal{I}^4_1 \mathcal{I}^4_2   ,   \mathrm{log} \big[  \big\{ \mathcal{I}^1_1 \mathcal{I}^1_2 + \mathcal{I}^3_1 \mathcal{I}^3_2   ,       \mathcal{I}^2_1 \mathcal{I}^2_2  + \mathcal{I}^4_1 \mathcal{I}^4_2  +   \mathcal{I}^1_1 \mathcal{I}^3_2 + \mathcal{I}^3_1 \mathcal{I}^4_2  +    \mathcal{I}^3_1 \mathcal{I}^2_2  + \mathcal{I}^4_1 \mathcal{I}^4_2   \big\} \big]  \bigg]\bigg] \\  + \underset{i \neq  j \in V ( \textbf{Z}^2 ) }{\underset{a,a^{\prime} \in \textbf{R}}{\mathrm{span}}} \bigg[\bigg[    \mathcal{I}^1_1 \mathcal{I}^3_2     ,   \mathrm{log} \big[  \big\{ \mathcal{I}^1_1 \mathcal{I}^1_2 + \mathcal{I}^3_1 \mathcal{I}^3_2   ,       \mathcal{I}^2_1 \mathcal{I}^2_2  + \mathcal{I}^4_1 \mathcal{I}^4_2  +   \mathcal{I}^1_1 \mathcal{I}^3_2 + \mathcal{I}^3_1 \mathcal{I}^4_2  +    \mathcal{I}^3_1 \mathcal{I}^2_2  + \mathcal{I}^4_1 \mathcal{I}^4_2   \big\} \big]  \bigg]\bigg] \\  + \underset{i \neq  j \in V ( \textbf{Z}^2 ) }{\underset{a,a^{\prime} \in \textbf{R}}{\mathrm{span}}}  \bigg[\bigg[   \mathcal{I}^3_1 \mathcal{I}^4_2     ,   \mathrm{log} \big[  \big\{ \mathcal{I}^1_1 \mathcal{I}^1_2 + \mathcal{I}^3_1 \mathcal{I}^3_2   +        \mathcal{I}^2_1 \mathcal{I}^2_2  + \mathcal{I}^4_1 \mathcal{I}^4_2  +   \mathcal{I}^1_1 \mathcal{I}^3_2 + \mathcal{I}^3_1 \mathcal{I}^4_2  +    \mathcal{I}^3_1 \mathcal{I}^2_2  + \mathcal{I}^4_1 \mathcal{I}^4_2   \big\} \big]  \bigg]\bigg] \\  + \underset{i \neq  j \in V ( \textbf{Z}^2 ) }{\underset{a,a^{\prime} \in \textbf{R}}{\mathrm{span}}} \bigg[\bigg[         \mathcal{I}^3_1 \mathcal{I}^2_2   ,   \mathrm{log} \big[  \big\{ \mathcal{I}^1_1 \mathcal{I}^1_2 + \mathcal{I}^3_1 \mathcal{I}^3_2   ,       \mathcal{I}^2_1  \mathcal{I}^2_2  + \mathcal{I}^4_1 \mathcal{I}^4_2  +   \mathcal{I}^1_1 \mathcal{I}^3_2 + \mathcal{I}^3_1 \mathcal{I}^4_2  +   \mathcal{I}^3_1 \mathcal{I}^2_2  + \mathcal{I}^4_1 \mathcal{I}^4_2   \big\} \big]  \bigg]\bigg] \\  + \underset{i \neq  j \in V ( \textbf{Z}^2 ) }{\underset{a,a^{\prime} \in \textbf{R}}{\mathrm{span}}}  \bigg[\bigg[     \mathcal{I}^4_1 \mathcal{I}^4_2  ,   \mathrm{log} \big[  \big\{ \mathcal{I}^1_1 \mathcal{I}^1_2 + \mathcal{I}^3_1 \mathcal{I}^3_2   ,       \mathcal{I}^2_1  \mathcal{I}^2_2  + \mathcal{I}^4_1 \mathcal{I}^4_2  +   \mathcal{I}^1_1 \mathcal{I}^3_2 + \mathcal{I}^3_1 \mathcal{I}^4_2  +    \mathcal{I}^3_1 \mathcal{I}^2_2  + \mathcal{I}^4_1 \mathcal{I}^4_2   \big\} \big]  \bigg]\bigg]   \approx 0 \text{, }
\end{align*} }

\noindent namely, whether higher-order terms in the expansion of the second block operator, with its natural logarithm, vanishes. Therefore, it suffices to demonstrate, as a starting point, that the above superposition of Poisson brackets vanishing implies that $\big\{ \Phi , \bar{\Phi} \big\}$ also vanishes. In particular, one can first demonstrate that a superposition of the above form of Poisson brackets approximately vanishes, from which,

{\small \begin{align*}
     \underset{i \neq  j \in V ( \textbf{Z}^2 ) }{\underset{a,a^{\prime} \in \textbf{R}}{\mathrm{span}}}  \bigg[\bigg[  \mathcal{I}^1_1 \mathcal{I}^1_2  ,   \mathrm{log} \big[  \big\{ \mathcal{I}^1_1 \mathcal{I}^1_2 + \mathcal{I}^3_1 \mathcal{I}^3_2   +        \mathcal{I}^2_1 \mathcal{I}^2_2  + \mathcal{I}^4_1 \mathcal{I}^4_2 +  \mathcal{I}^1_1 \mathcal{I}^3_2 + \mathcal{I}^3_1 \mathcal{I}^4_2  +    \mathcal{I}^3_1 \mathcal{I}^2_2  + \mathcal{I}^4_1 \mathcal{I}^4_2  + \cdots  \big\} \big]  \bigg]\bigg] \\ +   \underset{i \neq  j \in V ( \textbf{Z}^2 ) }{\underset{a,a^{\prime} \in \textbf{R}}{\mathrm{span}}}  \bigg[\bigg[  \mathcal{I}^3_1 \mathcal{I}^3_2 ,  \mathrm{log} \big[  \big\{ \mathcal{I}^1_1 \mathcal{I}^1_2 + \mathcal{I}^3_1 \mathcal{I}^3_2  +        \mathcal{I}^2_1  \mathcal{I}^2_2  + \mathcal{I}^4_1 \mathcal{I}^4_2 +   \mathcal{I}^1_1 \mathcal{I}^3_2 + \mathcal{I}^3_1 \mathcal{I}^4_2  +   \mathcal{I}^3_1 \mathcal{I}^2_2  + \mathcal{I}^4_1 \mathcal{I}^4_2  + \cdots   \big\} \big]  \bigg]\bigg] \\  +  \underset{i \neq  j \in V ( \textbf{Z}^2 ) }{\underset{a,a^{\prime} \in \textbf{R}}{\mathrm{span}}} \bigg[\bigg[           \mathcal{I}^2_1 \mathcal{I}^2_2  ,  \mathrm{log} \big[  \big\{ \mathcal{I}^1_1 \mathcal{I}^1_2 + \mathcal{I}^3_1 \mathcal{I}^3_2   +        \mathcal{I}^2_1 \mathcal{I}^2_2  + \mathcal{I}^4_1 \mathcal{I}^4_2  +  \mathcal{I}^1_1 \mathcal{I}^3_2 + \mathcal{I}^3_1 \mathcal{I}^4_2  +   \mathcal{I}^3_1 \mathcal{I}^2_2  + \mathcal{I}^4_1 \mathcal{I}^4_2    + \cdots \big\} \big]  \bigg]\bigg] \\ +  \underset{i \neq  j \in V ( \textbf{Z}^2 ) }{\underset{a,a^{\prime} \in \textbf{R}}{\mathrm{span}}}  \bigg[\bigg[   \mathcal{I}^4_1 \mathcal{I}^4_2   ,   \mathrm{log} \big[  \big\{ \mathcal{I}^1_1 \mathcal{I}^1_2 + \mathcal{I}^3_1 \mathcal{I}^3_2   ,       \mathcal{I}^2_1 \mathcal{I}^2_2  + \mathcal{I}^4_1 \mathcal{I}^4_2  +   \mathcal{I}^1_1 \mathcal{I}^3_2 + \mathcal{I}^3_1 \mathcal{I}^4_2  +    \mathcal{I}^3_1 \mathcal{I}^2_2  + \mathcal{I}^4_1 \mathcal{I}^4_2    + \cdots \big\} \big]  \bigg]\bigg] \\  + \underset{i \neq  j \in V ( \textbf{Z}^2 ) }{\underset{a,a^{\prime} \in \textbf{R}}{\mathrm{span}}} \bigg[\bigg[    \mathcal{I}^1_1 \mathcal{I}^3_2     ,   \mathrm{log} \big[  \big\{ \mathcal{I}^1_1 \mathcal{I}^1_2 + \mathcal{I}^3_1 \mathcal{I}^3_2   ,       \mathcal{I}^2_1 \mathcal{I}^2_2  + \mathcal{I}^4_1 \mathcal{I}^4_2  +   \mathcal{I}^1_1 \mathcal{I}^3_2 + \mathcal{I}^3_1 \mathcal{I}^4_2  +    \mathcal{I}^3_1 \mathcal{I}^2_2  + \mathcal{I}^4_1 \mathcal{I}^4_2    + \cdots \big\} \big]  \bigg]\bigg] \\  + \underset{i \neq  j \in V ( \textbf{Z}^2 ) }{\underset{a,a^{\prime} \in \textbf{R}}{\mathrm{span}}}  \bigg[\bigg[   \mathcal{I}^3_1 \mathcal{I}^4_2     ,   \mathrm{log} \big[  \big\{ \mathcal{I}^1_1 \mathcal{I}^1_2 + \mathcal{I}^3_1 \mathcal{I}^3_2   +        \mathcal{I}^2_1 \mathcal{I}^2_2  + \mathcal{I}^4_1 \mathcal{I}^4_2  +   \mathcal{I}^1_1 \mathcal{I}^3_2 + \mathcal{I}^3_1 \mathcal{I}^4_2  +    \mathcal{I}^3_1 \mathcal{I}^2_2  + \mathcal{I}^4_1 \mathcal{I}^4_2    + \cdots \big\} \big]  \bigg]\bigg] \\   + \underset{i \neq  j \in V ( \textbf{Z}^2 ) }{\underset{a,a^{\prime} \in \textbf{R}}{\mathrm{span}}} \bigg[\bigg[         \mathcal{I}^3_1 \mathcal{I}^2_2   ,   \mathrm{log} \big[  \big\{ \mathcal{I}^1_1 \mathcal{I}^1_2 + \mathcal{I}^3_1 \mathcal{I}^3_2   ,       \mathcal{I}^2_1  \mathcal{I}^2_2  + \mathcal{I}^4_1 \mathcal{I}^4_2  +   \mathcal{I}^1_1 \mathcal{I}^3_2 + \mathcal{I}^3_1 \mathcal{I}^4_2  +   \mathcal{I}^3_1 \mathcal{I}^2_2  + \mathcal{I}^4_1 \mathcal{I}^4_2    + \cdots \big\} \big]  \bigg]\bigg] \\ + \underset{i \neq  j \in V ( \textbf{Z}^2 ) }{\underset{a,a^{\prime} \in \textbf{R}}{\mathrm{span}}}  \bigg[\bigg[     \mathcal{I}^4_1 \mathcal{I}^4_2  ,   \mathrm{log} \big[  \big\{ \mathcal{I}^1_1 \mathcal{I}^1_2 + \mathcal{I}^3_1 \mathcal{I}^3_2   ,       \mathcal{I}^2_1  \mathcal{I}^2_2  + \mathcal{I}^4_1 \mathcal{I}^4_2  +   \mathcal{I}^1_1 \mathcal{I}^3_2 + \mathcal{I}^3_1 \mathcal{I}^4_2  +    \mathcal{I}^3_1 \mathcal{I}^2_2  + \mathcal{I}^4_1 \mathcal{I}^4_2    + \cdots \big\} \big]  \bigg]\bigg] \\    +    {\underset{a,a^{\prime} \in \textbf{R}}{\mathrm{span}}}  \bigg[\bigg[     \textit{Higher order Ising-type block operators}  ,   \mathrm{log} \big[  \big\{ \mathcal{I}^1_1 \mathcal{I}^1_2 + \mathcal{I}^3_1 \mathcal{I}^3_2   ,       \mathcal{I}^2_1  \mathcal{I}^2_2  + \mathcal{I}^4_1 \mathcal{I}^4_2  +   \mathcal{I}^1_1 \mathcal{I}^3_2 \\ + \mathcal{I}^3_1 \mathcal{I}^4_2   +    \mathcal{I}^3_1 \mathcal{I}^2_2  + \mathcal{I}^4_1 \mathcal{I}^4_2    + \cdots \big\} \big]  \bigg]\bigg]                   
     \approx 0 \\ \tag{1} \text{, }
\end{align*} }

\noindent is also expected to hold.

\bigskip

\noindent \textit{Proof of Theorem}. It suffices to demonstrate that several computations with the Poisson bracket can be performed. As mentioned earlier in the previous section, the arguments that enter within each Poisson bracket take the form,

\begin{align*}
  \underset{N \in \textbf{Z}}{\sum} \big[  x_1 x_{N-1} \delta_{a,a^{\prime}+1}   x_1 x_N \delta_{a,a^{\prime}+1} \big[    x_1 x_{N-1} \delta_{a,a^{\prime}+1} q^a \frac{x_1}{x_N} \delta_{a,a^{\prime}+1}  + q^a \frac{x_1}{x_{N-1}} \delta_{a,a^{\prime}+1} \frac{x_1 x_N}{y^2} \delta_{a,a^{\prime}-1} \end{align*}

        \begin{align*} + x_1 x_{N-2} \delta_{a,a^{\prime}+1}  x_1 x_N \delta_{a,a^{\prime}+1} + q^{-a} \frac{x_1}{x_{N-2}} \delta_{a,a^{\prime}-1} q^a \frac{x_1}{x_N} \delta_{a,a^{\prime}-1}  \big]    \big]  \text{, } \\  \\    \underset{N \in \textbf{Z}}{\sum} \big[  x_1 x_{N-1} \delta_{a,a^{\prime}+1} x_1 x_N \delta_{a,a^{\prime}+1 } \big[ x_1 x_{N-1} \delta_{a,a^{\prime}+1} q^a \frac{x_1}{x_N} \delta_{a,a^{\prime}+1} + q^a \frac{x_1}{x_{N-1}} \delta_{a,a^{\prime}+1}  \frac{x_1 x_N}{y^2} \delta_{a,a^{\prime}-1} \\ + x_1 x_{N-2} \delta_{a,a^{\prime}+1}         q^{-a} \frac{x_1}{x_N}  \delta_{a,a^{\prime}+1}  + q^{-a} \frac{x_1}{x_{N-2}} \delta_{a,a^{\prime}-1} \frac{x_1 x_N}{y^2} \delta_{a,a^{\prime}-1 }  \big]     \big]    \text{, }  \end{align*}

\begin{align*}  \underset{N \in \textbf{Z}}{\sum} \big[ q^{-a} \frac{x_1}{x_{N-1}} \delta_{a,a^{\prime}-1} q^a \frac{x_1}{x_N} \delta_{a,a^{\prime}+1}      \big[     x_1 x_{N-1} \delta_{a,a^{\prime}+1} q^{-a} \frac{x_1}{x_{N+1}} \delta_{a,a^{\prime}-1} + q^{-a} \frac{x_1}{x_{N-1}} \delta_{a,a^{\prime}+1} \frac{x_1 x_{N+1}}{y^2}  \delta_{a,a^{\prime}-1} \\ + x_1 x_{N-2} \delta_{a,a^{\prime}+1}   x_1 x_{N+2} \delta_{a,a^{\prime}+1} + q^{-a} \frac{x_1}{x_{N-2}}  \delta_{a,a^{\prime}-1} q^a \frac{x_1}{x_{N+2}} \delta_{a,a^{\prime}+1}   \big] \big]     \text{, }  \\ \\   \underset{N \in \textbf{Z}}{\sum}   \big[   q^{-a} \frac{x_1}{x_{N-1}}  \delta_{a,a^{\prime}-1}  q^a \frac{x_1}{x_N}  \delta_{a,a^{\prime}+1}  \big[ x_1 x_{N-1} \delta_{a,a^{\prime}+1} q^a \frac{x_1}{x_N} \delta_{a,a^{\prime}+1} + q^a \frac{x_1}{x_{N-1}} \delta_{a,a^{\prime}+1} \frac{x_1 x_N}{y^2} \delta_{a,a^{\prime}-1}\\  + x_1 x_{N-2} \delta_{a,a^{\prime}+1}  x_1 x_N \delta_{a,a^{\prime}+1}  + q^{-a} \frac{x_1}{x_{N-2}} \delta_{a,a^{\prime}+1} q^{-a} \frac{x_1}{x_N} \delta_{a,a^{\prime}+1}  \big] \big]   \text{, } 
\end{align*}

\noindent for the first group of terms, 

\begin{align*}
       \underset{N \in \textbf{Z}}{\sum}   \big[    x_1 x_{N-2} \delta_{a,a^{\prime}+1} x_1 x_{N+1} \delta_{a,a^{\prime}+1 }     \big[       x_1 x_{N-1} \delta_{a,a^{\prime}+1} q^{-a} \frac{x_1}{x_{N+1}} \delta_{a,a^{\prime}-1}     +  q^{-a} \frac{x_1}{x_{N-1}} \\ \times   \delta_{a,a^{\prime}+1} \frac{x_1 x_{N+1}}{y^2} \delta_{a,a^{\prime}-1}   +    x_1 x_{N-2} \delta_{a,a^{\prime}+1}   x_1 x_{N+2} \delta_{a,a^{\prime}+1} + q^{-a} \frac{x_1}{x_{N-2}} \delta_{a,a^{\prime}-1} q^a  \\ \times  \frac{x_1}{x_{N+2}} \delta_{a,a^{\prime}+1}           \big] \big]      \text{, } \end{align*}

       \begin{align*} b\underset{N \in \textbf{Z}}{\sum}   \big[     x_1 x_{N-2} \delta_{a,a^{\prime}+1} x_1 x_{N+1} \delta_{a,a^{\prime}+1}    \big[ x_1 x_{N-1} \delta_{a,a^{\prime}+1}  q^a \frac{x_1}{x_N} \delta_{a,a^{\prime}+1} +  q^a  \frac{x_1}{x_{N-1}} \delta_{a,a^{\prime}+1} \\ \times  \frac{x_1 x_N}{y^2} \delta_{a,a^{\prime}-1}   + x_1 x_{N-2} \delta_{a,a^{\prime}+1}  x_1 x_N \delta_{a,a^{\prime}+1} + q^{-a} \frac{x_1}{x_{N-2}} \delta_{a,a^{\prime}-1} q^a \\ \times  \frac{x_1}{x_N} \delta_{a,a^{\prime}+1}  \big] \big] \text{, } \end{align*}

       \begin{align*}   \underset{N \in \textbf{Z}}{\sum}   \big[    x_1 x_{N-2} \delta_{a,a^{\prime}+1} x_1 x_{N+1} \delta_{a,a^{\prime}+1}       \big[ x_1 x_{N-1} \delta_{a,a^{\prime}+1}  q^a \frac{x_1}{x_N} \delta_{a,a^{\prime}+1}  +  q^a \frac{x_1}{x_{N-1}} \delta_{a,a^{\prime}+1} \\  \times  \frac{x_1 x_N}{y^2} \delta_{a,a^{\prime}-1}     + x_1 x_{N-2} \delta_{a,a^{\prime}+1}  q^{-a} \frac{x_1}{x_N} \delta_{a,a^{\prime}-1} + q^{-a} \frac{x_1}{x_{N-2}} \delta_{a,a^{\prime}-1} \\   \times  \frac{x_1 x_N}{y^2} \delta_{a,a^{\prime}-1 }   \big]  \big]  \text{, } \\ \\  \underset{N \in \textbf{Z}}{\sum}   \big[  q^{-a} \frac{x_1}{x_{N-2}} \delta_{a,a^{\prime}-1} q^a \frac{x_1}{x_{N+2}} \delta_{a,a^{\prime}+1} \big[ x_1 x_{N-1} \delta_{a,a^{\prime}+1} q^{-a} \frac{x_1}{x_{N-1} } \delta_{a,a^{\prime}-1} + q^{-a} \frac{x_1}{x_{N-1}} \\ \times  \delta_{a,a^{\prime}+1}   \frac{x_1 x_{N+1}}{y^2} \delta_{a,a^{\prime}-1} +  x_1 x_{N-2} 
 \delta_{a,a^{\prime}+1}   x_1 x_{N+2} \delta_{a,a^{\prime}+1} + q^{-a} \frac{x_1}{x_{N-2} } \delta_{a,a^{\prime}-1} q^a \\ \times   \frac{x_1}{x_{N+2}} \delta_{a,a^{\prime}+1}  \big]  \big]   \text{, } \end{align*}

\noindent for the next group of terms,

 \begin{align*}
 \underset{N \in \textbf{Z}}{\sum}   \big[   q^{-a} \frac{x_1}{x_{N-2}} \delta_{a,a^{\prime}-1} q^a \frac{x_1}{x_{N+2}} \delta_{a,a^{\prime}+1}     \big[   x_1 x_{N-1} \delta_{a,a^{\prime}+1}  q^a \frac{x_1}{x_N} \delta_{a,a^{\prime}+1} + q^a \frac{x_1}{x_{N-1}} \\  \times \delta_{a,a^{\prime}+1} \frac{x_1 x_N}{y^2} \delta_{a,a^{\prime}-1}      +  x_1 x_{N-2} \delta_{a,a^{\prime}+1}  x_1 x_N \delta_{a,a^{\prime}+1} + q^{-a} \frac{x_1}{x_{N-2}} \delta_{a,a^{\prime}-1}  q^a \frac{x_1}{x_N} \delta_{a,a^{\prime}+1}     \big] \big]  \text{, } \\ \\  \underset{N \in \textbf{Z}}{\sum}   \big[  q^{-a} \frac{x_1}{x_{N-2}} \delta_{a,a^{\prime}-1}     q^a \frac{x_1}{x_{N+2}} \delta_{a,a^{\prime}+1}         \big[   x_1 x_{N-1} \delta_{a,a^{\prime}+1} q^a \frac{x_1}{x_N} \delta_{a,a^{\prime}+1} + q^a \frac{x_1}{x_{N-1}} \delta_{a,a^{\prime}+1}  \\ \times  \frac{x_1 x_N}{y^2} \delta_{a,a^{\prime}-1} + x_1 x_{N-2} \delta_{a,a^{\prime}+1} q^{-a} \frac{x_1}{x_N} \delta_{a,a^{\prime}-1}    + q^{-a} \frac{x_1}{x_{N-2}} \delta_{a,a^{\prime}-1} \frac{x_1 x_N}{y^2 } \delta_{a,a^{\prime}-1}   \big] \big] \text{, } \end{align*}

 \begin{align*} 
 \underset{N \in \textbf{Z}}{\sum}   \big[        x_1 x_{N-2} \delta_{a,a^{\prime}+1}   q^{-a} \frac{x_1}{x_{N+1}} \delta_{a,a^{\prime}-1}  \big[       q^{-a} \frac{x_1}{x_{N-1}} \delta_{a,a^{\prime}-1} q^a \frac{x_1}{x_{N+1}} \delta_{a,a^{\prime}+1}  + \frac{x_1 x_{N-1}}{y^2} \delta_{a,a^{\prime}-1}  \\ \times  \frac{x_1 x_{N+1}}{y^2} \delta_{a,a^{\prime}-1 }           +     x_1 x_{N-2} \delta_{a,a^{\prime}+1}  q^{-a} \frac{x_1}{x_{N+1}} \delta_{a,a^{\prime}-1} + q^{-a} \frac{x_1}{x_{N-2}} \delta_{a,a^{\prime}-1} \frac{x_1 x_{N+1}}{y^2} \delta_{a,a^{\prime}-1}  \big] \big]  \text{, }\\ \\ \underset{N \in \textbf{Z}}{\sum}   \big[  x_1 x_{N-2}  \delta_{a,a^{\prime}+1} q^{-a} \frac{x_1}{x_{N+1}} \delta_{a,a^{\prime}-1} \big[ q^{-a} \frac{x_1}{x_{N-1}} \delta_{a,a^{\prime}-1} q^a \frac{x_1}{x_N} \delta_{a,a^{\prime}+1} + \frac{x_1 x_{N-1}}{y^2} \delta_{a,a^{\prime}-1}  \\ \times  \frac{x_1 x_N}{y^2} \delta_{a,a^{\prime}-1}  + x_1 x_{N-2} \delta_{a,a^{\prime}+1}  q^a \frac{x_1}{x_N} \delta_{a,a^{\prime}+1}  + q^a \frac{x_1}{x_{N-2}} \delta_{a,a^{\prime}+1} \frac{x_1 x_N}{y^2} \delta_{a,a^{\prime}-1}   \big] \big]   \text{, } \end{align*}

\noindent for the next group of terms, and,

 \begin{align*}
 \underset{N \in \textbf{Z}}{\sum}   \big[  x_1 x_{N-2} \delta_{a,a^{\prime}+1} q^{-a} \frac{x_1}{x_{N+1}} \delta_{a,a^{\prime}-1}   \big[ q^{-a} \frac{x_1}{x_{N-1}} \delta_{a,a^{\prime}-1} q^a \frac{x_1}{x_N} \delta_{a,a^{\prime}+1}   + \frac{x_1 x_{N-1}}{y^2}  \delta_{a,a^{\prime}-1} \frac{x_1 x_N}{y^2} \delta_{a,a^{\prime}-1} \\  + q^{-a} \frac{x_1}{x_{N-2}} \delta_{a,a^{\prime}-1} q^a \frac{x_1}{x_N} \delta_{a,a^{\prime}+1} + \frac{x_1 x_{N-2}}{y^2} \delta_{a,a^{\prime}-1} \frac{x_1 x_{N+1}}{y^2} \delta_{a,a^{\prime}-1}  \big] \big]  \text{, } \end{align*}

 \begin{align*}
 \underset{N \in \textbf{Z}}{\sum}   \big[ q^a \frac{x_1}{x_{N-1}} \delta_{a,a^{\prime}+1} \frac{x_1 x_{N+1}}{y^2} \delta_{a,a^{\prime}-1}   \big[   x_1 x_{N-1} \delta_{a,a^{\prime}+1} q^a \frac{x_1}{x_N} \delta_{a,a^{\prime}+1} + q^a \frac{x_1}{x_{N+1}} \delta_{a,a^{\prime}+1} \frac{x_1 x_N}{y^2} \delta_{a,a^{\prime}-1} \\ + x_1 x_{N-2} \delta_{a,a^{\prime}+1}  x_1 x_N \delta_{a,a^{\prime}+1}      + q^a \frac{x_1}{x_{N-2}} \delta_{a,a^{\prime}-1} q^a \frac{x_1}{x_N} \delta_{a,a^{\prime}+1}  \big] \big] \text{, } \end{align*} \begin{align*}
 \underset{N \in \textbf{Z}}{\sum}   \big[ q^a \frac{x_1}{x_{N-1}} \delta_{a,a^{\prime}+1} \frac{x_1 x_{N+1}}{y^2} \delta_{a,a^{\prime}-1} \big[ q^{-a} \frac{x_1}{x_{N-1}} \delta_{a,a^{\prime}-1} q^a \frac{x_1}{x_N} \delta_{a,a^{\prime}+1} + \frac{x_1 x_{N-1}}{y^2} \delta_{a,a^{\prime}-1} \frac{x_1 x_N}{y^2} \delta_{a,a^{\prime}-1}  \\ + x_1 x_{N-2} \delta_{a,a^{\prime}+1}  q^a \frac{x_1}{x_N} \delta_{a,a^{\prime}+1} + q^a \frac{x_1}{x_{N-2}} \delta_{a,a^{\prime}+1 } \frac{x_1 x_N}{y^2} \delta_{a,a^{\prime}-1}  \big] \big] \text{, }
\end{align*}

\noindent for the last group of terms.

\bigskip

\noindent After having demonstrated that the first argument of the Poisson bracket of the action-angle coordinates take the form as indicated from the summations over $\textbf{Z}$ above, we turn towards arguing that the superposition of Poisson brackets given by $(1)$ approximately vanishes. To this end, in the forthcoming computations we isolate each Poisson bracket from lower-order block operators of the Ising-type transfer matrix. First, observe,

\begin{align*} \underset{i \neq  j \in V ( \textbf{Z}^2 ) }{\underset{a,a^{\prime} \in \textbf{R}}{\mathrm{span}}}  \bigg[\bigg[  \mathcal{I}^1_1 \mathcal{I}^1_2  ,   \mathrm{log} \big[  \big\{ \mathcal{I}^1_1 \mathcal{I}^1_2 + \mathcal{I}^3_1 \mathcal{I}^3_2   +        \mathcal{I}^2_1 \mathcal{I}^2_2  + \mathcal{I}^4_1 \mathcal{I}^4_2 +  \mathcal{I}^1_1 \mathcal{I}^3_2 + \mathcal{I}^3_1 \mathcal{I}^4_2  +    \mathcal{I}^3_1 \mathcal{I}^2_2  + \mathcal{I}^4_1 \mathcal{I}^4_2   \big\} \big]  \bigg]\bigg] \\ \\  <       \underset{i \neq  j \in V ( \textbf{Z}^2 ) }{\underset{a,a^{\prime} \in \textbf{R}}{\mathrm{span}}}  \bigg[\bigg[  \mathcal{I}^1_1 \mathcal{I}^1_2  ,   \mathrm{log} \big[  \big\{ \mathcal{I}^1_1 \mathcal{I}^1_2   \big\} \big]  \bigg]\bigg]  \\  \\ <   \underset{i \neq  j \in V ( \textbf{Z}^2 ) }{\underset{a,a^{\prime} \in \textbf{R}}{\mathrm{span}}}  \bigg[\bigg[  \mathcal{I}^1_1 \mathcal{I}^1_2  ,  \bigg[ \mathrm{log} \big[  \mathcal{I}^1_1 \mathcal{I}^1_2  \big] + \mathrm{log} \big[  \mathcal{I}^3_1 \mathcal{I}^3_2  \big] + \mathrm{log} \big[          \mathcal{I}^2_1 \mathcal{I}^2_2  \big] + \mathrm{log} \big[  \mathcal{I}^4_1 \mathcal{I}^4_2 \big] + \mathrm{log} \big[   \mathcal{I}^1_1 \mathcal{I}^3_2  \big] \end{align*}

\begin{align*}    + \mathrm{log} \big[ \mathcal{I}^3_1 \mathcal{I}^4_2  \big]  + \mathrm{log} \big[   \mathcal{I}^3_1 \mathcal{I}^2_2   \big] + \mathrm{log} \big[ \mathcal{I}^4_1 \mathcal{I}^4_2    \big]  \bigg] \bigg]\bigg]  \\ \\ \overset{(\mathrm{BL})}{=}        \underset{i \neq  j \in V ( \textbf{Z}^2 ) }{\underset{a,a^{\prime} \in \textbf{R}}{\mathrm{span}}}  \bigg[\bigg[  \mathcal{I}^1_1 \mathcal{I}^1_2 , \mathrm{log} \big[  \mathcal{I}^1_1 \mathcal{I}^1_2  \big] \bigg]\bigg]    + \underset{i \neq  j \in V ( \textbf{Z}^2 ) }{\underset{a,a^{\prime} \in \textbf{R}}{\mathrm{span}}}  \bigg[\bigg[  \mathcal{I}^1_1 \mathcal{I}^1_2 ,   \mathrm{log} \big[  \mathcal{I}^3_1 \mathcal{I}^3_2  \big]  \bigg]\bigg]   + \underset{i \neq  j \in V ( \textbf{Z}^2 ) }{\underset{a,a^{\prime} \in \textbf{R}}{\mathrm{span}}}  \bigg[\bigg[  \mathcal{I}^1_1 \mathcal{I}^1_2 \\  ,  \mathrm{log} \big[          \mathcal{I}^2_1 \mathcal{I}^2_2  \big] \bigg]\bigg]   + \underset{i \neq  j \in V ( \textbf{Z}^2 ) }{\underset{a,a^{\prime} \in \textbf{R}}{\mathrm{span}}}  \bigg[\bigg[  \mathcal{I}^1_1 \mathcal{I}^1_2 , \mathrm{log} \big[  \mathcal{I}^4_1 \mathcal{I}^4_2 \big]  \bigg]\bigg]   + \underset{i \neq  j \in V ( \textbf{Z}^2 ) }{\underset{a,a^{\prime} \in \textbf{R}}{\mathrm{span}}}  \bigg[\bigg[  \mathcal{I}^1_1 \mathcal{I}^1_2 ,\mathrm{log} \big[  \mathcal{I}^1_1 \mathcal{I}^3_2 \big]  \bigg]\bigg] \\  + \underset{i \neq  j \in V ( \textbf{Z}^2 ) }{\underset{a,a^{\prime} \in \textbf{R}}{\mathrm{span}}}  \bigg[\bigg[  \mathcal{I}^1_1 \mathcal{I}^1_2  , \mathrm{log} \big[ \mathcal{I}^3_1 \mathcal{I}^2_2 \big] \bigg]\bigg]     + \underset{i \neq  j \in V ( \textbf{Z}^2 ) }{\underset{a,a^{\prime} \in \textbf{R}}{\mathrm{span}}}  \bigg[\bigg[  \mathcal{I}^1_1 \mathcal{I}^1_2  , \mathrm{log } \big[ \mathcal{I}^3_1 \mathcal{I}^2_2 \big] \bigg]\bigg]  \\  +   \underset{i \neq  j \in V ( \textbf{Z}^2 ) }{\underset{a,a^{\prime} \in \textbf{R}}{\mathrm{span}}}  \bigg[\bigg[  \mathcal{I}^1_1 \mathcal{I}^1_2   ,  \mathrm{log} \big[ \mathcal{I}^4_1 \mathcal{I}^4_2 \big]      \bigg]\bigg]               \\ \\   =    {\underset{a,a^{\prime} \in \textbf{R}}{\mathrm{span}}}  \bigg[    \bigg[\bigg[  \mathcal{I}^1_1 \mathcal{I}^1_2 , \mathrm{log} \big[  \mathcal{I}^1_1 \mathcal{I}^1_2  \big] \bigg]\bigg]    +   \bigg[\bigg[  \mathcal{I}^1_1 \mathcal{I}^1_2 ,   \mathrm{log} \big[  \mathcal{I}^3_1 \mathcal{I}^3_2  \big]  \bigg]\bigg]   +   \bigg[\bigg[  \mathcal{I}^1_1 \mathcal{I}^1_2  ,  \mathrm{log} \big[          \mathcal{I}^2_1 \mathcal{I}^2_2  \big] \bigg]\bigg] \\  +   \bigg[\bigg[  \mathcal{I}^1_1 \mathcal{I}^1_2 , \mathrm{log} \big[  \mathcal{I}^4_1 \mathcal{I}^4_2 \big]  \bigg]\bigg]   +  \bigg[\bigg[  \mathcal{I}^1_1 \mathcal{I}^1_2 ,\mathrm{log} \big[  \mathcal{I}^1_1 \mathcal{I}^3_2 \big]  \bigg]\bigg]   +   \bigg[\bigg[  \mathcal{I}^1_1 \mathcal{I}^1_2  , \mathrm{log} \big[ \mathcal{I}^3_1 \mathcal{I}^2_2 \big] \bigg]\bigg]   \\   +   \bigg[\bigg[  \mathcal{I}^1_1 \mathcal{I}^1_2  , \mathrm{log } \big[ \mathcal{I}^3_1 \mathcal{I}^2_2 \big] \bigg]\bigg]  +   \bigg[\bigg[  \mathcal{I}^1_1 \mathcal{I}^1_2   ,  \mathrm{log} \big[ \mathcal{I}^4_1 \mathcal{I}^4_2 \big]      \bigg]\bigg]          \bigg]         \\ \\   <          {\underset{a,a^{\prime} \in \textbf{R}}{\mathrm{span}}}  \bigg[    \bigg[\bigg[  \mathcal{I}^1_1 \mathcal{I}^1_2 , \mathrm{log} \big[   \mathcal{I}^1_2  \big]^2 \bigg]\bigg]    +   \bigg[\bigg[  \mathcal{I}^1_1 \mathcal{I}^1_2 ,   \mathrm{log} \big[  \mathcal{I}^3_2  \big]^2  \bigg]\bigg]   +   \bigg[\bigg[  \mathcal{I}^1_1 \mathcal{I}^1_2  ,  \mathrm{log} \big[           \mathcal{I}^2_2  \big]^2 \bigg]\bigg] \\  +   \bigg[\bigg[  \mathcal{I}^1_1 \mathcal{I}^1_2 , \mathrm{log} \big[   \mathcal{I}^4_2 \big]^2  \bigg]\bigg]   +  \bigg[\bigg[  \mathcal{I}^1_1 \mathcal{I}^1_2 ,\mathrm{log} \big[  \mathcal{I}^1_2 \mathcal{I}^3_2 \big]  \bigg]\bigg]   +   \bigg[\bigg[  \mathcal{I}^1_1 \mathcal{I}^1_2  , \mathrm{log} \big[ \mathcal{I}^3_2 \mathcal{I}^2_2 \big] \bigg]\bigg]   \\   +   \bigg[\bigg[  \mathcal{I}^1_1 \mathcal{I}^1_2  , \mathrm{log } \big[ \mathcal{I}^3_2 \mathcal{I}^2_2 \big] \bigg]\bigg]  +   \bigg[\bigg[  \mathcal{I}^1_1 \mathcal{I}^1_2   ,  \mathrm{log} \big[  \mathcal{I}^4_2 \big]^2      \bigg]\bigg]          \bigg]           \\ \\ <   8  \underset{i \neq  j \in V ( \textbf{Z}^2 ) }{\underset{a,a^{\prime} \in \textbf{R}}{\mathrm{span}}} \bigg[       \bigg[\bigg[\mathcal{I}^1_1 \mathcal{I}^1_2 ,  \bigg[   \underset{i \neq j : 1 \leq i \leq 4}{\underset{1 \leq j \leq 4}{\mathrm{sup}}}    \bigg[\bigg[         4 \mathrm{log} \big[ \mathcal{I}^j_2 \big]^2  +          3  \mathrm{log} \big[ \mathcal{I}^i_2 \mathcal{I}^j_2  \big]           \bigg]\bigg]  \bigg]        \bigg]\bigg]   \bigg] \end{align*}

\begin{align*}   
      <   8  \underset{i \neq  j \in V ( \textbf{Z}^2 ) }{\underset{a,a^{\prime} \in \textbf{R}}{\mathrm{span}}}   \bigg[       \bigg[\bigg[\big[ \mathcal{I}^1_2 \big]^2  ,  \bigg[   \underset{i \neq j : 1 \leq i \leq 4}{\underset{1 \leq j \leq 4}{\mathrm{sup}}}    \bigg[\bigg[         4 \mathrm{log} \big[ \mathcal{I}^j_2 \big]^2  +          3  \mathrm{log} \big[ \mathcal{I}^i_2 \mathcal{I}^j_2  \big]           \bigg]\bigg]  \bigg]        \bigg]\bigg]   \bigg]   \\ \\  \overset{(\mathrm{BL})}{=}   8  \underset{i \neq  j \in V ( \textbf{Z}^2 ) }{\underset{a,a^{\prime} \in \textbf{R}}{\mathrm{span}}}  \bigg[       \bigg[\bigg[\big[ \mathcal{I}^1_2 \big]^2  ,  \bigg[   \underset{i \neq j : 1 \leq i \leq 4}{\underset{1 \leq j \leq 4}{\mathrm{sup}}}    \big\{          4 \mathrm{log} \big[ \mathcal{I}^j_2 \big]^2  \big\}  \bigg]        \bigg]\bigg]  +    \bigg[\bigg[\big[ \mathcal{I}^1_2 \big]^2  ,  \bigg[   \underset{i \neq j : 1 \leq i \leq 4}{\underset{1 \leq j \leq 4}{\mathrm{sup}}}    \big\{            3  \mathrm{log} \big[ \mathcal{I}^i_2 \mathcal{I}^j_2  \big]           \big\}  \bigg]        \bigg]\bigg]  \bigg]                  \\  <    8  \underset{i \neq  j \in V ( \textbf{Z}^2 ) }{\underset{a,a^{\prime} \in \textbf{R}}{\mathrm{span}}}  \bigg[       \bigg[\bigg[\big[ \mathcal{I}^1_2 \big]^2  ,  \bigg[   \underset{i \neq j : 1 \leq i \leq 4}{\underset{1 \leq j \leq 4}{\mathrm{sup}}}    \big\{          4 \mathrm{log} \big[ \mathcal{I}^j_2 \big]^2  \big\}  \bigg]        \bigg]\bigg]  +    \bigg[\bigg[\big[ \mathcal{I}^1_2 \big]^2  ,  \bigg[   \underset{i \neq j : 1 \leq i \leq 4}{\underset{1 \leq j \leq 4}{\mathrm{sup}}}    \big\{            3  \mathrm{log} \big[ \mathcal{I}^j_2  \big]           \big\}  \bigg]        \bigg]\bigg]  \bigg]  \\   \overset{(\mathrm{AC})}{=}      8  \underset{i \neq  j \in V ( \textbf{Z}^2 ) }{\underset{a,a^{\prime} \in \textbf{R}}{\mathrm{span}}}   \bigg[       \bigg[\bigg[\big[ \mathcal{I}^1_2 \big]^2  ,  \bigg[   \underset{i \neq j : 1 \leq i \leq 4}{\underset{1 \leq j \leq 4}{\mathrm{sup}}}    \big\{          4 \mathrm{log} \big[ \mathcal{I}^j_2 \big]^2  \big\}  \bigg]        \bigg]\bigg]  -    \bigg[\bigg[\bigg[   \underset{i \neq j : 1 \leq i \leq 4}{\underset{1 \leq j \leq 4}{\mathrm{sup}}}    \big\{            3  \mathrm{log} \big[ \mathcal{I}^j_2  \big]           \big\}  \bigg]  ,   \big[ \mathcal{I}^1_2 \big]^2       \bigg]\bigg]  \bigg]   \\ \approx           8  \underset{i \neq  j \in V ( \textbf{Z}^2 ) }{\underset{a,a^{\prime} \in \textbf{R}}{\mathrm{span}}}  \bigg[    \big\{ 0 , 0 \big\}  \bigg]  \approx 0                 , 
\end{align*}

\noindent corresponding to the first Poisson bracket,

\begin{align*}    \underset{i \neq  j \in V ( \textbf{Z}^2 ) }{\underset{a,a^{\prime} \in \textbf{R}}{\mathrm{span}}}  \bigg[\bigg[  \mathcal{I}^3_1 \mathcal{I}^3_2 ,  \mathrm{log} \big[  \big\{ \mathcal{I}^1_1 \mathcal{I}^1_2 + \mathcal{I}^3_1 \mathcal{I}^3_2  +        \mathcal{I}^2_1  \mathcal{I}^2_2  + \mathcal{I}^4_1 \mathcal{I}^4_2 +   \mathcal{I}^1_1 \mathcal{I}^3_2 + \mathcal{I}^3_1 \mathcal{I}^4_2  +   \mathcal{I}^3_1 \mathcal{I}^2_2  + \mathcal{I}^4_1 \mathcal{I}^4_2   \big\} \big]  \bigg]\bigg] \\ \\  < \underset{i \neq  j \in V ( \textbf{Z}^2 ) }{\underset{a,a^{\prime} \in \textbf{R}}{\mathrm{span}}}  \bigg[\bigg[  \mathcal{I}^3_1 \mathcal{I}^3_2 ,  \mathrm{log} \big[  \big\{ \mathcal{I}^1_1 \mathcal{I}^1_2      \big\} \big]  \bigg]\bigg] \\ \\ <  \underset{i \neq  j \in V ( \textbf{Z}^2 ) }{\underset{a,a^{\prime} \in \textbf{R}}{\mathrm{span}}}  \bigg[\bigg[  \mathcal{I}^3_1 \mathcal{I}^3_2 , \mathrm{log} \big[ \mathcal{I}^1_1 \mathcal{I}^1_2 \big] \bigg]\bigg]   + \underset{i \neq  j \in V ( \textbf{Z}^2 ) }{\underset{a,a^{\prime} \in \textbf{R}}{\mathrm{span}}}  \bigg[\bigg[  \mathcal{I}^3_1 \mathcal{I}^3_2 , \mathrm{log} \big[ \mathcal{I}^3_1 \mathcal{I}^3_2 \big] \bigg]\bigg]    + \underset{i \neq  j \in V ( \textbf{Z}^2 ) }{\underset{a,a^{\prime} \in \textbf{R}}{\mathrm{span}}}  \bigg[\bigg[  \mathcal{I}^3_1 \mathcal{I}^3_2 , \mathrm{log} \big[ \mathcal{I}^2_1 \mathcal{I}^2_2 \big] \bigg]\bigg]  \\     + \underset{i \neq  j \in V ( \textbf{Z}^2 ) }{\underset{a,a^{\prime} \in \textbf{R}}{\mathrm{span}}}  \bigg[\bigg[  \mathcal{I}^3_1 \mathcal{I}^3_2 , \mathrm{log} \big[ \mathcal{I}^4_1 \mathcal{I}^4_2 \big] \bigg]\bigg]     + \underset{i \neq  j \in V ( \textbf{Z}^2 ) }{\underset{a,a^{\prime} \in \textbf{R}}{\mathrm{span}}}  \bigg[\bigg[  \mathcal{I}^3_1 \mathcal{I}^3_2 , \mathrm{log} \big[ \mathcal{I}^1_1 \mathcal{I}^3_2 \big] \bigg]\bigg]    + \underset{i \neq  j \in V ( \textbf{Z}^2 ) }{\underset{a,a^{\prime} \in \textbf{R}}{\mathrm{span}}}  \bigg[\bigg[  \mathcal{I}^3_1 \mathcal{I}^3_2 ,  \mathrm{log} \big[ \mathcal{I}^1_1 \mathcal{I}^3_2 \big] \bigg]\bigg]  \end{align*}

\begin{align*}   + \underset{i \neq  j \in V ( \textbf{Z}^2 ) }{\underset{a,a^{\prime} \in \textbf{R}}{\mathrm{span}}}  \bigg[\bigg[  \mathcal{I}^3_1 \mathcal{I}^3_2 , \mathrm{log} \big[ \mathcal{I}^3_1 \mathcal{I}^4_2 \big] \bigg]\bigg]    + \underset{i \neq  j \in V ( \textbf{Z}^2 ) }{\underset{a,a^{\prime} \in \textbf{R}}{\mathrm{span}}}  \bigg[\bigg[  \mathcal{I}^3_1 \mathcal{I}^3_2 ,  \mathrm{log} \big[ \mathcal{I}^3_1 \mathcal{I}^2_2 \big] \bigg]\bigg]    + \underset{i \neq  j \in V ( \textbf{Z}^2 ) }{\underset{a,a^{\prime} \in \textbf{R}}{\mathrm{span}}}  \bigg[\bigg[  \mathcal{I}^3_1 \mathcal{I}^3_2 ,  \mathrm{log} \big[ \mathcal{I}^4_1 \mathcal{I}^4_2 \big] \bigg]\bigg] \\  \\ =  \underset{i \neq  j \in V ( \textbf{Z}^2 ) }{\underset{a,a^{\prime} \in \textbf{R}}{\mathrm{span}}} \bigg[  \bigg[\bigg[  \mathcal{I}^3_1 \mathcal{I}^3_2 , \mathrm{log} \big[ \mathcal{I}^1_1 \mathcal{I}^1_2 \big] \bigg]\bigg]   +  \bigg[\bigg[  \mathcal{I}^3_1 \mathcal{I}^3_2 , \mathrm{log} \big[ \mathcal{I}^3_1 \mathcal{I}^3_2 \big] \bigg]\bigg]    +  \bigg[\bigg[  \mathcal{I}^3_1 \mathcal{I}^3_2 , \mathrm{log} \big[ \mathcal{I}^2_1 \mathcal{I}^2_2 \big] \bigg]\bigg] \\   +  \bigg[\bigg[  \mathcal{I}^3_1 \mathcal{I}^3_2 , \mathrm{log} \big[ \mathcal{I}^4_1 \mathcal{I}^4_2 \big] \bigg]\bigg]     +  \bigg[\bigg[  \mathcal{I}^3_1 \mathcal{I}^3_2 , \mathrm{log} \big[ \mathcal{I}^1_1 \mathcal{I}^3_2 \big] \bigg]\bigg] +  \bigg[\bigg[  \mathcal{I}^3_1 \mathcal{I}^3_2 ,  \mathrm{log} \big[ \mathcal{I}^1_1 \mathcal{I}^3_2 \big] \bigg]\bigg]  \\   +  \bigg[\bigg[  \mathcal{I}^3_1 \mathcal{I}^3_2 , \mathrm{log} \big[ \mathcal{I}^3_1 \mathcal{I}^4_2 \big] \bigg]\bigg]      + \bigg[\bigg[  \mathcal{I}^3_1 \mathcal{I}^3_2 ,  \mathrm{log} \big[ \mathcal{I}^3_1 \mathcal{I}^2_2 \big] \bigg]\bigg]    + \bigg[\bigg[  \mathcal{I}^3_1 \mathcal{I}^3_2 ,  \mathrm{log} \big[ \mathcal{I}^4_1 \mathcal{I}^4_2 \big] \bigg]\bigg] \bigg] \\  <   8 {\underset{a,a^{\prime} \in \textbf{R}}{\mathrm{span}}}  \bigg[       \bigg[\bigg[\mathcal{I}^3_1 \mathcal{I}^3_2 ,  \bigg[   \underset{i \neq j : 1 \leq i \leq 4}{\underset{1 \leq j \leq 4}{\mathrm{sup}}}    \bigg[\bigg[         4 \mathrm{log} \big[ \mathcal{I}^j_2 \big]^2  +          3  \mathrm{log} \big[ \mathcal{I}^i_2 \mathcal{I}^j_2  \big]           \bigg]\bigg]  \bigg]        \bigg]\bigg]   \bigg] \\ 
      <   8 {\underset{a,a^{\prime} \in \textbf{R}}{\mathrm{span}}}  \bigg[       \bigg[\bigg[\big[ \mathcal{I}^3_2 \big]^2  ,  \bigg[   \underset{i \neq j : 1 \leq i \leq 4}{\underset{1 \leq j \leq 4}{\mathrm{sup}}}    \bigg[\bigg[         4 \mathrm{log} \big[ \mathcal{I}^j_2 \big]^2  +          3  \mathrm{log} \big[ \mathcal{I}^i_2 \mathcal{I}^j_2  \big]           \bigg]\bigg]  \bigg]        \bigg]\bigg]   \bigg]   \\ \\  \overset{(\mathrm{BL})}{=}   8 {\underset{a,a^{\prime} \in \textbf{R}}{\mathrm{span}}}  \bigg[       \bigg[\bigg[\big[ \mathcal{I}^3_2 \big]^2  ,  \bigg[   \underset{i \neq j : 1 \leq i \leq 4}{\underset{1 \leq j \leq 4}{\mathrm{sup}}}    \big\{          4 \mathrm{log} \big[ \mathcal{I}^j_2 \big]^2  \big\}  \bigg]        \bigg]\bigg]  +    \bigg[\bigg[\big[ \mathcal{I}^1_2 \big]^2  ,  \bigg[   \underset{i \neq j : 1 \leq i \leq 4}{\underset{1 \leq j \leq 4}{\mathrm{sup}}}    \big\{            3  \mathrm{log} \big[ \mathcal{I}^i_2 \mathcal{I}^j_2  \big]           \big\}  \bigg]        \bigg]\bigg]  \bigg]                 \end{align*}

      \begin{align*} <    8  \underset{i \neq  j \in V ( \textbf{Z}^2 ) }{\underset{a,a^{\prime} \in \textbf{R}}{\mathrm{span}}}   \bigg[       \bigg[\bigg[\big[ \mathcal{I}^3_2 \big]^2  ,  \bigg[   \underset{i \neq j : 1 \leq i \leq 4}{\underset{1 \leq j \leq 4}{\mathrm{sup}}}    \big\{          4 \mathrm{log} \big[ \mathcal{I}^j_2 \big]^2  \big\}  \bigg]        \bigg]\bigg]  +    \bigg[\bigg[\big[ \mathcal{I}^3_2 \big]^2  ,  \bigg[   \underset{i \neq j : 1 \leq i \leq 4}{\underset{1 \leq j \leq 4}{\mathrm{sup}}}    \big\{            3  \mathrm{log} \big[ \mathcal{I}^j_2  \big]           \big\}  \bigg]        \bigg]\bigg]  \bigg]  \\   \overset{(\mathrm{AC})}{=}      8  \underset{i \neq  j \in V ( \textbf{Z}^2 ) }{\underset{a,a^{\prime} \in \textbf{R}}{\mathrm{span}}}  \bigg[       \bigg[\bigg[\big[ \mathcal{I}^3_2 \big]^2  ,  \bigg[   \underset{i \neq j : 1 \leq i \leq 4}{\underset{1 \leq j \leq 4}{\mathrm{sup}}}    \big\{          4 \mathrm{log} \big[ \mathcal{I}^j_2 \big]^2  \big\}  \bigg]        \bigg]\bigg]  -    \bigg[\bigg[\bigg[   \underset{i \neq j : 1 \leq i \leq 4}{\underset{1 \leq j \leq 4}{\mathrm{sup}}}    \big\{            3  \mathrm{log} \big[ \mathcal{I}^j_2  \big]           \big\}  \bigg]  ,   \big[ \mathcal{I}^3_2 \big]^2       \bigg]\bigg]  \bigg]   \\ \approx           8  \underset{i \neq  j \in V ( \textbf{Z}^2 ) }{\underset{a,a^{\prime} \in \textbf{R}}{\mathrm{span}}}   \bigg[    \big\{ 0 , 0 \big\}  \bigg]  \approx 0     ,    \end{align*}

\noindent corresponding to the second Poisson bracket,


\begin{align*}   \underset{i \neq  j \in V ( \textbf{Z}^2 ) }{\underset{a,a^{\prime} \in \textbf{R}}{\mathrm{span}}} \bigg[\bigg[           \mathcal{I}^2_1 \mathcal{I}^2_2  ,  \mathrm{log} \big[  \big\{ \mathcal{I}^1_1 \mathcal{I}^1_2 + \mathcal{I}^3_1 \mathcal{I}^3_2   +        \mathcal{I}^2_1 \mathcal{I}^2_2  + \mathcal{I}^4_1 \mathcal{I}^4_2  +  \mathcal{I}^1_1 \mathcal{I}^3_2 + \mathcal{I}^3_1 \mathcal{I}^4_2  +   \mathcal{I}^3_1 \mathcal{I}^2_2  + \mathcal{I}^4_1 \mathcal{I}^4_2   \big\} \big]  \bigg]\bigg] \\ \\ <  \underset{i \neq  j \in V ( \textbf{Z}^2 ) }{\underset{a,a^{\prime} \in \textbf{R}}{\mathrm{span}}}  \bigg[\bigg[            \mathcal{I}^2_1 \mathcal{I}^2_2  ,   \mathrm{log} \big[  \big\{ \mathcal{I}^1_1 \mathcal{I}^1_2    \big\} \big]   \bigg]\bigg] \\  \\ < \underset{i \neq  j \in V ( \textbf{Z}^2 ) }{\underset{a,a^{\prime} \in \textbf{R}}{\mathrm{span}}}  \bigg[\bigg[            \mathcal{I}^2_1 \mathcal{I}^2_2  ,  \mathrm{log} \big[  \big\{ \mathcal{I}^1_1 \mathcal{I}^1_2  \big] \bigg]\bigg] + \underset{i \neq  j \in V ( \textbf{Z}^2 ) }{\underset{a,a^{\prime} \in \textbf{R}}{\mathrm{span}}} \bigg[\bigg[\mathcal{I}^2_1 \mathcal{I}^2_2 ,  \mathrm{log} \big[ \mathcal{I}^3_1 \mathcal{I}^3_2  \big] \bigg]\bigg] + \underset{i \neq  j \in V ( \textbf{Z}^2 ) }{\underset{a,a^{\prime} \in \textbf{R}}{\mathrm{span}}} \bigg[\bigg[\mathcal{I}^1_1 \mathcal{I}^1_2 , \mathrm{log} \big[          \mathcal{I}^2_1 \mathcal{I}^2_2     \big] \bigg]\bigg] \\  + \underset{i \neq  j \in V ( \textbf{Z}^2 ) }{\underset{a,a^{\prime} \in \textbf{R}}{\mathrm{span}}} \bigg[\bigg[\mathcal{I}^1_1 \mathcal{I}^1_2 , \mathrm{log} \big[ \mathcal{I}^4_1 \mathcal{I}^4_2   \big] \bigg]\bigg]      + \underset{i \neq  j \in V ( \textbf{Z}^2 ) }{\underset{a,a^{\prime} \in \textbf{R}}{\mathrm{span}}} \bigg[\bigg[\mathcal{I}^1_1 \mathcal{I}^1_2 , \mathrm{log} \big[ \mathcal{I}^1_1 \mathcal{I}^3_2  \big] \bigg]\bigg]      + \underset{i \neq  j \in V ( \textbf{Z}^2 ) }{\underset{a,a^{\prime} \in \textbf{R}}{\mathrm{span}}} \bigg[\bigg[\mathcal{I}^1_1 \mathcal{I}^1_2 , \mathrm{log} \big[ \mathcal{I}^3_1 \mathcal{I}^4_2  \big] \bigg]\bigg]      \\    + \underset{i \neq  j \in V ( \textbf{Z}^2 ) }{\underset{a,a^{\prime} \in \textbf{R}}{\mathrm{span}}} \bigg[\bigg[\mathcal{I}^1_1 \mathcal{I}^1_2 ,  \mathrm{log} \big[ \mathcal{I}^3_1 \mathcal{I}^2_2  \big] \bigg]\bigg]   +  \underset{i \neq  j \in V ( \textbf{Z}^2 ) }{\underset{a,a^{\prime} \in \textbf{R}}{\mathrm{span}}} \bigg[\bigg[\mathcal{I}^1_1 \mathcal{I}^1_2   , \mathrm{log} \big[ \mathcal{I}^4_1 \mathcal{I}^4_2   \big\} \big]   \bigg]\bigg]  \\ \\  = \underset{i \neq  j \in V ( \textbf{Z}^2 ) }{\underset{a,a^{\prime} \in \textbf{R}}{\mathrm{span}}}  \bigg[ \bigg[\bigg[            \mathcal{I}^2_1 \mathcal{I}^2_2  ,  \mathrm{log} \big[  \big\{ \mathcal{I}^1_1 \mathcal{I}^1_2  \big] \bigg]\bigg] + \bigg[\bigg[\mathcal{I}^2_1 \mathcal{I}^2_2 ,  \mathrm{log} \big[ \mathcal{I}^3_1 \mathcal{I}^3_2  \big] \bigg]\bigg]  + \bigg[\bigg[\mathcal{I}^1_1 \mathcal{I}^1_2 , \mathrm{log} \big[          \mathcal{I}^2_1 \mathcal{I}^2_2     \big] \bigg]\bigg]  \\   +  \bigg[\bigg[\mathcal{I}^1_1 \mathcal{I}^1_2 , \mathrm{log} \big[ \mathcal{I}^4_1 \mathcal{I}^4_2   \big] \bigg]\bigg]      +  \bigg[\bigg[\mathcal{I}^1_1 \mathcal{I}^1_2 , \mathrm{log} \big[ \mathcal{I}^1_1 \mathcal{I}^3_2  \big] \bigg]\bigg]     +  \bigg[\bigg[\mathcal{I}^1_1 \mathcal{I}^1_2 , \mathrm{log} \big[ \mathcal{I}^3_1 \mathcal{I}^4_2  \big] \bigg]\bigg]   \\    +  \bigg[\bigg[\mathcal{I}^1_1 \mathcal{I}^1_2 ,  \mathrm{log} \big[ \mathcal{I}^3_1 \mathcal{I}^2_2  \big] \bigg]\bigg]  +  \bigg[\bigg[\mathcal{I}^1_1 \mathcal{I}^1_2   , \mathrm{log} \big[ \mathcal{I}^4_1 \mathcal{I}^4_2   \big\} \big]   \bigg]\bigg]  \bigg] \\          <   8  \underset{i \neq  j \in V ( \textbf{Z}^2 ) }{\underset{a,a^{\prime} \in \textbf{R}}{\mathrm{span}}}  \bigg[       \bigg[\bigg[\mathcal{I}^3_1 \mathcal{I}^3_2 ,  \bigg[   \underset{i \neq j : 1 \leq i \leq 4}{\underset{1 \leq j \leq 4}{\mathrm{sup}}}    \bigg[\bigg[         4 \mathrm{log} \big[ \mathcal{I}^j_2 \big]^2  +          3  \mathrm{log} \big[ \mathcal{I}^i_2 \mathcal{I}^j_2  \big]           \bigg]\bigg]  \bigg]        \bigg]\bigg]   \bigg] \\ 
      <   8  \underset{i \neq  j \in V ( \textbf{Z}^2 ) }{\underset{a,a^{\prime} \in \textbf{R}}{\mathrm{span}}}  \bigg[       \bigg[\bigg[\big[ \mathcal{I}^2_2 \big]^2  ,  \bigg[   \underset{i \neq j : 1 \leq i \leq 4}{\underset{1 \leq j \leq 4}{\mathrm{sup}}}    \bigg[\bigg[         4 \mathrm{log} \big[ \mathcal{I}^j_2 \big]^2  +          3  \mathrm{log} \big[ \mathcal{I}^i_2 \mathcal{I}^j_2  \big]           \bigg]\bigg]  \bigg]        \bigg]\bigg]   \bigg]   \\  \overset{(\mathrm{BL})}{=}   8  \underset{i \neq  j \in V ( \textbf{Z}^2 ) }{\underset{a,a^{\prime} \in \textbf{R}}{\mathrm{span}}} \bigg[       \bigg[\bigg[\big[ \mathcal{I}^2_2 \big]^2  ,  \bigg[   \underset{i \neq j : 1 \leq i \leq 4}{\underset{1 \leq j \leq 4}{\mathrm{sup}}}    \big\{          4 \mathrm{log} \big[ \mathcal{I}^j_2 \big]^2  \big\}  \bigg]        \bigg]\bigg]  +    \bigg[\bigg[\big[ \mathcal{I}^2_2 \big]^2  ,  \bigg[   \underset{i \neq j : 1 \leq i \leq 4}{\underset{1 \leq j \leq 4}{\mathrm{sup}}}    \big\{            3  \mathrm{log} \big[ \mathcal{I}^i_2 \mathcal{I}^j_2  \big]           \big\}  \bigg]        \bigg]\bigg]  \bigg]                   \end{align*}

      \begin{align*}    <    8  \underset{i \neq  j \in V ( \textbf{Z}^2 ) }{\underset{a,a^{\prime} \in \textbf{R}}{\mathrm{span}}}  \bigg[       \bigg[\bigg[\big[ \mathcal{I}^2_2 \big]^2  ,  \bigg[   \underset{i \neq j : 1 \leq i \leq 4}{\underset{1 \leq j \leq 4}{\mathrm{sup}}}    \big\{          4 \mathrm{log} \big[ \mathcal{I}^j_2 \big]^2  \big\}  \bigg]        \bigg]\bigg]  +    \bigg[\bigg[\big[ \mathcal{I}^2_2 \big]^2  ,  \bigg[   \underset{i \neq j : 1 \leq i \leq 4}{\underset{1 \leq j \leq 4}{\mathrm{sup}}}    \big\{            3  \mathrm{log} \big[ \mathcal{I}^j_2  \big]           \big\}  \bigg]        \bigg]\bigg]  \bigg]  \\   \overset{(\mathrm{AC})}{=}      8  \underset{i \neq  j \in V ( \textbf{Z}^2 ) }{\underset{a,a^{\prime} \in \textbf{R}}{\mathrm{span}}} \bigg[       \bigg[\bigg[\big[ \mathcal{I}^2_2 \big]^2  ,  \bigg[   \underset{i \neq j : 1 \leq i \leq 4}{\underset{1 \leq j \leq 4}{\mathrm{sup}}}    \big\{          4 \mathrm{log} \big[ \mathcal{I}^j_2 \big]^2  \big\}  \bigg]        \bigg]\bigg]  -    \bigg[\bigg[\bigg[   \underset{i \neq j : 1 \leq i \leq 4}{\underset{1 \leq j \leq 4}{\mathrm{sup}}}    \big\{            3  \mathrm{log} \big[ \mathcal{I}^j_2  \big]           \big\}  \bigg]  ,   \big[ \mathcal{I}^3_2 \big]^2       \bigg]\bigg]  \bigg]   \\ \approx           8  \underset{i \neq  j \in V ( \textbf{Z}^2 ) }{\underset{a,a^{\prime} \in \textbf{R}}{\mathrm{span}}}   \bigg[    \big\{ 0 , 0 \big\}  \bigg]  \approx 0     , \end{align*}

\noindent corresponding to the third Poisson bracket,

\begin{align*} \underset{i \neq  j \in V ( \textbf{Z}^2 ) }{\underset{a,a^{\prime} \in \textbf{R}}{\mathrm{span}}}  \bigg[\bigg[   \mathcal{I}^4_1 \mathcal{I}^4_2   ,   \mathrm{log} \big[  \big\{ \mathcal{I}^1_1 \mathcal{I}^1_2 + \mathcal{I}^3_1 \mathcal{I}^3_2   ,       \mathcal{I}^2_1 \mathcal{I}^2_2  + \mathcal{I}^4_1 \mathcal{I}^4_2  +   \mathcal{I}^1_1 \mathcal{I}^3_2 + \mathcal{I}^3_1 \mathcal{I}^4_2  +    \mathcal{I}^3_1 \mathcal{I}^2_2  + \mathcal{I}^4_1 \mathcal{I}^4_2   \big\} \big]  \bigg]\bigg] \\ \\ <  \underset{i \neq  j \in V ( \textbf{Z}^2 ) }{\underset{a,a^{\prime} \in \textbf{R}}{\mathrm{span}}}  \bigg[\bigg[  \mathcal{I}^4_1 \mathcal{I}^4_2   ,   \mathrm{log} \big[  \big\{ \mathcal{I}^1_1 \mathcal{I}^1_2   \big\} \big] \bigg]\bigg]  \\ \\  <      \underset{i \neq  j \in V ( \textbf{Z}^2 ) }{\underset{a,a^{\prime} \in \textbf{R}}{\mathrm{span}}}  \bigg[\bigg[  \mathcal{I}^4_1 \mathcal{I}^4_2   , \mathrm{log} \big[ \mathcal{I}^1_1 \mathcal{I}^1_2    \big] \bigg]\bigg]   +  \underset{i \neq  j \in V ( \textbf{Z}^2 ) }{\underset{a,a^{\prime} \in \textbf{R}}{\mathrm{span}}}  \bigg[\bigg[  \mathcal{I}^4_1 \mathcal{I}^4_2 ,\mathrm{log} \big[ \mathcal{I}^3_1 \mathcal{I}^3_2            \big] \bigg]\bigg]    +   \underset{i \neq  j \in V ( \textbf{Z}^2 ) }{\underset{a,a^{\prime} \in \textbf{R}}{\mathrm{span}}}  \bigg[\bigg[  \mathcal{I}^4_1 \mathcal{I}^4_2   ,\mathrm{log} \big[ \mathcal{I}^2_1 \mathcal{I}^2_2     \big] \bigg]\bigg] \\  +  \underset{i \neq  j \in V ( \textbf{Z}^2 ) }{\underset{a,a^{\prime} \in \textbf{R}}{\mathrm{span}}}  \bigg[\bigg[  \mathcal{I}^4_1 \mathcal{I}^4_2  ,\mathrm{log} \big[ \mathcal{I}^4_1 \mathcal{I}^4_2      \big] \bigg]\bigg]   +   \underset{i \neq  j \in V ( \textbf{Z}^2 ) }{\underset{a,a^{\prime} \in \textbf{R}}{\mathrm{span}}}  \bigg[\bigg[  \mathcal{I}^4_1 \mathcal{I}^4_2    ,\mathrm{log} \big[ \mathcal{I}^1_1 \mathcal{I}^3_2   \big] \bigg]\bigg]   +  \underset{i \neq  j \in V ( \textbf{Z}^2 ) }{\underset{a,a^{\prime} \in \textbf{R}}{\mathrm{span}}}  \bigg[\bigg[  \mathcal{I}^4_1 \mathcal{I}^4_2  ,\mathrm{log} \big[\mathcal{I}^3_1 \mathcal{I}^4_2        \big] \bigg]\bigg]   \\   +  \underset{i \neq  j \in V ( \textbf{Z}^2 ) }{\underset{a,a^{\prime} \in \textbf{R}}{\mathrm{span}}}  \bigg[\bigg[  \mathcal{I}^4_1 \mathcal{I}^4_2  ,\mathrm{log} \big[  \mathcal{I}^3_1 \mathcal{I}^2_2   \big]   \bigg]\bigg]  +  \underset{i \neq  j \in V ( \textbf{Z}^2 ) }{\underset{a,a^{\prime} \in \textbf{R}}{\mathrm{span}}}  \bigg[\bigg[  \mathcal{I}^4_1 \mathcal{I}^4_2  ,\mathrm{log} \big[    \mathcal{I}^4_1 \mathcal{I}^4_2  \big] \bigg]\bigg]  \\ \\  =     \underset{i \neq  j \in V ( \textbf{Z}^2 ) }{\underset{a,a^{\prime} \in \textbf{R}}{\mathrm{span}}} \bigg[  \bigg[\bigg[  \mathcal{I}^4_1 \mathcal{I}^4_2   , \mathrm{log} \big[ \mathcal{I}^1_1 \mathcal{I}^1_2    \big] \bigg]\bigg]   +   \bigg[\bigg[  \mathcal{I}^4_1 \mathcal{I}^4_2 ,\mathrm{log} \big[ \mathcal{I}^3_1 \mathcal{I}^3_2            \big] \bigg]\bigg]    +    \bigg[\bigg[  \mathcal{I}^4_1 \mathcal{I}^4_2   ,\mathrm{log} \big[ \mathcal{I}^2_1 \mathcal{I}^2_2     \big] \bigg]\bigg] \\ +   \bigg[\bigg[  \mathcal{I}^4_1 \mathcal{I}^4_2  ,\mathrm{log} \big[ \mathcal{I}^4_1 \mathcal{I}^4_2      \big] \bigg]\bigg]   +     \bigg[\bigg[  \mathcal{I}^4_1 \mathcal{I}^4_2    ,\mathrm{log} \big[ \mathcal{I}^1_1 \mathcal{I}^3_2   \big] \bigg]\bigg]   +  \bigg[\bigg[  \mathcal{I}^4_1 \mathcal{I}^4_2  ,\mathrm{log} \big[\mathcal{I}^3_1 \mathcal{I}^4_2        \big] \bigg]\bigg]    \end{align*}

\begin{align*}   +    \bigg[\bigg[  \mathcal{I}^4_1 \mathcal{I}^4_2  ,\mathrm{log} \big[  \mathcal{I}^3_1 \mathcal{I}^2_2   \big]   \bigg]\bigg]  +   \bigg[\bigg[  \mathcal{I}^4_1 \mathcal{I}^4_2  ,\mathrm{log} \big[    \mathcal{I}^4_1 \mathcal{I}^4_2  \big] \bigg]\bigg]  \bigg] \\ \\       <   8  \underset{i \neq  j \in V ( \textbf{Z}^2 ) }{\underset{a,a^{\prime} \in \textbf{R}}{\mathrm{span}}}  \bigg[       \bigg[\bigg[\mathcal{I}^4_1 \mathcal{I}^4_2 ,  \bigg[   \underset{i \neq j : 1 \leq i \leq 4}{\underset{1 \leq j \leq 4}{\mathrm{sup}}}    \bigg[\bigg[         4 \mathrm{log} \big[ \mathcal{I}^j_2 \big]^2  +          3  \mathrm{log} \big[ \mathcal{I}^i_2 \mathcal{I}^j_2  \big]           \bigg]\bigg]  \bigg]        \bigg]\bigg]   \bigg] \\ 
      <   8  \underset{i \neq  j \in V ( \textbf{Z}^2 ) }{\underset{a,a^{\prime} \in \textbf{R}}{\mathrm{span}}}  \bigg[       \bigg[\bigg[\big[ \mathcal{I}^4_2 \big]^2  ,  \bigg[   \underset{i \neq j : 1 \leq i \leq 4}{\underset{1 \leq j \leq 4}{\mathrm{sup}}}    \bigg[\bigg[         4 \mathrm{log} \big[ \mathcal{I}^j_2 \big]^2  +          3  \mathrm{log} \big[ \mathcal{I}^i_2 \mathcal{I}^j_2  \big]           \bigg]\bigg]  \bigg]        \bigg]\bigg]   \bigg]  \\  \overset{(\mathrm{BL})}{=}   8  \underset{i \neq  j \in V ( \textbf{Z}^2 ) }{\underset{a,a^{\prime} \in \textbf{R}}{\mathrm{span}}}  \bigg[       \bigg[\bigg[\big[ \mathcal{I}^4_2 \big]^2  ,  \bigg[   \underset{i \neq j : 1 \leq i \leq 4}{\underset{1 \leq j \leq 4}{\mathrm{sup}}}    \big\{          4 \mathrm{log} \big[ \mathcal{I}^j_2 \big]^2  \big\}  \bigg]        \bigg]\bigg]  +    \bigg[\bigg[\big[ \mathcal{I}^4_2 \big]^2  ,  \bigg[   \underset{i \neq j : 1 \leq i \leq 4}{\underset{1 \leq j \leq 4}{\mathrm{sup}}}    \big\{            3  \mathrm{log} \big[ \mathcal{I}^i_2 \mathcal{I}^j_2  \big]           \big\}  \bigg]        \bigg]\bigg]  \bigg]                 \\  <    8  \underset{i \neq  j \in V ( \textbf{Z}^2 ) }{\underset{a,a^{\prime} \in \textbf{R}}{\mathrm{span}}}  \bigg[       \bigg[\bigg[\big[ \mathcal{I}^4_2 \big]^2  ,  \bigg[   \underset{i \neq j : 1 \leq i \leq 4}{\underset{1 \leq j \leq 4}{\mathrm{sup}}}    \big\{          4 \mathrm{log} \big[ \mathcal{I}^j_2 \big]^2  \big\}  \bigg]        \bigg]\bigg]  +    \bigg[\bigg[\big[ \mathcal{I}^4_2 \big]^2  ,  \bigg[   \underset{i \neq j : 1 \leq i \leq 4}{\underset{1 \leq j \leq 4}{\mathrm{sup}}}    \big\{            3  \mathrm{log} \big[ \mathcal{I}^j_2  \big]           \big\}  \bigg]        \bigg]\bigg]  \bigg]  \\ \overset{(\mathrm{AC})}{=}      8  \underset{i \neq  j \in V ( \textbf{Z}^2 ) }{\underset{a,a^{\prime} \in \textbf{R}}{\mathrm{span}}}  \bigg[       \bigg[\bigg[\big[ \mathcal{I}^4_2 \big]^2  ,  \bigg[   \underset{i \neq j : 1 \leq i \leq 4}{\underset{1 \leq j \leq 4}{\mathrm{sup}}}    \big\{          4 \mathrm{log} \big[ \mathcal{I}^j_2 \big]^2  \big\}  \bigg]        \bigg]\bigg]  -    \bigg[\bigg[\bigg[   \underset{i \neq j : 1 \leq i \leq 4}{\underset{1 \leq j \leq 4}{\mathrm{sup}}}    \big\{            3  \mathrm{log} \big[ \mathcal{I}^j_2  \big]           \big\}  \bigg]  ,   \big[ \mathcal{I}^4_2 \big]^2       \bigg]\bigg]  \bigg]   \\ \approx           8  \underset{i \neq  j \in V ( \textbf{Z}^2 ) }{\underset{a,a^{\prime} \in \textbf{R}}{\mathrm{span}}}   \bigg[    \big\{ 0 , 0 \big\}  \bigg]  \approx 0     , \end{align*}

\noindent corresponding to the fourth Poisson bracket,

\begin{align*} \underset{i \neq  j \in V ( \textbf{Z}^2 ) }{\underset{a,a^{\prime} \in \textbf{R}}{\mathrm{span}}} \bigg[\bigg[    \mathcal{I}^1_1 \mathcal{I}^3_2     ,   \mathrm{log} \big[  \big\{ \mathcal{I}^1_1 \mathcal{I}^1_2 + \mathcal{I}^3_1 \mathcal{I}^3_2   ,       \mathcal{I}^2_1 \mathcal{I}^2_2  + \mathcal{I}^4_1 \mathcal{I}^4_2  +   \mathcal{I}^1_1 \mathcal{I}^3_2 + \mathcal{I}^3_1 \mathcal{I}^4_2  +    \mathcal{I}^3_1 \mathcal{I}^2_2  + \mathcal{I}^4_1 \mathcal{I}^4_2   \big\} \big]  \bigg]\bigg] \\ <   \underset{i \neq  j \in V ( \textbf{Z}^2 ) }{\underset{a,a^{\prime} \in \textbf{R}}{\mathrm{span}}}  \bigg[\bigg[  \mathcal{I}^1_1 \mathcal{I}^3_2     ,  \mathrm{log} \big[  \big\{ \mathcal{I}^1_1 \mathcal{I}^1_2   \big\} \big]  \bigg]\bigg]  \\ \\ <       \underset{i \neq  j \in V ( \textbf{Z}^2 ) }{\underset{a,a^{\prime} \in \textbf{R}}{\mathrm{span}}}  \bigg[\bigg[  \mathcal{I}^1_1 \mathcal{I}^3_2     , \mathrm{log} \big[ \mathcal{I}^1_1 \mathcal{I}^1_2  \big] \bigg]\bigg]       + \underset{i \neq  j \in V ( \textbf{Z}^2 ) }{\underset{a,a^{\prime} \in \textbf{R}}{\mathrm{span}}}  \bigg[\bigg[  \mathcal{I}^1_1 \mathcal{I}^3_2     , \mathrm{log} \big[  \mathcal{I}^3_1 \mathcal{I}^3_2            \big] \bigg]\bigg]       + \underset{i \neq  j \in V ( \textbf{Z}^2 ) }{\underset{a,a^{\prime} \in \textbf{R}}{\mathrm{span}}}  \bigg[\bigg[  \mathcal{I}^1_1 \mathcal{I}^3_2     , \mathrm{log} \big[ \mathcal{I}^2_1 \mathcal{I}^2_2      \big] \bigg]\bigg]      \end{align*}

\begin{align*}   + \underset{i \neq  j \in V ( \textbf{Z}^2 ) }{\underset{a,a^{\prime} \in \textbf{R}}{\mathrm{span}}}  \bigg[\bigg[  \mathcal{I}^1_1 \mathcal{I}^3_2     , \mathrm{log} \big[ \mathcal{I}^4_1 \mathcal{I}^4_2       \big] \bigg]\bigg]       + \underset{i \neq  j \in V ( \textbf{Z}^2 ) }{\underset{a,a^{\prime} \in \textbf{R}}{\mathrm{span}}}  \bigg[\bigg[  \mathcal{I}^1_1 \mathcal{I}^3_2     , \mathrm{log} \big[ \mathcal{I}^1_1 \mathcal{I}^3_2     \big] \bigg]\bigg]       + \underset{i \neq  j \in V ( \textbf{Z}^2 ) }{\underset{a,a^{\prime} \in \textbf{R}}{\mathrm{span}}}  \bigg[\bigg[  \mathcal{I}^1_1 \mathcal{I}^3_2     ,  \mathrm{log} \big[  \mathcal{I}^3_1 \mathcal{I}^4_2      \big] \bigg]\bigg]      \\ + \underset{i \neq  j \in V ( \textbf{Z}^2 ) }{\underset{a,a^{\prime} \in \textbf{R}}{\mathrm{span}}}  \bigg[\bigg[  \mathcal{I}^1_1 \mathcal{I}^3_2     ,   \mathrm{log} \big[  \mathcal{I}^3_1 \mathcal{I}^2_2    \big] \bigg]\bigg]     + \underset{i \neq  j \in V ( \textbf{Z}^2 ) }{\underset{a,a^{\prime} \in \textbf{R}}{\mathrm{span}}}  \bigg[\bigg[  \mathcal{I}^1_1 \mathcal{I}^3_2     ,   \mathrm{log} \big[    \mathcal{I}^4_1 \mathcal{I}^4_2  \big] \bigg]\bigg]   \\  \\ =   \underset{i \neq  j \in V ( \textbf{Z}^2 ) }{\underset{a,a^{\prime} \in \textbf{R}}{\mathrm{span}}} \bigg[  \bigg[\bigg[  \mathcal{I}^1_1 \mathcal{I}^3_2     , \mathrm{log} \big[ \mathcal{I}^1_1 \mathcal{I}^1_2  \big] \bigg]\bigg]       +   \bigg[\bigg[  \mathcal{I}^1_1 \mathcal{I}^3_2     , \mathrm{log} \big[  \mathcal{I}^3_1 \mathcal{I}^3_2            \big] \bigg]\bigg]       +  \bigg[\bigg[  \mathcal{I}^1_1 \mathcal{I}^3_2     , \mathrm{log} \big[ \mathcal{I}^2_1 \mathcal{I}^2_2      \big] \bigg]\bigg]       \\  +  \bigg[\bigg[  \mathcal{I}^1_1 \mathcal{I}^3_2     , \mathrm{log} \big[ \mathcal{I}^4_1 \mathcal{I}^4_2       \big] \bigg]\bigg]       +  \bigg[\bigg[  \mathcal{I}^1_1 \mathcal{I}^3_2     , \mathrm{log} \big[ \mathcal{I}^1_1 \mathcal{I}^3_2     \big] \bigg]\bigg]       +  \bigg[\bigg[  \mathcal{I}^1_1 \mathcal{I}^3_2     ,  \mathrm{log} \big[  \mathcal{I}^3_1 \mathcal{I}^4_2      \big] \bigg]\bigg]      \\ +  \bigg[\bigg[  \mathcal{I}^1_1 \mathcal{I}^3_2     ,   \mathrm{log} \big[  \mathcal{I}^3_1 \mathcal{I}^2_2    \big] \bigg]\bigg]     +  \bigg[\bigg[  \mathcal{I}^1_1 \mathcal{I}^3_2     ,   \mathrm{log} \big[    \mathcal{I}^4_1 \mathcal{I}^4_2  \big] \bigg]\bigg]   \bigg] \\    <   8  \underset{i \neq  j \in V ( \textbf{Z}^2 ) }{\underset{a,a^{\prime} \in \textbf{R}}{\mathrm{span}}}  \bigg[       \bigg[\bigg[\big[ \mathcal{I}^3_2 \big]^2  ,  \bigg[   \underset{i \neq j : 1 \leq i \leq 4}{\underset{1 \leq j \leq 4}{\mathrm{sup}}}    \bigg[\bigg[         4 \mathrm{log} \big[ \mathcal{I}^j_2 \big]^2  +          3  \mathrm{log} \big[ \mathcal{I}^i_2 \mathcal{I}^j_2  \big]           \bigg]\bigg]  \bigg]        \bigg]\bigg]   \bigg]  \\  \overset{(\mathrm{BL})}{=}   8  \underset{i \neq  j \in V ( \textbf{Z}^2 ) }{\underset{a,a^{\prime} \in \textbf{R}}{\mathrm{span}}} \bigg[       \bigg[\bigg[\big[ \mathcal{I}^3_2 \big]^2  ,  \bigg[   \underset{i \neq j : 1 \leq i \leq 4}{\underset{1 \leq j \leq 4}{\mathrm{sup}}}    \big\{          4 \mathrm{log} \big[ \mathcal{I}^j_2 \big]^2  \big\}  \bigg]        \bigg]\bigg]  +    \bigg[\bigg[\big[ \mathcal{I}^3_2 \big]^2  ,  \bigg[   \underset{i \neq j : 1 \leq i \leq 4}{\underset{1 \leq j \leq 4}{\mathrm{sup}}}    \big\{            3  \mathrm{log} \big[ \mathcal{I}^i_2 \mathcal{I}^j_2  \big]           \big\}  \bigg]        \bigg]\bigg]  \bigg]                 \\  <    8  \underset{i \neq  j \in V ( \textbf{Z}^2 ) }{\underset{a,a^{\prime} \in \textbf{R}}{\mathrm{span}}}  \bigg[       \bigg[\bigg[\big[ \mathcal{I}^3_2 \big]^2  ,  \bigg[   \underset{i \neq j : 1 \leq i \leq 4}{\underset{1 \leq j \leq 4}{\mathrm{sup}}}    \big\{          4 \mathrm{log} \big[ \mathcal{I}^j_2 \big]^2  \big\}  \bigg]        \bigg]\bigg]  +    \bigg[\bigg[\big[ \mathcal{I}^3_2 \big]^2  ,  \bigg[   \underset{i \neq j : 1 \leq i \leq 4}{\underset{1 \leq j \leq 4}{\mathrm{sup}}}    \big\{            3  \mathrm{log} \big[ \mathcal{I}^j_2  \big]           \big\}  \bigg]        \bigg]\bigg]  \bigg]  \end{align*}

\begin{align*} \overset{(\mathrm{AC})}{=}      8  \underset{i \neq  j \in V ( \textbf{Z}^2 ) }{\underset{a,a^{\prime} \in \textbf{R}}{\mathrm{span}}}  \bigg[       \bigg[\bigg[\big[ \mathcal{I}^3_2 \big]^2  ,  \bigg[   \underset{i \neq j : 1 \leq i \leq 4}{\underset{1 \leq j \leq 4}{\mathrm{sup}}}    \big\{          4 \mathrm{log} \big[ \mathcal{I}^j_2 \big]^2  \big\}  \bigg]        \bigg]\bigg]  -    \bigg[\bigg[\bigg[   \underset{i \neq j : 1 \leq i \leq 4}{\underset{1 \leq j \leq 4}{\mathrm{sup}}}    \big\{            3  \mathrm{log} \big[ \mathcal{I}^j_2  \big]           \big\}  \bigg]  ,   \big[ \mathcal{I}^3_2 \big]^2       \bigg]\bigg]  \bigg]   \\ \approx           8  \underset{i \neq  j \in V ( \textbf{Z}^2 ) }{\underset{a,a^{\prime} \in \textbf{R}}{\mathrm{span}}}  \bigg[    \big\{ 0 , 0 \big\}  \bigg]  \approx 0           ,    \end{align*}

\noindent corresponding to the fifth Poisson bracket,

\begin{align*}\underset{i \neq  j \in V ( \textbf{Z}^2 ) }{\underset{a,a^{\prime} \in \textbf{R}}{\mathrm{span}}}  \bigg[\bigg[   \mathcal{I}^3_1 \mathcal{I}^4_2     ,   \mathrm{log} \big[  \big\{ \mathcal{I}^1_1 \mathcal{I}^1_2 + \mathcal{I}^3_1 \mathcal{I}^3_2   +        \mathcal{I}^2_1 \mathcal{I}^2_2  + \mathcal{I}^4_1 \mathcal{I}^4_2  +   \mathcal{I}^1_1 \mathcal{I}^3_2 + \mathcal{I}^3_1 \mathcal{I}^4_2  +    \mathcal{I}^3_1 \mathcal{I}^2_2  + \mathcal{I}^4_1 \mathcal{I}^4_2   \big\} \big]  \bigg]\bigg]\\ \\  <    \underset{i \neq  j \in V ( \textbf{Z}^2 ) }{\underset{a,a^{\prime} \in \textbf{R}}{\mathrm{span}}}  \bigg[\bigg[   \mathcal{I}^3_1 \mathcal{I}^4_2     ,   \mathrm{log} \big[  \big\{ \mathcal{I}^1_1 \mathcal{I}^1_2   \big\} \big]  \bigg]\bigg]       \\              \\ <    \underset{i \neq  j \in V ( \textbf{Z}^2 ) }{\underset{a,a^{\prime} \in \textbf{R}}{\mathrm{span}}}  \bigg[\bigg[  \mathcal{I}^3_1 \mathcal{I}^4_2     ,   \mathrm{log} \big[ \mathcal{I}^1_1 \mathcal{I}^1_2 \big] \bigg]\bigg] + \underset{i \neq  j \in V ( \textbf{Z}^2 ) }{\underset{a,a^{\prime} \in \textbf{R}}{\mathrm{span}}}  \bigg[\bigg[  \mathcal{I}^3_1 \mathcal{I}^4_2     ,   \mathrm{log} \big[  \mathcal{I}^3_1 \mathcal{I}^3_2            \big] \bigg]\bigg] + \underset{i \neq  j \in V ( \textbf{Z}^2 ) }{\underset{a,a^{\prime} \in \textbf{R}}{\mathrm{span}}}  \bigg[\bigg[  \mathcal{I}^3_1 \mathcal{I}^4_2     ,   \mathrm{log} \big[   \mathcal{I}^2_1 \mathcal{I}^2_2       \big] \bigg]\bigg]   \\  +  \underset{i \neq  j \in V ( \textbf{Z}^2 ) }{\underset{a,a^{\prime} \in \textbf{R}}{\mathrm{span}}}  \bigg[\bigg[  \mathcal{I}^3_1 \mathcal{I}^4_2     ,   \mathrm{log} \big[  \mathcal{I}^4_1 \mathcal{I}^4_2     \big] \bigg]\bigg] + \underset{i \neq  j \in V ( \textbf{Z}^2 ) }{\underset{a,a^{\prime} \in \textbf{R}}{\mathrm{span}}}  \bigg[\bigg[  \mathcal{I}^3_1 \mathcal{I}^4_2     ,   \mathrm{log} \big[ \mathcal{I}^1_1      \mathcal{I}^3_2    \big] \bigg]\bigg] + \underset{i \neq  j \in V ( \textbf{Z}^2 ) }{\underset{a,a^{\prime} \in \textbf{R}}{\mathrm{span}}}  \bigg[\bigg[  \mathcal{I}^3_1 \mathcal{I}^4_2     ,   \mathrm{log} \big[ \mathcal{I}^3_1  \mathcal{I}^4_2 \big] \bigg]\bigg]  \\ \underset{i \neq  j \in V ( \textbf{Z}^2 ) }{\underset{a,a^{\prime} \in \textbf{R}}{\mathrm{span}}}  \bigg[\bigg[  \mathcal{I}^3_1 \mathcal{I}^4_2     ,   \mathrm{log} \big[      \mathcal{I}^3_1 \mathcal{I}^2_2     \big] \bigg]\bigg] + \underset{i \neq  j \in V ( \textbf{Z}^2 ) }{\underset{a,a^{\prime} \in \textbf{R}}{\mathrm{span}}}  \bigg[\bigg[  \mathcal{I}^3_1 \mathcal{I}^4_2     ,   \mathrm{log} \big[ \mathcal{I}^4_1 \mathcal{I}^4_2   \big] \bigg]\bigg]   \\  \\ =   \underset{i \neq  j \in V ( \textbf{Z}^2 ) }{\underset{a,a^{\prime} \in \textbf{R}}{\mathrm{span}}} \bigg[  \bigg[\bigg[  \mathcal{I}^3_1 \mathcal{I}^4_2     ,   \mathrm{log} \big[ \mathcal{I}^1_1 \mathcal{I}^1_2 \big] \bigg]\bigg] +  \bigg[\bigg[  \mathcal{I}^3_1 \mathcal{I}^4_2     ,   \mathrm{log} \big[  \mathcal{I}^3_1 \mathcal{I}^3_2            \big] \bigg]\bigg] +  \bigg[\bigg[  \mathcal{I}^3_1 \mathcal{I}^4_2     ,   \mathrm{log} \big[   \mathcal{I}^2_1 \mathcal{I}^2_2       \big] \bigg]\bigg]  \\   +    \bigg[\bigg[  \mathcal{I}^3_1 \mathcal{I}^4_2     ,   \mathrm{log} \big[  \mathcal{I}^4_1 \mathcal{I}^4_2     \big] \bigg]\bigg] +   \bigg[\bigg[  \mathcal{I}^3_1 \mathcal{I}^4_2     ,   \mathrm{log} \big[ \mathcal{I}^1_1      \mathcal{I}^3_2    \big] \bigg]\bigg] +         \bigg[\bigg[  \mathcal{I}^3_1 \mathcal{I}^4_2     ,   \mathrm{log} \big[ \mathcal{I}^3_1  \mathcal{I}^4_2 \big] \bigg]\bigg]  \\ +    \bigg[\bigg[  \mathcal{I}^3_1 \mathcal{I}^4_2     ,   \mathrm{log} \big[      \mathcal{I}^3_1 \mathcal{I}^2_2     \big] \bigg]\bigg] + \bigg[\bigg[  \mathcal{I}^3_1 \mathcal{I}^4_2     ,   \mathrm{log} \big[ \mathcal{I}^4_1 \mathcal{I}^4_2   \big] \bigg]\bigg]  \bigg] \\    <   8 \underset{i \neq  j \in V ( \textbf{Z}^2 ) }{\underset{a,a^{\prime} \in \textbf{R}}{\mathrm{span}}}   \bigg[       \bigg[\bigg[\big[ \mathcal{I}^4_2 \big]^2  ,  \bigg[   \underset{i \neq j : 1 \leq i \leq 4}{\underset{1 \leq j \leq 4}{\mathrm{sup}}}    \bigg[\bigg[         4 \mathrm{log} \big[ \mathcal{I}^j_2 \big]^2  +          3  \mathrm{log} \big[ \mathcal{I}^i_2 \mathcal{I}^j_2  \big]           \bigg]\bigg]  \bigg]        \bigg]\bigg]   \bigg]   \\ \overset{(\mathrm{BL})}{=}   8 \underset{i \neq  j \in V ( \textbf{Z}^2 ) }{\underset{a,a^{\prime} \in \textbf{R}}{\mathrm{span}}}  \bigg[       \bigg[\bigg[\big[ \mathcal{I}^4_2 \big]^2  ,  \bigg[   \underset{i \neq j : 1 \leq i \leq 4}{\underset{1 \leq j \leq 4}{\mathrm{sup}}}    \big\{          4 \mathrm{log} \big[ \mathcal{I}^j_2 \big]^2  \big\}  \bigg]        \bigg]\bigg]  +    \bigg[\bigg[\big[ \mathcal{I}^4_2 \big]^2  ,  \bigg[   \underset{i \neq j : 1 \leq i \leq 4}{\underset{1 \leq j \leq 4}{\mathrm{sup}}}    \big\{            3  \mathrm{log} \big[ \mathcal{I}^i_2 \mathcal{I}^j_2  \big]           \big\}  \bigg]        \bigg]\bigg]  \bigg]             \\  <    8 \underset{i \neq  j \in V ( \textbf{Z}^2 ) }{\underset{a,a^{\prime} \in \textbf{R}}{\mathrm{span}}}  \bigg[       \bigg[\bigg[\big[ \mathcal{I}^4_2 \big]^2  ,  \bigg[   \underset{i \neq j : 1 \leq i \leq 4}{\underset{1 \leq j \leq 4}{\mathrm{sup}}}    \big\{          4 \mathrm{log} \big[ \mathcal{I}^j_2 \big]^2  \big\}  \bigg]        \bigg]\bigg]  +    \bigg[\bigg[\big[ \mathcal{I}^4_2 \big]^2  ,  \bigg[   \underset{i \neq j : 1 \leq i \leq 4}{\underset{1 \leq j \leq 4}{\mathrm{sup}}}    \big\{            3  \mathrm{log} \big[ \mathcal{I}^j_2  \big]           \big\}  \bigg]        \bigg]\bigg]  \bigg]     \end{align*}

\begin{align*}  \overset{(\mathrm{AC})}{=}      8 \underset{i \neq  j \in V ( \textbf{Z}^2 ) }{\underset{a,a^{\prime} \in \textbf{R}}{\mathrm{span}}}  \bigg[       \bigg[\bigg[\big[ \mathcal{I}^4_2 \big]^2  ,  \bigg[   \underset{i \neq j : 1 \leq i \leq 4}{\underset{1 \leq j \leq 4}{\mathrm{sup}}}    \big\{          4 \mathrm{log} \big[ \mathcal{I}^j_2 \big]^2  \big\}  \bigg]        \bigg]\bigg]  -    \bigg[\bigg[\bigg[   \underset{i \neq j : 1 \leq i \leq 4}{\underset{1 \leq j \leq 4}{\mathrm{sup}}}    \big\{            3  \mathrm{log} \big[ \mathcal{I}^j_2  \big]           \big\}  \bigg]  ,   \big[ \mathcal{I}^4_2 \big]^2       \bigg]\bigg]  \bigg]   \\ \approx           8 \underset{i \neq  j \in V ( \textbf{Z}^2 ) }{\underset{a,a^{\prime} \in \textbf{R}}{\mathrm{span}}}   \bigg[    \big\{ 0 , 0 \big\}  \bigg]  \approx 0           ,  \end{align*}

\noindent corresponding to the sixth Poisson bracket,

\begin{align*} \underset{i \neq  j \in V ( \textbf{Z}^2 ) }{\underset{a,a^{\prime} \in \textbf{R}}{\mathrm{span}}} \bigg[\bigg[         \mathcal{I}^3_1 \mathcal{I}^2_2   ,   \mathrm{log} \big[  \big\{ \mathcal{I}^1_1 \mathcal{I}^1_2 + \mathcal{I}^3_1 \mathcal{I}^3_2   +        \mathcal{I}^2_1  \mathcal{I}^2_2  + \mathcal{I}^4_1 \mathcal{I}^4_2  +   \mathcal{I}^1_1 \mathcal{I}^3_2 + \mathcal{I}^3_1 \mathcal{I}^4_2  +   \mathcal{I}^3_1 \mathcal{I}^2_2  + \mathcal{I}^4_1 \mathcal{I}^4_2   \big\} \big]  \bigg]\bigg] \\ \\ < \underset{i \neq  j \in V ( \textbf{Z}^2 ) }{\underset{a,a^{\prime} \in \textbf{R}}{\mathrm{span}}}  \bigg[\bigg[    \mathcal{I}^3_1 \mathcal{I}^2_2   , \big[   \mathrm{log} \big[  \big\{ \mathcal{I}^1_1 \mathcal{I}^1_2  \big\} \big] \big]  \bigg]\bigg]    \\  < \underset{i \neq  j \in V ( \textbf{Z}^2 ) }{\underset{a,a^{\prime} \in \textbf{R}}{\mathrm{span}}}  \bigg[\bigg[    \mathcal{I}^3_1 \mathcal{I}^2_2   ,  \mathrm{log} \big[ \mathcal{I}^1_1 \mathcal{I}^1_2    \big] \bigg]\bigg] +   \underset{i \neq  j \in V ( \textbf{Z}^2 ) }{\underset{a,a^{\prime} \in \textbf{R}}{\mathrm{span}}}  \bigg[\bigg[    \mathcal{I}^3_1 \mathcal{I}^2_2   ,  \mathrm{log} \big[ \mathcal{I}^3_1 \mathcal{I}^3_2             \big] \bigg]\bigg]  +   \underset{i \neq  j \in V ( \textbf{Z}^2 ) }{\underset{a,a^{\prime} \in \textbf{R}}{\mathrm{span}}}  \bigg[\bigg[    \mathcal{I}^3_1 \mathcal{I}^2_2   ,  \mathrm{log} \big[ \mathcal{I}^2_1  \mathcal{I}^2_2   \big] \bigg]\bigg] \\  +   \underset{i \neq  j \in V ( \textbf{Z}^2 ) }{\underset{a,a^{\prime} \in \textbf{R}}{\mathrm{span}}}  \bigg[\bigg[    \mathcal{I}^3_1 \mathcal{I}^2_2   ,  \mathrm{log} \big[  \mathcal{I}^4_1 \mathcal{I}^4_2   \big] \bigg]\bigg]  +   \underset{i \neq  j \in V ( \textbf{Z}^2 ) }{\underset{a,a^{\prime} \in \textbf{R}}{\mathrm{span}}}  \bigg[\bigg[    \mathcal{I}^3_1 \mathcal{I}^2_2   ,  \mathrm{log} \big[   \mathcal{I}^1_1 \mathcal{I}^3_2    \big] \bigg]\bigg]  +   \underset{i \neq  j \in V ( \textbf{Z}^2 ) }{\underset{a,a^{\prime} \in \textbf{R}}{\mathrm{span}}}  \bigg[\bigg[    \mathcal{I}^3_1 \mathcal{I}^2_2   ,  \mathrm{log} \big[  \mathcal{I}^3_1 \mathcal{I}^4_2    \big] \bigg]\bigg] \\   +   \underset{i \neq  j \in V ( \textbf{Z}^2 ) }{\underset{a,a^{\prime} \in \textbf{R}}{\mathrm{span}}}  \bigg[\bigg[    \mathcal{I}^3_1 \mathcal{I}^2_2   ,  \mathrm{log} \big[   \mathcal{I}^3_1 \mathcal{I}^2_2  \big] \bigg]\bigg]  +   \underset{i \neq  j \in V ( \textbf{Z}^2 ) }{\underset{a,a^{\prime} \in \textbf{R}}{\mathrm{span}}}  \bigg[\bigg[    \mathcal{I}^3_1 \mathcal{I}^2_2   ,  \mathrm{log} \big[   \mathcal{I}^4_1 \mathcal{I}^4_2  \big] \bigg]\bigg] \\  = \underset{i \neq  j \in V ( \textbf{Z}^2 ) }{\underset{a,a^{\prime} \in \textbf{R}}{\mathrm{span}}} \bigg[  \bigg[\bigg[    \mathcal{I}^3_1 \mathcal{I}^2_2   ,  \mathrm{log} \big[ \mathcal{I}^1_1 \mathcal{I}^1_2    \big] \bigg]\bigg] +  \bigg[\bigg[    \mathcal{I}^3_1 \mathcal{I}^2_2   ,  \mathrm{log} \big[ \mathcal{I}^3_1 \mathcal{I}^3_2             \big] \bigg]\bigg]  +   \bigg[\bigg[    \mathcal{I}^3_1 \mathcal{I}^2_2   ,  \mathrm{log} \big[ \mathcal{I}^2_1  \mathcal{I}^2_2   \big] \bigg]\bigg]  \\ +   \bigg[\bigg[    \mathcal{I}^3_1 \mathcal{I}^2_2   ,  \mathrm{log} \big[  \mathcal{I}^4_1 \mathcal{I}^4_2   \big] \bigg]\bigg]  +   \bigg[\bigg[    \mathcal{I}^3_1 \mathcal{I}^2_2   ,  \mathrm{log} \big[   \mathcal{I}^1_1 \mathcal{I}^3_2    \big] \bigg]\bigg]  +    \bigg[\bigg[    \mathcal{I}^3_1 \mathcal{I}^2_2   ,  \mathrm{log} \big[  \mathcal{I}^3_1 \mathcal{I}^4_2    \big] \bigg]\bigg]  \\ +   \bigg[\bigg[    \mathcal{I}^3_1 \mathcal{I}^2_2   ,  \mathrm{log} \big[   \mathcal{I}^3_1 \mathcal{I}^2_2  \big] \bigg]\bigg]  +    \bigg[\bigg[    \mathcal{I}^3_1 \mathcal{I}^2_2   ,  \mathrm{log} \big[   \mathcal{I}^4_1 \mathcal{I}^4_2  \big] \bigg]\bigg] \bigg] \\     <   8 \underset{i \neq  j \in V ( \textbf{Z}^2 ) }{\underset{a,a^{\prime} \in \textbf{R}}{\mathrm{span}}} \bigg[       \bigg[\bigg[\big[ \mathcal{I}^2_2 \big]^2  ,  \bigg[   \underset{i \neq j : 1 \leq i \leq 4}{\underset{1 \leq j \leq 4}{\mathrm{sup}}}    \bigg[\bigg[         4 \mathrm{log} \big[ \mathcal{I}^j_2 \big]^2  +          3  \mathrm{log} \big[ \mathcal{I}^i_2 \mathcal{I}^j_2  \big]           \bigg]\bigg]  \bigg]        \bigg]\bigg]   \bigg]     \end{align*}

\begin{align*}   \overset{(\mathrm{BL})}{=}   8 \underset{i \neq  j \in V ( \textbf{Z}^2 ) }{\underset{a,a^{\prime} \in \textbf{R}}{\mathrm{span}}}  \bigg[       \bigg[\bigg[\big[ \mathcal{I}^2_2 \big]^2  ,  \bigg[   \underset{i \neq j : 1 \leq i \leq 4}{\underset{1 \leq j \leq 4}{\mathrm{sup}}}    \big\{          4 \mathrm{log} \big[ \mathcal{I}^j_2 \big]^2  \big\}  \bigg]        \bigg]\bigg]  +    \bigg[\bigg[\big[ \mathcal{I}^2_2 \big]^2  ,  \bigg[   \underset{i \neq j : 1 \leq i \leq 4}{\underset{1 \leq j \leq 4}{\mathrm{sup}}}    \big\{            3  \mathrm{log} \big[ \mathcal{I}^i_2 \mathcal{I}^j_2  \big]           \big\}  \bigg]        \bigg]\bigg]  \bigg]               \\  <    8 \underset{i \neq  j \in V ( \textbf{Z}^2 ) }{\underset{a,a^{\prime} \in \textbf{R}}{\mathrm{span}}}  \bigg[       \bigg[\bigg[\big[ \mathcal{I}^2_2 \big]^2  ,  \bigg[   \underset{i \neq j : 1 \leq i \leq 4}{\underset{1 \leq j \leq 4}{\mathrm{sup}}}    \big\{          4 \mathrm{log} \big[ \mathcal{I}^j_2 \big]^2  \big\}  \bigg]        \bigg]\bigg]  +    \bigg[\bigg[\big[ \mathcal{I}^2_2 \big]^2  ,  \bigg[   \underset{i \neq j : 1 \leq i \leq 4}{\underset{1 \leq j \leq 4}{\mathrm{sup}}}    \big\{            3  \mathrm{log} \big[ \mathcal{I}^j_2  \big]           \big\}  \bigg]        \bigg]\bigg]  \bigg]  \\ \overset{(\mathrm{AC})}{=}      8 \underset{i \neq  j \in V ( \textbf{Z}^2 ) }{\underset{a,a^{\prime} \in \textbf{R}}{\mathrm{span}}}  \bigg[       \bigg[\bigg[\big[ \mathcal{I}^2_2 \big]^2  ,  \bigg[   \underset{i \neq j : 1 \leq i \leq 4}{\underset{1 \leq j \leq 4}{\mathrm{sup}}}    \big\{          4 \mathrm{log} \big[ \mathcal{I}^j_2 \big]^2  \big\}  \bigg]        \bigg]\bigg]  -    \bigg[\bigg[\bigg[   \underset{i \neq j : 1 \leq i \leq 4}{\underset{1 \leq j \leq 4}{\mathrm{sup}}}    \big\{            3  \mathrm{log} \big[ \mathcal{I}^j_2  \big]           \big\}  \bigg]  ,   \big[ \mathcal{I}^2_2 \big]^2       \bigg]\bigg]  \bigg]   \\ \approx           8 \underset{i \neq  j \in V ( \textbf{Z}^2 ) }{\underset{a,a^{\prime} \in \textbf{R}}{\mathrm{span}}} \bigg[    \big\{ 0 , 0 \big\}  \bigg]  \approx 0            ,  \end{align*}

\noindent corresponding to the seventh Poisson bracket, and, lastly,

\begin{align*} \underset{i \neq  j \in V ( \textbf{Z}^2 ) }{\underset{a,a^{\prime} \in \textbf{R}}{\mathrm{span}}}  \bigg[\bigg[     \mathcal{I}^4_1 \mathcal{I}^4_2  ,   \mathrm{log} \big[  \big\{ \mathcal{I}^1_1 \mathcal{I}^1_2 + \mathcal{I}^3_1 \mathcal{I}^3_2   ,       \mathcal{I}^2_1  \mathcal{I}^2_2  + \mathcal{I}^4_1 \mathcal{I}^4_2  +   \mathcal{I}^1_1 \mathcal{I}^3_2 + \mathcal{I}^3_1 \mathcal{I}^4_2  +    \mathcal{I}^3_1 \mathcal{I}^2_2  + \mathcal{I}^4_1 \mathcal{I}^4_2   \big\} \big]  \bigg]\bigg]  \\ \\ <             \underset{i \neq  j \in V ( \textbf{Z}^2 ) }{\underset{a,a^{\prime} \in \textbf{R}}{\mathrm{span}}}  \bigg[\bigg[     \mathcal{I}^4_1 \mathcal{I}^4_2  ,   \mathrm{log} \big[  \big\{ \mathcal{I}^1_1 \mathcal{I}^1_2   \big\} \big]   \bigg]\bigg]        \\ <  \underset{i \neq  j \in V ( \textbf{Z}^2 ) }{\underset{a,a^{\prime} \in \textbf{R}}{\mathrm{span}}}  \bigg[\bigg[ \mathcal{I}^4_1 \mathcal{I}^4_2  ,  \mathrm{log} \big[ \mathcal{I}^1_1 \mathcal{I}^1_2     \big] \bigg]\bigg] +   \underset{i \neq  j \in V ( \textbf{Z}^2 ) }{\underset{a,a^{\prime} \in \textbf{R}}{\mathrm{span}}}  \bigg[\bigg[ \mathcal{I}^4_1 \mathcal{I}^4_2  ,  \mathrm{log} \big[ \mathcal{I}^3_1 \mathcal{I}^3_2           \big] \bigg]\bigg] +  \underset{i \neq  j \in V ( \textbf{Z}^2 ) }{\underset{a,a^{\prime} \in \textbf{R}}{\mathrm{span}}}  \bigg[\bigg[ \mathcal{I}^4_1 \mathcal{I}^4_2  ,  \mathrm{log} \big[ \mathcal{I}^2_1  \mathcal{I}^2_2    \big] \bigg]\bigg]  \\ + \underset{i \neq  j \in V ( \textbf{Z}^2 ) }{\underset{a,a^{\prime} \in \textbf{R}}{\mathrm{span}}}  \bigg[\bigg[ \mathcal{I}^4_1 \mathcal{I}^4_2  ,  \mathrm{log} \big[ \mathcal{I}^4_1 \mathcal{I}^4_2        \big] \bigg]\bigg] + \underset{i \neq  j \in V ( \textbf{Z}^2 ) }{\underset{a,a^{\prime} \in \textbf{R}}{\mathrm{span}}}  \bigg[\bigg[ \mathcal{I}^4_1 \mathcal{I}^4_2  ,  \mathrm{log} \big[ \mathcal{I}^1_1 \mathcal{I}^3_2   \big] \bigg]\bigg] +  \underset{i \neq  j \in V ( \textbf{Z}^2 ) }{\underset{a,a^{\prime} \in \textbf{R}}{\mathrm{span}}}  \bigg[\bigg[ \mathcal{I}^4_1 \mathcal{I}^4_2  ,  \mathrm{log} \big[  \mathcal{I}^3_1 \mathcal{I}^4_2        \big] \bigg]\bigg] \\  +   \underset{i \neq  j \in V ( \textbf{Z}^2 ) }{\underset{a,a^{\prime} \in \textbf{R}}{\mathrm{span}}}  \bigg[\bigg[ \mathcal{I}^4_1 \mathcal{I}^4_2  ,  \mathrm{log} \big[       \mathcal{I}^3_1 \mathcal{I}^2_2     \big] \bigg]\bigg]   +   \underset{i \neq  j \in V ( \textbf{Z}^2 ) }{\underset{a,a^{\prime} \in \textbf{R}}{\mathrm{span}}}  \bigg[\bigg[ \mathcal{I}^4_1 \mathcal{I}^4_2  ,  \mathrm{log} \big[         \mathcal{I}^4_1 \mathcal{I}^4_2  \big] \bigg]\bigg]     \\ \\ =   
        \underset{i \neq  j \in V ( \textbf{Z}^2 ) }{\underset{a,a^{\prime} \in \textbf{R}}{\mathrm{span}}}  \bigg[ \bigg[\bigg[ \mathcal{I}^4_1 \mathcal{I}^4_2  ,  \mathrm{log} \big[ \mathcal{I}^1_1 \mathcal{I}^1_2     \big] \bigg]\bigg] +    \bigg[\bigg[ \mathcal{I}^4_1 \mathcal{I}^4_2  ,  \mathrm{log} \big[ \mathcal{I}^3_1 \mathcal{I}^3_2           \big] \bigg]\bigg] +  \bigg[\bigg[ \mathcal{I}^4_1 \mathcal{I}^4_2  ,  \mathrm{log} \big[ \mathcal{I}^2_1  \mathcal{I}^2_2    \big] \bigg]\bigg]  \end{align*}

\begin{align*}  +   \bigg[\bigg[ \mathcal{I}^4_1 \mathcal{I}^4_2  ,  \mathrm{log} \big[ \mathcal{I}^4_1 \mathcal{I}^4_2        \big] \bigg]\bigg] +   \bigg[\bigg[ \mathcal{I}^4_1 \mathcal{I}^4_2  ,  \mathrm{log} \big[ \mathcal{I}^1_1 \mathcal{I}^3_2   \big] \bigg]\bigg] +   \bigg[\bigg[ \mathcal{I}^4_1 \mathcal{I}^4_2  ,  \mathrm{log} \big[  \mathcal{I}^3_1 \mathcal{I}^4_2        \big] \bigg]\bigg] \\ +    \bigg[\bigg[ \mathcal{I}^4_1 \mathcal{I}^4_2  ,  \mathrm{log} \big[       \mathcal{I}^3_1 \mathcal{I}^2_2     \big] \bigg]\bigg]   +    \bigg[\bigg[ \mathcal{I}^4_1 \mathcal{I}^4_2  ,  \mathrm{log} \big[         \mathcal{I}^4_1 \mathcal{I}^4_2  \big] \bigg]\bigg]       \bigg]  \\    <   8 \underset{i \neq  j \in V ( \textbf{Z}^2 ) }{\underset{a,a^{\prime} \in \textbf{R}}{\mathrm{span}}}  \bigg[       \bigg[\bigg[\big[ \mathcal{I}^4_2 \big]^2  ,  \bigg[   \underset{i \neq j : 1 \leq i \leq 4}{\underset{1 \leq j \leq 4}{\mathrm{sup}}}    \bigg[\bigg[         4 \mathrm{log} \big[ \mathcal{I}^j_2 \big]^2  +          3  \mathrm{log} \big[ \mathcal{I}^i_2 \mathcal{I}^j_2  \big]           \bigg]\bigg]  \bigg]        \bigg]\bigg]   \bigg]  \\  \overset{(\mathrm{BL})}{=}   8 \underset{i \neq  j \in V ( \textbf{Z}^2 ) }{\underset{a,a^{\prime} \in \textbf{R}}{\mathrm{span}}}  \bigg[       \bigg[\bigg[\big[ \mathcal{I}^4_2 \big]^2  ,  \bigg[   \underset{i \neq j : 1 \leq i \leq 4}{\underset{1 \leq j \leq 4}{\mathrm{sup}}}    \big\{          4 \mathrm{log} \big[ \mathcal{I}^j_2 \big]^2  \big\}  \bigg]        \bigg]\bigg]  +    \bigg[\bigg[\big[ \mathcal{I}^4_2 \big]^2  ,  \bigg[   \underset{i \neq j : 1 \leq i \leq 4}{\underset{1 \leq j \leq 4}{\mathrm{sup}}}    \big\{            3  \mathrm{log} \big[ \mathcal{I}^i_2 \mathcal{I}^j_2  \big]           \big\}  \bigg]        \bigg]\bigg]  \bigg]                 \\  <    8 \underset{i \neq  j \in V ( \textbf{Z}^2 ) }{\underset{a,a^{\prime} \in \textbf{R}}{\mathrm{span}}}  \bigg[       \bigg[\bigg[\big[ \mathcal{I}^4_2 \big]^2  ,  \bigg[   \underset{i \neq j : 1 \leq i \leq 4}{\underset{1 \leq j \leq 4}{\mathrm{sup}}}    \big\{          4 \mathrm{log} \big[ \mathcal{I}^j_2 \big]^2  \big\}  \bigg]        \bigg]\bigg]  +    \bigg[\bigg[\big[ \mathcal{I}^4_2 \big]^2  ,  \bigg[   \underset{i \neq j : 1 \leq i \leq 4}{\underset{1 \leq j \leq 4}{\mathrm{sup}}}    \big\{            3  \mathrm{log} \big[ \mathcal{I}^j_2  \big]           \big\}  \bigg]        \bigg]\bigg]  \bigg]  \\ \overset{(\mathrm{AC})}{=}      8 \underset{i \neq  j \in V ( \textbf{Z}^2 ) }{\underset{a,a^{\prime} \in \textbf{R}}{\mathrm{span}}}  \bigg[       \bigg[\bigg[\big[ \mathcal{I}^4_2 \big]^2  ,  \bigg[   \underset{i \neq j : 1 \leq i \leq 4}{\underset{1 \leq j \leq 4}{\mathrm{sup}}}    \big\{          4 \mathrm{log} \big[ \mathcal{I}^j_2 \big]^2  \big\}  \bigg]        \bigg]\bigg]  -    \bigg[\bigg[\bigg[   \underset{i \neq j : 1 \leq i \leq 4}{\underset{1 \leq j \leq 4}{\mathrm{sup}}}    \big\{            3  \mathrm{log} \big[ \mathcal{I}^j_2  \big]           \big\}  \bigg]  ,   \big[ \mathcal{I}^4_2 \big]^2       \bigg]\bigg]  \bigg]   \\ \approx           8 {\underset{a,a^{\prime} \in \textbf{R}}{\mathrm{span}}}  \bigg[    \big\{ 0 , 0 \big\}  \bigg]  \approx 0            ,    \end{align*}

\noindent corresponding to the eighth Poisson bracket. Hence, the reverse direction of the iff correspondence follows, as,

\begin{align*}
 \underset{i \neq  j \in V ( \textbf{Z}^2 ) }{\underset{a,a^{\prime} \in \textbf{R}}{\mathrm{span}}}  \bigg[\bigg[  \mathcal{I}^1_1 \mathcal{I}^1_2  ,   \mathrm{log} \big[  \big\{ \mathcal{I}^1_1 \mathcal{I}^1_2 + \mathcal{I}^3_1 \mathcal{I}^3_2   +        \mathcal{I}^2_1 \mathcal{I}^2_2  + \mathcal{I}^4_1 \mathcal{I}^4_2 +  \mathcal{I}^1_1 \mathcal{I}^3_2 + \mathcal{I}^3_1 \mathcal{I}^4_2  ,   \mathcal{I}^3_1 \mathcal{I}^2_2  + \mathcal{I}^4_1 \mathcal{I}^4_2   \big\} \big]  \bigg]\bigg] \\  +   \underset{i \neq  j \in V ( \textbf{Z}^2 ) }{\underset{a,a^{\prime} \in \textbf{R}}{\mathrm{span}}}  \bigg[\bigg[  \mathcal{I}^3_1 \mathcal{I}^3_2 ,  \mathrm{log} \big[  \big\{ \mathcal{I}^1_1 \mathcal{I}^1_2 + \mathcal{I}^3_1 \mathcal{I}^3_2  +        \mathcal{I}^2_1  \mathcal{I}^2_2  + \mathcal{I}^4_1 \mathcal{I}^4_2 +   \mathcal{I}^1_1 \mathcal{I}^3_2 + \mathcal{I}^3_1 \mathcal{I}^4_2  +   \mathcal{I}^3_1 \mathcal{I}^2_2  + \mathcal{I}^4_1 \mathcal{I}^4_2   \big\} \big]  \bigg]\bigg] \\  +  \underset{i \neq  j \in V ( \textbf{Z}^2 ) }{\underset{a,a^{\prime} \in \textbf{R}}{\mathrm{span}}} \bigg[\bigg[           \mathcal{I}^2_1 \mathcal{I}^2_2  ,  \mathrm{log} \big[  \big\{ \mathcal{I}^1_1 \mathcal{I}^1_2 + \mathcal{I}^3_1 \mathcal{I}^3_2   +        \mathcal{I}^2_1 \mathcal{I}^2_2  + \mathcal{I}^4_1 \mathcal{I}^4_2  +  \mathcal{I}^1_1 \mathcal{I}^3_2 + \mathcal{I}^3_1 \mathcal{I}^4_2  +   \mathcal{I}^3_1 \mathcal{I}^2_2  + \mathcal{I}^4_1 \mathcal{I}^4_2   \big\} \big]  \bigg]\bigg] \\  +  \underset{i \neq  j \in V ( \textbf{Z}^2 ) }{\underset{a,a^{\prime} \in \textbf{R}}{\mathrm{span}}}  \bigg[\bigg[   \mathcal{I}^4_1 \mathcal{I}^4_2   ,   \mathrm{log} \big[  \big\{ \mathcal{I}^1_1 \mathcal{I}^1_2 + \mathcal{I}^3_1 \mathcal{I}^3_2   ,       \mathcal{I}^2_1 \mathcal{I}^2_2  + \mathcal{I}^4_1 \mathcal{I}^4_2  +   \mathcal{I}^1_1 \mathcal{I}^3_2 + \mathcal{I}^3_1 \mathcal{I}^4_2  +    \mathcal{I}^3_1 \mathcal{I}^2_2  + \mathcal{I}^4_1 \mathcal{I}^4_2   \big\} \big]  \bigg]\bigg] \end{align*}

\begin{align*}   + \underset{i \neq  j \in V ( \textbf{Z}^2 ) }{\underset{a,a^{\prime} \in \textbf{R}}{\mathrm{span}}} \bigg[\bigg[    \mathcal{I}^1_1 \mathcal{I}^3_2     ,   \mathrm{log} \big[  \big\{ \mathcal{I}^1_1 \mathcal{I}^1_2 + \mathcal{I}^3_1 \mathcal{I}^3_2   ,       \mathcal{I}^2_1 \mathcal{I}^2_2  + \mathcal{I}^4_1 \mathcal{I}^4_2  +   \mathcal{I}^1_1 \mathcal{I}^3_2 + \mathcal{I}^3_1 \mathcal{I}^4_2  +    \mathcal{I}^3_1 \mathcal{I}^2_2  + \mathcal{I}^4_1 \mathcal{I}^4_2   \big\} \big]  \bigg]\bigg] \\  + \underset{i \neq  j \in V ( \textbf{Z}^2 ) }{\underset{a,a^{\prime} \in \textbf{R}}{\mathrm{span}}}  \bigg[\bigg[   \mathcal{I}^3_1 \mathcal{I}^4_2     ,   \mathrm{log} \big[  \big\{ \mathcal{I}^1_1 \mathcal{I}^1_2 + \mathcal{I}^3_1 \mathcal{I}^3_2   +        \mathcal{I}^2_1 \mathcal{I}^2_2  + \mathcal{I}^4_1 \mathcal{I}^4_2  +   \mathcal{I}^1_1 \mathcal{I}^3_2 + \mathcal{I}^3_1 \mathcal{I}^4_2  +    \mathcal{I}^3_1 \mathcal{I}^2_2  + \mathcal{I}^4_1 \mathcal{I}^4_2   \big\} \big]  \bigg]\bigg] \\  + \underset{i \neq  j \in V ( \textbf{Z}^2 ) }{\underset{a,a^{\prime} \in \textbf{R}}{\mathrm{span}}} \bigg[\bigg[         \mathcal{I}^3_1 \mathcal{I}^2_2   ,   \mathrm{log} \big[  \big\{ \mathcal{I}^1_1 \mathcal{I}^1_2 + \mathcal{I}^3_1 \mathcal{I}^3_2   ,       \mathcal{I}^2_1  \mathcal{I}^2_2  + \mathcal{I}^4_1 \mathcal{I}^4_2  +   \mathcal{I}^1_1 \mathcal{I}^3_2 + \mathcal{I}^3_1 \mathcal{I}^4_2  +   \mathcal{I}^3_1 \mathcal{I}^2_2  + \mathcal{I}^4_1 \mathcal{I}^4_2   \big\} \big]  \bigg]\bigg] \\ + \underset{i \neq  j \in V ( \textbf{Z}^2 ) }{\underset{a,a^{\prime} \in \textbf{R}}{\mathrm{span}}}  \bigg[\bigg[     \mathcal{I}^4_1 \mathcal{I}^4_2  ,   \mathrm{log} \big[  \big\{ \mathcal{I}^1_1 \mathcal{I}^1_2 + \mathcal{I}^3_1 \mathcal{I}^3_2   ,       \mathcal{I}^2_1  \mathcal{I}^2_2  + \mathcal{I}^4_1 \mathcal{I}^4_2  +   \mathcal{I}^1_1 \mathcal{I}^3_2 + \mathcal{I}^3_1 \mathcal{I}^4_2  +    \mathcal{I}^3_1 \mathcal{I}^2_2  + \mathcal{I}^4_1 \mathcal{I}^4_2   \big\} \big]  \bigg]\bigg]    \approx 0  . 
\end{align*}

\noindent ($\Longrightarrow$) Suppose that $\big[ \big[ \Phi^{\mathrm{Ising}}_{\text{Lower order}} , \bar{\Phi^{\mathrm{Ising}}_{\text{Lower order}}} \big] \big]  \approx 0$. To demonstrate that (1) holds, given the assumption that the Poisson bracket of the \textit{lower} order expansion of the action-angle coordinates with its complex conjugate vanishes, it suffices to argue,

\begin{align*}
\underset{\text{representation of \textbf{T}}}{\underset{\text{of the finite-dimensional}}{\underset{\text{Rows and columns}}{\sum}}} \bigg[ 
\underset{i \neq  j \in V ( \textbf{Z}^2 ) }{\underset{a,a^{\prime} \in \textbf{R}}{\mathrm{span}}}   \bigg[\bigg[     \textit{Higher order Ising-type block operators}  ,   \mathrm{log} \big[  \big\{ \mathcal{I}^1_1 \mathcal{I}^1_2 + \mathcal{I}^3_1 \mathcal{I}^3_2  \\  ,       \mathcal{I}^2_1  \mathcal{I}^2_2   + \mathcal{I}^4_1 \mathcal{I}^4_2   +   \mathcal{I}^1_1 \mathcal{I}^3_2  + \mathcal{I}^3_1 \mathcal{I}^4_2   +    \mathcal{I}^3_1 \mathcal{I}^2_2  + \mathcal{I}^4_1 \mathcal{I}^4_2    + \cdots \big\} \big]  \bigg]\bigg]   \bigg]    \approx 0 . 
\end{align*}

\noindent The desired superposition of Poisson brackets, corresponding to,

\begin{align*}
{\underset{a,a^{\prime} \in \textbf{R}}{\mathrm{span}}}   \bigg[\bigg[ \mathcal{I}^1_5 \mathcal{I}^2_5   ,  \mathrm{log} \big[  \big\{ \mathcal{I}^1_1 \mathcal{I}^1_2 + \mathcal{I}^3_1 \mathcal{I}^3_2   ,       \mathcal{I}^2_1  \mathcal{I}^2_2  + \mathcal{I}^4_1 \mathcal{I}^4_2  +   \mathcal{I}^1_1 \mathcal{I}^3_2  + \mathcal{I}^3_1 \mathcal{I}^4_2   +    \mathcal{I}^3_1 \mathcal{I}^2_2  + \mathcal{I}^4_1 \mathcal{I}^4_2    + \cdots \big\} \big] \bigg]\bigg]    ,
\end{align*}

\noindent can be analyzed through the following rearrangements,

\begin{align*}
    {\underset{a,a^{\prime} \in \textbf{R}}{\mathrm{span}}}   \bigg\{\mathcal{I}^1_5   \mathcal{I}^2_5  ,  \mathrm{log} \big[  \big\{ \mathcal{I}^1_1 \mathcal{I}^1_2 + \mathcal{I}^3_1 \mathcal{I}^3_2   ,       \mathcal{I}^2_1  \mathcal{I}^2_2  + \mathcal{I}^4_1 \mathcal{I}^4_2  +   \mathcal{I}^1_1 \mathcal{I}^3_2  + \mathcal{I}^3_1 \mathcal{I}^4_2   +    \mathcal{I}^3_1 \mathcal{I}^2_2  + \mathcal{I}^4_1 \mathcal{I}^4_2    + \cdots \big\} \big] \bigg]\bigg]   \\ \\ <   {\underset{a,a^{\prime} \in \textbf{R}}{\mathrm{span}}}   \bigg[\bigg[\mathcal{I}^1_5   \mathcal{I}^2_5  ,  \mathrm{log} \big[   \mathcal{I}^1_1 \mathcal{I}^1_2  \big] \bigg]\bigg] \end{align*}

\begin{align*}        < {\underset{a,a^{\prime} \in \textbf{R}}{\mathrm{span}}}   \bigg[\bigg[\mathcal{I}^1_5   \mathcal{I}^2_5  ,  \mathrm{log} \big[   \mathcal{I}^1_1 \mathcal{I}^1_2  \big] \bigg]\bigg]  +   {\underset{a,a^{\prime} \in \textbf{R}}{\mathrm{span}}}   \bigg[\bigg[ \mathcal{I}^1_5   \mathcal{I}^2_5  ,  \mathrm{log} \big[   \mathcal{I}^3_1 \mathcal{I}^3_2  \big] \bigg]\bigg]   + \cdots + {\underset{a,a^{\prime} \in \textbf{R}}{\mathrm{span}}}   \bigg[\bigg[\mathcal{I}^1_5   \mathcal{I}^2_5   ,  \mathrm{log} \big[   \mathcal{I}^4_N \mathcal{I}^4_N  \big] \bigg]\bigg]                     \\ \\ \equiv  {\underset{a,a^{\prime} \in \textbf{R}}{\mathrm{span}}}  \bigg[ \bigg[\bigg[ \mathcal{I}^1_5   \mathcal{I}^2_5   ,  \mathrm{log} \big[   \mathcal{I}^1_1 \mathcal{I}^1_2  \big] \bigg]\bigg]  +     \bigg[\bigg[\mathcal{I}^1_5   \mathcal{I}^2_5   ,  \mathrm{log} \big[   \mathcal{I}^3_1 \mathcal{I}^3_2  \big] \bigg]\bigg]   + \cdots +   \bigg[\bigg[\mathcal{I}^1_5   \mathcal{I}^2_5  ,  \mathrm{log} \big[   \mathcal{I}^4_N \mathcal{I}^4_N  \big] \bigg]\bigg]    \bigg] \\ \\      <  \big( N + 3 \big) \bigg[  \underset{i \neq  j \in V ( \textbf{Z}^2 ) }{\underset{a,a^{\prime} \in \textbf{R}}{\mathrm{span}}}  \bigg[       \bigg[\bigg[\big[ \mathcal{I}^2_5 \big]^2  ,  \bigg[   \underset{i \neq j : 1 \leq i \leq 4}{\underset{1 \leq j \leq 4}{\mathrm{sup}}}    \bigg[\bigg[         4 \mathrm{log} \big[ \mathcal{I}^j_2 \big]^2  +          3  \mathrm{log} \big[ \mathcal{I}^i_2 \mathcal{I}^j_2  \big]           \bigg]\bigg]  \bigg]        \bigg]\bigg]   \bigg] \\   +             \underset{i \neq  j \in V ( \textbf{Z}^2 ) }{\underset{a,a^{\prime} \in \textbf{R}}{\mathrm{span}}} \bigg[   \bigg[\bigg[\big[ \mathcal{I}^3_5 \big]^2  ,  \bigg[   \underset{i \neq j : 1 \leq i \leq 4}{\underset{1 \leq j \leq 4}{\mathrm{sup}}}    \bigg[\bigg[         4 \mathrm{log} \big[ \mathcal{I}^j_2 \big]^2  +          3  \mathrm{log} \big[ \mathcal{I}^i_2 \mathcal{I}^j_2  \big]           \bigg]\bigg]  \bigg]        \bigg]\bigg] \bigg]   + \dots \\ +    \underset{i \neq  j \in V ( \textbf{Z}^2 ) }{\underset{a,a^{\prime} \in \textbf{R}}{\mathrm{span}}} \bigg[   \bigg[\bigg[\big[ \mathcal{I}^4_N \big]^2  ,  \bigg[   \underset{i \neq j : 1 \leq i \leq 4}{\underset{1 \leq j \leq 4}{\mathrm{sup}}}    \bigg[\bigg[         4 \mathrm{log} \big[ \mathcal{I}^j_2 \big]^2  +          3  \mathrm{log} \big[ \mathcal{I}^i_2 \mathcal{I}^j_2  \big]           \bigg]\bigg]  \bigg]        \bigg]\bigg]  \bigg]  \bigg]    \\ \\ 
  = \big( N + 3 \big) \underset{i \neq  j \in V ( \textbf{Z}^2 ) }{\underset{a,a^{\prime} \in \textbf{R}}{\mathrm{span}}}   \bigg[   \bigg[       \bigg[\bigg[\big[ \mathcal{I}^2_5 \big]^2  ,  \bigg[   \underset{i \neq j : 1 \leq i \leq 4}{\underset{1 \leq j \leq 4}{\mathrm{sup}}}    \bigg[\bigg[         4 \mathrm{log} \big[ \mathcal{I}^j_2 \big]^2  +          3  \mathrm{log} \big[ \mathcal{I}^i_2 \mathcal{I}^j_2  \big]           \bigg]\bigg]  \bigg]        \bigg]\bigg]   \bigg] \\   +             \bigg[   \bigg[\bigg[\big[ \mathcal{I}^3_5 \big]^2  ,  \bigg[   \underset{i \neq j : 1 \leq i \leq 4}{\underset{1 \leq j \leq 4}{\mathrm{sup}}}    \bigg[\bigg[         4 \mathrm{log} \big[ \mathcal{I}^j_2 \big]^2  +          3  \mathrm{log} \big[ \mathcal{I}^i_2 \mathcal{I}^j_2  \big]           \bigg]\bigg]  \bigg]        \bigg]\bigg] \bigg]   + \dots         
\\  +    \bigg[   \bigg[\bigg[\big[ \mathcal{I}^4_N \big]^2  ,  \bigg[   \underset{i \neq j : 1 \leq i \leq 4}{\underset{1 \leq j \leq 4}{\mathrm{sup}}}    \bigg[\bigg[         4 \mathrm{log} \big[ \mathcal{I}^j_2 \big]^2  +          3  \mathrm{log} \big[ \mathcal{I}^i_2 \mathcal{I}^j_2  \big]           \bigg]\bigg]  \bigg]        \bigg]\bigg]  \bigg]  \bigg] .             
\end{align*}

\noindent It suffices to demonstrate that if,

\begin{align*}
    \underset{i \neq  j \in V ( \textbf{Z}^2 ) }{\underset{a,a^{\prime} \in \textbf{R}}{\mathrm{span}}}   \bigg[\bigg[\big[ \mathcal{I}^2_5 \big]^2  ,  \bigg[   \underset{i \neq j : 1 \leq i \leq 4}{\underset{1 \leq j \leq 4}{\mathrm{sup}}}    \bigg[\bigg[         4 \mathrm{log} \big[ \mathcal{I}^j_2 \big]^2  +          3  \mathrm{log} \big[ \mathcal{I}^i_2 \mathcal{I}^j_2  \big]           \bigg]\bigg]  \bigg]        \bigg]\bigg]  \approx 0  , \end{align*}

\begin{align*}   \underset{i \neq  j \in V ( \textbf{Z}^2 ) }{\underset{a,a^{\prime} \in \textbf{R}}{\mathrm{span}}}  \bigg[\bigg[\big[ \mathcal{I}^3_5 \big]^2  ,  \bigg[   \underset{i \neq j : 1 \leq i \leq 4}{\underset{1 \leq j \leq 4}{\mathrm{sup}}}    \bigg[\bigg[         4 \mathrm{log} \big[ \mathcal{I}^j_2 \big]^2  +          3  \mathrm{log} \big[ \mathcal{I}^i_2 \mathcal{I}^j_2  \big]           \bigg]\bigg]  \bigg]        \bigg]\bigg] \approx 0 , \\ \\ \underset{i \neq  j \in V ( \textbf{Z}^2 ) }{\underset{a,a^{\prime} \in \textbf{R}}{\mathrm{span}}}  \bigg[\bigg[\big[ \mathcal{I}^4_N \big]^2  ,  \bigg[   \underset{i \neq j : 1 \leq i \leq 4}{\underset{1 \leq j \leq 4}{\mathrm{sup}}}    \bigg[\bigg[         4 \mathrm{log} \big[ \mathcal{I}^j_2 \big]^2  +          3  \mathrm{log} \big[ \mathcal{I}^i_2 \mathcal{I}^j_2  \big]           \bigg]\bigg]  \bigg]        \bigg]\bigg] \approx 0  , 
\end{align*}

\noindent then countably many linear combinations between the above terms also vanish, hence implying that $\big[ \big[ \Phi^{\mathrm{Ising-type}} , \bar{\Phi^{\mathrm{Ising-type}}} \big] \big]  \approx 0$. To demonstrate that the remaining direction of the iff correspondence holds, observe,

\begin{align*}
    \underset{i \neq  j \in V ( \textbf{Z}^2 ) }{\underset{a,a^{\prime} \in \textbf{R}}{\mathrm{span}}}   \bigg[\bigg[\big[ \mathcal{I}^2_5 \big]^2  ,  \bigg[   \underset{i \neq j : 1 \leq i \leq 4}{\underset{1 \leq j \leq 4}{\mathrm{sup}}}    \bigg[\bigg[         4 \mathrm{log} \big[ \mathcal{I}^j_2 \big]^2  +          3  \mathrm{log} \big[ \mathcal{I}^i_2 \mathcal{I}^j_2  \big]           \bigg]\bigg]  \bigg]        \bigg]\bigg]  \\ < \underset{i \neq  j \in V ( \textbf{Z}^2 ) }{\underset{a,a^{\prime} \in \textbf{R}}{\mathrm{span}}}   \bigg[\bigg[\big[ \mathcal{I}^2_N \big]^2  ,  \bigg[   \underset{i \neq j : 1 \leq i \leq 4}{\underset{1 \leq j \leq 4}{\mathrm{sup}}}    \bigg[\bigg[         4 \mathrm{log} \big[ \mathcal{I}^j_2 \big]^2  +          3  \mathrm{log} \big[ \mathcal{I}^i_2 \mathcal{I}^j_2  \big]           \bigg]\bigg]  \bigg]        \bigg]\bigg]   
\\ 
     <    \underset{i \neq  j \in V ( \textbf{Z}^2 ) }{\underset{a,a^{\prime} \in \textbf{R}}{\mathrm{span}}}   \bigg[\bigg[\big[ \mathcal{I}^2_N \big]^2  ,  \bigg[   \underset{i \neq j : 1 \leq i \leq 4}{\underset{1 \leq j \leq 4}{\mathrm{sup}}}    \bigg[\bigg[         4 \mathrm{log} \big[ \mathcal{I}^j_2 \big]^2          \bigg]\bigg]  \bigg]        \bigg]\bigg]   \\    <    \underset{i \neq  j \in V ( \textbf{Z}^2 ) }{\underset{a,a^{\prime} \in \textbf{R}}{\mathrm{span}}}   \bigg[\bigg[\big[ \mathcal{I}^2_N \big]^2  ,  \bigg[   \underset{i \neq j : 1 \leq i \leq 4}{\underset{1 \leq j \leq 4}{\mathrm{sup}}}    \bigg[\bigg[         4 \mathrm{log} \big[ \mathcal{I}^j_1 \big]^2          \bigg]\bigg]  \bigg]        \bigg]\bigg] \\   \equiv     \underset{i \neq  j \in V ( \textbf{Z}^2 ) }{\underset{a,a^{\prime} \in \textbf{R}}{\mathrm{span}}}   \bigg[\bigg[\big[ \mathcal{I}^2_N \big]^2  ,  \bigg[   \underset{i \neq j : 1 \leq i \leq 4}{\underset{1 \leq j \leq 4}{\mathrm{sup}}}    \bigg[\bigg[         4 \mathrm{log} \big[ \mathcal{I}^j_1 \big]^3 - 4 \mathrm{log} \big[ \mathcal{I}^j_1  \big]          \bigg]\bigg]  \bigg]        \bigg]\bigg]     \\ \overset{(\mathrm{BL})}{=}          \underset{i \neq  j \in V ( \textbf{Z}^2 ) }{\underset{a,a^{\prime} \in \textbf{R}}{\mathrm{span}}} \bigg[  \bigg[\bigg[\big[ \mathcal{I}^2_N \big]^2  ,  \bigg[   \underset{i \neq j : 1 \leq i \leq 4}{\underset{1 \leq j \leq 4}{\mathrm{sup}}}    \bigg[\bigg[         4 \mathrm{log} \big[ \mathcal{I}^j_1 \big]^3        \bigg]\bigg]  \bigg]        \bigg]\bigg]    \\ -   \bigg[\bigg[\big[ \mathcal{I}^2_N \big]^2  ,  \bigg[   \underset{i \neq j : 1 \leq i \leq 4}{\underset{1 \leq j \leq 4}{\mathrm{sup}}}    \bigg[\bigg[         4 \mathrm{log} \big[ \mathcal{I}^j_1  \big]          \bigg]\bigg]  \bigg]        \bigg]\bigg]     \bigg]    \end{align*}

\begin{align*}      \overset{(\mathrm{AC})}{=}          \underset{i \neq  j \in V ( \textbf{Z}^2 ) }{\underset{a,a^{\prime} \in \textbf{R}}{\mathrm{span}}} \bigg[  \bigg[\bigg[\big[ \mathcal{I}^2_N \big]^2  ,  \bigg[   \underset{i \neq j : 1 \leq i \leq 4}{\underset{1 \leq j \leq 4}{\mathrm{sup}}}    \bigg[\bigg[         4 \mathrm{log} \big[ \mathcal{I}^j_1 \big]^3        \bigg]\bigg]  \bigg]        \bigg]\bigg]  \\  +    \bigg[\bigg[  \bigg[   \underset{i \neq j : 1 \leq i \leq 4}{\underset{1 \leq j \leq 4}{\mathrm{sup}}}    \bigg[\bigg[         4 \mathrm{log} \big[ \mathcal{I}^j_1  \big]          \bigg]\bigg]  \bigg]   ,  \big[ \mathcal{I}^2_N \big]^2      \bigg]\bigg]     \bigg]    \\   \approx            \underset{i \neq  j \in V ( \textbf{Z}^2 ) }{\underset{a,a^{\prime} \in \textbf{R}}{\mathrm{span}}} \bigg[  \big\{ 0 , 0 \big\} \bigg] \\ \approx 0                   , 
\end{align*}

\noindent For the remaining terms,

\begin{align*}
     \underset{i \neq  j \in V ( \textbf{Z}^2 ) }{\underset{a,a^{\prime} \in \textbf{R}}{\mathrm{span}}}  \bigg[\bigg[  \mathcal{I}^1_1 \mathcal{I}^1_2  ,   \mathrm{log} \big[  \big\{ \mathcal{I}^1_1 \mathcal{I}^1_2 + \mathcal{I}^3_1 \mathcal{I}^3_2   +        \mathcal{I}^2_1 \mathcal{I}^2_2  + \mathcal{I}^4_1 \mathcal{I}^4_2 +  \mathcal{I}^1_1 \mathcal{I}^3_2 + \mathcal{I}^3_1 \mathcal{I}^4_2  +    \mathcal{I}^3_1 \mathcal{I}^2_2  + \mathcal{I}^4_1 \mathcal{I}^4_2  + \cdots  \big\} \big]  \bigg]\bigg] \\  \approx 0 \\ \\    \underset{i \neq  j \in V ( \textbf{Z}^2 ) }{\underset{a,a^{\prime} \in \textbf{R}}{\mathrm{span}}}  \bigg[\bigg[  \mathcal{I}^3_1 \mathcal{I}^3_2 ,  \mathrm{log} \big[  \big\{ \mathcal{I}^1_1 \mathcal{I}^1_2 + \mathcal{I}^3_1 \mathcal{I}^3_2  +        \mathcal{I}^2_1  \mathcal{I}^2_2  + \mathcal{I}^4_1 \mathcal{I}^4_2 +   \mathcal{I}^1_1 \mathcal{I}^3_2 + \mathcal{I}^3_1 \mathcal{I}^4_2  +   \mathcal{I}^3_1 \mathcal{I}^2_2  + \mathcal{I}^4_1 \mathcal{I}^4_2  + \cdots   \big\} \big]  \bigg]\bigg] \\ \approx 0 \\ \\   \underset{i \neq  j \in V ( \textbf{Z}^2 ) }{\underset{a,a^{\prime} \in \textbf{R}}{\mathrm{span}}} \bigg[\bigg[           \mathcal{I}^2_1 \mathcal{I}^2_2  ,  \mathrm{log} \big[  \big\{ \mathcal{I}^1_1 \mathcal{I}^1_2 + \mathcal{I}^3_1 \mathcal{I}^3_2   +        \mathcal{I}^2_1 \mathcal{I}^2_2  + \mathcal{I}^4_1 \mathcal{I}^4_2  +  \mathcal{I}^1_1 \mathcal{I}^3_2 + \mathcal{I}^3_1 \mathcal{I}^4_2  +   \mathcal{I}^3_1 \mathcal{I}^2_2  + \mathcal{I}^4_1 \mathcal{I}^4_2    + \cdots \big\} \big]  \bigg]\bigg] \\   \approx 0 \\ \\   \underset{i \neq  j \in V ( \textbf{Z}^2 ) }{\underset{a,a^{\prime} \in \textbf{R}}{\mathrm{span}}}  \bigg[\bigg[   \mathcal{I}^4_1 \mathcal{I}^4_2   ,   \mathrm{log} \big[  \big\{ \mathcal{I}^1_1 \mathcal{I}^1_2 + \mathcal{I}^3_1 \mathcal{I}^3_2   ,       \mathcal{I}^2_1 \mathcal{I}^2_2  + \mathcal{I}^4_1 \mathcal{I}^4_2  +   \mathcal{I}^1_1 \mathcal{I}^3_2 + \mathcal{I}^3_1 \mathcal{I}^4_2  +    \mathcal{I}^3_1 \mathcal{I}^2_2  + \mathcal{I}^4_1 \mathcal{I}^4_2    + \cdots \big\} \big]  \bigg]\bigg] \\  \approx 0 \end{align*}

     \begin{align*}   \underset{i \neq  j \in V ( \textbf{Z}^2 ) }{\underset{a,a^{\prime} \in \textbf{R}}{\mathrm{span}}} \bigg[\bigg[    \mathcal{I}^1_1 \mathcal{I}^3_2     ,   \mathrm{log} \big[  \big\{ \mathcal{I}^1_1 \mathcal{I}^1_2 + \mathcal{I}^3_1 \mathcal{I}^3_2   ,       \mathcal{I}^2_1 \mathcal{I}^2_2  + \mathcal{I}^4_1 \mathcal{I}^4_2  +   \mathcal{I}^1_1 \mathcal{I}^3_2 + \mathcal{I}^3_1 \mathcal{I}^4_2  +    \mathcal{I}^3_1 \mathcal{I}^2_2  + \mathcal{I}^4_1 \mathcal{I}^4_2    + \cdots \big\} \big]  \bigg]\bigg] \\  \approx 0 \\ \\     \underset{i \neq  j \in V ( \textbf{Z}^2 ) }{\underset{a,a^{\prime} \in \textbf{R}}{\mathrm{span}}}  \bigg[\bigg[   \mathcal{I}^3_1 \mathcal{I}^4_2     ,   \mathrm{log} \big[  \big\{ \mathcal{I}^1_1 \mathcal{I}^1_2 + \mathcal{I}^3_1 \mathcal{I}^3_2   +        \mathcal{I}^2_1 \mathcal{I}^2_2  + \mathcal{I}^4_1 \mathcal{I}^4_2  +   \mathcal{I}^1_1 \mathcal{I}^3_2 + \mathcal{I}^3_1 \mathcal{I}^4_2  +    \mathcal{I}^3_1 \mathcal{I}^2_2  + \mathcal{I}^4_1 \mathcal{I}^4_2    + \cdots \big\} \big]  \bigg]\bigg] \\  \approx 0 \\ \\  \underset{i \neq  j \in V ( \textbf{Z}^2 ) }{\underset{a,a^{\prime} \in \textbf{R}}{\mathrm{span}}} \bigg[\bigg[         \mathcal{I}^3_1 \mathcal{I}^2_2   ,   \mathrm{log} \big[  \big\{ \mathcal{I}^1_1 \mathcal{I}^1_2 + \mathcal{I}^3_1 \mathcal{I}^3_2   ,       \mathcal{I}^2_1  \mathcal{I}^2_2  + \mathcal{I}^4_1 \mathcal{I}^4_2  +   \mathcal{I}^1_1 \mathcal{I}^3_2 + \mathcal{I}^3_1 \mathcal{I}^4_2  +   \mathcal{I}^3_1 \mathcal{I}^2_2  + \mathcal{I}^4_1 \mathcal{I}^4_2    + \cdots \big\} \big]  \bigg]\bigg] \\  \approx 0 \\ \\ \underset{i \neq  j \in V ( \textbf{Z}^2 ) }{\underset{a,a^{\prime} \in \textbf{R}}{\mathrm{span}}}  \bigg[\bigg[     \mathcal{I}^4_1 \mathcal{I}^4_2  ,   \mathrm{log} \big[  \big\{ \mathcal{I}^1_1 \mathcal{I}^1_2 + \mathcal{I}^3_1 \mathcal{I}^3_2   ,       \mathcal{I}^2_1  \mathcal{I}^2_2  + \mathcal{I}^4_1 \mathcal{I}^4_2  +   \mathcal{I}^1_1 \mathcal{I}^3_2 + \mathcal{I}^3_1 \mathcal{I}^4_2  +    \mathcal{I}^3_1 \mathcal{I}^2_2  + \mathcal{I}^4_1 \mathcal{I}^4_2    + \cdots \big\} \big]  \bigg]\bigg] \\ \approx 0 ,  
     \end{align*}

     \noindent can be shown to hold by applying identical computations with the Poisson bracket, as it was demonstrated above to conclude that,

\begin{align*}
   \big( N + 3 \big) \underset{i \neq  j \in V ( \textbf{Z}^2 ) }{\underset{a,a^{\prime} \in \textbf{R}}{\mathrm{span}}}   \bigg[   \bigg[       \bigg[\bigg[\big[ \mathcal{I}^2_5 \big]^2  ,  \bigg[   \underset{i \neq j : 1 \leq i \leq 4}{\underset{1 \leq j \leq 4}{\mathrm{sup}}}    \bigg[\bigg[         4 \mathrm{log} \big[ \mathcal{I}^j_2 \big]^2  +          3  \mathrm{log} \big[ \mathcal{I}^i_2 \mathcal{I}^j_2  \big]           \bigg]\bigg]  \bigg]        \bigg]\bigg]   \bigg] \end{align*}

   \begin{align*} +             \bigg[   \bigg[\bigg[\big[ \mathcal{I}^3_5 \big]^2  ,  \bigg[   \underset{i \neq j : 1 \leq i \leq 4}{\underset{1 \leq j \leq 4}{\mathrm{sup}}}    \bigg[\bigg[         4 \mathrm{log} \big[ \mathcal{I}^j_2 \big]^2  +          3  \mathrm{log} \big[ \mathcal{I}^i_2 \mathcal{I}^j_2  \big]           \bigg]\bigg]  \bigg]        \bigg]\bigg] \bigg]   + \dots \\ +    \bigg[   \bigg[\bigg[\big[ \mathcal{I}^4_N \big]^2  ,  \bigg[   \underset{i \neq j : 1 \leq i \leq 4}{\underset{1 \leq j \leq 4}{\mathrm{sup}}}    \bigg[\bigg[         4 \mathrm{log} \big[ \mathcal{I}^j_2 \big]^2  +          3  \mathrm{log} \big[ \mathcal{I}^i_2 \mathcal{I}^j_2  \big]           \bigg]\bigg]  \bigg]        \bigg]\bigg]  \bigg]  \bigg] \approx 0 .             
\end{align*}

\noindent Hence,

{\small \begin{align*}
     \underset{i \neq  j \in V ( \textbf{Z}^2 ) }{\underset{a,a^{\prime} \in \textbf{R}}{\mathrm{span}}}  \bigg[\bigg[  \mathcal{I}^1_1 \mathcal{I}^1_2  ,   \mathrm{log} \big[  \big\{ \mathcal{I}^1_1 \mathcal{I}^1_2 + \mathcal{I}^3_1 \mathcal{I}^3_2   +        \mathcal{I}^2_1 \mathcal{I}^2_2  + \mathcal{I}^4_1 \mathcal{I}^4_2 +  \mathcal{I}^1_1 \mathcal{I}^3_2 + \mathcal{I}^3_1 \mathcal{I}^4_2  +    \mathcal{I}^3_1 \mathcal{I}^2_2  + \mathcal{I}^4_1 \mathcal{I}^4_2  + \cdots  \big\} \big]  \bigg]\bigg] \\ +   \underset{i \neq  j \in V ( \textbf{Z}^2 ) }{\underset{a,a^{\prime} \in \textbf{R}}{\mathrm{span}}}  \bigg[\bigg[  \mathcal{I}^3_1 \mathcal{I}^3_2 ,  \mathrm{log} \big[  \big\{ \mathcal{I}^1_1 \mathcal{I}^1_2 + \mathcal{I}^3_1 \mathcal{I}^3_2  +        \mathcal{I}^2_1  \mathcal{I}^2_2  + \mathcal{I}^4_1 \mathcal{I}^4_2 +   \mathcal{I}^1_1 \mathcal{I}^3_2 + \mathcal{I}^3_1 \mathcal{I}^4_2  +   \mathcal{I}^3_1 \mathcal{I}^2_2  + \mathcal{I}^4_1 \mathcal{I}^4_2  + \cdots   \big\} \big]  \bigg]\bigg] \\  +  \underset{i \neq  j \in V ( \textbf{Z}^2 ) }{\underset{a,a^{\prime} \in \textbf{R}}{\mathrm{span}}} \bigg[\bigg[           \mathcal{I}^2_1 \mathcal{I}^2_2  ,  \mathrm{log} \big[  \big\{ \mathcal{I}^1_1 \mathcal{I}^1_2 + \mathcal{I}^3_1 \mathcal{I}^3_2   +        \mathcal{I}^2_1 \mathcal{I}^2_2  + \mathcal{I}^4_1 \mathcal{I}^4_2  +  \mathcal{I}^1_1 \mathcal{I}^3_2 + \mathcal{I}^3_1 \mathcal{I}^4_2  +   \mathcal{I}^3_1 \mathcal{I}^2_2  + \mathcal{I}^4_1 \mathcal{I}^4_2    + \cdots \big\} \big]  \bigg]\bigg] \\ +  \underset{i \neq  j \in V ( \textbf{Z}^2 ) }{\underset{a,a^{\prime} \in \textbf{R}}{\mathrm{span}}}  \bigg[\bigg[   \mathcal{I}^4_1 \mathcal{I}^4_2   ,   \mathrm{log} \big[  \big\{ \mathcal{I}^1_1 \mathcal{I}^1_2 + \mathcal{I}^3_1 \mathcal{I}^3_2   ,       \mathcal{I}^2_1 \mathcal{I}^2_2  + \mathcal{I}^4_1 \mathcal{I}^4_2  +   \mathcal{I}^1_1 \mathcal{I}^3_2 + \mathcal{I}^3_1 \mathcal{I}^4_2  +    \mathcal{I}^3_1 \mathcal{I}^2_2  + \mathcal{I}^4_1 \mathcal{I}^4_2    + \cdots \big\} \big]  \bigg]\bigg] \\  + \underset{i \neq  j \in V ( \textbf{Z}^2 ) }{\underset{a,a^{\prime} \in \textbf{R}}{\mathrm{span}}} \bigg[\bigg[    \mathcal{I}^1_1 \mathcal{I}^3_2     ,   \mathrm{log} \big[  \big\{ \mathcal{I}^1_1 \mathcal{I}^1_2 + \mathcal{I}^3_1 \mathcal{I}^3_2   ,       \mathcal{I}^2_1 \mathcal{I}^2_2  + \mathcal{I}^4_1 \mathcal{I}^4_2  +   \mathcal{I}^1_1 \mathcal{I}^3_2 + \mathcal{I}^3_1 \mathcal{I}^4_2  +    \mathcal{I}^3_1 \mathcal{I}^2_2  + \mathcal{I}^4_1 \mathcal{I}^4_2    + \cdots \big\} \big]  \bigg]\bigg] \\    + \underset{i \neq  j \in V ( \textbf{Z}^2 ) }{\underset{a,a^{\prime} \in \textbf{R}}{\mathrm{span}}}  \bigg[\bigg[   \mathcal{I}^3_1 \mathcal{I}^4_2     ,   \mathrm{log} \big[  \big\{ \mathcal{I}^1_1 \mathcal{I}^1_2 + \mathcal{I}^3_1 \mathcal{I}^3_2   +        \mathcal{I}^2_1 \mathcal{I}^2_2  + \mathcal{I}^4_1 \mathcal{I}^4_2  +   \mathcal{I}^1_1 \mathcal{I}^3_2 + \mathcal{I}^3_1 \mathcal{I}^4_2  +    \mathcal{I}^3_1 \mathcal{I}^2_2  + \mathcal{I}^4_1 \mathcal{I}^4_2    + \cdots \big\} \big]  \bigg]\bigg] \\   + \underset{i \neq  j \in V ( \textbf{Z}^2 ) }{\underset{a,a^{\prime} \in \textbf{R}}{\mathrm{span}}} \bigg[\bigg[         \mathcal{I}^3_1 \mathcal{I}^2_2   ,   \mathrm{log} \big[  \big\{ \mathcal{I}^1_1 \mathcal{I}^1_2 + \mathcal{I}^3_1 \mathcal{I}^3_2   ,       \mathcal{I}^2_1  \mathcal{I}^2_2  + \mathcal{I}^4_1 \mathcal{I}^4_2  +   \mathcal{I}^1_1 \mathcal{I}^3_2 + \mathcal{I}^3_1 \mathcal{I}^4_2  +   \mathcal{I}^3_1 \mathcal{I}^2_2  + \mathcal{I}^4_1 \mathcal{I}^4_2    + \cdots \big\} \big]  \bigg]\bigg] \\  + \underset{i \neq  j \in V ( \textbf{Z}^2 ) }{\underset{a,a^{\prime} \in \textbf{R}}{\mathrm{span}}}  \bigg[\bigg[     \mathcal{I}^4_1 \mathcal{I}^4_2  ,   \mathrm{log} \big[  \big\{ \mathcal{I}^1_1 \mathcal{I}^1_2 + \mathcal{I}^3_1 \mathcal{I}^3_2   ,       \mathcal{I}^2_1  \mathcal{I}^2_2  + \mathcal{I}^4_1 \mathcal{I}^4_2  +   \mathcal{I}^1_1 \mathcal{I}^3_2 + \mathcal{I}^3_1 \mathcal{I}^4_2  +    \mathcal{I}^3_1 \mathcal{I}^2_2  + \mathcal{I}^4_1 \mathcal{I}^4_2    + \cdots \big\} \big]  \bigg]\bigg] \\    +    {\underset{a,a^{\prime} \in \textbf{R}}{\mathrm{span}}}  \bigg[\bigg[     \textit{Higher order Ising-type block operators}  ,   \mathrm{log} \big[  \big\{ \mathcal{I}^1_1 \mathcal{I}^1_2 + \mathcal{I}^3_1 \mathcal{I}^3_2   ,       \mathcal{I}^2_1  \mathcal{I}^2_2  + \mathcal{I}^4_1 \mathcal{I}^4_2  +   \mathcal{I}^1_1 \mathcal{I}^3_2 \\  + \mathcal{I}^3_1 \mathcal{I}^4_2   +    \mathcal{I}^3_1 \mathcal{I}^2_2  + \mathcal{I}^4_1 \mathcal{I}^4_2    + \cdots \big\} \big]  \bigg]\bigg]                   
     \approx 0 \text{. } \\ \tag{1} 
\end{align*} }

\bigskip

\noindent With each direction of the correspondence established, we conclude the argument. \boxed{}

\section{Declarations}

\subsection{Ethics approval and consent to participate}

The author consents to participate in the peer review process.

\subsection{Consent for publication}

The author consents to submit the following work for publication.

\subsection{Availability of data and materials}

Not applicable

\subsection{Conflict of interest}

The author declares no competing interests.

\subsection{Funding}

Not applicable





\end{document}